\documentclass[longauth]{aa}
\usepackage{graphicx}
\usepackage{lscape}
\usepackage[varg]{txfonts}
\usepackage{amsmath, amssymb}
\usepackage[utf8]{inputenc}
\usepackage{booktabs}
\usepackage{hyperref}
\usepackage{multirow}

\usepackage{CJK}
\usepackage[whole]{bxcjkjatype}
\usepackage{CJKutf8}

\usepackage{listings}
\usepackage{textcomp}
\usepackage{natbib}

\usepackage[dvipsnames]{xcolor}
\definecolor{myblue}{HTML}{1F77B4}
\definecolor{mygreen}{HTML}{2CA02C}
\definecolor{myred}{HTML}{D62728}
\definecolor{mymagenta}{HTML}{D33682}
\definecolor{codepurple}{HTML}{C42043}

\usepackage{tikz}

\usepackage{pifont}

\newcommand{\cmark}{\color{mygreen} \ding{51}}%
\newcommand{\xmark}{\color{myred}\ding{56}}%
\newcommand{\qmark}{\color{black}{\bf ?}}%

\hypersetup{
unicode=true,           
colorlinks=true,        
linkcolor=myred,        
citecolor=myblue,       
filecolor=magenta,      
urlcolor=mygreen        
}

\newcommand{\gaia}{\textit{Gaia}}

\newcommand{\hst}{\textit{HST}}
\newcommand{\swift}{\textit{Swift}}
\newcommand{\xmm}{\textit{XMM-Newton}}

\newcommand{\missing}[1]{\textcolor{red}{\textbf{#1}}}
\renewcommand{\missing}[1]{#1}
\newcommand{\nodata}{\dots}
\newcommand{\kms}{~\rm km\,s^{-1}}
\newcommand{\program}[1]{\textsc{#1}}
\newcommand{\tmax}{$t_{\rm max}$}
\newcommand{\sn}{SN\,2018ibb}

\newcommand{\mycomment}[1]{}

\newcommand{\berkeley}{Department of Astronomy, University of California, Berkeley, CA 94720-3411, USA}
\newcommand{\bham}{Birmingham Institute for Gravitational Wave Astronomy and School of Physics and Astronomy, University of Birmingham, Birmingham B15 2TT, UK}
\newcommand{\caltechobs}{The Caltech Optical Observatories, California Institute of Technology, Pasadena, CA 91125, USA}
\newcommand{\caltechastro}{Division of Physics, Mathematics and Astronomy, California Institute of Technology, Pasadena, CA 91125, USA}
\newcommand{\cardiff}{Cardiff Hub for Astrophysics Research and Technology, School of Physics \& Astronomy, Cardiff University, Queens Buildings, The Parade, Cardiff, CF24 3AA, UK}
\newcommand{\cfa}{Center for Astrophysics, Harvard \& Smithsonian, 60 Garden Street, Cambridge, MA 02138-1516, USA}
\newcommand{\cnrs}{Universit\'e de Lyon, Universit\'e Claude-Bernard Lyon 1, CNRS/IN2P3, IP2I Lyon, 69622 Villeurbanne, France}
\newcommand{\cornell}{Department of Astronomy, Cornell University, Ithaca, NY 14853, USA}
\newcommand{\dawn}{Cosmic Dawn Center (DAWN), Denmark; Niels Bohr Institute, Copenhagen University, Jagtvej 128, 2200, Copenhagen N, Denmark}
\newcommand{\dtu}{DTU Space, National Space Institute, Technical University of Denmark, Elektrovej 327, 2800 Kgs. Lyngby, Denmark}
\newcommand{\eso}{European Southern Observatory, Alonso de C\'ordova 3107, Casilla 19, Santiago, Chile}
\newcommand{\gemini}{Gemini Observatory / NSF's NOIRLab, Casilla 603, La Serena, Chile}
\newcommand{\gsi}{GSI Helmholtzzentrum für Schwerionenforschung, Planckstra{\ss}e 1, 64291 Darmstadt, Germany}
\newcommand{\fairfax}{George Mason University, Department of Physics \& Astronomy, MS 3F3, 4400 University Dr., Fairfax, VA 22030, USA}
\newcommand{\finca}{Finnish Centre for Astronomy with ESO (FINCA), 20014 University of Turku, Finland}
\newcommand{\hits}{Heidelberger Institut f\"ur Theoretische Studien, Schloss-Wolfsbrunnenweg 35, 69118 Heidelberg, Germany}
\newcommand{\ice}{Institute of Space Sciences (ICE, CSIC), Campus UAB, Carrer de Can Magrans, s/n, 08193 Barcelona, Spain}
\newcommand{\ieec}{Institut d’Estudis Espacials de Catalunya (IEEC), 08034 Barcelona, Spain}
\newcommand{\inafbo}{INAF - Osservatorio di Astrofisica e Scienza dello Spazio, via Piero Gobetti 93/3, 40129 Bologna, Italy}
\newcommand{\inafmi}{INAF - Istituto di Astrofisica Spaziale e Fisica cosmica Milano (IASF), via Alfonso Corti 12, 20133 Milano, Italy}

\newcommand{\inafpa}{INAF - Osservatorio Astronomico di Padova, vicolo dell'Osservatorio 5, 35122 Padova, Italy}
\newcommand{\inafteramo}{INAF - Osservatorio Astronomico d'Abruzzo, Via M. Maggini snc, 64100 Teramo, Italy}
\newcommand{\ipac}{IPAC, California Institute of Technology, 1200 E. California Blvd, Pasadena, CA 91125, USA}
\newcommand{\lbt}{Large Binocular Telescope Observatory, University of Arizona, 933 N. Cherry Ave., Tucson, AZ 85721, USA}
\newcommand{\lcogt}{Las Cumbres Observatory, 6740 Cortona Dr. Suite 102, Goleta, CA, 93117, USA}
\newcommand{\ljmu}{Astrophysics Research Institute, Liverpool John Moores University, Liverpool Science Park, 146 Brownlow Hill, Liverpool L3 5RF, UK}
\newcommand{\mas}{Millennium Institute of Astrophysics MAS, Nuncio Monsenor Sotero Sanz 100, Off. 104, Providencia, Santiago, Chile}
\newcommand{\mitinst}{MIT-Kavli Institute for Astrophysics and Space Research, 77 Massachusetts Ave., Cambridge, MA 02139, USA}
\newcommand{\mpa}{Max-Planck-Institut f{\"u}r Astrophysik, Karl-Schwarzschild Stra{\ss}e 1, 85748 Garching, Germany}
\newcommand{\mpe}{Max-Planck-Institut f{\"u}r Extraterrestrische Physik, Giessenbachstra{\ss}e 1, 85748, Garching, Germany}
\newcommand{\nordita}{Nordita, Stockholm University and KTH Royal Institute of Technology Hannes Alfv\'ens v\"ag 12, 106 91 Stockholm, Sweden}
\newcommand{\okcastro}{The Oskar Klein Centre, Department of Astronomy, Stockholm University, Albanova University Center, 106 91 Stockholm, Sweden}
\newcommand{\okcphys}{The Oskar Klein Centre, Department of Physics, Stockholm University, Albanova University Center, 106 91 Stockholm, Sweden}
\newcommand{\oxford}{Department of Physics, University of Oxford, Denys Wilkinson Building, Keble Road, Oxford OX1 3RH, UK}
\newcommand{\qub}{Astrophysics Research Centre, School of Mathematics and Physics, Queen's University Belfast, BT7 1NN, UK}
\newcommand{\tsinghua}{Physics Department and Tsinghua Center for Astrophysics (THCA), Tsinghua University, Beijing, 100084, China}
\newcommand{\tum}{Technische Universit{\"a}t M{\"u}nchen, TUM School of Natural Sciences, Physik-Department, James-Franck-Stra{\ss}e 1, 85748 Garching, Germany}
\newcommand{\tuorla}{Tuorla Observatory, Department of Physics and Astronomy, 20014 University of Turku, Finland}
\newcommand{\turku}{Tuorla Observatory, Department of Physics and Astronomy, University of Turku, 20014, Finland}
\newcommand{\ucdavis}{Department of Physics, University of California, 1 Shields Avenue, Davis, CA 95616-5270, USA}
\newcommand{\ucsb}{Department of Physics, University of California, Santa Barbara, CA 93106-9530, USA}
\newcommand{\ucsc}{Department of Astronomy and Astrophysics, University of California, Santa Cruz, CA 95064}
\newcommand{\vienna}{Department of Astrophysics, University of Vienna, T\"urkenschanzstra{\ss}e 17, 1180 Vienna, Austria}
\newcommand{\warsaw}{Astronomical Observatory, University of Warsaw, Al. Ujazdowskie 4, 00-478 Warszawa, Poland}
\newcommand{\wis}{Department of Particle Physics and Astrophysics, Weizmann Institute of Science, 234 Herzl St, 76100 Rehovot, Israel}
\newcommand{\lpl}{Lunar and Planetary Laboratory, University of Arizona, Tucson, AZ 85721, USA}
\newcommand{\iafi}{The NSF AI Institute for Artificial Intelligence and Fundamental Interactions, USA}

\usepackage{natbib}
\begin{document} 

\title{
1100~days in the life of the supernova 2018ibb --- \\
The best pair-instability supernova candidate, to date
}

\author{
    Steve Schulze             \inst{1}\thanks{\email{steve.schulze@fysik.su.se}}     \href{https://orcid.org/0000-0001-6797-1889}{\includegraphics[scale=0.5]{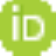}} \and
    Claes Fransson            \inst{2}    \and 
    Alexandra Kozyreva        \inst{3}    \href{https://orcid.org/0000-0001-9598-8821}{\includegraphics[scale=0.5]{ORCIDiD_icon16x16-eps-converted-to.pdf}}\and
    Ting-Wan Chen             \inst{4,5,6} \href{https://orcid.org/0000-0002-1066-6098}{\includegraphics[scale=0.5]{ORCIDiD_icon16x16-eps-converted-to.pdf}}\and
    Ofer Yaron                \inst{7}     \and 
    Anders Jerkstrand         \inst{2}    \href{https://orcid.org/0000-0001-8005-4030}{\includegraphics[scale=0.5]{ORCIDiD_icon16x16-eps-converted-to.pdf}}\and
    Avishay Gal-Yam           \inst{7}     \and 
    Jesper Sollerman          \inst{2}    \href{https://orcid.org/0000-0003-1546-6615}{\includegraphics[scale=0.5]{ORCIDiD_icon16x16-eps-converted-to.pdf}}\and
    Lin Yan                   \inst{8}    \and 
    Tuomas Kangas             \inst{9,10} \href{https://orcid.org/0000-0002-5477-0217}{\includegraphics[scale=0.5]{ORCIDiD_icon16x16-eps-converted-to.pdf}}\and
    Giorgos Leloudas          \inst{11}    \href{https://orcid.org/0000-0002-8597-0756}{\includegraphics[scale=0.5]{ORCIDiD_icon16x16-eps-converted-to.pdf}}\and
    Conor~M.~B. Omand         \inst{2}    \href{https://orcid.org/0000-0002-9646-8710}{\includegraphics[scale=0.5]{ORCIDiD_icon16x16-eps-converted-to.pdf}}\and
    Stephen~J. Smartt         \inst{12,13} \href{https://orcid.org/0000-0002-8229-1731}{\includegraphics[scale=0.5]{ORCIDiD_icon16x16-eps-converted-to.pdf}}\and
    Yi Yang \begin{CJK*}{UTF8}{gbsn}(杨轶)\end{CJK*} \inst{14}            \href{https://orcid.org/0000-0002-6535-8500}{\includegraphics[scale=0.5]{ORCIDiD_icon16x16-eps-converted-to.pdf}}\and
    Matt Nicholl              \inst{15,13}    \href{https://orcid.org/0000-0002-2555-3192}{\includegraphics[scale=0.5]{ORCIDiD_icon16x16-eps-converted-to.pdf}} \and
    Nikhil Sarin              \inst{16,1}  \href{https://orcid.org/0000-0003-2700-1030}{\includegraphics[scale=0.5]{ORCIDiD_icon16x16-eps-converted-to.pdf}}\and
    Yuhan Yao                 \inst{17}    \href{https://orcid.org/0000-0001-6747-8509}{\includegraphics[scale=0.5]{ORCIDiD_icon16x16-eps-converted-to.pdf}}\and
    Thomas~G. Brink           \inst{14}     \href{https://orcid.org/0000-0001-5955-2502}{\includegraphics[scale=0.5]{ORCIDiD_icon16x16-eps-converted-to.pdf}}\and
    Amir Sharon               \inst{7}     \and 
    Andrea Rossi              \inst{18}    \and 
    Ping Chen                 \inst{7}     \href{https://orcid.org/0000-0003-0853-6427}{\includegraphics[scale=0.5]{ORCIDiD_icon16x16-eps-converted-to.pdf}}\and
    Zhihao Chen               \inst{19}     \href{https://orcid.org/0000-0001-5175-4652}{\includegraphics[scale=0.5]{ORCIDiD_icon16x16-eps-converted-to.pdf}}\and
    Aleksandar Cikota         \inst{20}     \href{https://orcid.org/0000-0001-7101-9831}{\includegraphics[scale=0.5]{ORCIDiD_icon16x16-eps-converted-to.pdf}}\and
    Kishalay De\thanks{NASA Einstein Fellow}          \inst{21}    \and
    Andrew~J. Drake           \inst{17} \and
    Alexei V. Filippenko           \inst{14}     \and 
    Christoffer Fremling      \inst{8}    \and 
    Laurane Fr\'eour          \inst{22}    \and
    Johan~P.~U. Fynbo         \inst{23}    \href{https://orcid.org/0000-0002-8149-8298}{\includegraphics[scale=0.5]{ORCIDiD_icon16x16-eps-converted-to.pdf}}\and
    Anna~Y.~Q. Ho             \inst{24}    \href{https://orcid.org/0000-0002-9017-3567}{\includegraphics[scale=0.5]{ORCIDiD_icon16x16-eps-converted-to.pdf}}\and
    Cosimo Inserra            \inst{25}    \and 
    Ido Irani                 \inst{7}     \href{https://orcid.org/0000-0002-7996-8780}{\includegraphics[scale=0.5]{ORCIDiD_icon16x16-eps-converted-to.pdf}}\and
    Hanindyo Kuncarayakti     \inst{26,27} \href{https://orcid.org/0000-0002-1132-1366}{\includegraphics[scale=0.5]{ORCIDiD_icon16x16-eps-converted-to.pdf}} \and
    Ragnhild Lunnan           \inst{2}    \href{https://orcid.org/0000-0001-9454-4639}{\includegraphics[scale=0.5]{ORCIDiD_icon16x16-eps-converted-to.pdf}}\and
    Paolo Mazzali             \inst{28,5}  \and 
    Eran O. Ofek                 \inst{7} \and
    Eliana Palazzi            \inst{18}    \href{https://orcid.org/0000-0002-8691-7666}{\includegraphics[scale=0.5]{ORCIDiD_icon16x16-eps-converted-to.pdf}}\and
    Daniel~A. Perley          \inst{28}    \href{https://orcid.org/0000-0001-8472-1996}{\includegraphics[scale=0.5]{ORCIDiD_icon16x16-eps-converted-to.pdf}}\and
    Miika Pursiainen          \inst{11}    \href{https://orcid.org/0000-0002-8597-0756}{\includegraphics[scale=0.5]{ORCIDiD_icon16x16-eps-converted-to.pdf}}\and
    Barry Rothberg            \inst{29,30} \and 
    Luke~J. Shingles          \inst{31} \href{https://orcid.org/0000-0002-5738-1612}{\includegraphics[scale=0.5]{ORCIDiD_icon16x16-eps-converted-to.pdf}} \and
    Ken Smith                 \inst{13}    \and 
    Kirsty Taggart            \inst{32}    \href{https://orcid.org/0000-0002-5748-4558}{\includegraphics[scale=0.5]{ORCIDiD_icon16x16-eps-converted-to.pdf}}\and
    Leonardo Tartaglia        \inst{33,34} \href{https://orcid.org/0000-0003-3433-1492}{\includegraphics[scale=0.5]{ORCIDiD_icon16x16-eps-converted-to.pdf}}\and
    WeiKang Zheng             \inst{14}     \and 
    Joseph~P. Anderson        \inst{35,36} \href{https://orcid.org/0000-0003-0227-3451}{\includegraphics[scale=0.5]{ORCIDiD_icon16x16-eps-converted-to.pdf}} \and
    Letizia Cassara           \inst{37}     \and
    Eric Christensen          \inst{47} \and
    S.~George Djorgovski      \inst{17} \and
    Llu\'is Galbany           \inst{38,39} \href{https://orcid.org/0000-0002-1296-6887}{\includegraphics[scale=0.5]{ORCIDiD_icon16x16-eps-converted-to.pdf}} \and 
    Anamaria Gkini            \inst{2} \and
    Matthew~J. Graham         \inst{17}    \and 
    Mariusz Gromadzki         \inst{40} \href{https://orcid.org/0000-0002-1650-1518}{\includegraphics[scale=0.5]{ORCIDiD_icon16x16-eps-converted-to.pdf}} \and
    Steven L. Groom           \inst{41}    \href{https://orcid.org/0000-0001-5668-3507}{\includegraphics[scale=0.5]{ORCIDiD_icon16x16-eps-converted-to.pdf}} \and
    Daichi Hiramatsu          \inst{42,48} \href{https://orcid.org/0000-0002-1125-9187}{\includegraphics[scale=0.5]{ORCIDiD_icon16x16-eps-converted-to.pdf}}\and
    D. Andrew Howell          \inst{43,44} \and 
    Mansi M. Kasliwal         \inst{17}    \href{https://orcid.org/0000-0002-5619-4938}{\includegraphics[scale=0.5]{ORCIDiD_icon16x16-eps-converted-to.pdf}} \and
    Curtis McCully            \inst{43} \and
    Tom\'as E. M\"uller-Bravo \inst{38,39} \href{https://orcid.org/0000-0003-3939-7167}{\includegraphics[scale=0.5]{ORCIDiD_icon16x16-eps-converted-to.pdf}}\and
    Simona Paiano             \inst{37}     \and 
    Emmanouela Paraskeva      \inst{45} \and 
    Priscila J. Pessi         \inst{2} \href{https://orcid.org/0000-0002-8041-8559}{\includegraphics[scale=0.5]{ORCIDiD_icon16x16-eps-converted-to.pdf}} \and 
    David Polishook           \inst{7}     \href{https://orcid.org/0000-0002-6977-3146}{\includegraphics[scale=0.5]{ORCIDiD_icon16x16-eps-converted-to.pdf}}\and
    Arne Rau                  \inst{6} \and
    Mickael Rigault        \inst{46}    \href{https://orcid.org/0000-0002-8121-2560}{\includegraphics[scale=0.5]{ORCIDiD_icon16x16-eps-converted-to.pdf}} \and
    Ben Rusholme              \inst{41}    \href{https://orcid.org/0000-0001-7648-4142}{\includegraphics[scale=0.5]{ORCIDiD_icon16x16-eps-converted-to.pdf}} 
}

\institute{
\okcphys      \and 
\okcastro     \and 
\hits         \and 
\tum          \and 
\mpa          \and 
\mpe          \and 
\wis          \and 
\caltechobs   \and 
\finca        \and 
\turku        \and 
\dtu          \and 
\oxford       \and 
\qub          \and 
\berkeley     \and 
\bham         \and 
\nordita      \and 
\caltechastro \and 
\inafbo       \and 
\tsinghua     \and 
\gemini       \and 
\mitinst      \and 
\vienna       \and 
\dawn         \and 
\cornell      \and 
\cardiff      \and 
\tuorla       \and 
\finca        \and 
\ljmu         \and 
\lbt          \and 
\fairfax      \and 
\gsi          \and 
\ucsc         \and 
\inafpa       \and 
\inafteramo   \and 
\eso          \and 
\mas          \and 
\inafmi       \and 
\ice          \and 
\ieec         \and 
\warsaw       \and 
\ipac         \and 
\cfa          \and 
\lcogt        \and 
\ucsb         \and 
\ucdavis      \and 
\cnrs         \and 
\lpl          \and 
\iafi              
}

\date{Received 10 May 2023; accepted 12 October 2023}

\abstract
{Stars with zero-age main sequence masses between 140 and $260~M_\odot$ are thought to explode as pair-instability supernovae (PISNe). During their thermonuclear runaway, PISNe can produce up to several tens of solar masses of radioactive nickel, resulting in luminous transients similar to some superluminous supernovae (SLSNe). Yet, no unambiguous PISN has been discovered so far. \sn\ is a hydrogen-poor SLSN at $z=0.166$ that evolves extremely slowly compared to the hundreds of known SLSNe. Between mid 2018 and early 2022, we monitored its photometric and spectroscopic evolution from the UV to the near-infrared (NIR) with 2--10\,m class telescopes. \sn\ radiated $>3\times10^{51}~\rm erg$ during its evolution, and its bolometric light curve reached $>2\times10^{44}~\rm erg\,s^{-1}$ at its peak. The long-lasting rise of $>93$ rest-frame days implies a long diffusion time, which requires a very high total ejected mass. The PISN mechanism naturally provides both the energy source ($^{56}$Ni) and the long diffusion time. Theoretical models of PISNe make clear predictions as to their photometric and spectroscopic properties. \sn\ complies with most tests on the light curves, nebular spectra and host galaxy, and potentially all tests with the interpretation we propose. Both the light curve and the spectra require 25--44~$M_\odot$ of freshly nucleosynthesised $^{56}$Ni, pointing to the explosion of a metal-poor star with a helium core mass of 120--130~$M_\odot$ at the time of death. This interpretation is also supported by the tentative detection of [\ion{Co}{ii}]\,$\lambda$\,1.025$\mu$m, which has never been observed in any other PISN candidate or SLSN before. We observe a significant excess in the blue part of the optical spectrum during the nebular phase, which is in tension with predictions of existing PISN models. However, we have compelling observational evidence for an eruptive mass-loss episode of the progenitor of \sn\ shortly before the explosion, and our dataset reveals that the interaction of the SN ejecta with this oxygen-rich circumstellar material contributed to the observed emission. That may explain this specific discrepancy with PISN models. Powering by a central engine, such as a magnetar or a black hole, can be excluded with high confidence. This makes \sn\ by far the best candidate for being a PISN, to date.}


\authorrunning{Schulze, et al.}
\titlerunning{1100 days in the life of the PISN candidate 2018ibb}

\maketitle

\section{Introduction} \label{sec:intro}

Observations of stellar nurseries \citep[e.g.][]{Krumholz2019a}, as well as massive stars \citep[e.g.][]{Crowther2007a} and their fates \citep[e.g.][]{Filippenko1997a, GalYam2017a} have led to stellar evolution models of an ever-increasing complexity \citep[e.g.][]{, McKee2007a}. These models also predict the existence of stars with $\gtrsim100~M_\odot$ \citep[e.g.][]{Heger2002a, Heger2003a}, which may have no analogues in the local Universe \citep[][but see \citealt{Brands2022a}]{Mackey2003a, Bromm2004a, Langer2007a}, and exotic types of stellar explosions \citep{Fowler1964a, Rakavy1967a, Woosley2007a, Sakstein2022a}. 

One of those predicted, yet not securely discovered object classes, is pair-instability supernovae (PISNe). This SN class is produced by the thermonuclear runaway of metal-poor stars with zero-age main sequence (ZAMS) masses between 140 and $260~M_\odot$ \citep[][]{Fowler1964a, Barkat1967a, Rakavy1967a}. When such a massive star dies, its helium core would have grown to 65--$130~M_\odot$ \citep{Heger2002a}. The combination of a relatively low matter density and high temperature leads to the production of $e^-e^+$ pairs, reducing the radiation pressure that supports the star against the gravitational collapse. As a result, implosive oxygen and silicon burning produce enough energy to revert the collapse and obliterate the entire star, leaving no remnant behind.

During the past 15 years, PISNe have been a focus of fundamental physics and SN science. Stars with helium-topped cores slightly less massive than $\sim65~M_\odot$ presumably leave black holes behind, and stars whose helium-topped cores exceed $\sim130~M_\odot$ are thought to collapse directly into black holes. In this paradigm, there should be a dearth of black holes with masses between $\sim50$ and $\sim120~M_\odot$ \citep{Farmer2019a, Renzo2020a}. Observations by the Laser Interferometer Gravitational-Wave Observatory (LIGO) and the Virgo interferometer found tentative evidence for the existence of such a drop in the black-hole mass function \citep[][]{Abbott2020a}. A more recent study by the LIGO-Virgo collaboration using the larger Gravitational-Wave Transient Catalog 3 shows that the evidence of a mass gap at $\sim50~M_\odot$ is inconclusive \citep{Abbott2022a}. However, this could be due to the inclusion of binary black holes that formed through dynamical channels involving repeated mergers rather than evidence for the lack of a mass gap \citep[e.g.][and references therein]{Belczynski2020a, Gerosa2021a}. A further consequence of the PISN model is its peculiar nucleosynthetic pattern. In the case of an extremely metal-poor star being formed from the gas of PISN ejecta, its chemical composition should show a strong variance between elements with an odd and even nuclear charge \citep{Heger2002a, Umeda2002a}. \citet{Xing2023a} recently reported the discovery of a very metal-poor star with such a chemical signature \citep[for additional candidates see also][]{Aoki2014a, Salvadori2019a}, lending support to the first stars having been very massive and exploding as PISNe.

Finding PISNe is one of the main challenges in the SN field. PISN models predict that up to $\sim57~M_\odot$ of radioactive $^{56}$Ni are produced during the thermonuclear runaway \citep{Heger2002a}. Such high Ni-yield PISNe are thought to produce long-lived (rise times $>80$~days), luminous ($M_{\rm peak}<-21$~mag) transients \citep{Kasen2011a, Kozyreva2017a} in the regime of superluminous supernovae \citep[SLSNe;][]{Quimby2011a, GalYam2012a, GalYam2019a}. Although the powering mechanism of SLSNe is debated \citep{GalYam2009a, Blinnikov2010a, Inserra2013a}, numerous studies of both H-poor and H-rich SLSNe have revealed that nickel is not the primary source of energy \citep[e.g.][]{Chatzopoulos2012b, Chen2013a, Inserra2013a, Nicholl2017a, Inserra2018a, Moriya2018a, GalYam2019a, Inserra2019a, Kangas2022a, Chen2023b}. Yet, a few SLSNe had markedly broad and luminous light curves similar to predictions of PISN models, for example, SN\,1999as, SN\,2007bi, PTF12dam, PS1-14bj, and SN\,2015bn \citep{Hatano2001a, GalYam2009a, Nicholl2013a, Chen2015a, Lunnan2016a,  Nicholl2016a, Kozyreva2017a}. However, the published candidates either had incomplete datasets, not long enough rise times, too high ejecta velocities, too blue spectra, or exploded in galaxies with a metallicity that is too high to conclusively argue for the discovery of a PISN \citep[e.g.][]{Nicholl2013a, Jerkstrand2017a}. 

Not all PISNe are expected to be superluminous. The stars at the low-mass end of the PISN range ($M({\rm ZAMS})\gtrsim140~M_\odot$ equivalent to a He-core mass of $\gtrsim65~M_\odot$ at the end of their evolution) are thought to produce a small-to-no-mass of $^{56}$Ni \citep{Kasen2011a, Kozyreva2014a, Gilmer2017a}.
The bright supernova OGLE-2014-SN-073 is a candidate for a low-mass PISN, as its light curve would require only $\sim1~M_\odot$ of $^{56}$Ni \citep[][but see also \citealt{Dessart2018a} and \citealt{Moriya2018c} for alternative interpretations]{Terreran2017a, Kozyreva2018a}. However, also in that case, there is still some tension between observations and theoretical models due to the unknown explosion date and the lack of sufficiently early data to search for the luminous shock-breakout predicted by PISN models \citep{Kozyreva2018a}.
The Type I SN 2016iet is another candidate for a lower-mass PISN. \citet{Gomez2019a} concluded that the light curve could be powered by a few $M_\odot$ of $^{56}$Ni. However, the SN was discovered only around maximum light. Furthermore, its nebular spectra are very different to the PISN model spectra from \citet{Mazzali2019a} as pointed out by \citet{Gomez2019a}. While the inferred nickel mass of a few $M_\odot$ and ejecta mass of $\sim64~M_\odot$ could point to a PISN explosion, other powering mechanisms are also possible \citep{Gomez2019a}.

In June 2018, the Zwicky Transient Facility \citep[ZTF;][]{Bellm2019a, Graham2019a} started to survey the northern sky every 2--3 nights in two filters and detects thousands of supernovae every year \citep{Fremling2020a, Perley2020a}. Until autumn 2021, we carried out a systematic survey for SLSNe in ZTF \citep{Chen2023a, Chen2023b}. \sn, the slowest evolving SLSN in our sample, has several properties that match predictions of PISN models. Between mid 2018 and early 2022, we built a comprehensive photometric and spectroscopic dataset covering the evolution from \tmax$-$93 to \tmax$+$1000~rest-frame days to scrutinise SLSN and PISN models. In this paper, we present this dataset along with our conclusions on \sn's source of energy and progenitor. The paper is structured as follows: we report the SN discovery in Section \ref{sec:discovery} and describe the observations in Section \ref{sec:obs}. In Section \ref{sec:results}, we derive the properties of \sn's light curve, spectra and host galaxy and in Section \ref{sec:discussion} we contrast SLSN and PISN models with our dataset. Finally, in Section \ref{sec:conclusion} we summarise our findings and present our conclusions on the nature of \sn\ and its connection to PISNe.

Throughout the paper, we provide all uncertainties at $1\sigma$ confidence. The photometry is reported in the AB system. We assume $\Lambda$CDM cosmology with $H_0=67.8~{\rm km\,s}^{-1}\,{\rm Mpc}^{-1}$, $\Omega_{\rm M}=0.308$, and $\Omega_\Lambda=0.692$ \citep{Planck2016a}. Phase information is reported in the rest-frame with respect to the $g|r$-band maximum (\tmax) at MJD=58455.

\section{Discovery}\label{sec:discovery}

\begin{figure*}
    \centering
    \includegraphics[width=0.75\textwidth]{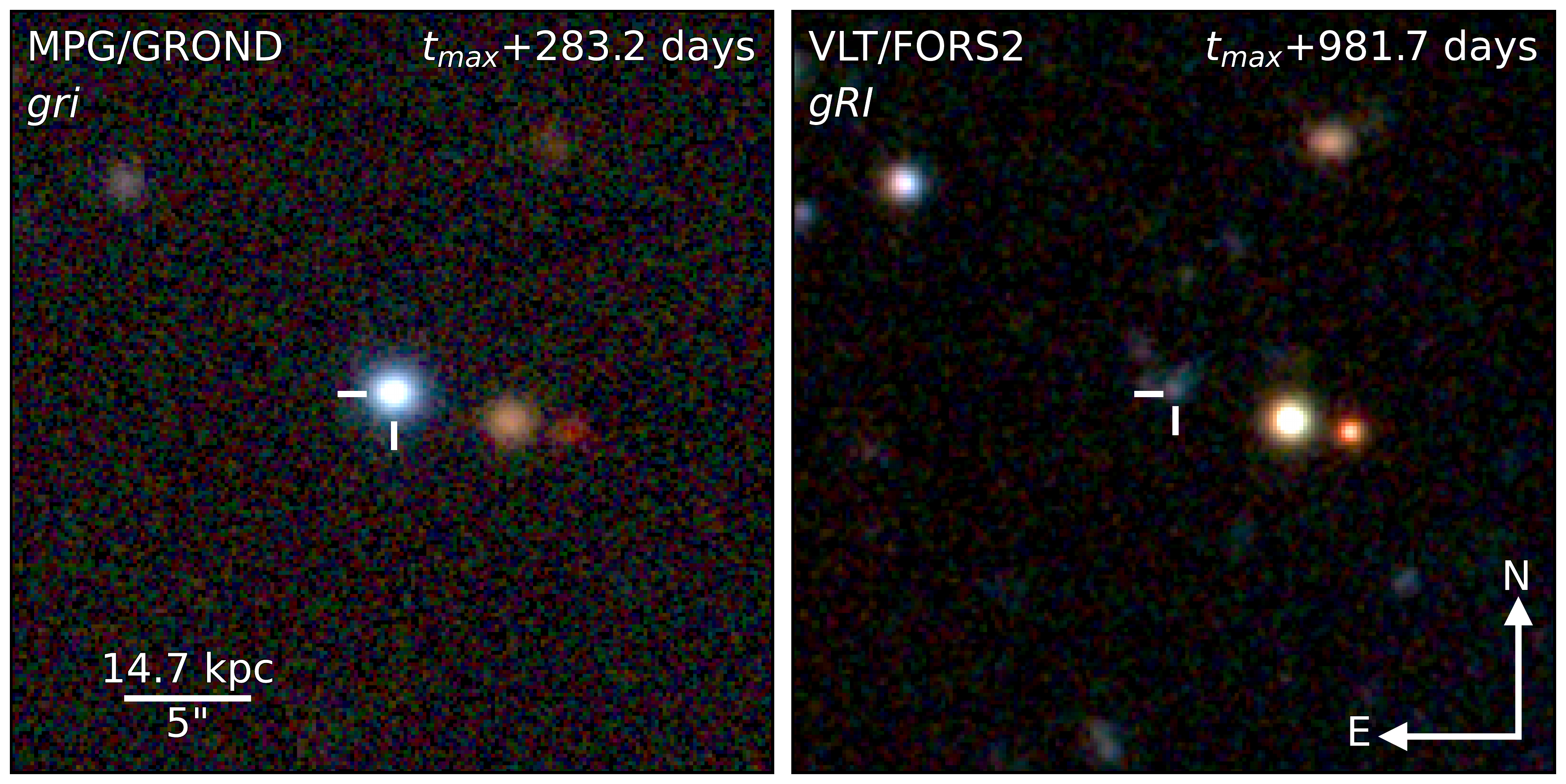}
    \caption{False-colour image of the field when \sn\ was bright (left) and after it had faded below the host level (right). The SN position, marked by the crosshair, is located $\sim1$~kpc from the centre of its star-forming dwarf host galaxy ($M^{\rm host}_r\sim-15.4$~mag, $M_\star\sim10^{7.6}~M_\odot$). For more information about the host, see Section \ref{res:host}. The false-colour image was built with \program{STIFF} version 2.4.0 \citep{Bertin2012a}.}
    \label{fig:rgb}
\end{figure*}

\sn, located at $\alpha$ = 04:38:56.950, $\delta=-$20:39:44.10 (J2000), was discovered by the Asteroid Terrestrial-impact Last Alert System \citep[ATLAS;][]{Tonry2011a, Smith2020a} survey as ATLAS18unu on 10 September 2018 with an apparent magnitude of $o=18.89$~mag \citep[wavelength range 5600--8200~\AA;][]{Tonry2018a}. Later detections were reported by the public northern sky survey of the Zwicky Transient Facility \citep{Bellm2019b} on 16 November 2018 (internal name: ZTF18acenqto), the Pan-STARRS Survey for Transients \citep{Huber2015a} on 8 January 2019 (internal name: PS19crg) and the \gaia\ Photometric Science Alerts transient survey \citep{Hodgkin2021a} on 4 July 2019 (internal name: Gaia19cvo). A false-colour image of the field when \sn\ was bright and after it had faded is shown in Figure \ref{fig:rgb}. \citet{Fremling2018a} initially classified \sn\ as a Type Ia SN on 5 December 2018 but retracted this classification on 6 December 2018 and set a new classification to `supernova' on 6 December 2018 \citep{Fremling2018b}. \citet{Pursiainen2018b} obtained a spectrum with the 3.58\,m New Technology Telescope at La Silla Observatory (Chile) as a part of the Extended Public ESO Spectroscopic Survey of Transient Objects \citep[ePESSTO;][]{Smartt2015a} on 14 December 2018 and classified \sn\ as a H-poor SLSN at $z=0.16$.

\section{Observations and data reduction} \label{sec:obs}

\subsection{Supernova photometry}\label{sec:obs:imaging}

Our imaging campaign had three tiers:
\textit{i}) all-sky surveys with sufficient depth and cadence to monitor the evolution from \tmax$-$93 to \tmax+306~days;
\textit{ii}) dedicated follow-up campaigns to expand the wavelength coverage to the UV and near-IR and to extend the light curve coverage to \tmax+1000 days; and
\textit{iii}) smaller targeted campaigns to mitigate data gaps, expand the wavelength coverage to the near-IR, and ensure a good flux calibration of the photometric and spectroscopic data. Owing to the large number of facilities involved in this effort, we present the details of each campaign and the data reduction in Appendix \ref{appendix:photometry}.

The ground-based photometry was calibrated with field stars from PanSTARRS1 \citep[PS1]{Chambers2016a}, the Dark Energy Survey \citep[DES;][]{DES2005a}, the Dark Energy Spectroscopic Instrument (DESI) Legacy Imaging survey \citep[LS;][]{Dey2018a}, and the Two Micron All-Sky Survey \citep[2MASS;][]{Skrutskie2006a}. We applied known colour equations between PS1/DES and Bessell/GROND/SDSS/ZTF filters \citep{Finkbeiner2016a, DrlicaWagner2018a, Greiner2008a, Medford2020a} and Lupton\footnote{\href{https://www.sdss.org/dr12/algorithms/sdssubvritransform}{https://www.sdss.org/dr12/algorithms/sdssubvritransform}}, to account for differences in the filter response function. We applied the offsets from \citet{Blanton2007a} to convert all measurements to the AB photometric system. The \swift/UVOT data were calibrated with zeropoints from the \swift\ pipeline and converted to the AB system following \citet{Breeveld2011a}. The \hst\ photometry was done with a custom-made aperture photometry tool, based on the python package \program{photutils} \citep{Bradley2020a} version 1.5, using an aperture with a diameter of 0\farcs5 and calibrated against tabulated zeropoints in \program{pysynphot} version 2.0.0 \citep{STScI2013a}.

SN spectra are characterised by strong absorption and emission features that evolve with time. This can lead to time-dependent colour terms between similar but not identical filters \citep[e.g.][]{Stritzinger2002a} and add a non-negligible systematic scatter to the light curves if these differences are not corrected. To illustrate this issue, we compute the synthetic magnitude in ZTF/$g$, GROND/$g$ and EFOSC2/$g$ at \tmax\ and \tmax+210~days\footnote{The GROND, ZTF and EFOSC2 $g$-band filters have an effective wavelength of $4504,\,4723, 5104$~\AA\ and width of $1373, 1282, 788$~\AA, respectively \citep[retrieved from the Spanish Virtual;][and references therein]{Rodrigo2021a}.}. At \tmax, the colour term between the EFOSC2/GROND and ZTF filters is $-0.01$ and $+0.04$ mag, respectively, but at \tmax+210~days the differences increased to $-0.13$ and $+0.12$ mag. Since the EFOSC2 and GROND data cover the late-time evolution, the differences in the filters would be well visible in the final light curve if they remained uncorrected.

To calibrate the various datasets into the same photometric system, we defined a set of reference filters consisting of the \swift\ filters, ZTF/$gr$, GROND/$izJH$ and 2MASS/$K$. Then, we extracted synthetic photometry of all ground-based filters used in our campaign from the Keck and VLT spectra (Section \ref{sec:obs:spectroscopy}), which were obtained in clear/photometric conditions, and measured the expected colours with respect to our reference filter system as a function of time. After applying this s-correction \citep{Stritzinger2002a}, we merged the different datasets to build a photometric sequence of \sn\ from \tmax$-93$ to \tmax+706~days. We omitted these corrections for the $BVJHK$ data because most observations in these filters were done with the same instrument. We also skipped applying an s-correction on the $HST$ observations in F336W at \tmax+1645 days as the SN was not detected anymore.
Table \ref{tab:phot} in Appendix \ref{appendix:photometry} summarises the homogenised SN photometry. The measurements are not corrected for Galactic extinction along the line of sight [$E(B-V)=0.03$ mag; \citealt{Schlafly2011a}], but this correction is applied to all derived properties and photometric data presented in this paper. 

The photometry is available on WISeREP\footnote{\href{https://www.wiserep.org}{https://www.wiserep.org}} \citep{Yaron2012a}. It is also available as a machine-readable table in the electronic version of this paper.

\subsection{Host galaxy photometry}

We obtained additional photometry with the ESO VLT, the 3.58\,m New Technology Telescope and the \textit{Hubble Space Telescope} approximately 1000~days after maximum (Appendix \ref{appendix:photometry}). The brightness of the host galaxy was measured with elliptical apertures encircling the entire host galaxy and calibrated in the same way as the SN photometry. The \hst\ photometry was done with a custom-made aperture photometry tool, based on the python package \program{photutils}, using an aperture comparable in area to the ground-based images and calibrated against tabulated zeropoints in \program{pysynphot}. In the $R$-band, we measure a brightness of $24.39\pm0.05$~mag. The brightness in the other filters is reported in  Table \ref{tab:host_phot}.

\subsection{Spectroscopy} \label{sec:obs:spectroscopy}

\begin{table}
    \caption{Photometry of the host galaxy}\label{tab:host_phot}
    \centering
    \begin{tabular}{lllc}
    \toprule
    Telescope & Instrument    & Filter    & Brightness\\
              &               &           & (mag)\\
    \midrule
    \hst & WFC3   &$F336W             $& $>26.04$\\
    NTT & EFOSC2 &$B                 $& $24.94\pm0.22$\\
    VLT & FORS2  &$g\_{\rm HIGH}     $& $24.95\pm0.05$\\
    VLT & FORS2  &$R\_{\rm SPECIAL}  $& $24.39\pm0.05$\\
    VLT & FORS2  &$I\_{\rm BESSELL}   $& $24.32\pm0.10$\\
    VLT & FORS2  &$z\_{\rm SPECIAL}  $& $23.78\pm0.14$\\
    \bottomrule
    \end{tabular}
    \tablefoot{All measurements are reported in the AB system and not corrected for reddening. Non-detections are reported at $3\sigma$ confidence.
    }
\end{table}

\begin{table*}
    \caption{Log of spectroscopic observations}\label{tab:spec_log}
    \scriptsize
    \centering
    \begin{tabular}{lccccccc}
    \toprule
    MJD         & Phase     & Telescope/Instrument  & Disperser     & Slit         & Wavelength    & Spectral    & Exposure\\
                & (day)     &                       &               & width ($''$) & range (\AA)   & resolution  & time (s)\\
    \midrule
    
    58453.349 & -1.4  & Keck-I/LRIS                &  400/3400 + 400/8500                   & 1.0             & 3076 -- 9350   & 600/1200       & 300/300       \\
    58461.248 & 5.4   & P60/SEDm                   &  \dots                                 & IFU             & 4650 -- 9200   & 100            & 2250          \\
    58464.228 & 7.9   & P60/SEDm                   &   \dots                                & IFU             & 3950 -- 9200   & 100            & 2250          \\
    58465.246 & 8.8   & P200/DBSP                  & 600/316                                & 1.5             & 3500 -- 10,000 & 1000/1000      & 1200          \\
    58467.254 & 10.5  & NTT/EFOSC2                 & Gr\#13                                 & 1.0             & 3650 -- 9250   & 350            & 3600          \\
    58480.217 & 21.6  & P60/SEDm                   & \dots                                  & IFU             & 5500 -- 8850   & 100            & 1200          \\
    58483.292 & 24.3  & Lick/Kast                  & 600/4310 + 300/7500                    & 2.0             & 3500 -- 10,500 & 800            & 1960          \\
    58487.225 & 27.6  & Lick/Kast                  & 600/4310 + 300/7500                    & 2.0             & 3500 -- 10,500 & 800            & 2400          \\
    58490.934 & 30.8  & NOT/ALFOSC                 & Gr\#4                                  & 1.3             & 3600 -- 9600   & 280            & 1800          \\     
    58491.226 & 31.1  & NTT/EFOSC2                 & Gr\#11 + Gr\#16/OG530                  & 1.0             & 3345 -- 9995   & 460/460        & 1800/1800     \\
    58493.096 & 32.7  & VLT/X-shooter              & \dots                                  & 1.0/0.9/0.9     & 3000 -- 24,800 & 5400/8900/5600 & 1800          \\
    58509.904 & 47.1  & NOT/ALFOSC                 & Gr\#4 + WG345                          & 1.3             & 3800 -- 9450   & 280            & 600           \\     
    58510.179 & 47.3  & Lick/Kast                  & 600/4310 + 300/7500                    & 2.0             & 3500 -- 10,500 & 800            & 3600          \\
    58515.100 & 51.5  & NTT/EFOSC2                 & Gr\#11 + Gr\#16/OG530                  & 1.0/1.0         & 3345 -- 9995   & 460/460        & 1800/1800     \\
    58525.048 & 60.1  & VLT/X-shooter              & \dots                                  & 1.0/0.9/0.9     & 3000 -- 24,800 & 5400/8900/5600 & 2400          \\
    58536.929 & 70.3  & NOT/ALFOSC                 & Gr\#4                                  & 1.3             & 3900 -- 9600   & 280            & 2400          \\
    58541.083 & 73.8  & NTT/EFOSC2                 & Gr\#11 + Gr\#16/OG530                  & 1.0/1.0         & 3345 -- 9995   & 460/460        & 2200/2200     \\
    58550.037 & 81.5  & VLT/X-shooter              & \dots                                  & 1.0/0.9/0.9     & 3000 -- 24,800 & 5400/8900/5600 & 3600          \\
    58559.153 & 89.3  & Lick/Kast                  & 600/4310 + 300/7500                    & 2.0             & 3500 -- 10,500 & 800            & 2400          \\
    58565.000 & 94.3  & VLT/X-shooter              & \dots                                  & 1.0/0.9/0.9     & 3000 -- 24,800 & 5400/8900/5600 & 3600          \\
    58718.304 & 225.8 & NTT/EFOSC2                 & Gr\#13                                 & 1.0             & 3650 -- 9250   & 350            & 5400          \\
    58724.585 & 231.2 & Keck-I/LRIS                & 400/3400 + 400/8500                    & 1.0             & 3076 -- 9350   & 600/1200       & 300/300       \\
    58776.290 & 275.6 & NTT/EFOSC2                 & Gr\#13                                 & 1.5             & 3650 -- 9250   & 230            & 5400          \\
    58776.907 & 276.1 & LBT/MODS+LUCI\tablefootmark{\scriptsize a} &  G400L/G670L/G200      & 1.2/1.2/1.0     & 3200 -- 12,000 & 925/1150/1100  & 3049/3083/3800\\
    58789.287 & 286.7 & VLT/X-shooter\tablefootmark{\scriptsize b}              & \dots                                       & 1.0/1.0/0.9     & 3000 -- 20,700 & 5400/8900/5600 & 3600          \\
    58866.134 & 352.6 & VLT/X-shooter\tablefootmark{\scriptsize b}              & \dots                                       & 1.0/1.0/0.9     & 3000 -- 20,700 & 5400/8900/5600 & 3600          \\
    58876.648 & 361.6 & LBT/MODS+LUCI\tablefootmark{\scriptsize a} &  G400L/G670L/G200/G200 & 1.2/1.2/1.0/1.0 & 3200 -- 23,500 & 925/1150/1100/1100& 7200/7200/3400/2800\\
    58895.120 & 377.5 & VLT/X-shooter\tablefootmark{\scriptsize b}              & \dots                                    & 1.0/1.0/0.9     & 3000 -- 20,700 & 5400/8900/5600 & 3600          \\
    59110.570 & 562.3 & Keck-I/LRIS                & 600/4000 + 400/8500                    & 1.0             & 3400 -- 10,275 & 1000/1200      & 4935/4935     \\
    59113.780 & 565.0 & VLT/FORS2                  & 300V                                   & 1.0             & 3800 -- 9600   & 440            & 6600          \\
    59198.044 & 637.3 & VLT/FORS2                  & 300V                                   & 1.0             & 3800 -- 9600   & 440            & 14400         \\
    59607.089 & 988.1 & Gemini-S/GMOS              & R150/GG455                             & 1.0             & 5000 -- 10,000 & 310            & 4800          \\
    59608.380 & 989.2 & VLT/FORS2                  & 300V                                   & 1.0             & 3800 -- 9600   & 440            & 14400         \\
    
    \bottomrule
    \end{tabular}
    \tablefoot{The modified Julian dates quote the beginning of each spectroscopic observation. The phase is reported for the mid-exposure time in the rest-frame with respect to the $g|r$-band maximum at MJD=58455. For the multi-arm instruments Kast, LBT, LRIS and X-shooter, we report the exposure time and spectral resolution of each arm. The wavelength ranges and the values of the spectral resolutions are taken from instrument manuals and are reported in the observer frame. The spectral resolutions refer to the unbinned data.\\
    \tablefoottext{a}{The optical spectra were obtained with MODS and the grisms G400L and G670L. The NIR spectra were obtained with the G200 grating and the $zJ$ (\tmax+276.1~days) and $zJ$+$HK$ (\tmax+361.6~days) spectroscopy filters. The start time is the average of the start times of the visual and NIR spectra.}\\
    \tablefoottext{b}{The observation was done with a $K$-band blocking filter.}
    }
\end{table*}

We collected a series of spectra spanning from the time of maximum to \tmax+989.2 days. Similarly to the imaging campaign, we utilised a large number of 2--10\,m class telescopes. A brief summary of the observations is provided in Table \ref{tab:spec_log}. The details of the observations and data reduction are presented in Appendix \ref{appendix:spectroscopy}. All spectra were absolute-flux-calibrated with multi-band photometry. Since the photometry was not obtained contemporaneously with the spectroscopic observation, we linearly interpolated between adjacent observations.

The spectra obtained after August 2021 have an increasing contribution from the host galaxy. The host contamination was removed with the FORS2 spectrum from January 2022 (\tmax+989.2~days). The slit did not cover the entire host galaxy. We scaled the spectrum to the flux encircled by the slit. We note that to determine whether \sn\ contributed to the observed spectrum from January 2022, we compared the observed spectrum to the fit of the spectral energy distribution (SED) of the entire host galaxy. The continuum level of the January 2022 spectrum is fully consistent with the best fit to the host galaxy SED (Figure \ref{fig:host_contamination}). The only remaining SN feature is broad [\ion{O}{iii}]\,$\lambda\lambda$\,4959,5007 in emission, produced by the interaction of the SN ejecta with circumstellar material (Section \ref{sec:csm:oiii_oii}). Owing to that, we masked the region and estimated the host galaxy flux with linear interpolation. To recover the host-subtracted spectrum of \sn\ from the January 2022 epoch, we utilised the best fit to the galaxy SED.

All data were also corrected for Milky-Way (MW) extinction. We note that a few spectra were affected by adverse weather conditions. The absolute-flux calibrated spectra \textit{without} MW extinction correction are available on WISeREP. 

\subsection{Imaging polarimetry} \label{sec:obs:impol}

\begin{table*}[h!]
	\caption{Log of polarimetric observations \label{tab:impol}}
\centering
\begin{tabular}{ccccc|cccc}
\toprule
MJD  & Phase  &  Exposure  &  Mean airmass  &  Filter  & $q$    & $u$    & $p$    & $\theta$\\
     & (day) &  time (s)  &                &          & (\%) & (\%) & (\%) & ($^{\circ}$)\\
\midrule
58492.241 & 31.9 &  $4\times100$  &  1.60  &   $v$\_HIGH     &  $0.14\pm0.08$  & $-0.24\pm0.08$  &  $0.28\pm0.08$  &  $150.2\pm7.9 $ \\
58512.121 & 49.0 &  $4\times100$  &  1.15  &   $v$\_HIGH     &  $0.10\pm0.09$  & $-0.31\pm0.09$  &  $0.33\pm0.09$  &  $140.0\pm8.3 $ \\
58524.125 & 59.3 &  $4\times100$  &  1.35  &   $v$\_HIGH     &  $0.11\pm0.10$  & $-0.25\pm0.10$  &  $0.28\pm0.10$  &  $146.8\pm10.2$ \\
58565.007 & 94.4 &  $4\times250$  &  1.33  &   $v$\_HIGH     &  $0.21\pm0.07$  & $-0.09\pm0.07$  &  $0.23\pm0.07$  &  $168.5\pm5.2 $ \\
58565.020 & 94.4 &  $4\times250$  &  1.43  &   $R$\_SPECIAL  &  $0.45\pm0.07$  & $-0.16\pm0.07$  &  $0.48\pm0.07$  &  $170.4\pm4.0 $ \\
\bottomrule
\end{tabular}
\tablefoot{The first four columns summarise the observations. The last four columns report the debiased polarisation properties: Stokes parameters $q$ and $u$, the debiased polarisation level $p=\sqrt{q^2+u^2}$, and the position angle $\theta=1/2\,\arctan\left(u/q\right)$.
}
\end{table*}

To measure the ejecta geometry, we acquired four epochs of imaging polarimetry in the $v$\_HIGH filter with VLT/FORS2 between \tmax+31.9 and \tmax+94.4~days (Table~\ref{tab:impol}). In addition, we got one epoch with the $R$\_SPECIAL filter at \tmax+94.4~days. Each polarisation measurement required four exposures at four different retarder-plate angles: $0^\circ$, $22\fdg5$, $45^\circ$, and $67\fdg5$. The beam was split with a Wollaston prism into the ordinary (o) and the extraordinary (e) ray. The o-ray and the e-ray were placed at the 7th and the 8th multi-object spectroscopy (MOS) stripes, respectively.
 
We reduced the data in a standard manner using \program{IRAF} \citep{Tody1993a} tasks. The flux of the SN in the o-ray and e-ray were measured through aperture photometry at all four retarder-plate angles using the \program{DAOPHOT.PHOT} package \citep{Stetson_1987}. Stokes parameters and polarisation of the target were derived based on the FORS2 manual \citep{Anderson_etal_2018}, and the polarisation degrees were corrected for polarisation bias, caused by the non-negativity nature of the polarisation degree, following \citet{Wang_etal_1997}. The extracted, debiased polarisation properties are summarised in Table~\ref{tab:impol}.

These values need to be corrected for polarisation induced by dichroic extinction from non-spherical dust grains that aligned with the magnetic field of the interstellar medium of the Milky Way (MW) and the host galaxy. Following \citet{Serkowski_etal_1975}, the polarisation level from the Milky Way can be as high as $\lesssim9\%\times E(B-V)$. With a Galactic extinction of $E(B-V)=0.03$~mag towards \sn, the MW polarisation level could be up to 0.26\%. The determination of the interstellar polarisation from \sn's host galaxy is not feasible. We note that the polarisation degree is only $p\lesssim$0.3\% in $v$\_HIGH between \tmax+32 and \tmax+94 days (see Table \ref{tab:impol}). Such a low level of polarisation is very unlikely to be caused by a high intrinsic polarisation aligned and cancelled to a comparable level of significant interstellar polarisation. Therefore, without correcting for the polarisation from the host galaxy, the observations point to a high degree of spherical symmetry of \sn\ during the phase of our polarisation measurement.

\begin{table}
    \caption{Log of X-ray observations\label{tab:xray}}
    \scriptsize
    \centering
    \begin{tabular}{cccccc}
    \toprule
    MJD & Phase & Count rate                  & $F_{\rm X}$           & $L_{\rm X}$        \\
        & (day) & ($10^{-3}\rm ct\,s^{-1}$)   & ($10^{-15}\rm erg\,s^{-1}\,cm^{-2}$)   & ($10^{42}\rm erg\,s^{-1}$)\\
    \midrule
    \multicolumn{5}{c}{\textbf{\swift}}\\
    \midrule
    58469.56    & $ 12.5^{+65.5}_{ -4.1}$&$<0.7$&$ <27.0 $&$ <2.2$\\
    58567.93    & $ 96.9^{ +5.2}_{ -6.9}$&$<1.9$&$ <68.6 $&$ <5.6$\\
    58592.17    & $117.7^{ +5.9}_{ -6.0}$&$<2.3$&$ <82.7 $&$ <6.7$\\
    58688.35    & $200.1\pm24.2         $&$<0.6$&$ <22.9 $&$ <1.9$\\
    58741.72    & $245.9\pm12.8	        $&$<1.8$&$ <64.9 $&$ <5.3$\\
    \midrule
    \multicolumn{5}{c}{\textbf{\xmm}}\\
    \midrule
    58511.22 & $48.5 \pm0.3$ & $<9.1$  &$<17.1$ & $<1.4$\\
    58561.70 & $91.8 \pm0.3$ & $<10.2$ &$<19.2$ & $<1.6$\\
    58694.68 & $205.8 \pm0.3$ & $<19.9$ &$<37.4$ & $<3.0$\\
    58723.37 & $230.4 \pm0.3$ & $<8.7$  &$<16.4$ & $<1.3$\\
    \bottomrule
    \end{tabular}
    \tablefoot{The phases report the mid-exposure time.
    For \swift\ data, the modified Julian date reports the mid-exposure time. The phase error quotes the bin size after dynamically rebinning the data. For \xmm\ data, the modified Julian date reports the beginning of the observation. The total phase error corresponds to the on-source integration time. All limits are reported at $3\sigma$ confidence. The measurements are corrected for MW absorption and reported for the bandpass from 0.3 to 10~keV.
    }
\end{table}

\subsection{X-ray Observations} \label{sec:obs:xray}

\subsubsection{\swift/XRT}

While monitoring \sn\ with UVOT between \tmax+8.4 and \tmax+224~days, \swift\ also observed the field with the X-ray telescope XRT between 0.3 and 10 keV in photon-counting mode \citep{Burrows2005a}. We analysed these data with the online-tools of the UK \swift\ team\footnote{\href{https://www.swift.ac.uk/user_objects/}{https://www.swift.ac.uk/user\_objects}} that use the software package \program{HEASoft} version 6.26.1 and methods described in \citet{Evans2007a, Evans2009a}.

\sn\ evaded detection in all epochs. The median $3\sigma$ count-rate limit of all observing blocks is 0.005~count~s$^{-1}$ (0.3--10~keV). Using the dynamic rebinning option in the \swift~online tools pushes the $3\sigma$ count-rate limits to 0.002~count~s$^{-1}$ (median value). A list of the limits from the stacking analysis is shown in Table~\ref{tab:xray}. To convert the count-rate limits into a flux, we used \program{WebPIMMS}\footnote{\href{https://heasarc.gsfc.nasa.gov/cgi-bin/Tools/w3pimms/w3pimms.pl}{https://heasarc.gsfc.nasa.gov/cgi-bin/Tools/w3pimms/w3pimms.pl}} and assumed a power-law spectrum with a photon index\footnote{The photon index is defined as the power-law index of the photon flux density ($N(E)\propto E^{-\Gamma}$).} of $\Gamma=2$ and a Galactic neutral hydrogen column density of $1.97\times10^{20}$~cm$^{-2}$ \citep{HI4PI2016a}. The average energy conversion factor for the unabsorbed flux is $3.66\times10^{-11}\,\rm\left(erg\,s^{-1}\,cm^{-2}\right)/\left(ct\,s^{-1}\right)$.
The median count-rate limit corresponds to an unabsorbed flux of $<7.4\times10^{-14}~{\rm erg\,cm}^{-2}\,{\rm s}^{-1}$  between 0.3--10 keV and a luminosity of $<4.9\times10^{42}~{\rm erg\,s}^{-1}$. The flux and luminosity limits of the individual bins are shown in Table~\ref{tab:xray}.

\subsubsection{\xmm}

The field of \sn\ was also observed by \xmm\ \citep[][Principal Investigator: R. Margutti, University of California, Berkeley, USA]{Jansen2001a}. Four epochs were taken with the European Photon Imaging Camera (EPIC) with the pn \citep[][]{Strueder2001a} and MOS1|2 cameras \citep{Turner2001a} between 28 January 2019 and 28 August 2019 (\tmax+48.5 -- \tmax+230.4). We reduced the \xmm/EPIC pn data using the \xmm\ Science Analysis System\footnote{\href{https://www.cosmos.esa.int/web/xmm-newton/sas}{https://www.cosmos.esa.int/web/xmm-newton/sas}} (SAS) following standard procedures. We extracted the source using a circular region with a radius of $32^{\prime\prime}$, and the background from a source-free region on the same CCD. The MOS data are shallower than the pn data, so we omit reporting them in this paper.

All \xmm\ observations led to non-detections with count rate limits between 0.009 and 0.020 ct s$^{-1}$ between 0.3 and 10~keV. Using the same spectral model as for XRT and an energy conversion factor of $1.88\times10^{-12}\,\rm\left(erg\,s^{-1}\,cm^{-2}\right)/\left(ct\,s^{-1}\right)$, these limits translate to unabsorbed flux limits between 1.6 and $3.7\times10^{-14}~{\rm erg\,cm}^{-2}\,{\rm s}^{-1}$. Table~\ref{tab:xray} summarises the measurements.

\subsection{Radio observations}\label{sec:obs:radio}

The field was observed by the VLA Sky Survey \citep{Lacy2020a} between 2 and 4~GHz on 27 October 2020 (\tmax+595~days). No source was detected. The flux at the SN position is $-47 \pm 223$~$\mu$Jy, translating to a $3\sigma$ flux limit of 622~$\mu$Jy and a luminosity of $2\times10^{39}\,\rm erg\,s^{-1}$. \citet{Eftekhari2021a} presented sub-mm observations at 100~GHz obtained with the Atacama Large Millimeter Array on 24 December 2019 (\tmax+331~days). These authors also reported a non-detection with an r.m.s. of 19~$\mu$Jy, translating to a $3\sigma$ flux limit of 58~$\mu$Jy and a luminosity of $5\times10^{39}\,\rm erg\,s^{-1}$.

\section{Results}\label{sec:results}
\subsection{Redshift}

\begin{figure}
    \centering
    \includegraphics[width=1\columnwidth]{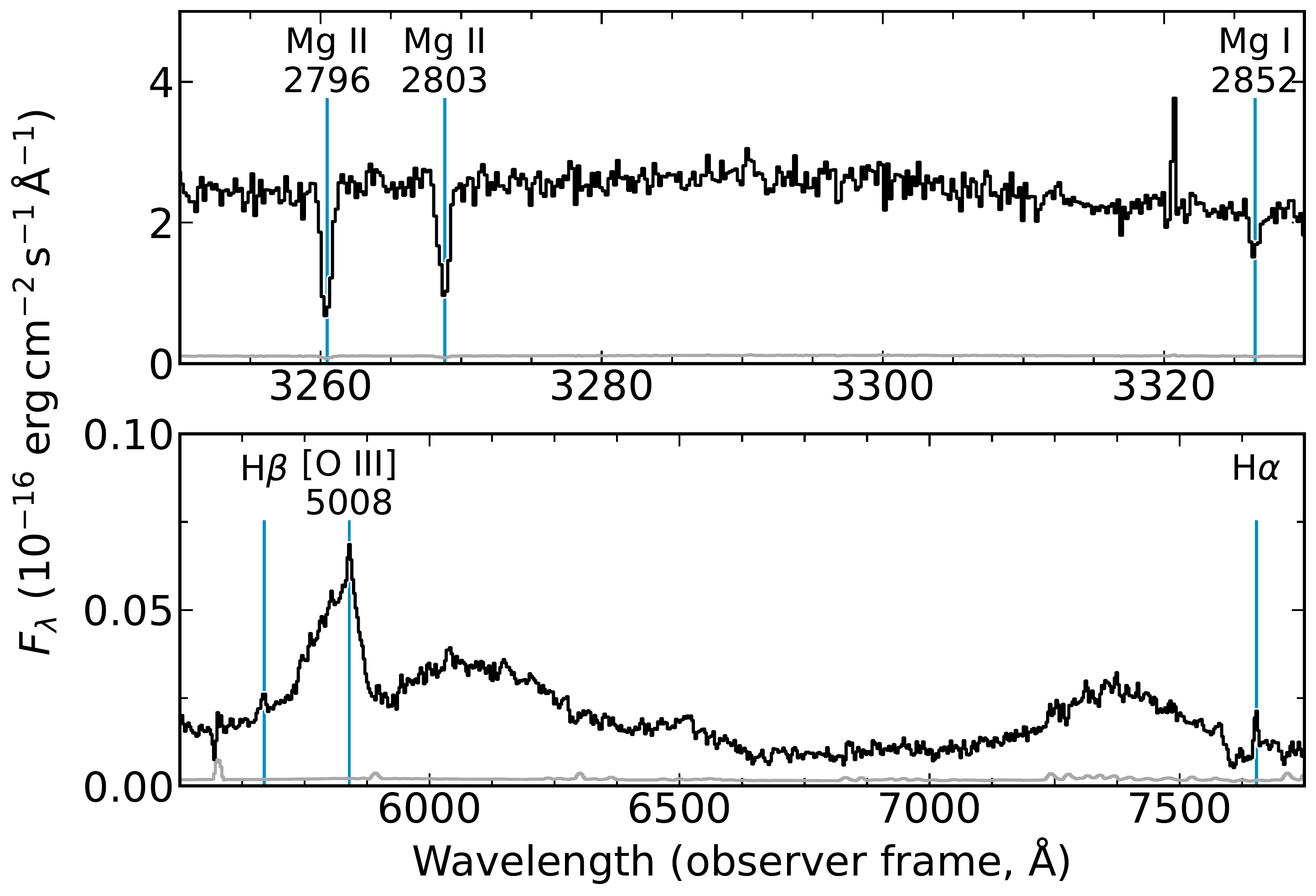}
    \caption{Galaxy absorption and emission lines at a common redshift of $z=0.166$ in the supernova spectra at \tmax+32.7~days (top) and  at \tmax+565.3~days (bottom). The error spectrum of each epoch is shown in grey.}
    \label{fig:spec:redshift}
\end{figure}

The X-shooter spectra between \tmax+32.7 and \tmax+94.3~days show narrow absorption lines of \ion{Mg}{i}\,$\lambda$\,2852 and \ion{Mg}{ii}\,$\lambda\lambda$\,2796,2803 from the host galaxy at a common redshift of $z=0.1660$ (Figure \ref{fig:spec:redshift}, top panel). The low-resolution FORS2 spectrum obtained at \tmax+565.3~days, shown in the bottom panel of Figure \ref{fig:spec:redshift}, reveals narrow emission lines from hydrogen and oxygen from the \ion{H}{ii} regions in the host galaxy at the same redshift as the absorption-line redshift. This redshift translates to a luminosity distance of 822.6~Mpc using the cosmological parameters from \citet{Planck2016a}.

\begin{figure*}
    \centering
    \includegraphics[width=1\textwidth]{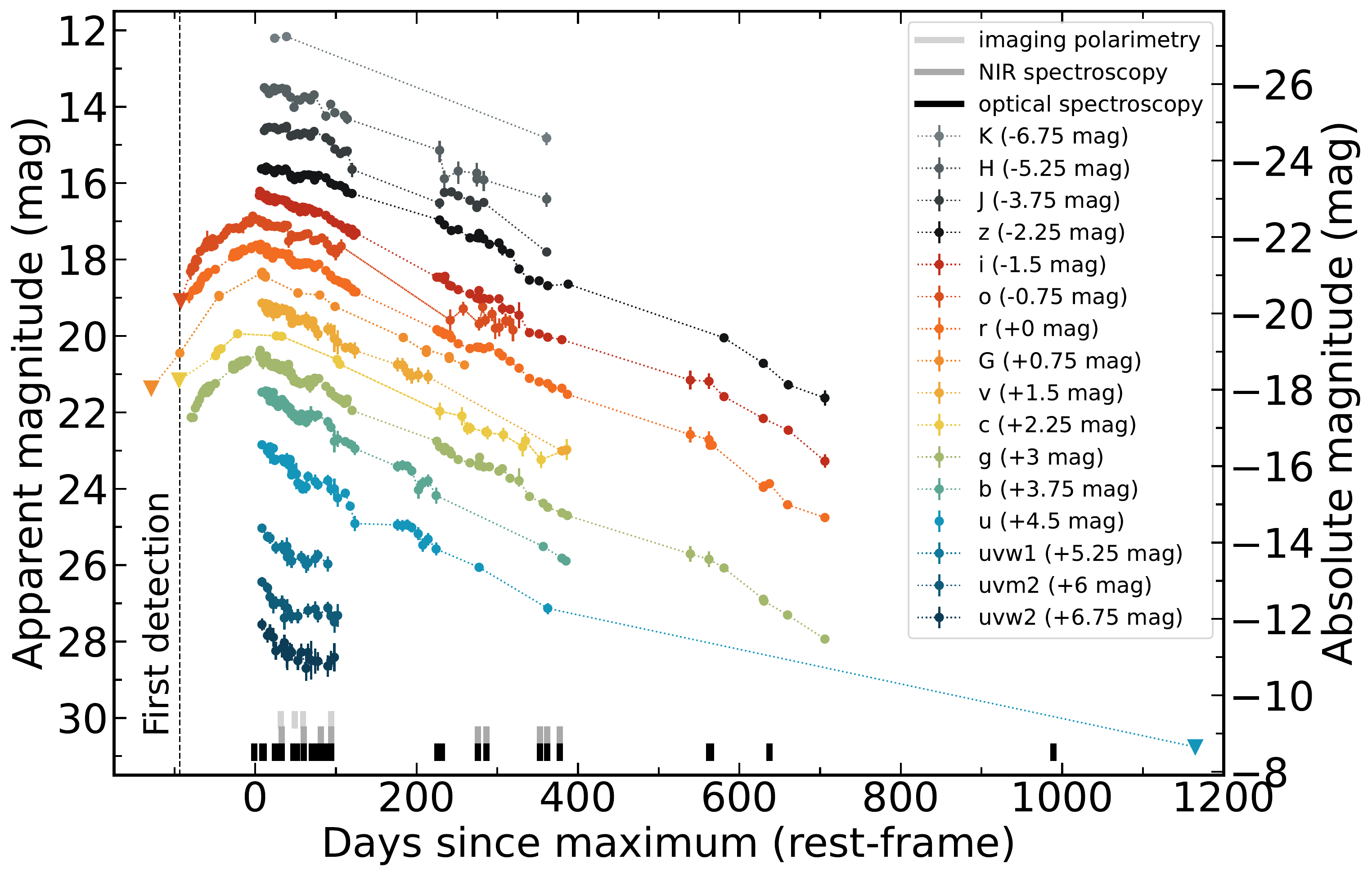}
    \caption{Multi-band light curve of \sn\ from 1800 to 18,500~\AA~(rest-frame) after correcting for the Galactic extinction. \sn\ was first detected by \gaia. The last non-detections before the first detection by \gaia\ and ATLAS are shown by the downward-pointing triangles. With a rise time of $>93$~rest-frame days, \sn\ is one of the slowest evolving SLSN known. The decline of 1.1~mag\,(100~days)$^{-1}$ is similar to the decay time of radioactive $^{56}$Co. After \tmax+575~days, the decline steepened to 1.5~mag\,(100~days)$^{-1}$. The light curve shows undulations up to \tmax+100~days and a longer-lasting bump at $\sim300$~rest-frame days. Vertical bars represent the epochs of spectroscopy and imaging polarimetry. The absolute magnitude is computed with $M=m - {\rm DM}(z) + 2.5\,\log \left(1+z\right)$, where DM is the distance modulus and $z$ the redshift.
    }
    \label{fig:lc:multiband}
\end{figure*}

\subsection{Light curve}\label{sec:lc}

\subsubsection{General properties}\label{sec:lc:general}

Figure \ref{fig:lc:multiband} shows the evolution of \sn\ from \tmax$-93$ to \tmax+706 rest-frame days. The early-time evolution was captured by the ATLAS, \gaia\ and ZTF surveys. Human scanners discovered \sn\ shortly before maximum light, which triggered our large monitoring campaign from UV to NIR wavelengths. The $g$, $r$ and $o$ band light curves cover the evolution from early to late times. We use these datasets to infer the time of maximum light and the rise and decline time scales. Fitting the light curves with 3rd order polynomials between ${\rm MJD}=58425$ and ${\rm MJD}=58485$ returns the time of maximum light at ${\rm MJD}=58458\pm2$, $58454\pm2$ and $58452\pm4$ in $g$, $r$ and $o$, respectively. Throughout the paper, we adopt the weighted mean MJD $58455\pm2$ as the time of maximum light. At the time of peak, \sn\ reached a brightness of $17.54\pm0.02$, $17.65\pm0.01$ and $17.92\pm0.04$ in $g$, $r$ and $o$ band, respectively (all corrected for MW extinction; Table \ref{tab:lc_prop}).\footnote{The host extinction is negligible (Section \ref{res:host}).} Using the Keck spectrum at \tmax$-$1.4~days, we infer k-corrected absolute magnitudes of $-21.79\pm0.02$, $-21.66\pm0.01$ and $-21.43\pm0.04$~mag in the aforementioned bands (Table \ref{tab:lc_prop}) and a k-corrected $g-r$ colour of $-0.12\pm0.02$~mag at peak (corrected for MW extinction), a typical luminosity and colour for a H-poor SLSN \citep{Nicholl2015a, deCia2018a, Lunnan2018b, Angus2019a, Chen2023a}.

Similar to \citet{Chen2023a}, we measure the rise and decline time scales from 10\% and 50\% peak flux to peak in all three bands. In the $g$ band, we obtain $t_{1/2,\,\rm rise}=52\pm1$~days, $t_{1/2,\,\rm decline}=88^{+1}_{-2}$~days, $t_{1/10,\,\rm rise}>79.3$~days, and $t_{1/10,\,\rm decline}=242\pm1$~days, that is, 1 mag (100 days)$^{-1}$ (all measured in the rest-frame). The light curve parameters in the other bands are summarised in Table \ref{tab:lc_prop}. Although the \gaia\ light curve is poorly sampled, the data are of sufficient quality to improve the lower limit on the rise timescale $t_{1/10,\,\rm rise}$. The \gaia\ alert database reports the first detection on ${\rm MJD}=58346.11$ (16 August 2018), 11.5 and 13.3 rest-frame days before the first ZTF and ATLAS\footnote{The last ATLAS non-detection is from 17 August 2018, that is, 1.3 rest-frame days after the first \gaia\ detection, reaching a limiting magnitude of $o\approx19.9$~mag at 3 sigma confidence.} detection, respectively. At the time of discovery, \sn\ had a brightness of 19.8 mag; around the time of maximum light, the brightness reached 17.7 mag. This sets a lower limit of $>93.4$~days on $t_{1/10,\,\rm rise}$.

\addtolength{\tabcolsep}{-2pt}   
\begin{table}
\caption{Light curve properties}
\label{tab:lc_prop}
\small
\centering
\begin{tabular}{cccc}
\toprule
\mycomment{
Band    & $m_{\rm peak}$ & $M_{\rm peak}$ & $\tau_{1/2,\rm rise}$ & $\tau_{1/2,\rm decline}$ & $\tau_{1/10,\rm rise}$ & $\tau_{1/10,\rm decline}$ \\
        & (mag) & (mag) & (day) & (day) & (day) & (day)\\
\midrule
$g$ &  $17.54\pm0.02$   & $-21.79\pm0.02$ & $52\pm1$ & $88^{+1}_{-2}$ & $>79.3$ & $242\pm1$\\
$r$ &  $17.65\pm0.01$   & $-21.66\pm0.01$ & $60\pm1$ & $93\pm1$       & $>82$   & $248\pm1$\\
$o$ &  $17.72\pm0.02$   & $-21.43\pm0.04$ & $67^{+2}_{-1}$ & $80^{+5}_{-4}$       & $>80$   & $>106$\\
}
Property                    & $g$             & $r$             & $o$             \\
\midrule
Peak time (MJD)                  & $58458\pm2$     & $58454 \pm 2$   & $58452 \pm 4$   \\
$m_{\rm peak}$ (mag)             & $17.54\pm0.02$  & $17.65\pm0.01$  & $17.72\pm0.02$  \\
$M_{\rm peak}$ (mag)             & $-21.80\pm0.02$ & $-21.66\pm0.01$ & $-21.62\pm0.02$ \\
$t_{1/2,\rm rise}$ (day)      & $52\pm1$        & $60\pm1$        & $64\pm1$        \\
$t_{1/2,\rm decline}$ (day)   & $88^{+1}_{-2}$  & $93\pm1$        & $95\pm3$        \\
$t_{1/e,\rm rise}$ (day)      & $68.3\pm0.4$    & $72.5\pm0.5$    & $73.8\pm0.5$\\
$t_{1/e,\rm decline}$ (day)   & $102\pm2$       & $117\pm1$       & $>107$\\
$t_{1/10,\rm rise}$ (day)     & $>79.3$         & $>82$           & $>80$           \\
$t_{1/10,\rm decline}$ (day)  & $242\pm1$       & $248\pm1$       & $235^{+1}_{-4}$ \\
\bottomrule
\end{tabular}
\tablefoot{All magnitudes are corrected for Galactic extinction. The absolute magnitudes include a k-correction inferred from the Keck spectrum at \tmax$-1.4$~days. All time scales are reported in the rest-frame. The uncertainties reflect the $1\sigma$ statistical errors.
}
\end{table}

Between July 2018 and the date of the first \gaia\ detection (16 August 2018), the field was visible to observing facilities in the southern hemisphere. We searched the data archives of the Australian Astronomical Observatory, the European Southern Observatory, the Gemini Observatory, and the Las Cumbres Observatory for serendipitous observations of this field but found no relevant data. We conclude that \sn's progenitor exploded $>93$ rest-frame days before the maximum light, but we have no firm constraint on the explosion date.\footnote{The \gaia\ alert database reports an observation on 5 July 2018 but no measurement. This could either mean \textit{i}) a non-detection (limiting magnitude $G=20.7$~mag) and hence imposing an upper limit of $<129$~days on $t_{1/10,\,\rm rise}$, \textit{ii}) the observation was not performed, or \textit{iii}) a problem in the data processing \citep{Hodgkin2021a}.}

We obtained a final epoch of photometry with \hst/WFC3 in $u$ band at \tmax+1165~days as a part of an \hst\ Snapshot programme to search for signs of late-time CSM interaction in SNe \citep{Fremling2021a}. \sn\ evaded detection. We place an upper limit of $u=26.2$ ($3\sigma$ confidence), shown as a downward pointing triangle in Figure \ref{fig:lc:multiband}.

\begin{figure}
    \centering
    \includegraphics[width=1\columnwidth]{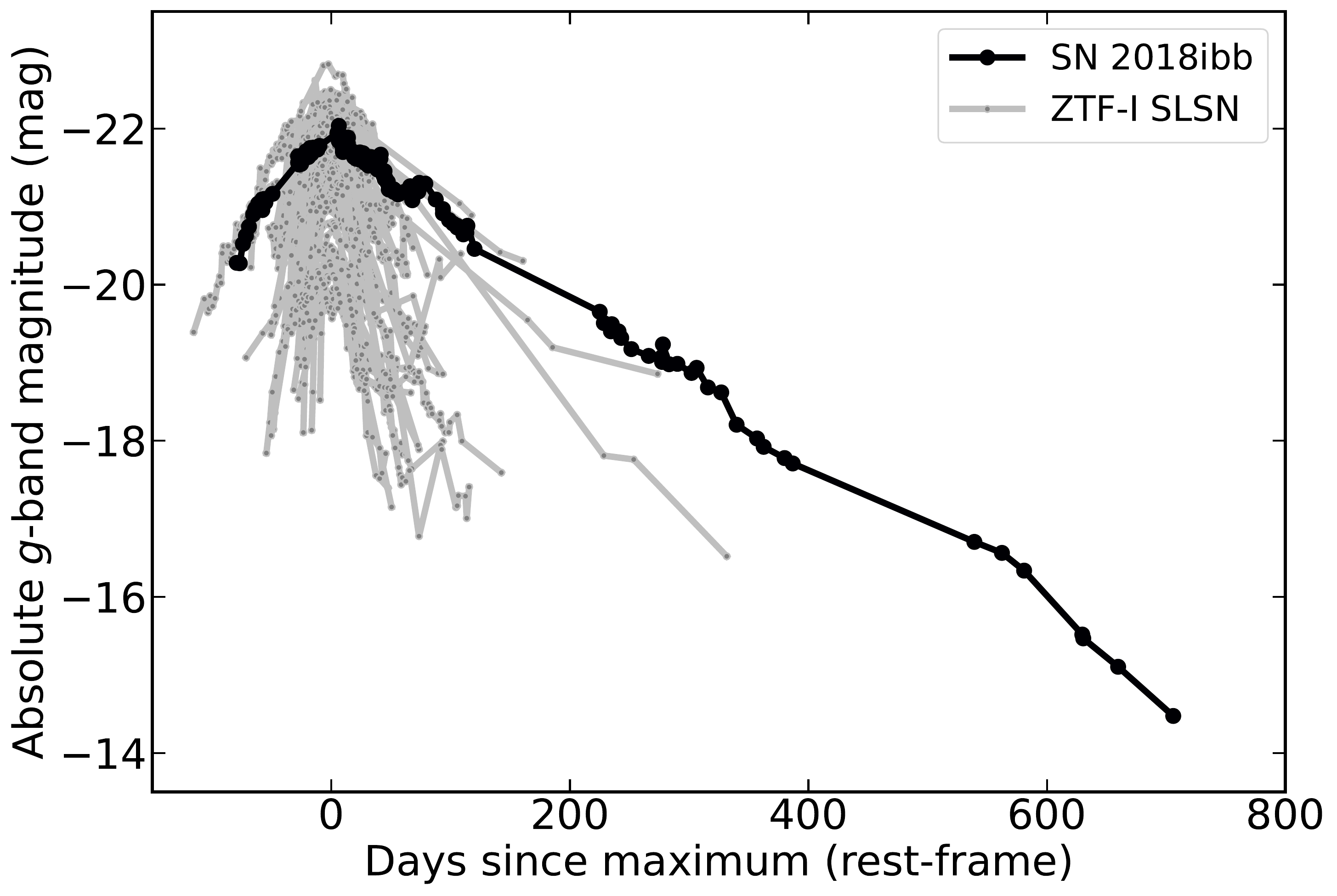}
    \caption{$g$-band light curve of \sn\ in the context of the homogeneous ZTF-I SLSN sample. \sn\ has a typical peak absolute magnitude. The rise of $>93$ rest-frame days is significantly longer than of the average ZTF SLSN. The long-lasting rise implies a long diffusion time, which requires a very high total ejected mass. The high peak luminosity requires a very energetic explosion. Both properties together hint to an explosion mechanism that might be different from that of regular SLSNe.
        }
    \label{fig:lc:comparison}
\end{figure}

\sn's light curve exhibits several peculiar properties. Figure \ref{fig:lc:comparison} compares the absolute magnitude versus the rest-frame phase of \sn\ to the light curves of the 78 H-poor SLSNe from ZTF-I presented in \citet{Chen2023a}. The absolute magnitude of all objects is computed with $M=m - {\rm DM}(z) + 2.5\,\log \left(1+z\right)$, where DM is the distance modulus and $z$ the redshift. \sn\ has the longest rise in the ZTF sample. The $g$-band rise time $t_{1/10,\,\rm rise}$ exceeds the sample mean value (41.9~days) by a factor of 2.1 times the sample standard deviation [$\sigma(t_{1/10,\,\rm rise}) = 17.8$~days; \citealt{Chen2023a}]. This factor could increase to even $4.9\sigma$ if the \gaia\ data are a good proxy of the rise time in ZTF/$g$. The light curve fades by 1.1~mag (100 days)$^{-1}$ for 500--600 days before the decline steepens to 1.5~mag (100 days)$^{-1}$. The decline time scale is slower than for any of the other H-poor SLSNe from the ZTF-I sample. The rise is even slower than any of the $>100$ H-poor SLSNe found by other surveys (as queried from the Transient Name Server\footnote{\href{https://www.wis-tns.org}{https://www.wis-tns.org}} and ADS Abstract Service\footnote{\href{https://ui.adsabs.harvard.edu}{https://ui.adsabs.harvard.edu}}). Only the H-poor SLSN PS1-14bj \citep{Lunnan2016a} had a longer rise. We discuss this in more detail in Section \ref{discussion:comparison_w_slow_slsn}.

A number of SNe have shown a pre-bump with observed peak luminosities between $M_g\sim-18$ and $-23$~mag \citep{Leloudas2012a, Nicholl2015b, Smith2016a, Angus2019a}. Such a bump is not visible in the light curve of \sn\ (Figures \ref{fig:lc:multiband} and \ref{fig:lc:comparison}). However, the progenitor of \sn\ exploded before the field came from behind the sun, precluding drawing a firm conclusion on the absence or presence of a pre-bump.

\subsubsection{Bolometric light curve}\label{sec:bb}\label{sec:lc:bolometric}

We compute the bolometric luminosity of \sn\ over a wavelength range from $\sim1800$ to $\sim14,300$~\AA\ (rest-frame), which is defined by the wavelength coverage of our photometric dataset. However, our dataset does not have the same wavelength coverage throughout the entire duration of the observations. In the following, we describe how the bolometric light curve is constructed and discuss the bolometric corrections that we derived for time intervals with incomplete spectral information.

\begin{figure}
    \centering
    \includegraphics[width=1\columnwidth]{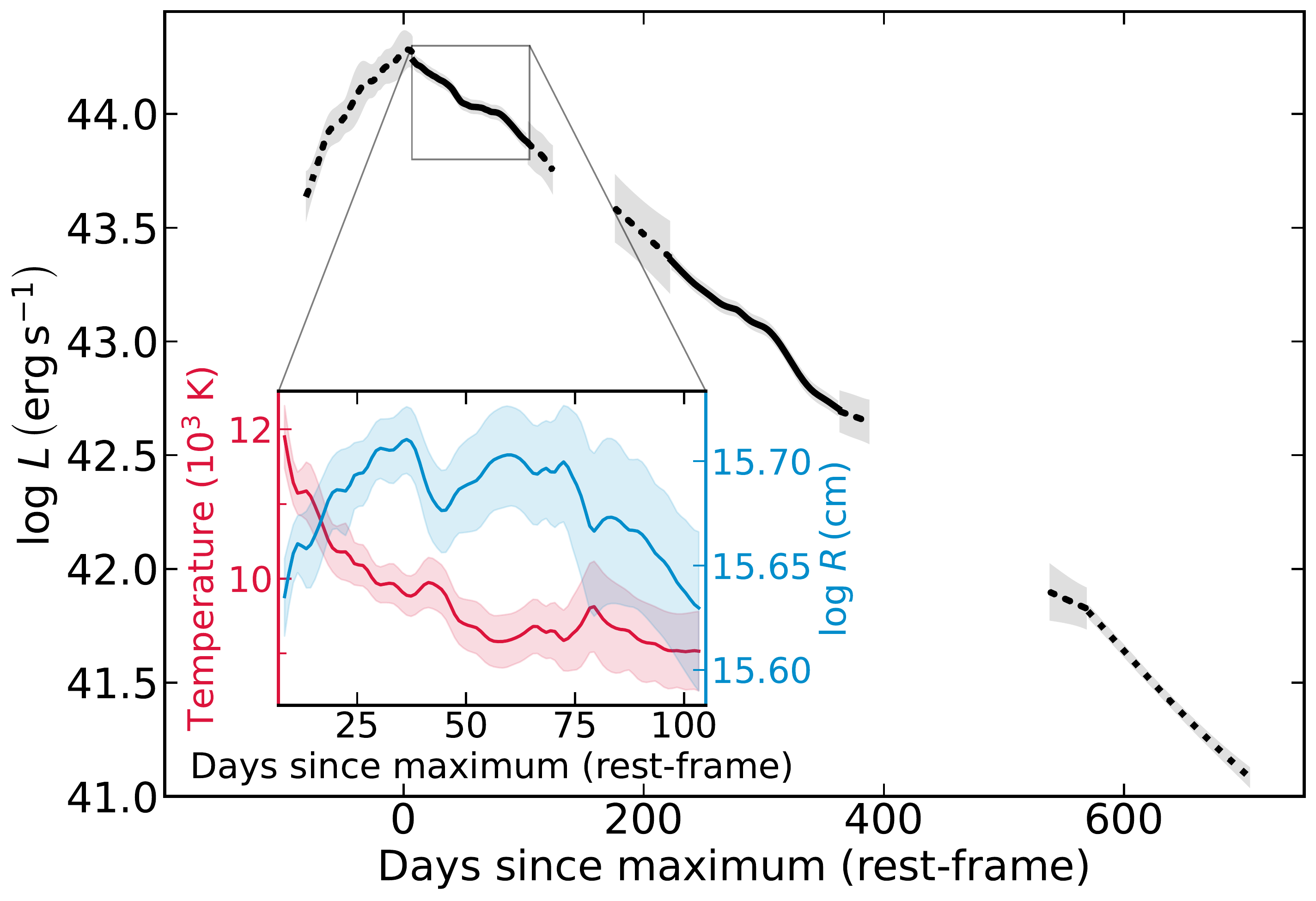}
    \caption{
    Bolometric light curve of \sn\ from 1800 to 14,300~\AA\ (rest-frame). The dotted lines indicate time segments with partial wavelength coverage. At peak \sn\ reached a luminosity of $>2\times10^{44}~\rm erg\,s^{-1}$. Integrating over the light curve from \tmax$-93$ to \tmax+706~days yields a radiated energy of $3>\times10^{51}~\rm erg$. Both values are conservative lower limits. The inset shows the evolution of the blackbody temperature and radius of the photospheric phase where photometry has been carried out from the $u$ to $H$ bands. The shaded regions indicate the $1\sigma$ statistical uncertainties.
    }
    \label{fig:lc:bolometric}
\end{figure}

The bolometric light curve is built as follows:
\textit{i}) correcting all photometric data for the MW extinction,
\textit{ii}) dividing the entire dataset into segments defined by the observing seasons,
\textit{iii}) interpolating the light curve in each band of each observing season with a Gaussian process with the python package \program{George} version 0.4.0 \citep{Ambikasaran2015a}\footnote{
We added a systematic error of 5\% to all optical and NIR filters and 10\% to all UV filters in quadrature to account for uncertainties in the flux calibration.},
\textit{iv}) constructing the spectral energy distributions for every time step,
\textit{v}) calculating the bolometric flux by numerical integration of each SED, and 
\textit{vi}) multiplying the bolometric flux by $4\pi\,d^2_L$, where $d_L$ is the luminosity distance, to obtain the bolometric luminosity.

\begin{figure*}
    \centering
    \includegraphics[width=1\textwidth]{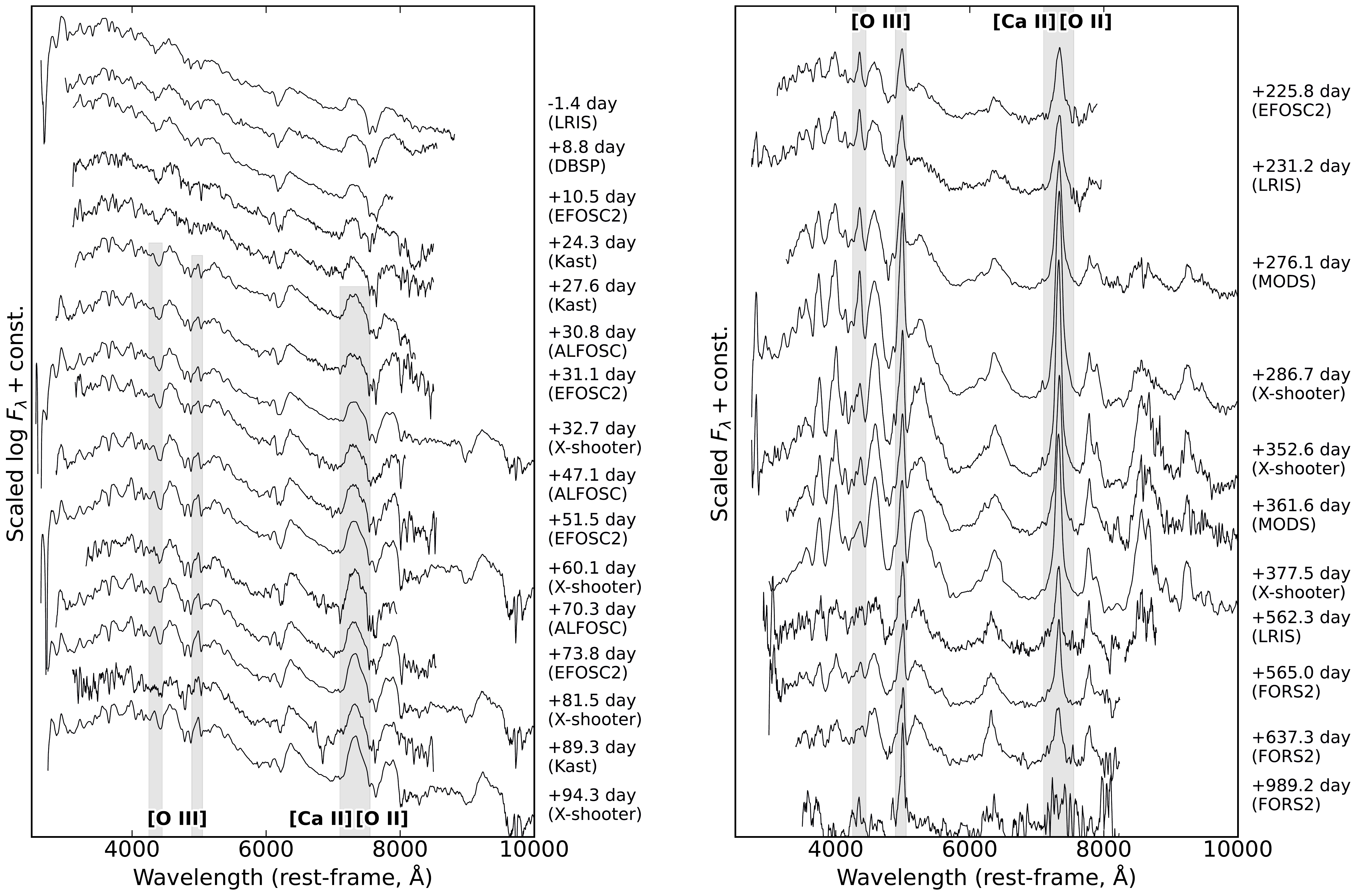}
    \caption{Spectroscopic sequence from 2500~\AA\ to 10,000~\AA\ and from the time of maximum to \tmax+1000~days (rebinned to 5~\AA\ and smoothed with a Savitzky-Golay filter). Spectra up to \tmax+100 days (left panel) are characterised by a blackbody continuum with superimposed absorption lines from the SN ejecta, expanding with a velocity of $\sim8500~\rm km\,s^{-1}$. Between \tmax+100 and \tmax+225~days (while \sn\ was behind the sun), the spectroscopic behaviour of \sn\ evolved drastically. The late-time spectra (right panel) are characterised in the blue ($<5000$~\AA) by a pseudo-continuum and emission lines produced by the interaction of the SN ejecta with circumstellar material and in the red ($>5000$~\AA) by nebular emission lines from the $^{56}$Ni-heated SN core. The regions with the fastest evolution are highlighted by the grey-shaded regions. Figure \ref{fig:spec:line_id} shows the identification of the most prominent features of the photospheric and nebular phases.
    Regions affected by strong telluric absorption were clipped. Their locations are indicated in Figure \ref{fig:spec:line_id}.
    }
    \label{fig:spec:seq_uvvis}
\end{figure*}

Our dataset has the best spectral coverage between \tmax\ and \tmax+375 days: 1800 to 14,300~\AA\ between \tmax\ to \tmax+100 days and 3000 to 14,300~\AA\ between \tmax+200 to \tmax+375 days. The bolometric light curve for these time intervals are shown as solid lines in Figure \ref{fig:lc:bolometric} and their $1\sigma$ confidence intervals as a shaded region. A tabulated version can be found in Table \ref{tab:bolometric}. Based on the blackbody fits to the data from $u$ to $H$ band, we estimate that $\lesssim3\%$ of the observed bolometric flux is emitted at longer wavelengths between \tmax\ and \tmax+100~days. Linearly extrapolating the observed SED from 1800 to 14,300~\AA\ towards shorter wavelengths yields a missing UV contribution of $\ll1\%$. Hence, we omit to correct the observed bolometric flux. At later phases, the spectrum does not resemble a blackbody anymore (Figure \ref{fig:spec:seq_uvvis}), and we cannot quantify the missing flux at longer and shorter wavelengths.

For the other epochs, we used these time intervals (\tmax\ to \tmax+100 days and \tmax+200 to \tmax+375 days) to estimate bolometric corrections. The pre-max dataset consists of photometry in ZTF $g$ and $r$, and ATLAS $c$ and $o$ filters, and the \gaia\ white band. We only use the ZTF data when computing the bolometric luminosity because the ATLAS and \gaia\ filters are too broad for building SEDs. At the time of the first epoch with coverage from $w2$ to $H$, $\sim26\%$ of the bolometric flux was emitted in $g+r$ band. We use this flux ratio as an estimate of the missing flux. Since SN ejecta cool with time, such a universal correction will progressively underestimate the bolometric flux towards earlier epochs. Between the first and second observing seasons, we continued the follow-up with \swift/UVOT in $ubv$ when \sn\ was no longer visible from the ground. Similar to the pre-max data, we chose time intervals with data from $w2$ to $H$ or $u$ to $H$ band to correct for the missing flux. At phases later than \tmax+500~days, photometric data are only available from $g$ to $z$ band. We omit to apply any bolometric correction for this time interval because we have no good estimate of the missing bolometric flux.

\sn\ reached a peak luminosity of $L_{\rm bol,\,peak}\geq2\times10^{44}~{\rm erg\,s}^{-1}$. Integrating the light curve from \tmax$-93$ to \tmax+706~days yields $\geq3\times10^{51}~{\rm erg}$ for the total radiated energy $E_{\rm rad}$. We emphasise that both values are strict lower limits. Our multi-band campaign only started when \sn\ peaked in the $g$ and $r$ bands, which was likely after the bolometric peak. 

Between \tmax\ and \tmax+100~days, the spectra of \sn\ are characterised by a cooling photosphere (Figure \ref{fig:spec:seq_uvvis}), and the spectral energy distributions from the $u$ to $H$ band are adequately fitted with a Planck function. The red and blue curves in the inset of Figure \ref{fig:lc:bolometric} show the evolution of the blackbody temperature and radius (see also Table \ref{tab:bolometric}), respectively. The photosphere has a temperature of 12,000~K at the time of maximum light and cools by 3000~K in 100~rest-frame days. During the same time interval, the location of the photosphere hardly changes from its mean value of $5\times10^{15}$~cm. The values of the blackbody radius and temperature are comparable to regular SLSNe \citep{Chen2023a} and the slow-evolving SLSN 2015bn \citep{Nicholl2016a}, which have observations in the UV. The blackbody temperature of \sn\ evolves slower than for regular SLSNe \citep{Chen2023a}, mirroring its slowly evolving light curve. We remark that including data at shorter wavelengths would have led to lower temperatures ($\approx0.1$~dex at \tmax) and larger radii ($\approx0.12$~dex at \tmax) due to absorption lines in the UV \citep{Yan2017a, Lunnan2018b, Angus2019a}. Owing to this, we omit these data to infer the blackbody radius and temperature.

\begin{figure*}
    \centering
    \includegraphics[width=1\textwidth]{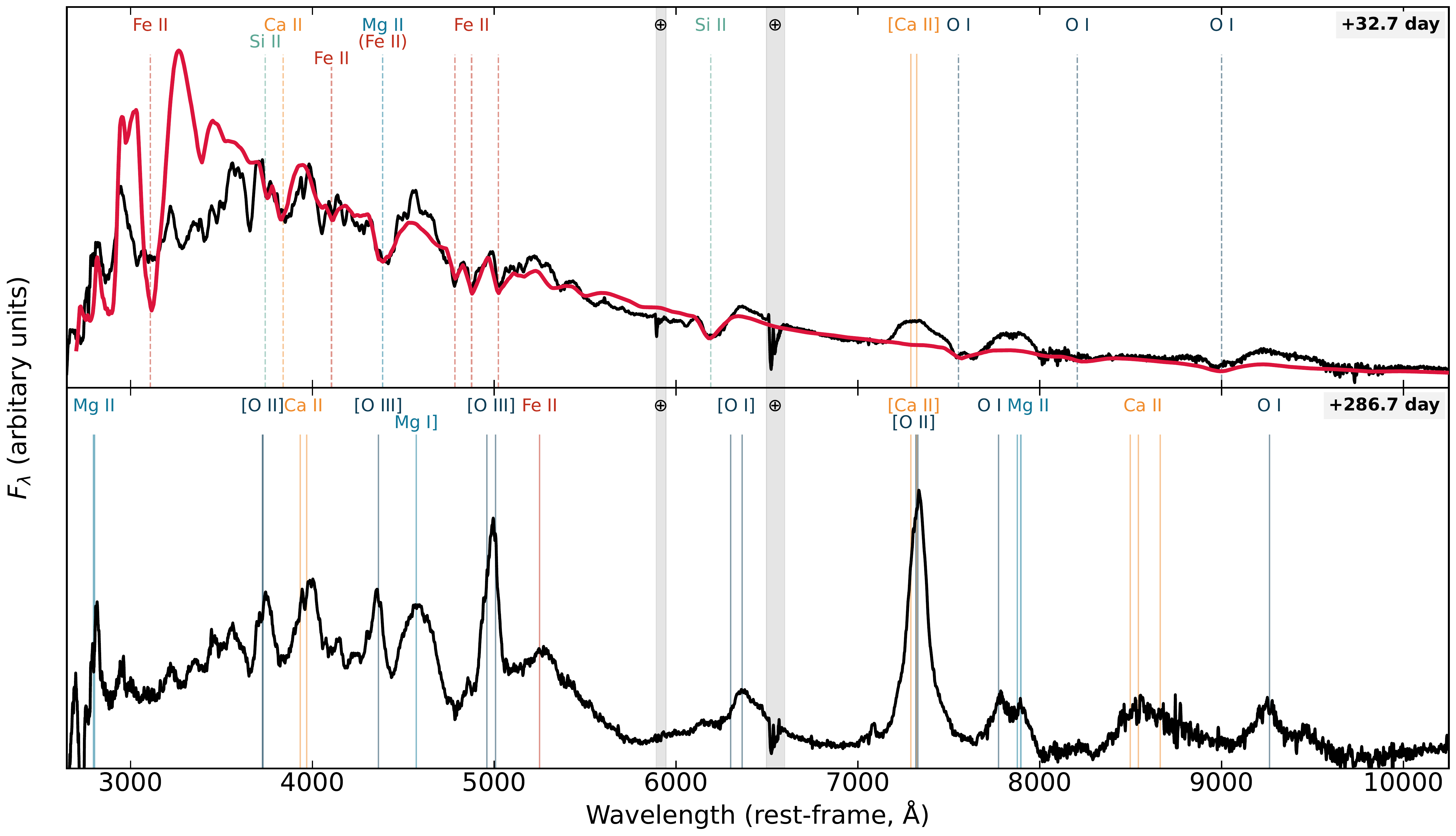}
    \caption{Line identification of the photospheric-phase spectrum (top) and nebular-phase spectrum (bottom).
    \textbf{Top:} The photospheric phase spectrum was fitted with the parameterised spectral synthesis code \program{SYNOW} (red curve). Most of the spectral features can be attributed to \ion{O}{i}, \ion{Mg}{ii}, \ion{Si}{ii}, \ion{Ca}{ii}, and \ion{Fe}{ii} as seen in other SLSNe during their cool photospheric phase \citep{GalYam2019a}. In addition to the absorption lines in the SN ejecta, the photospheric phase spectrum shows conspicuous [\ion{Ca}{ii}]\,$\lambda\lambda$\,7291,\,7324, a feature that gets dominated by [\ion{O}{ii}]\,$\lambda\lambda$7320,7330 at about \tmax+30~days.
    \textbf{Bottom}: The spectrum of the nebular phase consists of a blue pseudo-continuum and a series of allowed and forbidden emission lines from singly and doubly ionised oxygen, calcium, magnesium and iron. Remarkable is the presence of [\ion{O}{ii}] and [\ion{O}{iii}] in emission (as early as \tmax+30~days), indicating ionising radiation from shock interactions (Section \ref{sec:csm:oiii_oii}). SN absorption lines are indicated by dashed lines, and the locations mark the absorption trough minima (blueshifted by $8500~\kms$ from their rest wavelengths). SN emission lines are indicated by solid lines; their line centres are at the velocity coordinate $v=0$. Regions of strong atmospheric absorption are grey-shaded.}
    \label{fig:spec:line_id}
\end{figure*}

\subsection{Spectroscopy}
\subsubsection{Spectroscopic sequence}\label{res:spectroscopy}

Figure \ref{fig:spec:seq_uvvis} shows the spectral evolution between $\sim2800$~\AA\ to $\sim10,000$~\AA\ from the time of maximum to \tmax+990~days (all rest-frame). The spectra up to \tmax+100~days capture the photospheric phase. To identify the elements and ions responsible for the most prominent features, we model the  spectrum at \tmax+32.7~days with the spectrum synthesis code \program{SYNOW} \citep{Branch2005a}. The \program{SYNOW} fit, shown in the top panel of Figure \ref{fig:spec:line_id}, was obtained for a photospheric expansion velocity of 8000~km\,s$^{-1}$ (Section \ref{sec:velocities}) and for a blackbody temperature of 12,000~K (Section \ref{sec:bb}; a range in the order of $\pm500$ is applicable for both properties). The major ions that are securely identified and match the spectrum well are those of: \ion{O}{i}, \ion{Mg}{ii}, \ion{Si}{ii}, \ion{Ca}{ii}, and \ion{Fe}{ii} (the \ion{Mg}{ii} mainly improves the match of the feature around 4400~\AA, together with the \ion{Fe}{ii} line), in agreement with \citet{Konyves-Toth2021a}. Various additional iron group elements, such as \ion{Ti}{ii}, clearly help to lower the model flux on the blue side (3000--4000~\AA). However, we do not include those in the final \program{SYNOW} fit because the overall fit was not convincingly improved. Intriguingly, the spectrum shows narrow absorption lines from \ion{Fe}{II}\,$\lambda\lambda$\,4924,5018,5169 (half-width at zero intensity $\approx1500~\rm km\,s^{-1}$), reminiscent of the H-poor SLSNe 1999as and 2007bi. Modelling the photospheric-phase spectra of these two SLSNe revealed that such narrow features are challenging for existing SLSN models \citep{Moriya2019a}. These narrow features could point to a velocity cut in the density structure of the SN ejecta. This could be related to the density structure of the progenitor or possibly point to the deceleration of the outermost layer of the ejecta by the collision with dense circumstellar material \citep{Kasen2004a, Moriya2019a}. To differentiate between these scenarios, spectral modelling is required. This is beyond the scope of this paper. Absorption from \ion{O}{ii} between 3700~\AA\ and 4700~\AA, as seen in many SLSN spectra around peak \citep{Quimby2018a}, is not present. 

Owing to the limitations of the \program{SYNOW} approach, for instance, the simplifying underlying assumptions such as spherical, homologous expansion, resonant scattering line formation above a sharp blackbody spectrum-emitting photosphere, we perform this modelling only for the identification and verification of the major features. We avoid any fine-tuning of the different ion parameters and assessing the elemental abundances or relative mass fractions.

A complementary analysis with the National Institute of Standards and Technology (NIST) Atomic Spectra Database \citep{NIST_ASD}, following the methodology described in \citet{GalYam2019b}, which includes the same elements as above for relative intensities $\geq0.5$ in the range 2000--10,000~\AA\ (and $\geq0.2$ in the range 3000--6000~\AA\ for the \ion{Fe}{ii} lines), reveals additional possible identification of features that are not accounted for by the \program{SYNOW} fit. For instance, lines of \ion{Mg}{ii} and/or \ion{Si}{ii} may contribute to the small dip redwards of the $\sim7773$~\AA\ (rest-frame) \ion{O}{i} triplet. Also, numerous \ion{Fe}{ii} lines may contribute to the valley around 3000--3200~\AA\ as well as additional \ion{Mg}{ii} lines accounting for the dips around 4300~\AA. Remarkably, in addition to absorption lines from the SN ejecta, the first spectrum at \tmax$-1.4$~days shows conspicuous [\ion{Ca}{ii}]\,$\lambda\lambda$\,7291,7323 in emission. This is one of the strongest forbidden emission lines seen in nebular SN spectra \citep[][]{Filippenko1997a, GalYam2017a}. The only SLSNe that show [\ion{Ca}{ii}] during the photospheric phase are slow-evolving SLSNe \citep[e.g. SN\,2007bi, LSQ14an and  SN\,2015bn;][]{GalYam2009a, Nicholl2019a, Inserra2017a}.

\begin{figure*}[ht!]
    \centering
    \includegraphics[width=1\textwidth]{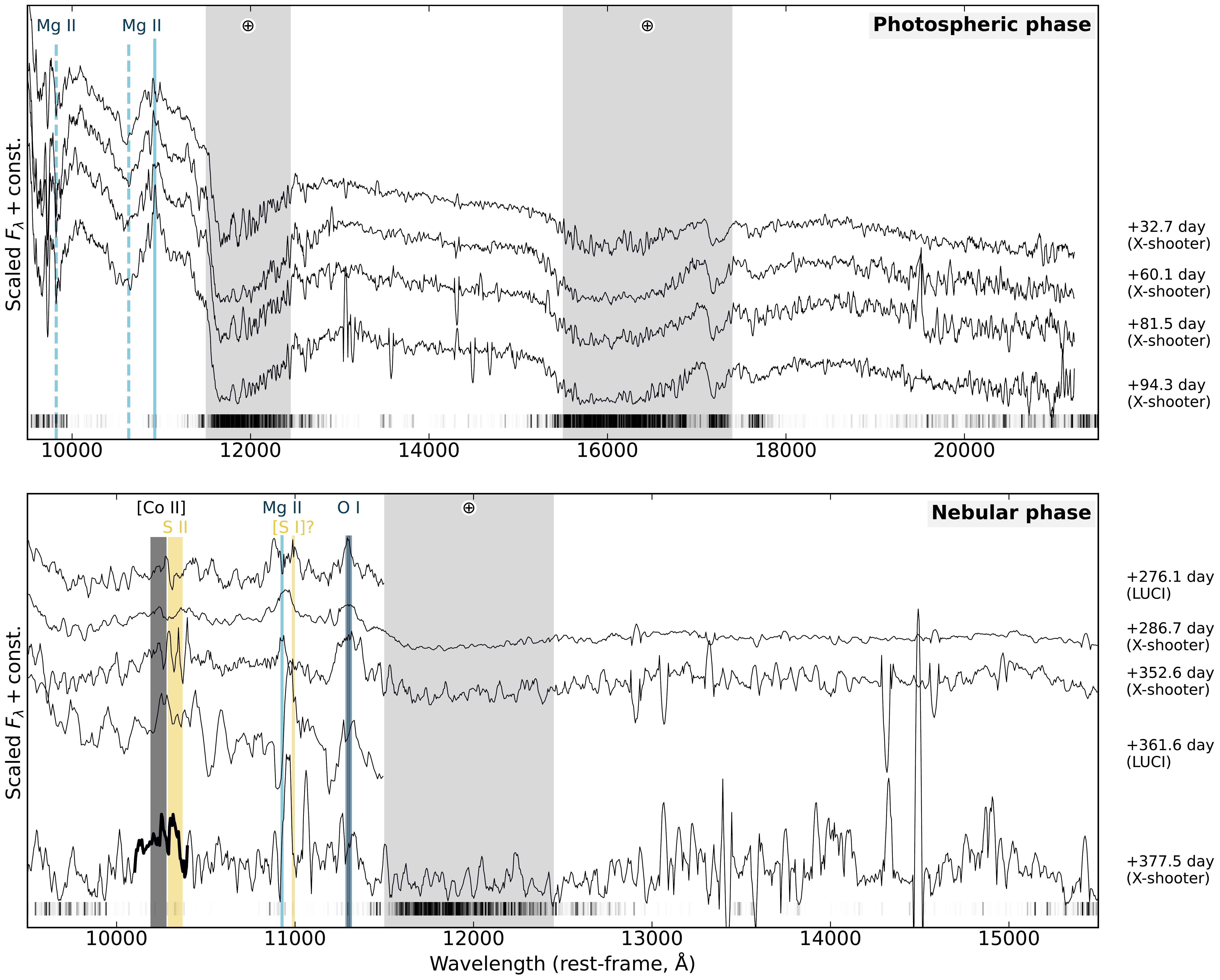}
    \caption{Spectroscopic sequence from $9500$ to $21,500$~\AA. The spectral sequence covers the evolution of the photospheric (top) and nebular (bottom) phases. The NIR spectra at $>1$~$\mu$m show only a few features in contrast to the optical spectra (Figure \ref{fig:spec:seq_uvvis}). The most prominent features are labelled. All spectra were rebinned to 5~\AA\ and smoothed with a Savitzky-Golay filter, except the spectrum at \tmax+361.6~days that was rebinned to 10~\AA. The grey scale at the bottom of each panel displays the strength of telluric features (white = transparent, black = opaque). In addition, regions of strong atmospheric absorption are grey-shaded.
    }
    \label{fig:spec:seq_nir}
\end{figure*}

During the first seasonal observing gap, the photosphere recedes and we start to see the core of the explosion. The nebular spectra (right panel in Figure \ref{fig:spec:seq_uvvis}) are dominated by emission lines with widths up to 10,000~km\,s$^{-1}$ and a  blue pseudo-continuum, similar to that seen in SNe Ia-CSM, Ibn, and Icn and some SNe IIn \citep[e.g.][]{Silverman2013a, Hosseinzadeh2017a, Gal-Yam2022a, Perley2022a}. Following previous observations of slow-evolving SLSNe \citep{Nicholl2016b, Lunnan2016a} and theoretical models by \citet{Jerkstrand2016a}, we identify the most conspicuous emission lines as allowed and forbidden transitions from neutral and ionised calcium, iron, magnesium, and oxygen (Figure \ref{fig:spec:line_id}).

Common to both the photospheric and the nebular phase is that the evolution is very slow with the exceptions of the regions at $\sim4360$, $\sim5000$ and 7300~\AA\ (highlighted in grey in Figure \ref{fig:spec:seq_uvvis}). At about 30~days after maximum, the region at $\sim5000$~\AA\ shows a rapidly growing emission feature. A weak emission line at $\sim4360$~\AA\ also emerges and reveals a similar trend to the $\sim5000$~\AA\ feature. Owing to this, we identify the two features as [\ion{O}{iii}]\,$\lambda$\,4363 and [\ion{O}{iii}]\,$\lambda\lambda$\,4959,5007, respectively. Most remarkably, the [\ion{O}{iii}]\,$\lambda\lambda$\,4959,5007 emission lines are present throughout the entire post-max evolution, even in the spectrum at \tmax+989.2 days after all other SN features faded below the detection threshold of the 4-hour VLT spectrum. This has never been observed in any SLSN before. Simultaneous with the rise of [\ion{O}{iii}], the centre of the 7300~\AA\ feature moves a few~\AA\ to longer wavelengths, the line profile changes from roughly top-hat to bell-shaped, the width decreases and the peak flux increases by a factor $\sim2$ within $<60$~days (Figure \ref{fig:spec:seq_uvvis}, left panel). This suggests that this line complex, commonly identified as [\ion{Ca}{ii}]\,$\lambda\lambda$\,7291,7324, gets dominated by [\ion{O}{ii}]\,$\lambda\lambda$\,7320,7330.

[\ion{O}{ii}] and even more so [\ion{O}{iii}] are not common features for SNe. [\ion{O}{iii}] was only observed in the slow-evolving H-poor SLSNe LSQ14an \citep{Inserra2017a} and PS1-14bj \citep{Lunnan2016a} during the photospheric phase and in SN\,2015bn in the nebular phase \citep{Nicholl2016b, Jerkstrand2017a}. Occasionally, it is also seen in regular H-poor and H-rich SNe predominantly years after the explosion (e.g. SNe II 1979C and 1980K, \citealt{Milisavljevic2009a} and \citealt{Fesen1999a}; SN IIb 1993J, \citealt{Milisavljevic2012a}; SNe IIn 1995N, 1996cr, 2010jl, \citealt{Fransson2002a, Bauer2008a, Milisavljevic2012a, Fransson2014a}; SN Ib 2012au, \citealt{Milisavljevic2018a}), and even more rarely during the photospheric phase of regular SNe \citep[e.g. Type Ic SN\,2021ocs;][]{Kuncarayakti2022a}. Possible mechanisms to produce [\ion{O}{ii}] and [\ion{O}{iii}] are
\textit{i}) excitation by CSM interaction \citep{Chevalier1994a},
\textit{ii}) photoionisation by the interaction of the pulsar wind nebula with the SN ejecta \citep{Chevalier1992a, Omand2023a},
and \textit{iii}) radioactivity \citep[for high ratios of deposited energy to O-density;][]{Jerkstrand2017a}.
In Section \ref{sec:csm:oiii_oii}, we show that [\ion{O}{ii}] and [\ion{O}{iii}] are produced by the interaction of the SN ejecta with circumstellar material.

Our series of NIR spectra (shown in Figure \ref{fig:spec:seq_nir}) covers the photospheric phase from \tmax+33 to \tmax+94~days, and the nebular phase from \tmax+276 to \tmax+378~days. The NIR spectra show a limited number of absorption and emission lines. The photospheric-phase spectra show two features at 1.093 and 1.13~$\mu$m. Following \citet{Jerkstrand2015a}, \citet{Hsiao2019a} and \citet{Shahbandeh2022a}, we tentatively identify the former as an absorption line of \ion{Mg}{ii}\,$\lambda$\,1.092~$\rm\mu m$ blueshifted by $\sim8500~\rm km\,s^{-1}$, and the latter as the recombination line \ion{O}{i}\,$\lambda$\,1.13~$\mu$m. These features can also be blended with emission lines from sulphur. The emission lines clearly stand out in the nebular-phase spectra. Our NIR spectra at \tmax+378~days show a prominent emission line at 1.025~$\rm \mu m$ that we tentatively identify as [\ion{Co}{ii}]\,$\lambda$\,1.025. This is the first time that a cobalt line has been detected in a SLSN spectrum. In Section \ref{disc:CoII}, we examine this detection in more detail. 

\subsubsection{Ejecta velocity}\label{sec:velocities}

\begin{figure}
    \centering
    \includegraphics[width=1\columnwidth]{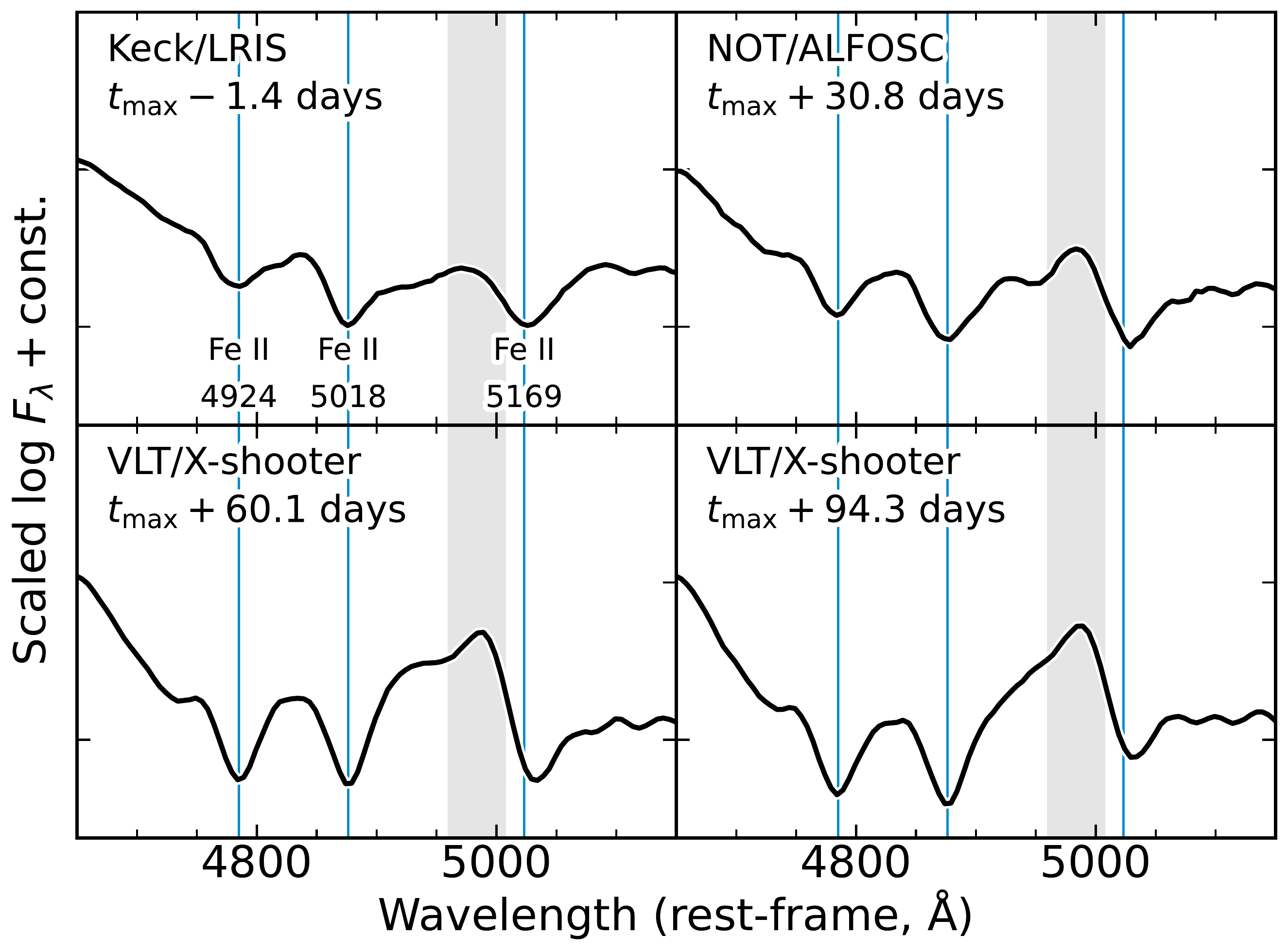}
    \caption{
    Zoom-in onto the \ion{Fe}{ii} absorption lines from the SN ejecta at selected epochs of the photospheric phase. The SN photosphere expands with a velocity of merely $\approx8500~\rm km\,s^{-1}$. There are no signs of deceleration between \tmax\ and \tmax+100~days. Starting at about \tmax+30~days, emission from [\ion{O}{iii}]\,$\lambda\lambda$\,4959, 5007 (grey shaded region), produced by the interaction of the SN ejecta with circumstellar material, contaminates the blue wing of \ion{Fe}{ii}\,$\lambda$\,5169.
    }
    \label{fig:spec:FeII}
\end{figure}

The photospheric-phase spectra of \sn\ show a large number of narrow absorption lines, mirroring a low ejecta velocity and the slow light curve evolution. The ejecta velocities are commonly measured from \ion{Fe}{ii}\,$\lambda$\,5169. Owing to the high velocities of SLSNe \citep[e.g.][]{Liu2017a, Chen2023b}, this line is usually blended with \ion{Fe}{ii}\,$\lambda$\,4924 and \ion{Fe}{ii}\,$\lambda$\,5018, necessitating template matching techniques to extract the velocities \citep{Modjaz2016a, Liu2017a}. However, the ejecta velocity of \sn\ is slow, and the \ion{Fe}{ii}\,$\lambda$\,5169 region is not blended and resolves into three absorption lines that we identify as \ion{Fe}{ii}\,$\lambda\lambda$\,4924, 5018 and 5169 (Figure \ref{fig:spec:FeII}). By measuring the minima of the three absorption lines, we extract a photospheric velocity of $\approx8500~\rm km\,s^{-1}$ that remains constant between \tmax\ and \tmax+100 days as demonstrated in Figure \ref{fig:spec:FeII} (all measurements are summarised in Table \ref{tab:velocities}).

The maximum ejecta velocity is best determined from the blue edge of the strong \ion{Mg}{ii}\,$\lambda\lambda$\,2796,2803 and \ion{Ca}{ii}\,$\lambda\lambda$\,3934, 3968 resonance lines. In Figure \ref{fig:spec:mgii_caii}, we show the regions around the two features at \tmax+32.7 days, centred on the blue doublet components. Because of the complexity of line features, we omit to subtract any continuum. For illustration purposes, we normalise the spectral regions so that the peak intensity and maximum absorption approximately match both lines. The blue components of the doublets exhibit complex profiles at low velocities because of the superposition with the wings of the red doublet components. The highest velocities are less affected by this. The \ion{Ca}{ii}\,$\lambda$\,3934 line gives the best estimate for the maximum ejecta velocity, $\sim 12,500 \kms$. This is consistent with the extent of the absorption component of \ion{Mg}{ii}\,$\lambda$\,2796, which, however, is more affected by other SN lines. The absorption minima of \ion{Mg}{ii}\,$\lambda$\,2796 and \ion{Ca}{ii}\,$\lambda$\,3934 are at $\sim8000 \kms$, but are affected by the doublet nature of the lines. Nonetheless, the locations of the absorption minima are consistent with the photospheric velocity determined from the absorption minima of \ion{Fe}{ii}\,$\lambda\lambda$\,4924,5018,5169.

To put the measurements in the context of other SLSNe, we first compare the velocity of \sn\ at maximum light to those of SLSNe in the ZTF-I sample \citep{Chen2023b}. The histogram in the top panel of Figure \ref{fig:spec:velocities} shows a kernel density estimate of the velocity distribution of the 27 SLSNe from the ZTF-I sample with \ion{Fe}{ii} velocities measured within $\pm20$ rest-frame days from maximum light. After bootstrapping the sample and propagating the measurement uncertainties with a Monte Carlo simulation, the median velocity of the ZTF-I sample is $14,800~\rm km\,s^{-1}$ and its $1\sigma$ confidence region extends from $10,500$ to $19,000~\rm km\,s^{-1}$. \sn\ lies in the bottom 8\% of this sample, but its velocity is not unparalleled. SN\,2019aamu had a lower photospheric velocity at peak, but the measurement is poorly constrained \citep[$5876^{+8110}_{-349}~\rm km\,s^{-1}$; ][]{Chen2023b}. 
In the bottom panel of Figure \ref{fig:spec:velocities}, we show the evolution of the \ion{Fe}{ii} velocities of \sn\ together with those of H-poor SLSNe from \citet{Liu2017a} (in grey). Within 50~days after maximum, the ejecta usually decelerate from $\sim15,000~\rm km\,s^{-1}$ to $\lesssim10,000~\rm km\,s^{-1}$, whereas \sn\ shows no evolution.

\subsubsection{A CSM shell around the progenitor of \sn}\label{sec:csm:shell}

The X-shooter spectra between \tmax+32.7~days and \tmax+94.3~days show two \ion{Mg}{ii} absorption line systems (Figure \ref{fig:spec:csm_shell}). The narrow component is associated with the gas in the SLSN host galaxy (Section \ref{res:host}). The lines of the broader component have a full width at half maximum (FWHM) of $406~{\rm km\,s}^{-1}$ and are blueshifted by $2918~{\rm km\,s}^{-1}$ (not varying between \tmax\ and \tmax+90~days; upper panels in Figure \ref{fig:spec:csm_shell}). They are significantly broader than expected for the interstellar medium in the dwarf host galaxy or any intervening dwarf galaxy\footnote{Based on the correlations between the stellar mass of galaxies and the width of galaxy absorption and emission lines \citep[][]{Kruehler2015a, Arabsalmani2018a}.}  but also significantly narrower than the narrowest SN features ($\sim1900~\rm km\,s^{-1}$; measured from \ion{Fe}{ii}). The equivalent widths are $2.00\pm0.09$ and $1.27\pm0.08$~\AA\ for \ion{Mg}{ii}\,$\lambda$\,2796 and \ion{Mg}{ii}\,$\lambda$\,2803, respectively. The observed line ratio is $1.57\pm0.12$ in tension with the predicted value of 2 for unsaturated lines. Assuming that the \ion{Mg}{ii} lines are unsaturated, we can convert their equivalent widths to a lower limit on the column density of singly ionised magnesium in the CSM shell. The rest-frame equivalent width ${\rm EW}_{\rm r}$ is related to the column density $N$, in units of atoms per $\rm cm^{2}$, via $N=1.13\times10^{20}~{\rm EW}_{\rm r}\,/\,\left(\lambda^2_{\rm r}\,f\right)$ where $\lambda_{\rm r}$ is the rest-frame wavelength, in units of \AA, and $f$ the oscillator strength \citep{Ellison2004a}. Using the oscillator strengths from \cite{Theodosiou1999a} for \ion{Mg}{ii}\,$\lambda$\,2796 and \ion{Mg}{ii}\,$\lambda$\,2803, we derive a lower limit of $N>5\times10^{13}\,\rm cm^{-2}$.

\begin{table}
    \caption{\ion{Fe}{ii} absorption line velocities during the photospheric phase\label{tab:velocities}}
    \centering
    \begin{tabular}{cccc}
    \toprule
    Phase & Velocity & Phase & Velocity\\
    (day)& $\left(\rm km\,s^{-1}\right)$ & (day)& $\left(\rm km\,s^{-1}\right)$\\
    \midrule
    -1.4	&$ 8489\pm 88 $ & 51.5	&$ 8371\pm 126$\\
    8.8	    &$ 8610\pm 28 $ & 60.1	&$ 8382\pm 211$\\
    10.5	&$ 8349\pm 64 $ & 70.3	&$ 8303\pm 198$\\
    30.8	&$ 8453\pm 121$ & 73.8	&$ 8313\pm 198$\\
    31.1	&$ 8426\pm 205$ & 81.5	&$ 8417\pm 200$\\
    32.7	&$ 8637\pm 168$ & 94.3	&$ 8431\pm 201$\\
    47.1	&$ 8433\pm 218$ \\
    \bottomrule
    \end{tabular}
\end{table}

\begin{figure}
    \centering
    \includegraphics[width=1\columnwidth]{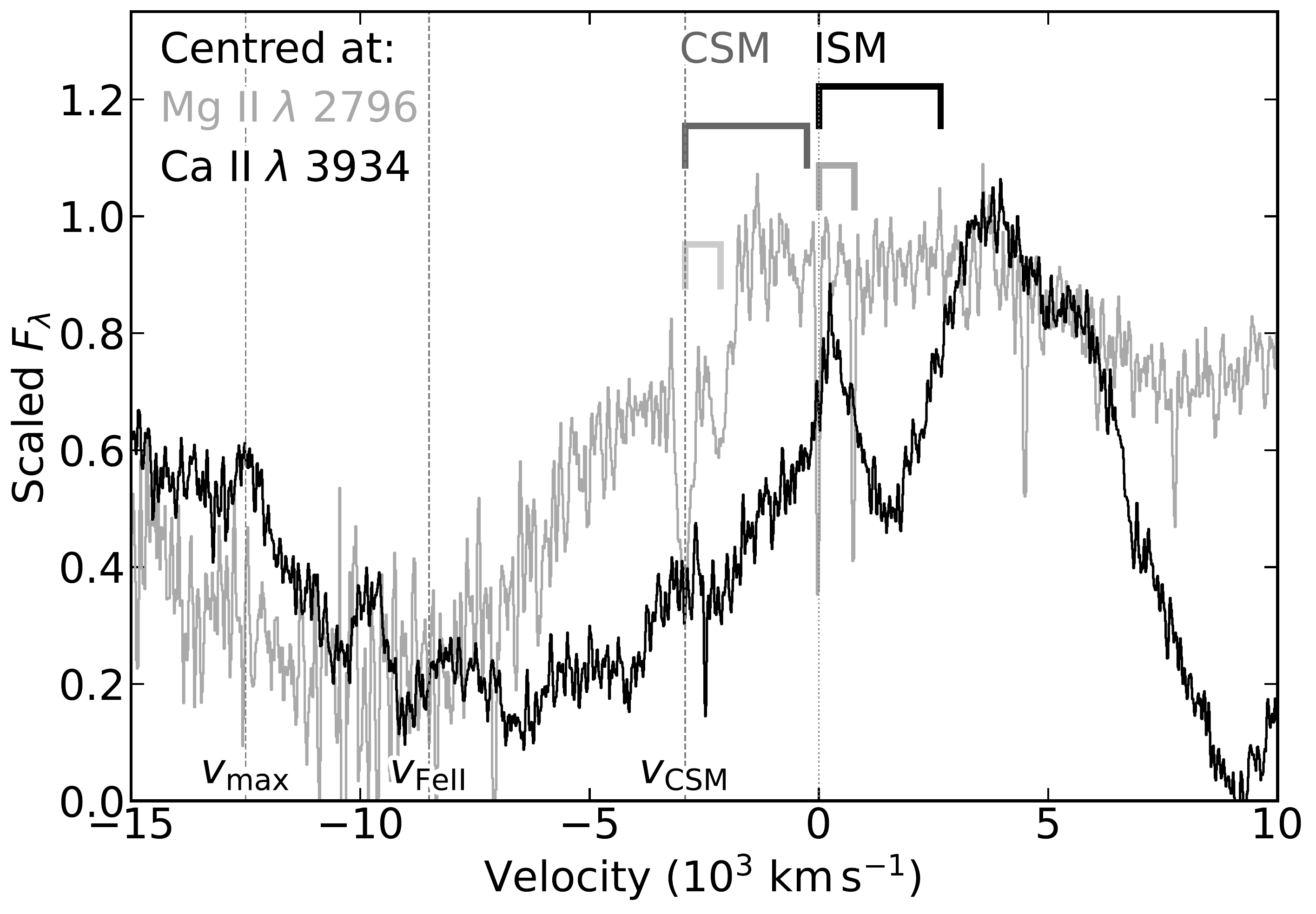}
    \caption{Maximum ejecta velocity. The extent of the \ion{Ca}{ii}\,$\lambda$\,3934 (black) absorption on the blue side can be traced to $\sim 12,500 \kms$ at \tmax+32.7 days. \ion{Mg}{ii}\,$\lambda$\,2796 (dark grey) has a comparable maximum velocity, albeit this region is affected by additional SN features. The location of the blue and red doublet components of \ion{Mg}{ii}\,$\lambda\lambda$\,2796, 2803 and \ion{Ca}{ii}\,$\lambda\lambda$\,3934, 3968 of the host galaxy ISM are indicated by brackets in a darker shade at the top of the figure. We also mark the position of the doublets of the CSM shell with brackets in a lighter shade. The CSM shell is detected through an additional \ion{Mg}{ii} absorption-line system blueshifted by $2918 \kms$. The CSM shell is not detected in \ion{Ca}{ii}.
    }
    \label{fig:spec:mgii_caii}
\end{figure}

The only other SLSN that showed such a blueshifted \ion{Mg}{ii} component was the H-poor SLSN iPTF16eh \citep{Lunnan2018a}. For that SLSN, the \ion{Mg}{ii} doublet was blueshifted by 3300~$\kms$. \citet{Lunnan2018a} also detected \ion{Mg}{ii} in emission between 100 and 300 days after maximum light. Moreover, the line centre of the emission lines moved from -1600 to +2900~$\kms$ during that time interval. These authors attributed the blueshifted \ion{Mg}{ii} absorption line system with a CSM shell expelled decades before the explosion and the time and frequency variable \ion{Mg}{ii} emission lines with a light echo from that shell. How such a light echo evolves depends mainly on its distance to the progenitor star. With that in mind, we analyse the Keck and X-shooter spectra between \tmax+230 and \tmax+378 days to constrain the properties of the CSM shell. Rebinning the spectra reveals \ion{Mg}{ii} in emission (Figures \ref{fig:spec:seq_uvvis} and \ref{fig:spec:line_id}). However, due to heavy rebinning, the information about the variability of the line centre was lost. We can, therefore, not ascertain whether the \ion{Mg}{ii} emission is connected with illuminated magnesium in the CSM shell or produced by the interaction of the SN ejecta with circumstellar material.

Motivated by the discovery of a CSM shell around \sn, we next search for corresponding \ion{Ca}{ii}\,$\lambda\lambda$\,3934,3969 absorption  in the X-shooter spectrum from \tmax+32.7  (Figure \ref{fig:spec:mgii_caii}). The search is aggravated by how the two \ion{Ca}{ii} doublets (CSM shell and SN ejecta) overlap in contrast to the \ion{Mg}{ii} doublets. Using the wavelength of \ion{Ca}{ii}\,$\lambda$\,3934 as the velocity reference, the blue doublet absorption of \ion{Ca}{ii} should be at the same velocity as the \ion{Mg}{ii} doublet ($-2918 \kms$). The red component of the \ion{Ca}{ii} doublet will, however, be displaced by 34.8~\AA, or 2653 $\kms$, to $\sim -265 \kms$. The position of a possible \ion{Ca}{ii} CSM component is marked with the light grey bracket in Figure \ref{fig:spec:mgii_caii}. We do indeed see a sharp drop in the \ion{Ca}{ii} profile at zero velocity, which could be the result of a red CSM absorption. For the blue component, it is more difficult because we do not know the line profile of the \ion{Ca}{ii} absorption from the SN ejecta. Therefore, it is difficult to assess the significance of this. However, we conclude that there is no evidence for \ion{Ca}{ii} absorption from the CSM shell. 

\subsubsection{Circumstellar interaction --- bumps and undulations in the light curve}\label{sec:csm:spectrum}

The multi-band light curves show a series of bumps and wiggles throughout the entire evolution of \sn\ (Figures \ref{fig:lc:multiband} and \ref{fig:lc:bolometric}). Between \tmax\ and \tmax+100 days, the bumps are well visible from $u$ to $H$ band (luminosity increases by a few 0.1~mag). The amplitudes of the bumps in \sn\ are comparable to the bumps seen in light curves of the other SLSNe  \citep[e.g.][]{Nicholl2016a, Inserra2017a, Fiore2021a, Hosseinzadeh2022a, Chen2023b}. Following the nomenclature in \citet{Chen2023b}, these bumps fall in the `weak' category. The bumps in \sn\ also introduce wiggles in the evolution of its blackbody radius and temperature (Figure \ref{fig:lc:bolometric}). These modulations are well within the measurement uncertainties of the long-term trends of these parameters, hindering a more in-depth analysis of these features.

The late-time photometric evolution of \sn\ reveals an increase in luminosity of 0.2~dex between \tmax+240 and \tmax+340~days (Figures \ref{fig:lc:multiband} and \ref{fig:lc:bolometric}) that is well isolated allowing for a more in-depth analysis. The bolometric light curve before and after this bump exhibits a decline rate of 1.18 mag (100 days)$^{-1}$. After subtracting the underlying fading light curve, we conclude that the light curve bump lasted for $\sim80$~days (measured between zero intensity) and reached its highest luminosity at \tmax+300 days. In total, $6.7\pm0.8\times10^{48}$~erg are radiated in excess to the $8.1\times10^{49}$~erg that \sn\ would have emitted without the bump during this time.

\begin{figure}
    \centering
    \includegraphics[width=1\columnwidth]{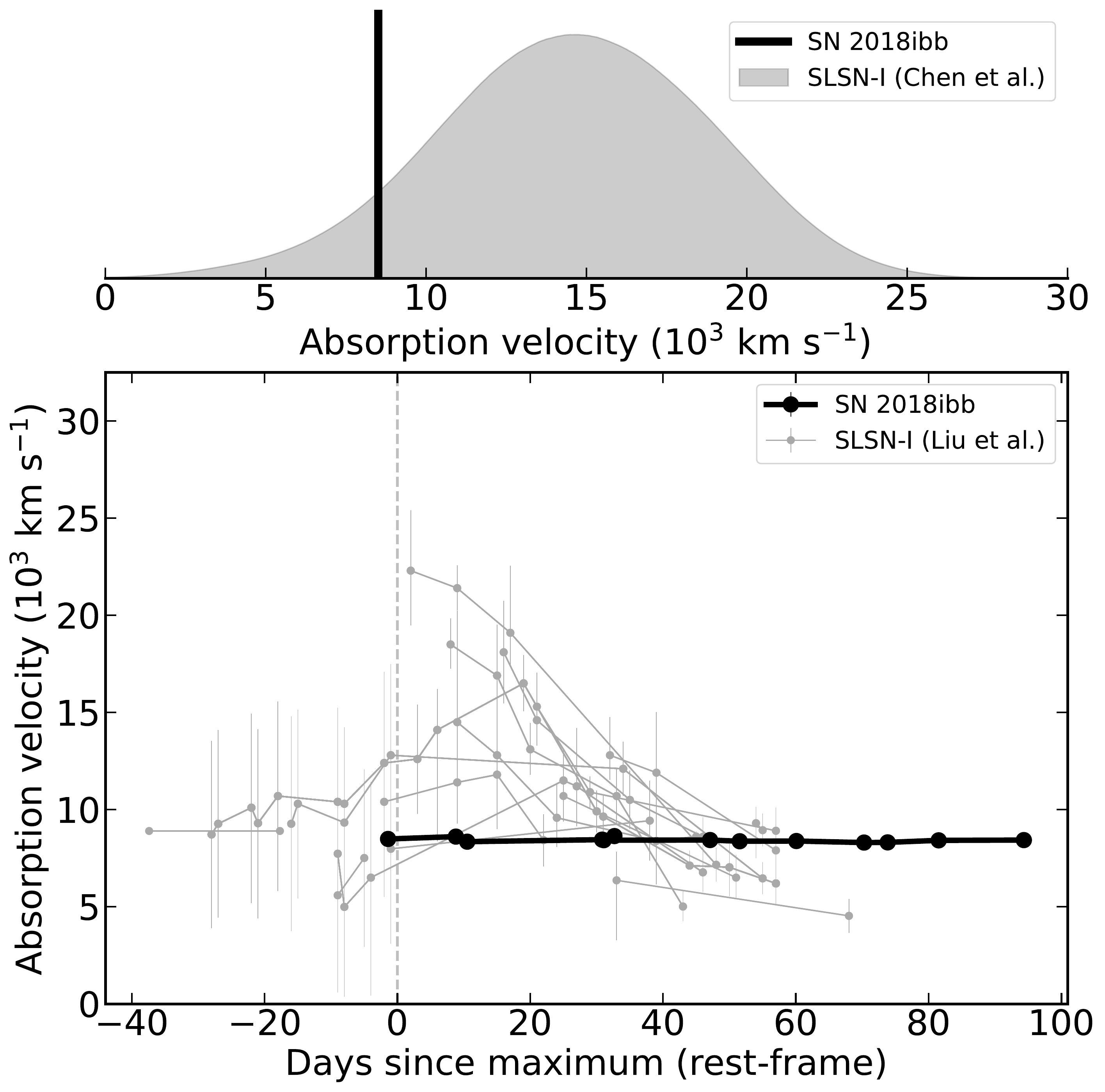}
    \caption{\ion{Fe}{ii} ejecta velocities of \sn\ and general SLSN samples (grey) at the time of maximum (top panel) and as a function of time (bottom panel). \sn\ has a remarkably low velocity at the time of maximum and an unprecedentedly flat velocity evolution, which is in stark contrast to known SLSNe.
    }
    \label{fig:spec:velocities}
\end{figure}

\begin{figure}
\centering
\includegraphics[width=1\columnwidth]{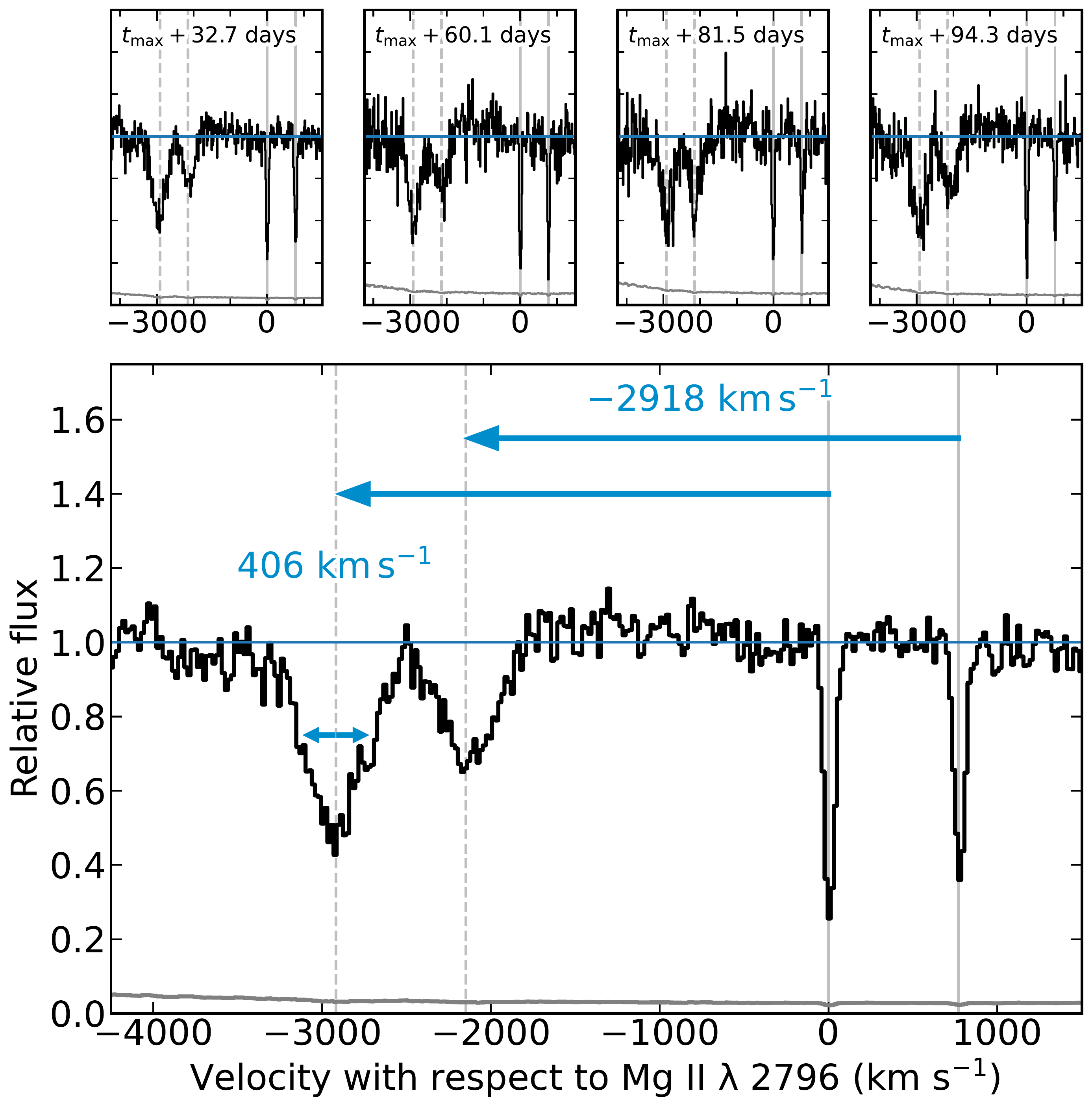}
\caption{Normalised X-shooter spectra from \tmax+32.7~days to \tmax+94.3~days (top panels) and their inverse-variance weighted co-added spectrum (bottom panel). The individual and stacked spectra show barely resolved, narrow absorption lines from the host ISM (marked by the solid vertical lines). In addition, a blue-shifted (2918~$\kms$) absorption line system is visible (marked by the dashed vertical lines). The FWHMs of the blue-shifted component are 406 km\,s$^{-1}$, significantly larger than the ISM lines but significantly smaller than the SN lines. This blue-shifted absorption-line system is connected with a shell of circumstellar material expelled by the progenitor star shortly before the explosion. No significant evolution in the position or shape of the absorption lines can be seen in the individual spectra (upper panels). The error spectrum is shown in grey.}
\label{fig:spec:csm_shell}
\end{figure}

\begin{figure}
    \centering
    \includegraphics[width=1\columnwidth]{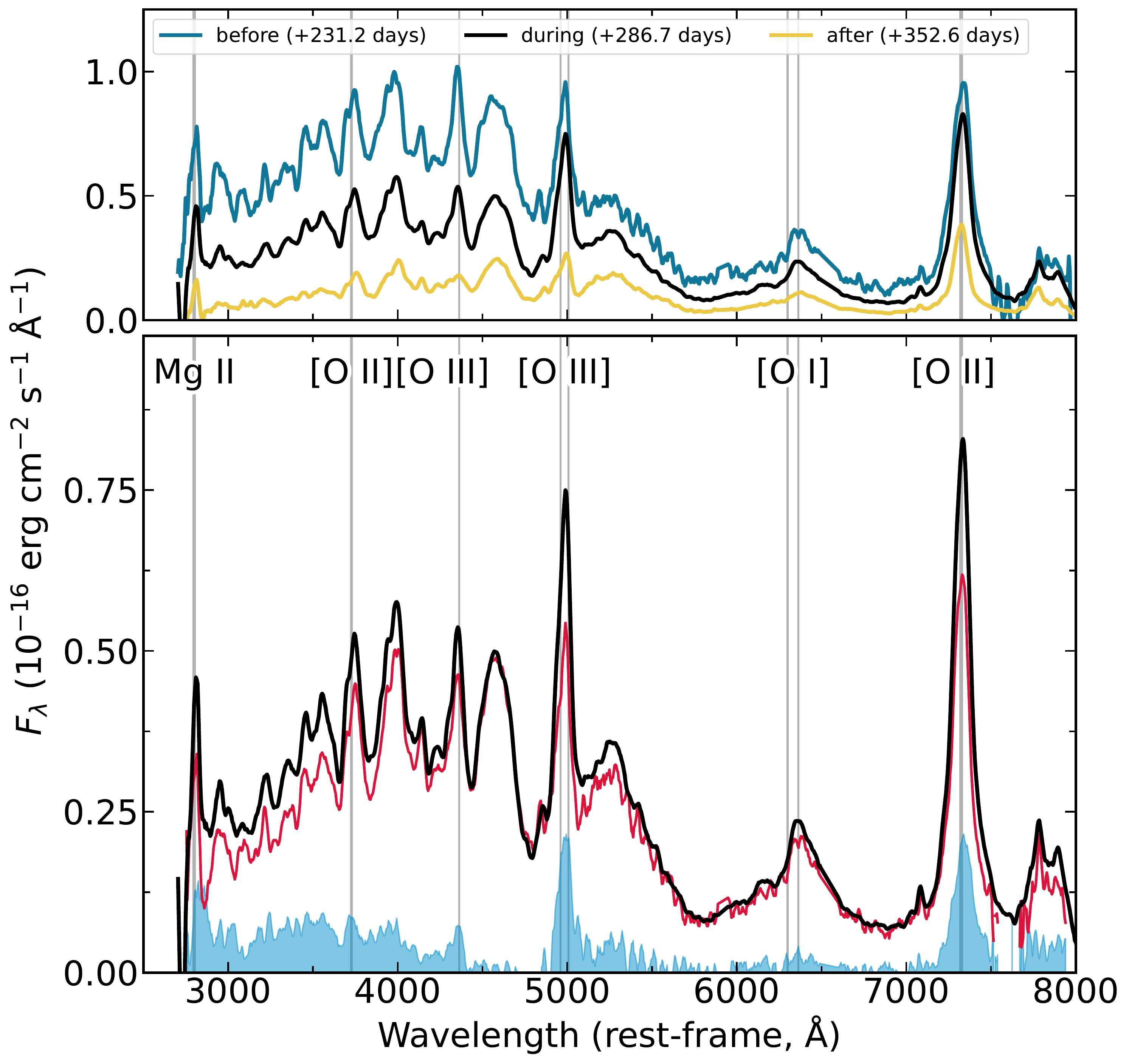}
    \caption{Impact of the light curve bump between \tmax+240 and \tmax+340~days on the SN spectrum at \tmax+286.7~days.
    \textbf{Top}: The spectra before, during and after the light curve bump. 
    \textbf{Bottom}: The observed spectrum at \tmax+286.7~days ($\approx13$~days before the peak of the bump) is shown in black. We estimate the `bump-free' spectrum of \sn\ at \tmax+286.7~days (red) based on the spectra obtained before and after the bump. The difference between the observed (black) and interpolated (red) spectra at \tmax+286.7~days is shown in blue. It reveals a series of emission lines that can be attributed to [\ion{O}{ii}] and [\ion{O}{iii}]. An excess bluewards of 5000~\AA\ is also visible, while no apparent residual can be seen at the location of [\ion{O}{i}].
    }
    \label{fig:spec:energy_injection}
\end{figure}

Figure \ref{fig:spec:energy_injection} presents the spectroscopic evolution of \sn\ during the bump phase. Assuming that all spectral features fade on exponential timescales similar to the multi-band and bolometric light curves, we use the spectra obtained before (blue) and after (yellow) the light curve bump to interpolate the spectrum at \tmax+286.7~days (black). Such an approach estimates the spectroscopic behaviour of \sn\ in the absence of the bump. The bottom panel of Figure \ref{fig:spec:energy_injection} shows the observed spectrum at \tmax+286.7~days in black and the estimated spectrum without the bump in red. The difference spectrum (blue) reveals substantially enhanced line fluxes in [\ion{O}{ii}] and [\ion{O}{iii}] but no change in [\ion{O}{i}]. The light-curve bump might also have increased the flux of the continuum level bluewards of 5000~\AA. Its shape is reminiscent of the blue pseudo-continuum seen in interaction-powered SNe \citep{Silverman2013a, Hosseinzadeh2017a, Gal-Yam2022a, Perley2022a}. Considering the similarity of the difference spectrum to the spectrum before and after the bump raises the question of whether a larger fraction of the emission bluewards of 5000~\AA\ in all nebular spectra is due to CSM interaction. We investigate that further in Section \ref{sec:spectra_discussion}.

\subsection{Radio and X-ray emission}

\begin{figure}
    \centering
    \includegraphics[width=1\columnwidth]{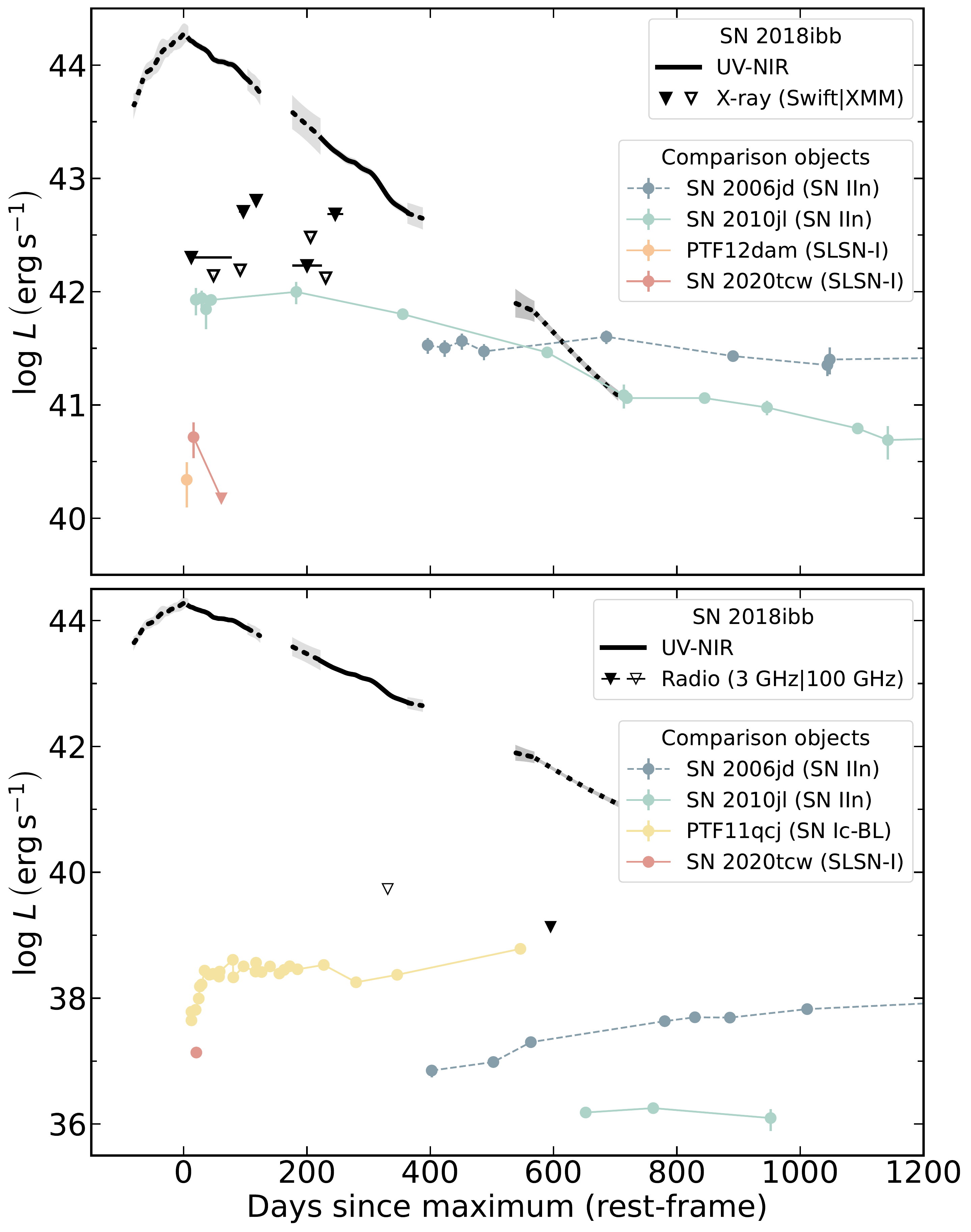}
    \caption{Thermal and non-thermal emission of \sn. Less than a few percent of the total radiated energy is emitted in the radio and X-rays. The luminosity limits lie in the ballpark of non-detections of other SLSNe, and they are a factor of 50 larger than the luminosity of the four H-poor SLSNe with either radio or X-ray detection. The limits of \sn\ are larger than the most luminous radio and X-ray SNe.}
    \label{fig:lc:non-thermal}
\end{figure}

The interaction of the SN ejecta with circumstellar material and heating of the SN ejecta by a central engine (e.g. magnetar or a black hole) can produce thermal X-ray emission and non-thermal radio emission \citep{Chevalier1992a, Chevalier1994a}. \sn\ was observed in the X-rays and radio between \tmax+13 and \tmax+246~days (Sections \ref{sec:obs:xray} and \ref{sec:obs:radio}). All observations led to non-detections with detection limits between 1 and $6\times10^{41}\,\rm erg\,s^{-1}$ in the X-rays and between $10^{39}$ and $10^{40}\,\rm erg\,s^{-1}$ in the radio. To put those measurements in the context of the UV-to-NIR bolometric light curve, we show the radio and X-ray measurements together with the bolometric light curve in Figure \ref{fig:lc:non-thermal}. From that, we conclude that $<2\%$ and $<10\%$ of the total emission are radiated in the radio and X-rays, respectively.

The non-detection limits are in the observed range of other SLSNe with X-ray and radio observations \citep{Levan2013a, Coppejans2018a, Margutti2018a, Law2019a, Eftekhari2021a, Murase2021a}. Only four SLSNe were detected at X-ray or radio frequencies: PTF10hgi \citep[radio;][]{Eftekhari2019a, Law2019a}, PTF12dam \citep[X-ray;][]{Margutti2018a, Eftekhari2021a}, SCP06P6 \citep[X-ray;][]{Levan2013a}, and SN\,2020tcw \citep[radio and X-ray;][]{Coppejans2021a, Matthews2021a}. Their measurements\footnote{PTF10hgi was detected in the radio $>7.5$~years after the SN explosion. Owing to this, we omit to show PTF10hgi in that figure.}, shown in Figure \ref{fig:lc:non-thermal}, are a factor of $>50$ smaller than the detection limits of \sn. 

To put the radio and X-ray properties of \sn\ in the context of interaction-powered SNe, we also show the light curves of the most luminous X-ray and radio SNe in Figure \ref{fig:lc:non-thermal}. The Type IIn SNe 2006jd and 2010jl are the most luminous X-ray SNe with absorption-corrected luminosities of $\sim10^{42}\rm\,erg\,s^{-1}$ \citep[][]{Chandra2012a, Chandra2015a}. The radio-loudest SNe \citep[e.g. SN Ic-BL PTF11qcj][]{Corsi2014a} reached luminosities of $\sim10^{38}\rm\,erg\,s^{-1}$, i.e. $\lesssim10$ times fainter than the limits for \sn. Their observed luminosities before correcting for host absorption can be significantly dimmer for hundreds of days \citep{Chandra2015a}.

In conclusion, the non-detection of \sn\ neither rules out CSM interaction nor a central engine as the dominant powering mechanism. Furthermore, the non-detection of \sn\ also agrees with theoretical models of magnetar- and interaction-powered SLSNe that predict no bright radio and X-ray emission for years after the SN explosion \citep{Murase2016a, Margalit2018a, Omand2018a}.

\subsection{Imaging polarimetry} \label{res:impol}

Our polarimetric observations between \tmax+31.9~days and \tmax+94.4~days revealed a polarisation signal of $0.27\pm0.04~\%$ in $V$ (weighted average of all epochs) and $0.48\pm0.07~\%$ in the $R$ band (Table \ref{tab:impol}). Dust grains in the Milky Way and the host galaxy could introduce a polarisation signal. As detailed in Section \ref{sec:obs:impol}, the polarisation level of the MW could be up to 0.26\%. The level of polarisation from the SN host galaxy is unknown, meaning that all reported measurements are upper limits.

Considering the observed low degree of polarisation and the consistent levels of Stokes parameters measured from \sn\ (Table~\ref{tab:impol}), we conclude that the continuum polarisation intrinsic to \sn\ is $\lesssim0.3\%$ in $V$ band between \tmax+31.9~days and \tmax+94.4~days. To convert this measurement into an asphericity of the ejecta, we assume an oblate ellipsoidal ejecta with a Thomson scattering atmosphere and a number density distribution of $N(r) \propto r^{-n}$, where $r$ is the ejecta radius and $n$ is the power-law index. Adopting $p \lesssim0.3\%$, we infer an axis ratio B/A (minor axis vs. major axis) of $\gtrsim0.9$ for an optical depth of $\tau=1$ and a power-law index of $n=2$, and B/A of $\gtrsim0.8$ for $\tau=5$ and $n=3$--5 \citep{Hoeflich_1991}. The degree of polarisation in the $R$ band is slightly higher ($p\approx0.5\%$). Therefore, we cannot exclude that the continuum polarisation is $p>0.3\%$. A polarisation degree $p \sim0.5\%$ implies an axis ratio B/A of $\sim0.88$ for $\tau=1$ and $n=2$ \citep{Hoeflich_1991}.

Therefore, we suggest that \sn's photosphere exhibits a high degree of spherical symmetry. \citet{Pursiainen2023a} analysed the data of the 16 SLSNe-I with polarimetric observations, including \sn. After correcting the phases of all objects for the diverse photometric decline rates, the properties of \sn\ are well within the observed distribution. While some of the events exhibit a non-zero level of polarisation at similar phases to \sn\ \citep[e.g. SN\,2015bn and SN\,2021fpl;][]{Leloudas2017a, Inserra2016a, Poidevin2022a}, most SLSNe show a consistently low polarisation degree at comparable normalised phases \citep[see figure 6 in ][]{Pursiainen2023a}.

The presence of any component in the atmosphere of \sn\ significantly deviating from spherical symmetry is thus unlikely within the photospheric phases covered by VLT polarimetry observations. Although Thomson scattering is wavelength independent, broad emission lines (see spectra in Figure \ref{fig:spec:seq_uvvis}), which are in general not polarised, may dominate the polarisation spectrum in the $V$ band and produce the apparent low polarisation values. Furthermore, iron-group elements in the ejecta (Figure \ref{fig:spec:line_id}) have a large number of bound-bound transitions in the blue and UV part of the spectrum, which can also depolarise the signal \citep[e.g.][]{Chornock2008a}, accounting for the slightly different polarisation levels measured in $V$ and $R$ bands.

\subsection{Host galaxy} \label{res:host}

\begin{table}
    \caption{Properties of the interstellar medium in the host galaxy}\label{tab:host_ism}
    \centering
    \begin{tabular}{lcc}
    \toprule
    Transition                 & ${\rm EW}_{\rm r}$  & Flux \\
                               & (\AA)                                         & $\left(10^{-18}\,{\rm erg\,cm}^{-2}\,{\rm s}^{-1}\right)$\\
    \midrule
    \multicolumn{3}{c}{\textbf{Absorption lines}}\\
    \midrule
    \ion{Mn}{ii}\,$\lambda$\,2594    & $0.18  \pm 0.13$  & \nodata     \\
                                     & ($<0.39$)         &             \\
    \ion{Fe}{ii}\,$\lambda$\,2600    & $0.07  \pm 0.13$  & \nodata     \\
                                     & ($<0.39$)         &             \\
    \ion{Mn}{ii}\,$\lambda$\,2606    & $-0.12 \pm 0.15$  & \nodata     \\
                                     & ($<0.45$)         &             \\
    \ion{Mg}{ii}\,$\lambda$\,2796    & $0.51  \pm 0.04$  & \nodata     \\
    \ion{Mg}{ii}\,$\lambda$\,2804    & $0.46  \pm 0.04$  & \nodata     \\
    \ion{Mg}{i}\,$\lambda$\,2852     & $0.14  \pm 0.04$  & \nodata     \\
                                     & ($<0.16$)         &             \\
    \ion{Ca}{ii}\,$\lambda$\,3934    & $0.03  \pm 0.01$  & \nodata     \\
    \ion{Ca}{ii}\,$\lambda$\,3969    & $0.01  \pm 0.01$  & \nodata     \\
                                     & ($<0.03$)         &             \\
    \midrule
    \multicolumn{3}{c}{\textbf{Emission lines}}\\
    \midrule
    H$\beta$                            & \nodata       & $3.68  \pm 0.78$ \\
    {[\ion{O}{iii}]}\,$\lambda$\,4363   & \nodata       & $0.25  \pm 0.16$ \\
                                        & \nodata       & ($<0.48$)        \\
    {[\ion{O}{iii}]}\,$\lambda$\,4959   & \nodata       & $1.96  \pm 0.81$ \\
                                        & \nodata       & ($<2.43$)        \\
    {[\ion{O}{iii}]}\,$\lambda$\,5007   & \nodata       & $12.96 \pm 1.11$ \\
    H$\alpha$                           & \nodata       & $10.43 \pm 0.76$ \\
    {[N\textsc{ii}]}\,$\lambda$\,6584   & \nodata       & $0.12  \pm 0.54$ \\
                                        & \nodata       & ($<1.62$)        \\
    \bottomrule
    \end{tabular}
    \tablefoot{We report rest-frame equivalent widths EW$_r$ for absorption lines and fluxes for emission lines. For lines detected with a significance of $<3\sigma$, we also report the $<3\sigma$ upper limits.
    The rest-frame equivalent widths are measured by averaging the measurements from the X-shooter spectra between \tmax+32.7~days and \tmax+94.3~days of each line species. The emission lines are measured from the FORS2 spectrum at \tmax+637~days. The emission-line fluxes are not corrected for reddening.
    }
\end{table}

\begin{figure}
    \centering
    \includegraphics[width=1\columnwidth]{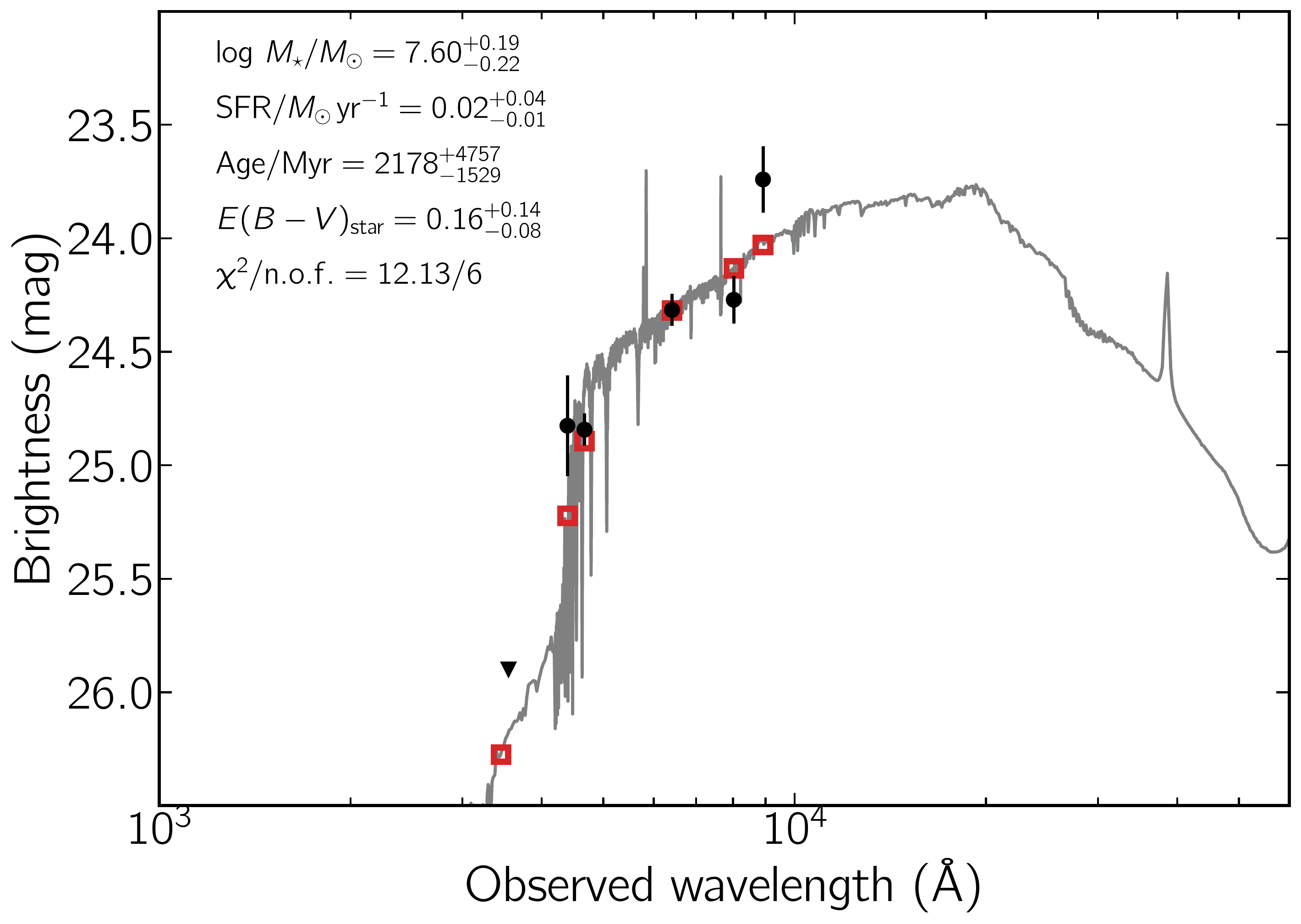}
    \caption{Spectral energy distribution of the host galaxy from 1000 to 60,000~\AA\ (black dots). The solid line displays the best-fitting model of the SED. The red squares represent the model-predicted magnitudes. The fitting parameters are shown in the upper-left corner. The abbreviation `n.o.f.' stands for 
    the number of filters.}
    \label{fig:host:sed}
\end{figure}

\sn's host galaxy was detected in several optical broad-band filters ($m_R \sim 24.4$~mag; Table \ref{tab:host_phot}). A false colour image of the field is shown in Figure \ref{fig:rgb}. The SN explosion site, marked by the crosshair, is $\approx1$~kpc from the centre of its host galaxy, a common offset for SLSNe \citep{Lunnan2014a, Schulze2018a, Schulze2021a}. To infer the mass and star-formation rate of the host, we model the observed spectral energy distribution (black data points in Figure \ref{fig:host:sed}) with the software package \program{Prospector} version 1.1 \citep{Johnson2021a}.\footnote{\program{Prospector} uses the \program{Flexible Stellar Population Synthesis} (\program{FSPS}) code \citep{Conroy2009a} to generate the underlying physical model and \program{python-fsps} \citep{ForemanMackey2014a} to interface with \program{FSPS} in \program{python}. The \program{FSPS} code also accounts for the contribution from the diffuse gas based on the \program{Cloudy} models from \citet{Byler2017a}. We use the dynamic nested sampling package \program{dynesty} \citep{Speagle2020a} to sample the posterior probability.} We assume a Chabrier IMF \citep{Chabrier2003a} and approximate the star formation history (SFH) by a linearly increasing SFH at early times followed by an exponential decline at late times [functional form $t \times \exp\left(-t/t_{1/e}\right)$, where $t$ is the age of the SFH episode and $t_{1/e}$ is the $e$-folding timescale]. The model is attenuated with the \citet{Calzetti2000a} model. The priors of the model parameters are set identical to those used by \citet{Schulze2021a}. The observed SED is adequately described by a galaxy model with a stellar mass of $\log\,M_\star/M_\odot=7.60^{+0.19}_{-0.22}$ and star-formation rate of $0.02^{+0.04}_{-0.01}~M_\odot\,{\rm yr}^{-1}$ (grey curve in Figure \ref{fig:host:sed}).

\begin{figure}
    \centering
    \includegraphics[width=1\columnwidth]{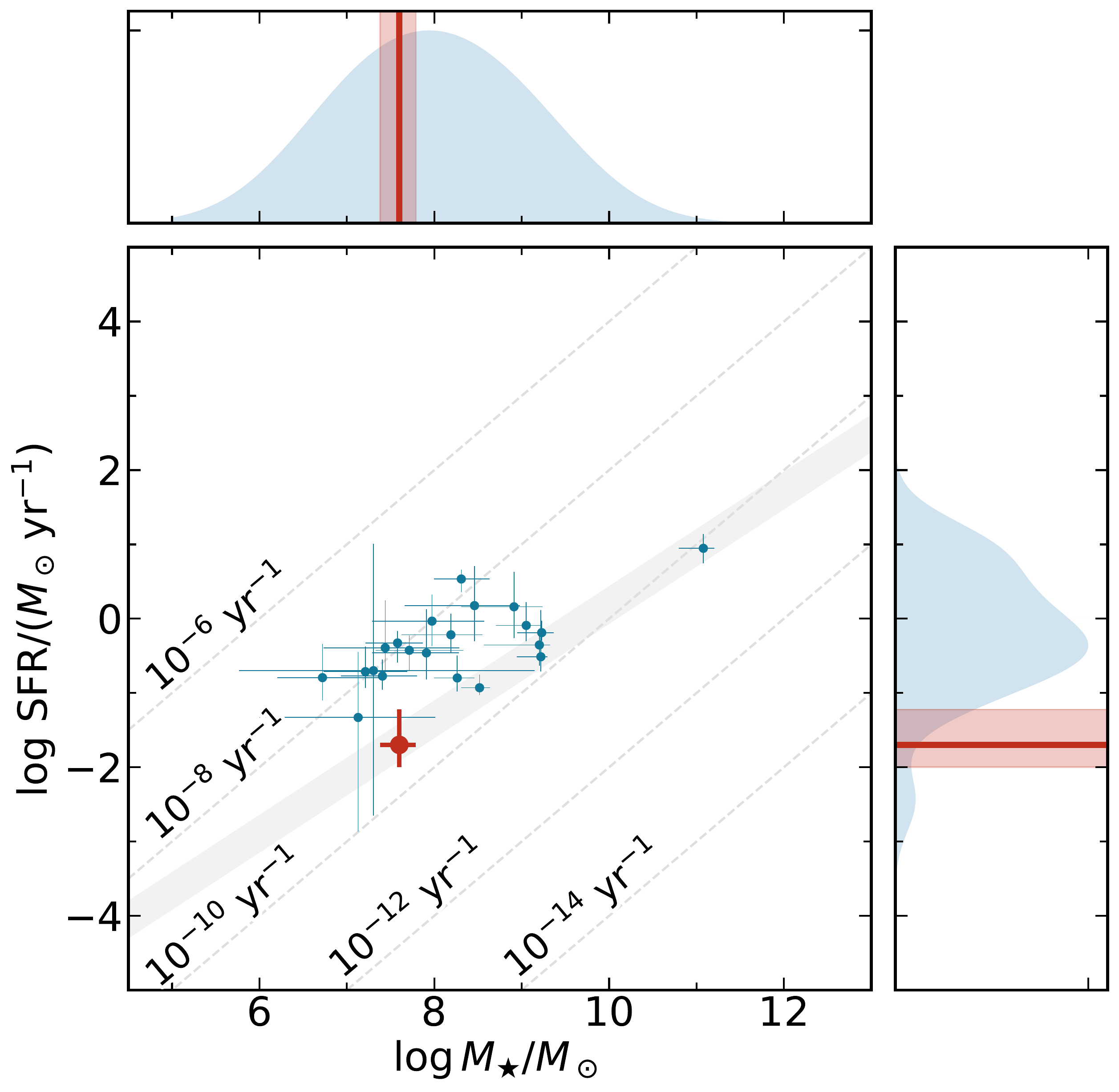}
    \caption{Star-formation rate and stellar mass of the host galaxy of \sn\ in the context of SLSN-I host galaxies from the PTF survey \citep{Schulze2021a}. The host galaxy of \sn\ lies in the expected parameter space of SLSN host galaxies but in the lower half of the mass and SFR distributions (kernel density estimates of the observed distributions are shown at the top and to the right of the figure). Its specific star-formation rate (SFR / mass) is comparable to the typical star-forming galaxies (grey band) but lower than for an average SLSN host galaxy.}
    \label{fig:host:sfr_mass}
\end{figure}

The mass and the star-formation rate of the host of \sn\ agree with the expected values of SLSNe-I host galaxies  at $z<0.3$ \citep{Leloudas2015a,Perley2016a,Chen2017a,Schulze2018a,Schulze2021a}, although both fall in the lower half of the distributions. The specific star-formation rate (SFR normalised by the stellar mass of the host) is comparable to a common star-forming galaxy of that stellar mass  \citep[grey band in Figure \ref{fig:host:sfr_mass};][]{Elbaz2007a} but in the lower half of the observed distribution of SLSN host galaxies \citep{Schulze2021a}. We caution that specific SFRs are notoriously difficult to measure (e.g. see figure 3 in \citealt{Schulze2021a}) as they rely on well-sampled SEDs from the UV to the NIR.

The X-shooter spectra up until $T_{\rm max} + 80$~days reveal narrow absorption lines from \ion{Mg}{i} and \ion{Mg}{ii} from the interstellar medium in the host galaxy but no absorption features from \ion{Ca}{ii}, \ion{Fe}{ii}, and \ion{Mn}{ii}, which have prominent features in the wavelength range accessible with X-shooter and are typically seen in low-mass star-forming galaxies, e.g. \citet{Prochaska2007a} and \citet{Fynbo2009a}. The equivalent widths of the detected lines and the upper limits of the strongest expected absorption features are reported in Table \ref{tab:host_ism}. The measurements of \ion{Mg}{i}\,$\lambda$\,2852 and \ion{Mg}{ii}\,$\lambda\lambda$\,2796,\,2804 are comparable to those of the SLSN host galaxies reported in \citet{Vreeswijk2014a}. Following the methodology of \citet{deUgartePostigo2012a}, we infer an absorption-line strength parameter of $\sim-3.5$ from \ion{Ca}{ii}, \ion{Mg}{i} and \ion{Mg}{ii}, putting the host of \sn\ at the low-metallicity end of the distribution (albeit the diagnostic is tailored to host galaxies of long-duration gamma-ray bursts, which are also connected with the death of very massive stars but which prefer galaxies with slightly higher metallicities and slightly older stellar populations than SLSNe-I; \citealt{Hjorth2012a, Leloudas2015a, Vergani2015a, Perley2016b, Schulze2018a}).

\sn's nebular spectra exhibit emission lines from hydrogen and oxygen from \ion{H}{ii} regions in the host galaxy. We measure their intensities by integrating over their line profiles. To apply emission-line diagnostics for measuring the oxygen abundance, we also need the flux of [\ion{N}{ii}]\,$\lambda$\,6584, which evaded detection. Using the H$\alpha$ line profile as a template of the [\ion{N}{ii}]\,$\lambda$\,6584 line profile, we measure the nominal flux and its uncertainty. Table \ref{tab:host_ism} summarises all measurements. Using the O3N2 metallicity indicator with the calibration from \citet{Marino2013a} yields a low oxygen abundance of $12 + \log\,\left({\rm O/H}\right) = 8.06^{+0.07}_{-0.11}$ in accordance with the low value from the absorption-line strength parameter. The oxygen abundance is comparable to the mean of SLSN host galaxies at similar redshifts \citep{Leloudas2015a, Perley2016a, Chen2017a}. The non-detection of [\ion{N}{ii}]\,$\lambda$\,6584 and [\ion{O}{iii}]\,$\lambda$\,4363 adds a systematic uncertainty to the inferred metallicity. However, the average metallicity of a galaxy is correlated with its mass. \citet{Andrews2013a} reported the mass-metallicity relation for star-forming galaxies between $z=0.027$ and 0.25, a stellar mass of $\log~M_\star/M_\odot=7.4$--10.5, and metallicities measured with the $T_e$ method. Using this mass-metallicity relation yields an oxygen abundance of $12+\log\,\rm O/H \approx 8.07^{+0.10}_{-0.12}$, identical to the value inferred from the galaxy emission lines. Assuming a solar oxygen abundance of 8.67 \citep{Asplund2009a}, the host galaxy metallicity is $0.25^{+0.07}_{-0.06}$ solar.

The flux ratio between H$\alpha$ and H$\beta$ is $2.76 \pm 0.62$, which is consistent within $1\sigma$ with the theoretically expected value of 2.86 for no extinction (assuming a temperature of $10^4$~K and an electron density of $10^2~{\rm cm}^{-3}$ for Case B recombination; \citealt{Osterbrock1989a}). We conclude that the host attenuation is negligible. The H$\alpha$ flux translates to a star-formation rate of ${\rm SFR} = 4.4 \pm 0.3 \times 10^{-3}~M_\odot\,{\rm yr}^{-1}$ using \citet{Kennicutt1998a} and the relation from \citet{Madau2014a} to convert from the Salpeter to the Chabrier IMF in the \citet{Kennicutt1998a} relation. This value is lower than the SFR estimated from the host SED fitting but consistent within $2\sigma$.

\section{Discussion}\label{sec:discussion}

\subsection{SN ejecta emission vs. CSM interaction\label{sec:csm:oiii_oii}}

In Section \ref{sec:csm:shell}, we have shown that the progenitor of \sn\ is embedded in circumstellar material ejected shortly before the explosion. In this section, we examine the line profiles and evolution of selected oxygen and metal lines to infer the physical conditions of the SN ejecta and the CSM.

\begin{figure}
\includegraphics[width=1\columnwidth]{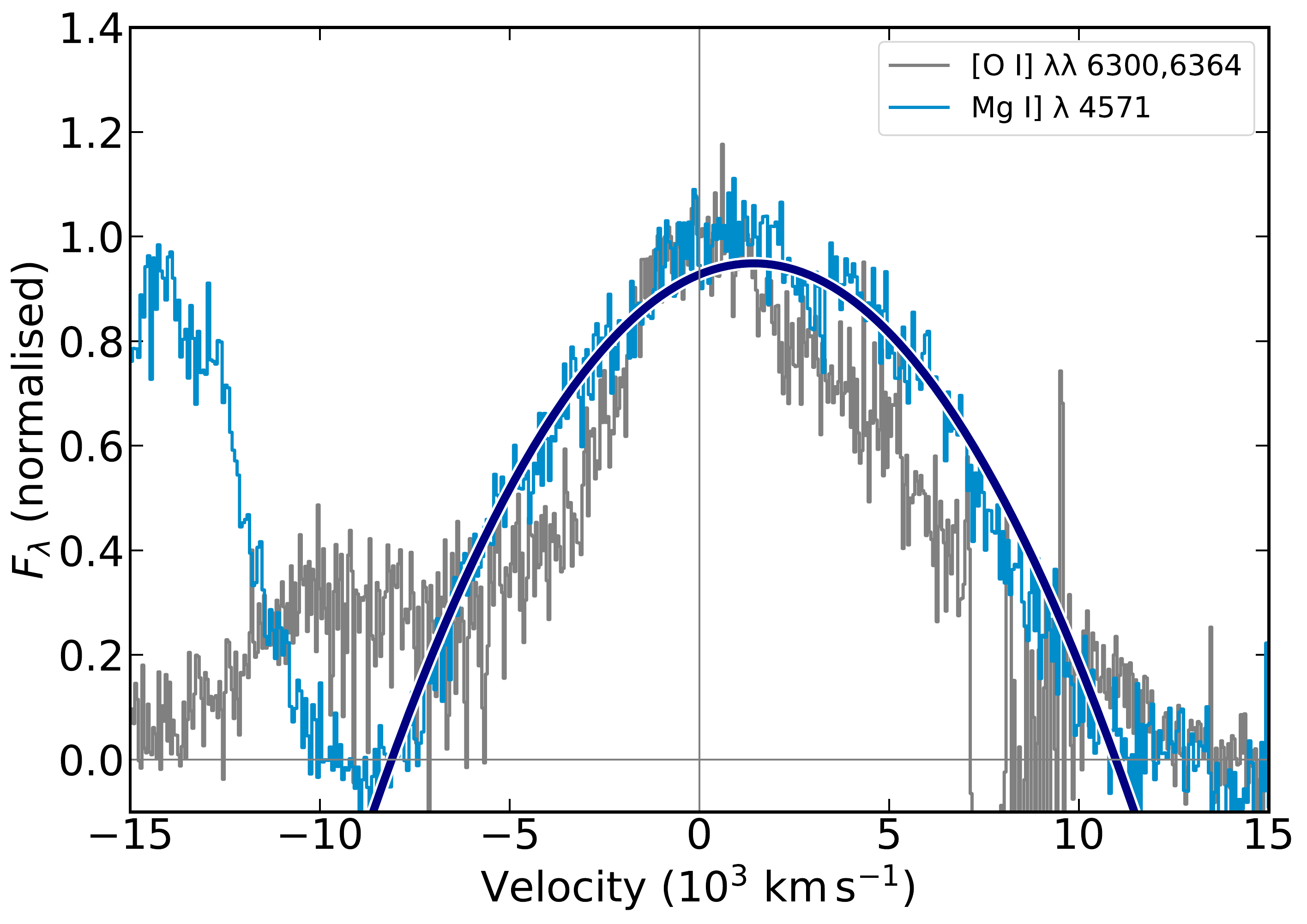}
\caption{[\ion{O}{i}]\,$\lambda\lambda$\,6300,6364  and \ion{Mg}{i}]\,$\lambda$\,4571 lines at \tmax+286.7~days. The \ion{Mg}{i}] line is well fitted with a parabolic shape (dark blue), expected from an optically thick expanding shell (e.g. swept up CSM and unshocked SN ejecta), while the [\ion{O}{i}] lines show a strong blue deficit because the line-forming region is still optically thick. The [\ion{O}{i}] doublet is centred on the 6364~\AA\ doublet component.}
\label{fig:oi_mgi}
\end{figure}

The line profiles are most clear in the nebular phase. Figure \ref{fig:oi_mgi} shows the continuum-subtracted \ion{Mg}{i}]\,$\lambda$\,4571 line with the [\ion{O}{i}] \,$\lambda\lambda$\, 6300, 6364 doublet at \tmax+286.7~days. Both lines extend to $\sim 10,000 \kms$. Their maximum velocity hardly changes up to the last well-observed epoch at \tmax+637.3~days. Its similarity to the maximum velocity of the \ion{Ca}{ii}\,$\lambda$\,3934 absorption line (Figure \ref{fig:spec:mgii_caii}) suggests \ion{Mg}{i}] and [\ion{O}{i}] are produced in the high-velocity ejecta. The \ion{Mg}{i}] line is well fitted with a parabolic line profile with similar maximum velocity, shown by the dark-blue line in Figure \ref{fig:oi_mgi}. This indicates emission from an optically thick shell with constant velocity \cite[e.g.][]{Fransson1984a}.

\begin{figure}
\includegraphics[width=1\columnwidth]{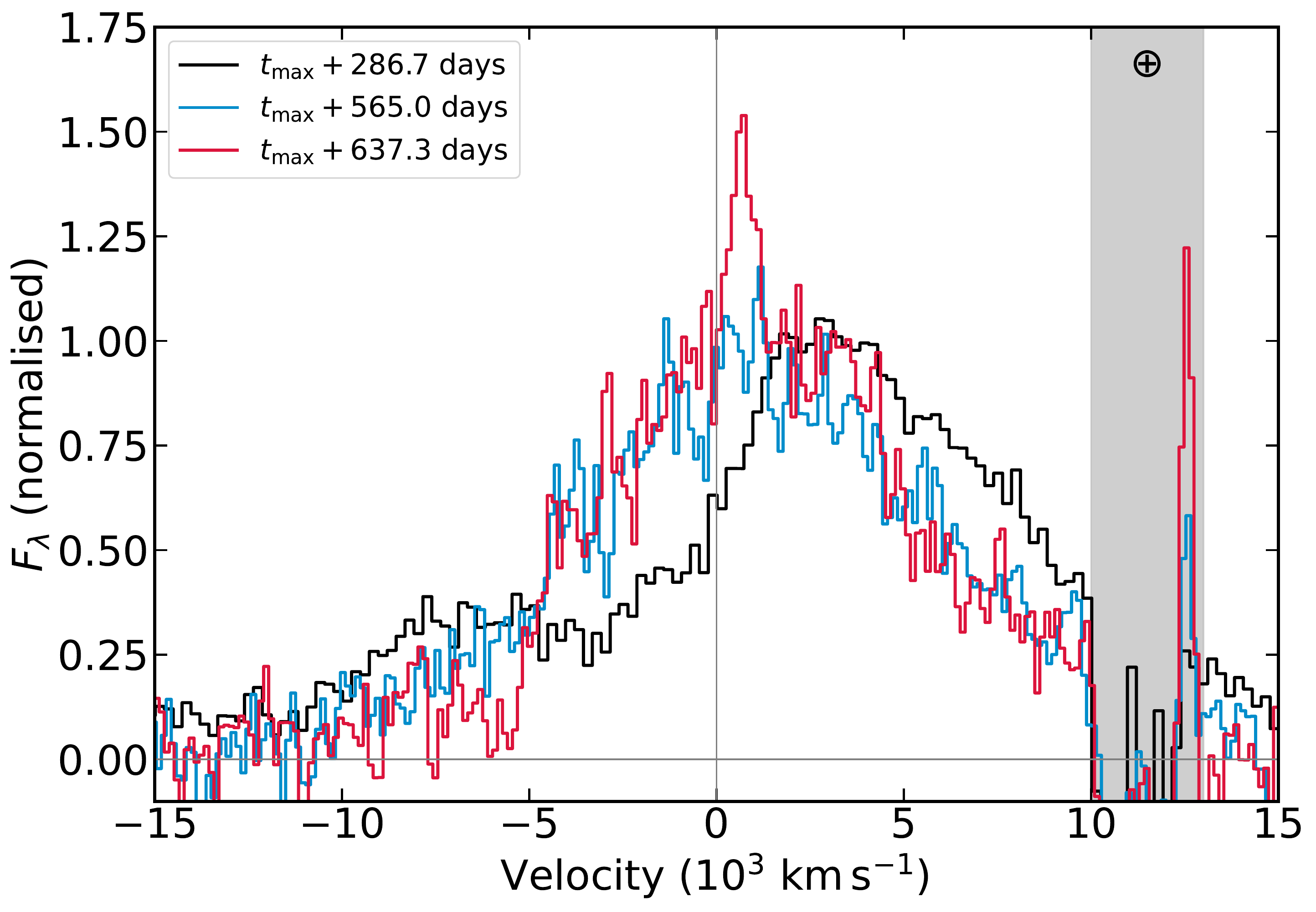}
\caption{Late-time evolution of the [\ion{O}{i}] \,$\lambda\lambda$\,6300,6364 doublet. Note the strong evolution on the blue side, while the red side of the lines is evolving slower. This indicates a transition from optically thick to optically thin [\ion{O}{i}] lines, implying that the scattering in the absorption part of the P-Cygni profile is decreasing. Regions of strong atmospheric absorption are grey-shaded.}
\label{fig:oi_evol}
\end{figure}

A similar parabolic line profile is consistent with the red side of the [\ion{O}{i}]\,$\lambda$\,6364 doublet component. However, the blue side of the doublet, dominated by the 6300~\AA\ component, lacks most of the emission compared to the \ion{Mg}{i}] line. By \tmax+637.3~days (Figure \ref{fig:oi_evol}), the blue doublet component has grown and is now the stronger of the two lines. The evolution of the [\ion{O}{i}] line profile may be explained if both doublet components are optically thick to at least \tmax+286.7~days.  In that case, the blue component will be scattered by the red component, which extends over most of the blue component (the velocity difference between the two components is $3016~\kms$). These photons will either be thermalised or emerge on the red side of the 6364~\AA\ doublet component. Emission from the front side of the ejecta with velocities $\lesssim -(v-3016 \kms$) are only partially scattered, and some of this emission may leak out, explaining the `bump' at $\sim -7500~\kms$. At \tmax+565.0~days, the blue doublet component has grown, and the blue wing is equally bright, or somewhat brighter, compared to the 6364~\AA\ component. This trend continues at \tmax+637.3~days. The expected 3:1 ratio is still not reached, indicating that the ejecta is not optically thin, yet.

Using the \citet{Sobolev1957a} theory for the line formation, we can estimate the optical depth $\tau$ for a given \ion{O}{i} density $n(\ion{O}{i})$ in the ejecta \citep[e.g.][]{Li1992a}. Assuming LTE among the ${}^3$P ground state levels, the optical depth of each line of the [\ion{O}{i}] doublet is given by
\begin{equation}
\tau = \frac{A\left({}^1{\rm D_2},{}^3{\rm P}_J\right)\,\lambda\left({}^1{\rm D_2},{}^3{\rm P}_J ^3\right)^3\,g\left({}^2{\rm D}_2 \right) }{8 \pi\,g_{\rm tot} }\,\ n({\rm \ion{O}{i}})\times t \nonumber
\end{equation}
where $A({}^1{\rm D_2},{}^3{\rm P}_J)$ with $J=2,1$ are the transition probabilities for the 6300~\AA\ and 6364~\AA\ lines, respectively, $\lambda\left({}^1{\rm D_2},{}^3{\rm P}_J ^3\right)$ is the wavelength of blue and red doublet component, respectively, $g_{\rm tot}=9$ is the total statistical weight to the ground multiplet, and $t$ is the time since the explosion, in units of day. Putting in the atomic constants, we get an optical depth for the 6364~\AA\ line of
\begin{equation}
\tau = 2.7 \   
\left( \frac{ n({\rm \ion{O}{i}})} {10^{10} \ {\rm cm}^{-3} } \right) 
\left(\frac{t}{300 \rm \ day}\right) \nonumber
\end{equation}
and a depth that is a factor of 2.9 larger for the 6300~\AA\ line. The typical \ion{O}{i} density needed to get an optically thick line is, therefore, $\ga 10^9$ cm$^{-3}$. This can be compared to the mean oxygen density of the core.  Assuming \ion{O}{i} is the dominant species of oxygen in the core, the number density is 
\begin{align}
n({\rm O}) &\approx 3 \times 10^7 \, f^{-1} \,\left(\frac{M({\rm O})}{30\,M_\odot}\right)\,\left(\frac{v_{\rm ej}}{10^4 \kms}\right)^{-3} 
 \,\left(\frac{t}{300 \ \rm day}\right)^{-3} \rm cm^{-3}
\label{eq:number_dens}
\end{align}
where $M(\rm O)$ is the mass of oxygen in the core, $f$ is the filling factor, and $v_{\rm ej}$ the ejecta velocity. To get an optically thick 6364~\AA\ line at \tmax+286.7~days, in other words a density $\ga 10^{10}$ cm$^{-3}$, requires a very small oxygen filling factor, $\lesssim 10^{-3}$ (i.e. a highly clumped medium), or an unphysically large oxygen mass of $\ga 10^3 M_\odot$.

In a CSM and PPISN interaction scenario, the continued high optical depth at late times could point to a highly compressed cool dense shell (CDS). A CDS will form from the compression behind the shock, which results from the interaction between the ejecta and the CSM \cite[for a discussion see, e.g.][]{Chevalier2017a}. If the density of the CSM is large, the forward shock will be radiative and dominate the emission, which will then have the composition of the CSM. In the opposite case, the reverse shock dominates with a composition typical of the outer ejecta. The latter case is more relevant for lower mass loss rates. In both cases, the density enhancement behind the cooling shock will be very large. Assuming an approximate pressure balance behind the shock, the density enhancement will be of the order of $T_{\rm shock}/T_{\rm ps} \approx 3/16\,\mu\,m_{\rm u}\,v_{\rm rel}^2/(k\,T_{\rm ps})$, where $\mu$ is the mean molecular weight ($\sim 1.7$ for a fully ionised oxygen gas), $m_{\rm u}$ the atomic mass unit, $v_{\rm rel}$ the relative velocity between the CSM shell and the ejecta, $T_{\rm shock}$ the temperature immediately behind the shock, and $T_{\rm ps}$  the post-shock temperature in the CDS ($\sim 10^4$ K). With $v_{\rm rel} \approx 5000 \kms$\footnote{The photospheric velocity of the ejecta is $<8500~\kms$ at late times (Section \ref{sec:velocities}), and the velocity of the CSM shell is $2918~\kms$ (Section \ref{sec:csm:shell}).}, the compression is of the order of $10^5$. Both $v_{\rm rel}$ and $T_{\rm ps}$ are uncertain and magnetic pressure could limit the compression. The CDS is also most likely unstable \citep{Chevalier1995}, leading to clumping of the shell and limiting of the compression. However, the estimate shows that a very large density could result in the CDS, making the line optically thick, equivalent to a low filling factor. In the PISN scenario,  strong clumping in the ejecta is needed. This is, however, not indicated from simulations of PISN models without CSM by \cite{Chen2020a}.

\begin{figure}
    \centering
    \includegraphics[width=1\columnwidth]{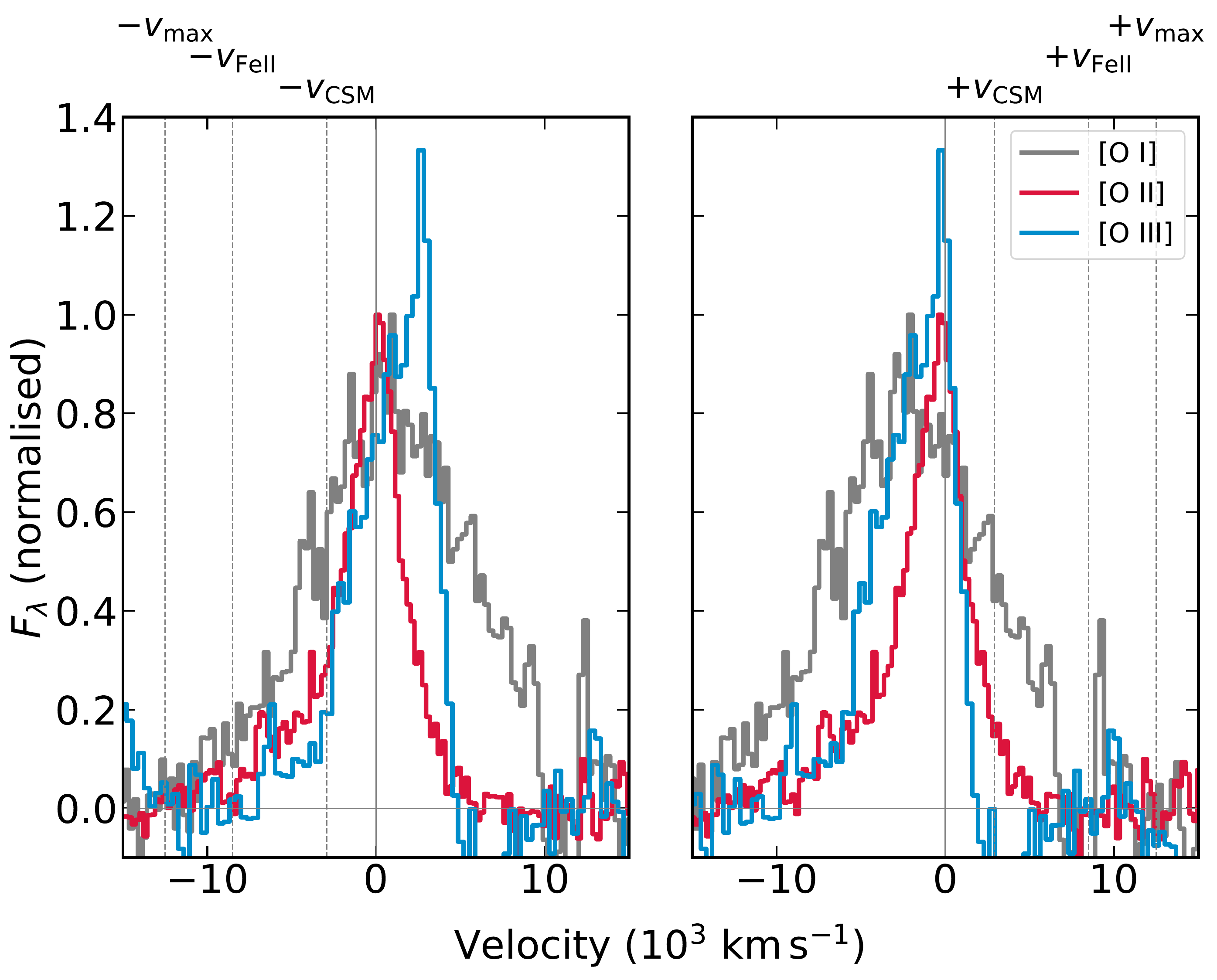}
    \caption{[\ion{O}{i}]\,$\lambda\lambda$\,6300,6364, [\ion{O}{ii}]\,$\lambda\lambda$\,7320,7330, and [\ion{O}{iii}]\,$\lambda\lambda$\,4959,5007 line profiles at \tmax+565.0 days. The velocity scale is centred on the blue doublet component in the left panel and on the red component in the right panel. [\ion{O}{ii}] and [\ion{O}{iii}] only reach out to approximately the velocity of the CSM shell ($v_{\rm CSM}$), much less than the photospheric velocity ($v_{\rm FeII}$) and the maximum velocity of the ejecta ($v_{\rm max}$). In contrast to that, the [\ion{O}{i}] profile extends to $\approx 12,500 \kms$. This points to [\ion{O}{ii}] and [\ion{O}{iii}] being produced close to the CSM shell whereas [\ion{O}{i}] is produced in the SN ejecta.
    }
    \label{fig:oi_oii_oiii_profiles}
\end{figure}

We now turn to the origin of the higher ionisation  [\ion{O}{ii}] and  [\ion{O}{iii}]  lines. Figure \ref{fig:oi_oii_oiii_profiles} shows the line profiles of the [\ion{O}{i}], [\ion{O}{ii}] and [\ion{O}{iii}] lines after subtracting the continuum. Owing to the doublet nature of the lines, we centre the line profiles on the blue component in the left panel and on the red component in the right panel. These line widths can be compared to the velocity of the CSM shell, the photospheric velocity, and the maximum velocity of the ejecta (vertical lines in Figure \ref{fig:oi_oii_oiii_profiles}). It is clear that the [\ion{O}{ii}] and [\ion{O}{iii}] line widths are closer to the velocity of the CSM shell than to the photospheric velocity of the SN ejecta; in contrast to the [\ion{O}{i}] line, which extends to the maximum velocity of the SN ejecta.

\begin{figure}[t]
\includegraphics[width=1\columnwidth]{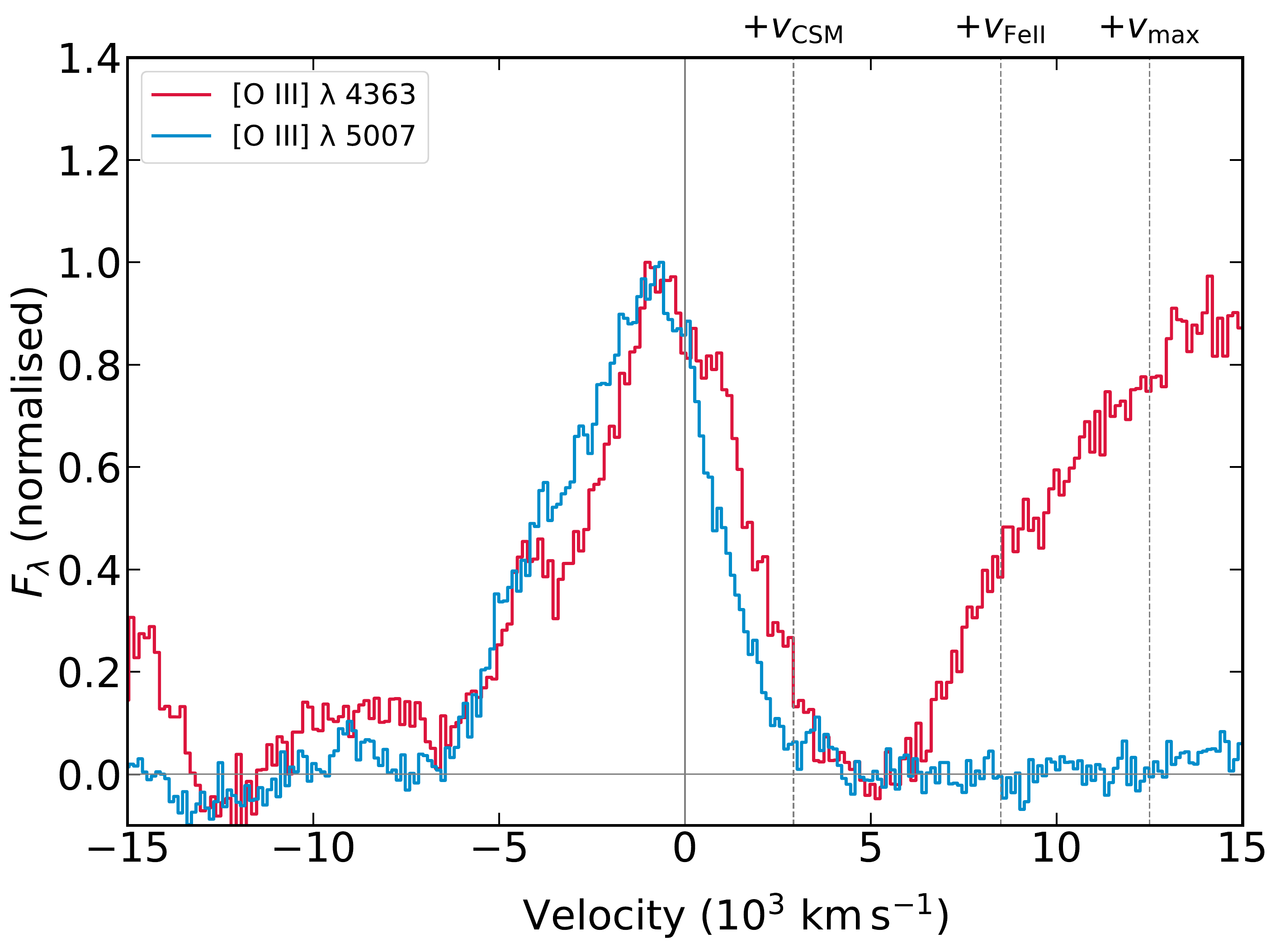}
\caption{Comparison of the [\ion{O}{iii}]\,$\lambda\lambda$\,4959,5007 lines with the [\ion{O}{iii}]$\lambda$ 4363 line. These lines are produced by the interaction of the SN ejecta with circumstellar material. Due to the occulation of the CSM by an optically thick SN ejecta, less of the `backside' of the CSM is seen. The velocities of the CSM shell ($v_{\rm CSM}$), the photospheric velocity ($v_{\rm FeII}$) and the maximum velocity of the ejecta ($v_{\rm max}$) are indicated.}
\label{fig:oiii}
\end{figure}

\begin{figure}
\includegraphics[width=1\columnwidth]{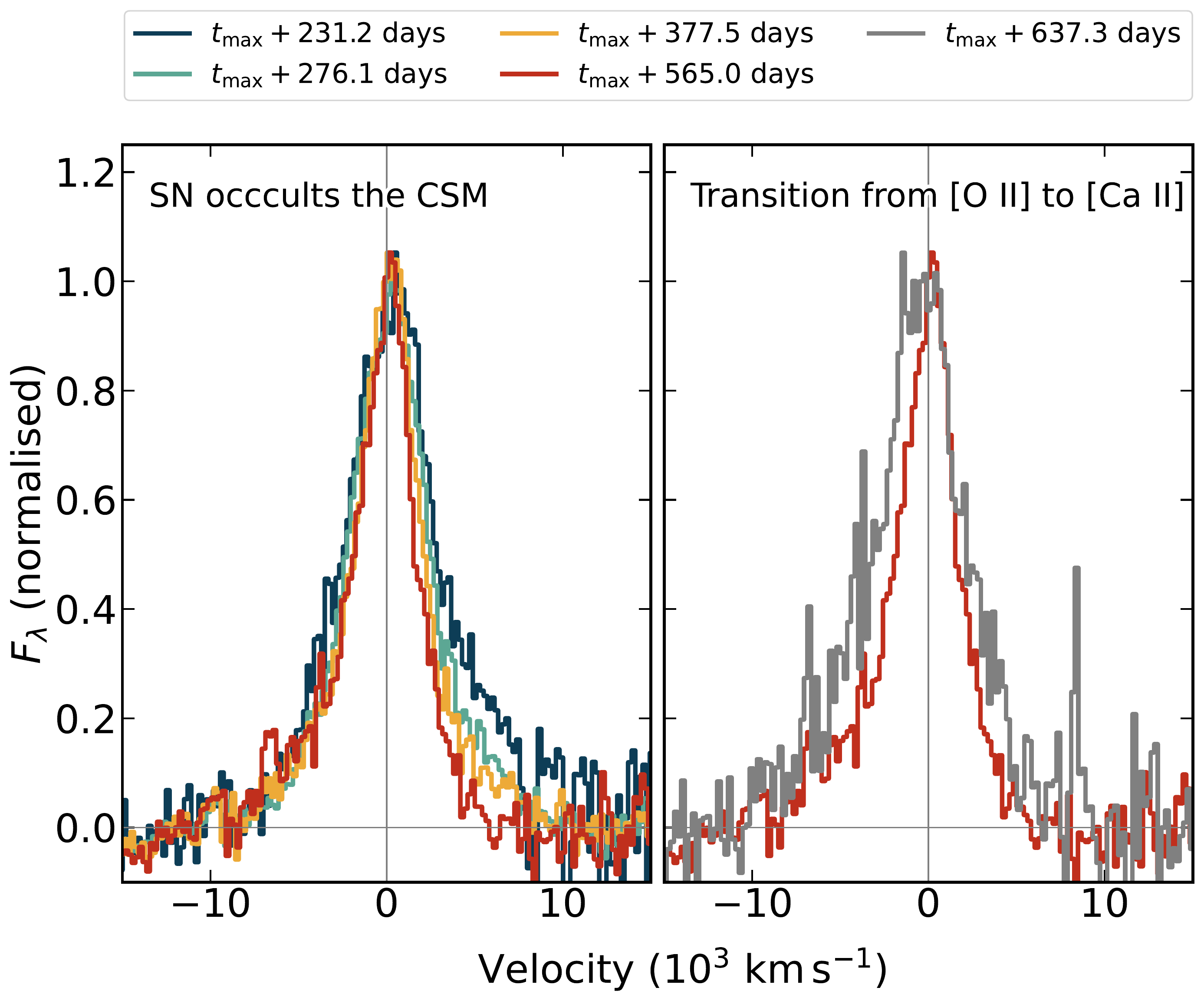}
    \caption{Evolution of the [\ion{O}{ii}]\,$\lambda\lambda$\,7320,7330 + [\ion{Ca}{ii}]\,$\lambda\lambda$\,7291,7324 line complex.
    \textbf{Left}: Up to \tmax+565.0~days, the line complex is dominated by [\ion{O}{ii}].
    The red wing narrows due to an increasing occultation of the CSM shell by the optically thick expanding SN photosphere.
    \textbf{Right}: Between \tmax+565.0~days and \tmax+637.2~days days, the line complex shifts to the blue, consistent with the [\ion{Ca}{ii}] line becoming more dominant. All profiles are centred on [\ion{O}{ii}]\,$\lambda$\,7320.}
\label{fig:oii7320}
\end{figure}

The differences in the origin of the forbidden oxygen lines are corroborated by the \ion{O}{i}-\ion{O}{iii} line profiles (Figures \ref{fig:oi_oii_oiii_profiles}, \ref{fig:oiii}). The asymmetric [\ion{O}{iii}]\,$\lambda$\,4363 and [\ion{O}{iii}]\,$\lambda\lambda$\,4959,5007 lines have little emission in the red wings indicative of emission from a thin shell, where most of the red emission is absorbed by the photosphere. Examples of this can be seen in figure 4b in \citet{Fransson1984a}. That scenario is also consistent with the evolution of the [\ion{O}{ii}]\,$\lambda\lambda$\,7320,7330 doublet. The 7300~\AA\ line, which may be a blend of the [\ion{O}{ii}] \,$\lambda\lambda$\,7320,7330 lines and [\ion{Ca}{ii}]\,$\lambda\lambda$\,7291,7324, is shown in Figure \ref{fig:oii7320}, centred on the [\ion{O}{ii}]\,$\lambda$\,7320 line. Focusing on the [\ion{O}{ii}] lines (left panel), the blue wing has a nearly constant line profile between \tmax+231.2 and \tmax+565.0~days. The red wing of the [\ion{O}{ii}]\,$\lambda$\,7330 doublet component gets considerably narrower during the same time interval. At \tmax+637.3~days (right panel), the entire 7300~\AA\ line profile changes quite dramatically, becoming flat-topped and broader. This is a result of the [\ion{Ca}{ii}]\,$\lambda \lambda$\,7291,7324 lines becoming strong, while the [\ion{O}{ii}] lines get weaker.

The increasing asymmetry of the [\ion{O}{ii}] doublet may be qualitatively understood by the CSM being occulted by the SN ejecta. Assuming that the optically thick SN ejecta with the velocity $v_{\rm ej}$ slams into the CSM shell with a low velocity (ideally $v\approx0$) located at a distance $R_{\rm s}$ from the progenitor star, the maximum velocity of the red wing $v_{\rm red}$ is (a pure geometric effect)

\begin{equation}
\begin{split}
v_{\rm red} &= v_s\,\left(1 - \left(v_{\rm ej}\,t/R_{\rm s}\right)^2\right) \approx v_{\rm s}\,\left(1 - \left(v_{\rm ej}\,t/v_{\rm s} \left(t+\Delta t\right)\right)^2\right) \\
&\approx v_{\rm s} \left(1 - \left(v_{\rm ej}\,t/v_{\rm s} \Delta t\right)^2\right) \nonumber 
\end{split}
\end{equation}
where $\Delta t$ is the time between the shell ejection and the explosion and $t$ the time since explosion. (We have assumed that $t \ll \Delta t$.) Because the ejecta with the CDS, which may define the photosphere, expands with a much higher velocity ($\sim 8,500 \kms$ vs. $\sim 3000 \kms$) a progressively increasing portion of the dense CSM will be occulted by the photosphere and less of the `backside' of the CSM will be seen. This would lead to the red side getting narrower with time. At the same time, an increasing portion of the dense CSM will be shocked,  leading to a decreasing luminosity from the dense CSM, including the [\ion{O}{ii-iii}] emission.

\begin{figure}
\includegraphics[width=1\columnwidth]{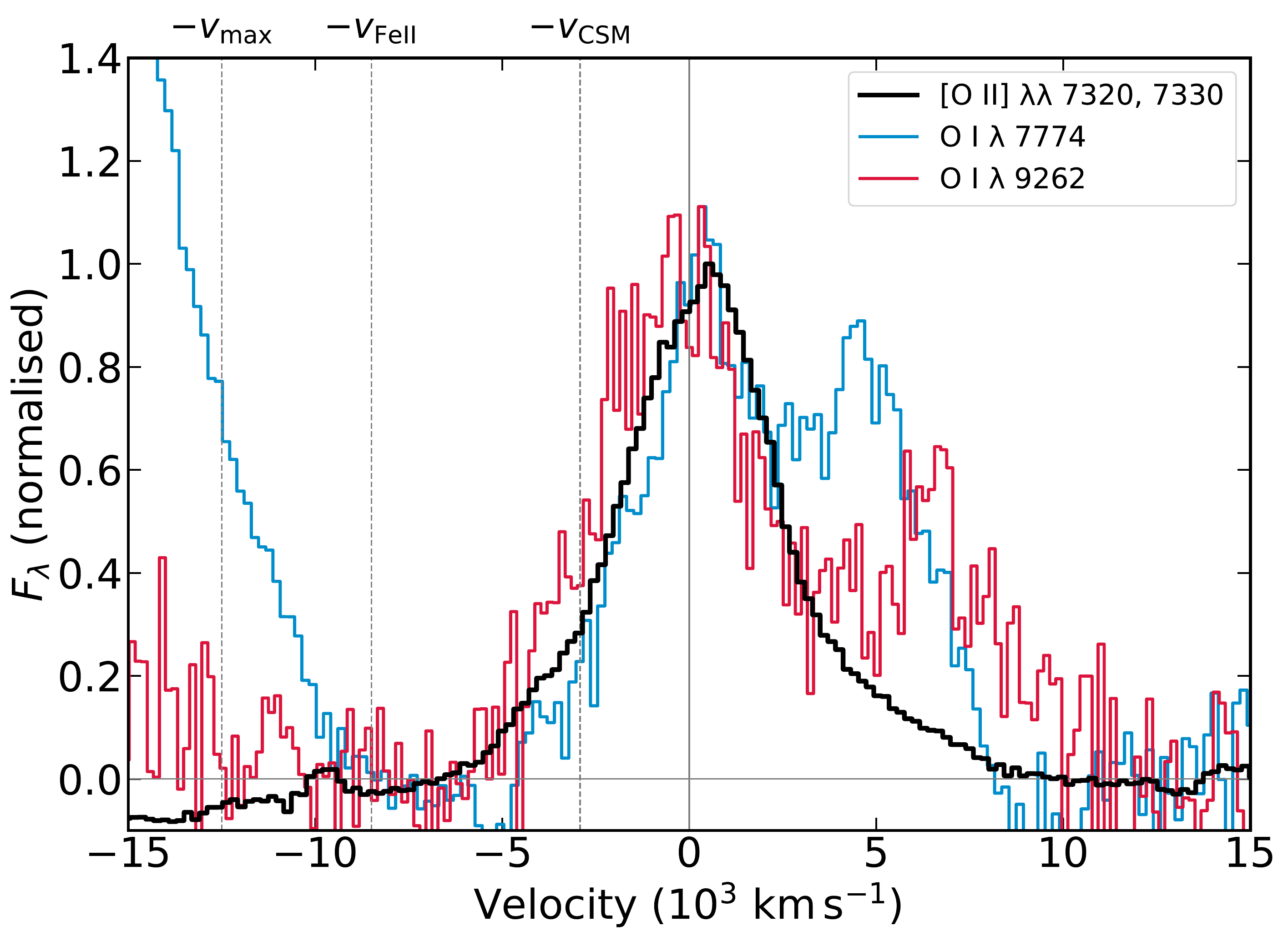}
\caption{Comparison of the  [\ion{O}{ii}]\,$\lambda\lambda$\,7320,7330 lines with the \ion{O}{i}\,$\lambda$\,7773 and \ion{O}{i}\,$\lambda$\,9263 recombination lines. The similar line profiles, extending to $\lesssim 5000 \kms$ indicates an origin in a highly processed CSM shell.}
\label{fig:oi_9263}
\end{figure}

The fact that the forbidden [\ion{O}{ii}] and [\ion{O}{iii}] lines are seen at about \tmax+30~days adds additional constraints on the physical conditions where they originate. The critical densities, above which collisional de-excitation becomes important, are less than $\sim 2\times 10^6\,\rm cm^{-3}$ for [\ion{O}{ii}] and [\ion{O}{iii}] \citep{Osterbrock2006a}. This is much lower than the densities expected in the ejecta (Equation \ref{eq:number_dens}). Therefore, the [\ion{O}{ii}] and [\ion{O}{iii}] lines would be severely suppressed if they were coming from the ejecta. Not only do [\ion{O}{ii}] and [\ion{O}{iii}] originate from the CSM, but also the recombination lines \ion{O}{i}\,$\lambda$\,7773 and \ion{O}{i}\,$\lambda$\,9263. The blue wings of their line profiles are similar to [\ion{O}{ii}]\,$\lambda$\,7320, extending to $\sim 5000 \kms$ (Figure \ref{fig:oi_9263}).

In summary, we propose a two-component scenario where the broad component, seen in particular in the [\ion{O}{i}]\,$\lambda\lambda$\,6300,6364 and \ion{Mg}{i}]\,$\lambda$\,4571 lines as well as the broad absorption in \ion{Mg}{ii}\,$\lambda$\,2800 and \ion{Ca}{ii}\,$\lambda$\,3934 come from either the CDS or possibly the unshocked ejecta. The low-velocity component seen in the [\ion{O}{iii}] lines, as well as  the [\ion{O}{ii}] and \ion{O}{i} recombination lines come from the CSM shell at $\sim 3000~\kms$. The fact that we see [\ion{O}{iii}]\,$\lambda\lambda$\,4959,5007 emission even at \tmax+989.2~days means that the dense CSM must extend out to at least a few $\approx 10^{17}$~cm. The velocity width of the \ion{Mg}{ii} absorption of the CSM shell of $406 \kms$ (Section \ref{sec:csm:shell}; Figure \ref{fig:spec:csm_shell}) may correspond to the velocity gradient over the CSM shell. That we do not see any change in the width with time suggests that this gradient must be small enough so that the velocity close to the shock is nearly constant. The origin of this gradient is not clear, though. One explanation could be that this is the result of a time-limited eruption, where a Hubble-like outflow is expected after a few dynamical time scales. This has been observed, for instance, in the Eta Carinae Homunculus nebula produced during the great eruption in 1843 \citep[e.g.][]{Smith2006a}. The absence of H and He lines throughout the entire evolution reveals (Figure \ref{fig:spec:seq_uvvis}) that the CSM shell must be processed gas from the stripped progenitor. Any hydrogen and helium must have been lost before this eruption and reside at much larger radii.

Among the $>200$ H-poor SLSNe known, \sn\ is only the seventh object with spectroscopic evidence of CSM interaction. In previous cases, CSM interaction did not manifest itself via [\ion{O}{iii}] in emission \citep[a possible candidate for CSM interaction with O-rich material is PS1-14bj;][]{Lunnan2016a}. iPTF16eh revealed CSM interaction through a light echo produced in a shell of H-poor and He-poor material \citep{Lunnan2016a}. Late-time spectra of iPTF10aagc, 13ehe, 15esb and 16bad \citep{Yan2015a, Yan2017a} and SN\,2018bsz \citep{Pursiainen2022a} showed broad Balmer emission lines, suggesting that their progenitors lost their hydrogen envelopes much closer to the time of the terminal explosion than \sn\ and iPTF16eh.

\subsection{Constraints on the powering mechanism and progenitor}\label{sec:discussion:lc_spec}

In the following, we contrast SLSN and PISN models with our photometric and spectroscopic datasets and discuss the most likely powering mechanism and progenitor of \sn.

\subsubsection{Modelling the bolometric light curve}\label{discussion:lightcurve:katz}

We first analyse the bolometric light curve. \citet{Katz2013a} proposed an exact method for testing whether a light curve is powered by the decay of radioactive material and, therefore, allows us to place an upper limit on any $^{56}$Ni produced during the explosion of \sn's progenitor. This method is independent of details in the radiative transport, including the highly uncertain opacity, the velocity distribution and the ejecta geometry. The method is described in detail in \citet{Wygoda2019a} and \citet{Sharon2020a}. In brief, the Katz integral is given by

\begin{eqnarray}
    QT &=& LT + ET~\rm with \nonumber \\
    QT &=& \int^t _0 dt'\,t'\,Q_{\rm dep}\left(t'\right),~
    LT = \int^t _0 dt'\,t'\,L\left(t'\right)\nonumber
\end{eqnarray}
and $ET$ is the integrated time-weighted luminosity that would be emitted if no $^{56}$Ni were produced. Assuming that there is no additional source of energy, $ET$ can be assumed to be negligible. The total energy deposition rate from radioactive decay of $^{56}$Ni, $Q_{\rm dep}$, is given by \citep{Jeffery1999a}
\begin{align}
    Q_{\rm dep}(t) &\approx Q_\gamma \left(1- e^{\left(-t_0/t\right)^2}\right) + Q_{e^+}(t) \nonumber
\end{align}
where $t_0$ is the $\gamma$-ray escape time. The deposition rates from $\gamma$-ray photons and positrons are
\begin{align} 
Q_\gamma &= \frac{M({\rm Ni})}{M_\odot}\,\left(6.45\,e^{-t / t_{1/2,\rm Ni}} + 1.38\,e^{-t / t_{1/2,\rm Co}}\right) \times10^{43}~\rm erg\,s^{-1} \nonumber\\
Q_{e^+} &= 4.64\frac{M({\rm Ni})}{M_\odot}\,\left(-e^{-t / t_{1/2,\rm Ni}} + e^{-t / t_{1/2,\rm Co}}\right) \times10^{41}~\rm erg\,s^{-1}\nonumber
\end{align}
where the mean lifetimes of $^{56}$Ni and $^{56}$Co are $t_{1/2,\rm Ni}=8.76$~days and $t_{1/2,\rm Co}=111.4$~days, respectively \citep{Junde1999a}.

\begin{figure}
    \centering
    \includegraphics[width=1\columnwidth, angle=0]{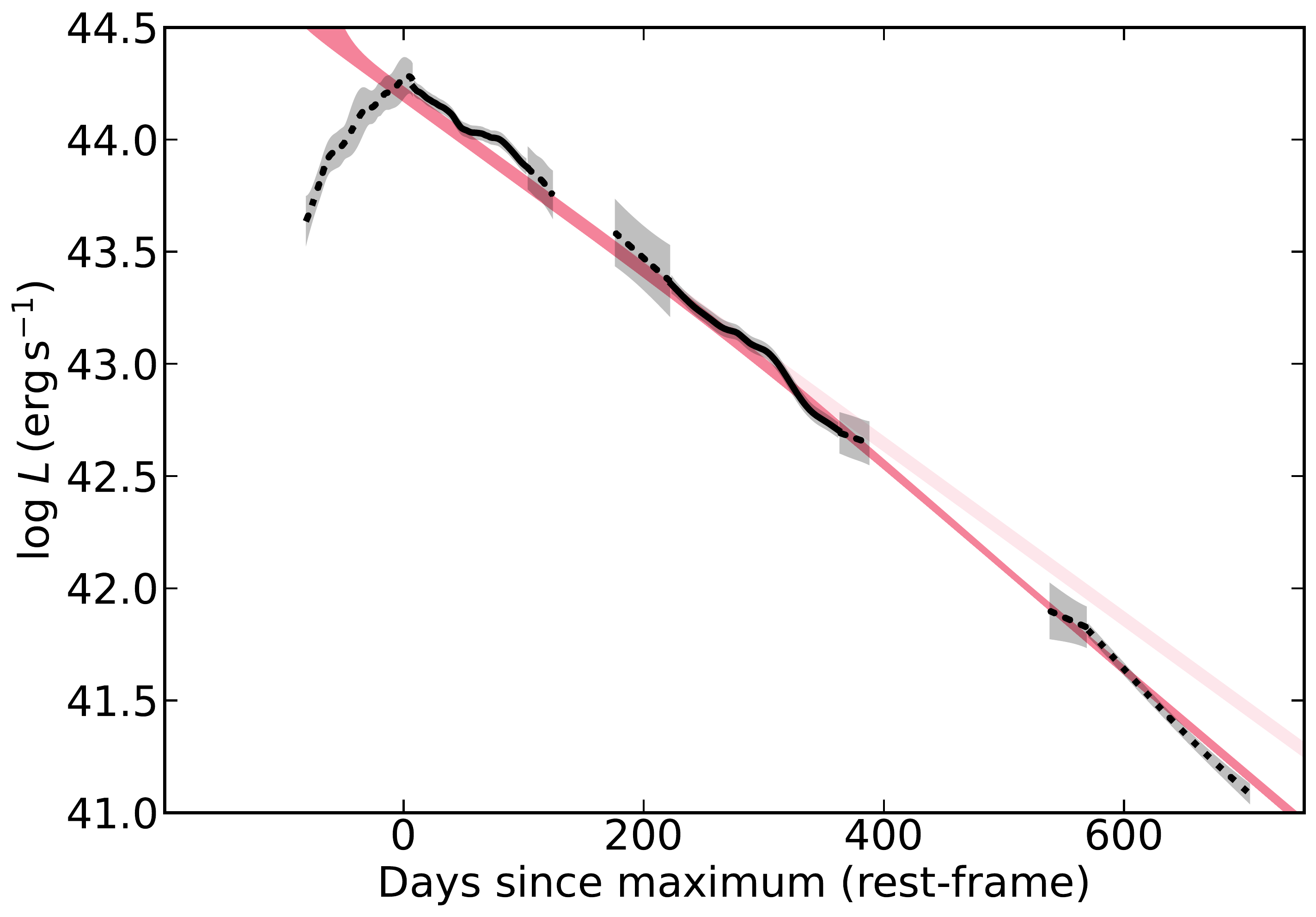}
    \caption{
    The bolometric light curve of \sn\ from 1800 to 14,300~\AA\ (rest-frame) and light-curve fits to the fading light curve using the \citet{Katz2013a} method. The entire fading light curve from up to 706~days after peak is fully consistent by being powered 24--$35~M_\odot$ of $^{56}$Ni (dark red), suggesting that \sn\ could be a pair-instability supernova. At about 300 days after peak (i.e. $\gtrsim400$~days after the explosion), the $\gamma$-ray trapping decreases with time. The loss of trapping is indicated by the difference between the light-red ($100\%$ trapping) and dark-red ($<100\%$ trapping) curves.
    }
    \label{fig:lc:katz}
\end{figure}

Since the explosion time is not well known, we vary the explosion time between 0 and 50 rest-frame days before the first detection and use the relation $L/LT = Q/QT$ to determine the $\gamma$-ray escape time. We measure a range of 600 to 700~rest-frame days for $t_0$. After the $\gamma$-ray escape time is determined, we infer the nickel mass by comparing the luminosity in the fitted range to the deposited radioactive energy. The best fit for each point in the $t_{\rm exp}$ grid is shown in Figure \ref{fig:lc:katz}. Indeed, the declining light curve is fully consistent with being powered by 24--$35~M_\odot$ of $^{56}$Ni. The upper bound could be even larger if the SN explosion happened more than 50 rest-frame days before the detection by \gaia.

The rise time of 90--140~days, the range of $\gamma$-ray escape times and the range of nickel masses are consistent with expectations from PISN models \citep{Kasen2011a, Kozyreva2015a, Kozyreva2017a} and SN Ia\footnote{Type Ia supernovae are a class of supernovae that occurs in binary systems in which one of the stars is a white dwarf \citep[e.g.][ and references therein]{Hillebrandt2000a}. This explosion produces $\sim0.6~M_\odot$ of $^{56}$Ni \citep[e.g.][and references therein]{Maoz2014a}.} \citep[another class of thermonuclear explosions; e.g.][]{Wygoda2019a, Sharon2020a}, after scaling their average nickel mass to the nickel mass of \sn. Both the excellent match with nickel powering and coverage of the fading light curve for 706 days is unprecedented for any of the $>200$ SLSNe known, suggesting that \sn\ could indeed be a PISN.

\subsubsection{Modelling the broadband light curve}\label{discussion:lightcurve:mosfit}

Next, we fit the multi-band light curve with the Modular Open-Source Fitter for Transients \program{MOSFiT} software tool \citep{Guillochon2018a}. In addition to the \program{MOSFiT} \program{nickel} model that is based on the parameterisation by \citet{Nadyozhin1994a}, we also select the central-engine models \program{slsn} (describing powering by a spin down of a magnetar; \citealt{Nicholl2017a}) and \program{fallback} (describing powering by a black hole accreting fallback material; \citealt{Moriya2018b}), and the \citet{Chatzopoulos2012a} model to characterise the powering by CSM interaction. We also utilise the more complex models \program{magni} combining powering by a magnetar and radioactive $^{56}$Ni \citep{Blanchard2019a} and \program{csmni} which combines powering by CSM interaction and $^{56}$Ni. In all models, the photosphere is assumed to have a blackbody spectral energy distribution at all times. While this approximation is adequate during the photospheric phase, it is inadequate at later times when the spectrum is dominated by emission lines and an interaction-powered pseudo-continuum. The spectral energy distribution of the model \program{slsn} is modified in the UV to account for absorption by the SN ejecta. As our dataset covers a very long time span, the trapping of $\gamma$-ray photons will eventually decrease, which accelerates the fading. All chosen models include a component to account for the loss of trapping \citep{Nicholl2017a}. 

\begin{figure*}
    \centering
    \includegraphics[width=1\textwidth]{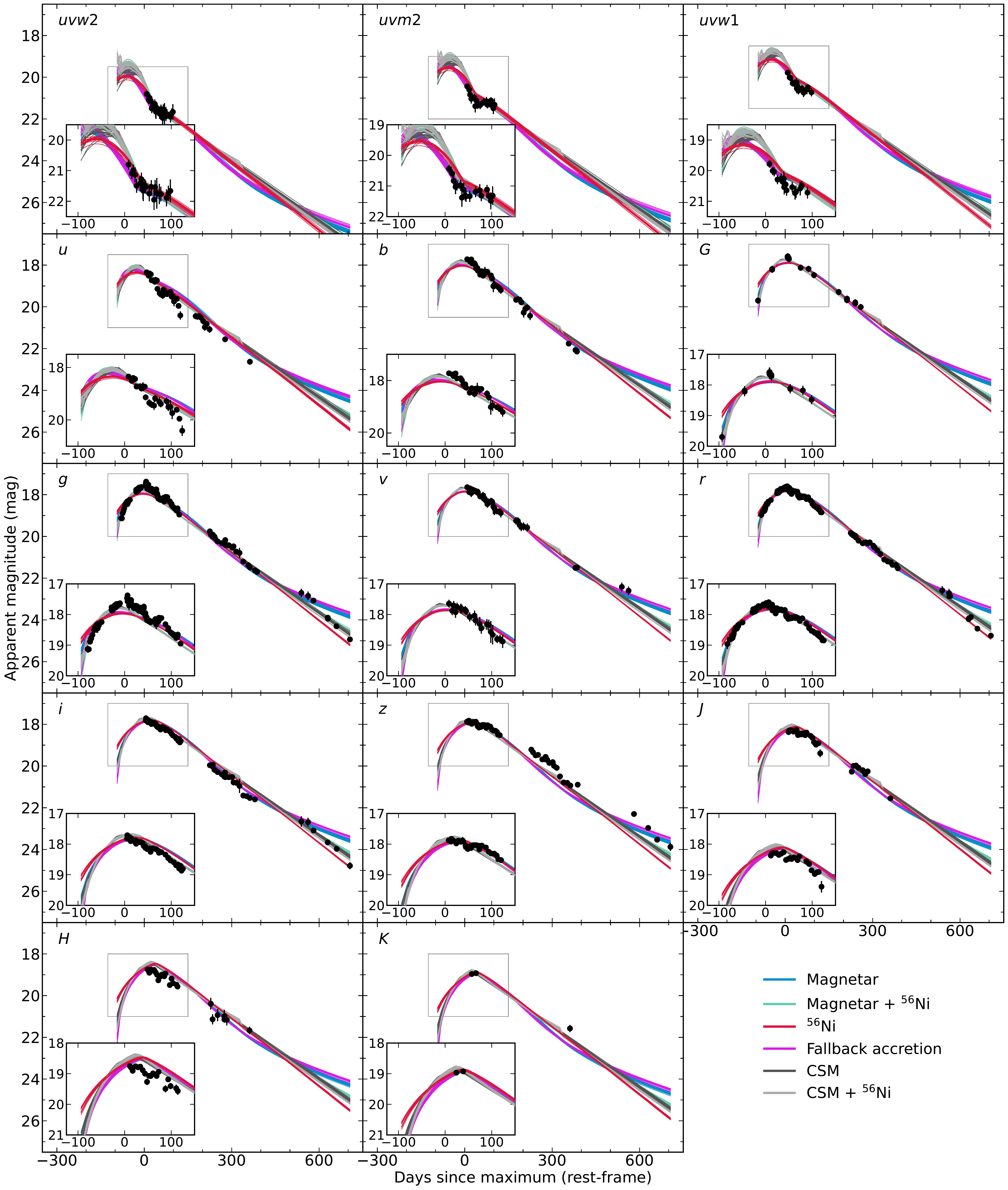}
    \caption{
    Modelling of the light curves from the rest-frame UV to the NIR with MOSFiT. All models provide an adequate description of the data up to \tmax+400~days, though with differences in the fit quality. At later times, the models diverge. The pure nickel model is the only model that captures the evolution after \tmax+500~days and has physically meaningful parameters. The central engine models (magnetar and fallback) predict a flattening of the light curve due to a power-law-shaped heating rate in contrast to powering by $^{56}$Ni that has an exponential energy deposition rate. The magnetar+nickel model also captures the full evolution. However, the inferred model parameters would be either physically implausible or require an exotic star, which we deem not viable. We note that the Keck photometry at \tmax+539~days and \tmax+562~days is not corrected for host contamination owing to the lack of Keck reference images to perform image subtraction. The expected host contribution is $\approx 10\%$.
}
    \label{fig:lc:mosfit}
\end{figure*}

The priors of the model parameters are chosen to cover a broad range of the physically allowed parameter spaces. Their ranges and shapes are similar to \citet{Nicholl2017a}, \citet{Kangas2022a} and \citet{Chen2023a}, and they are summarised in Table \ref{tab:mosfit}. For the pure nickel model, we set the opacities $\kappa$ and $\kappa_\gamma$ to 0.07 and $0.027~\rm g\,cm^{-2}$, respectively \citep[][and references therein]{Swartz1995a, Wang2015a}. For the models that include two sources of energy, in which one of the energy sources is $^{56}$Ni, we perform fits with unconstrained opacities as well as fits with the opacities set to $\kappa=0.07~\rm g\,cm^{-2}$ and $\kappa_\gamma=0.03~\rm g\,cm^{-2}$. The model parameters are inferred in a Bayesian way using the nested sampler \program{dynesty}. The fits of each model are shown in Figure \ref{fig:lc:mosfit}. As the fit covers a wide time interval, each panel in Figure \ref{fig:lc:mosfit} also contains a window zooming in onto the region of maximum light. The marginalised posteriors of the model parameters are summarised in Table \ref{tab:mosfit}.

Visually, all models capture the rise, peak and decline up to \tmax+400~days. There are noticeable differences between the fits and the data because of \textit{i}) not all models can be correct, \textit{ii}) the inherent assumptions of each model, and \textit{iii}) the assumption of a blackbody photosphere at all times. Owing to this, none of the models can capture the bumps and undulations (see inset in Figure \ref{fig:lc:mosfit}). The significant deviation in the $z$ band at $>$\tmax+200~days is due to the assumption of a blackbody photosphere. The late-time spectra reveal a blue pseudo-continuum with super-imposed emission lines. The luminescent [\ion{O}{ii}]\,$\lambda\lambda$\,7320,7330 emission lines are redshifted to the $z$ band and cause the apparent discrepancy between the data and the models (Figure \ref{fig:spec:seq_uvvis}).

Besides these general caveats, differences in the fit qualities between the models are visible. At epochs later than \tmax+500~days, the pure central engine models fail to describe the data. The discrepancies grow with time and reach $\sim2$~mag per band at \tmax+706~days. In contrast to that, all nickel and CSM models are able to describe the entire light curve from \tmax$-93$~days to \tmax+706~days. The fundamental difference in the late-time behaviour of the central-engine and nickel models lies in how the models deposit energy into the ejecta (and hence the SN luminosity). In the pure central engine models, the energy deposition evolves as a power law in time. This leads to a decrease in the decline rate of the SN brightness in the time vs. magnitude space. In contrast to that, radioactive material has an exponentially declining energy deposition rate. This results in a linear decline in the time vs. magnitude space, which fits our observations well. The loss of $\gamma$-ray trapping accelerates the fading independent of the powering mechanism, but it only modifies the light curve without altering its general shape. In other words, the loss of gamma-ray trapping cannot convert a power-law decline into an exponential decline \citep[e.g.][]{Chen2015a, Wang2015a, Nicholl2018a}.

The nickel models ($^{56}$Ni and $^{56}$Ni+CSM|magnetar) provide an adequate fit to the entire light curve. All models require $\approx35~M_\odot$ (spread in the median values: 20--$63~M_\odot$; statistical error of the individual measurements: 7--67\%; Table \ref{tab:mosfit}) of freshly synthesised nickel, consistent with our conclusions on the bolometric light curve (Section \ref{discussion:lightcurve:katz}). The pure $^{56}$Ni has a worse fit statistic because of the loss of $\gamma$-ray trapping. This leads to underpredicting the expected brightness. This is not a critical issue. The modelling assumes that the SED is still a blackbody, which is not the case anymore (Section \ref{sec:spectra_discussion}). Furthermore, fitting the bolometric light curve with $^{56}$Ni gave an excellent fit (Section \ref{discussion:lightcurve:katz}. Such large nickel masses can only be produced in a PISN explosion. PISN models predict no remnant after the entire star is obliterated \citep{Fowler1964a, Barkat1967a, Rakavy1967a}, eliminating the magnetar + $^{56}$Ni model. If we were to ignore stellar evolution theory, the rotational energy of the magnetar, which defines how much energy could be converted into radiation, would contribute $<1\%$ to the total radiated energy ($\sim1\times10^{50}~\rm erg$), whereas $>99\%$ of the radiated energy would come from the radioactive decay of $^{56}$Ni and its daughter products (Table \ref{tab:mosfit}). Moreover, the inferred spin period of $\sim15$~ms is much larger than the median spin period of $\sim2.6$~ms from the ZTF-I SLSN sample \citep[][see also \citealt{Nicholl2017a} and \citealt{Blanchard2020a}]{Chen2023a}. Even the slowest spinning SLSN magnetars never exceeded 6--7~ms \citep[][]{Nicholl2017a, Blanchard2020a, Chen2023b}.\footnote{All measurements are based on the fiducial assumption of dipole spin-down radiation. For a discussion on the impact of non-dipole radiation see \citet{Omand2023b}.} 

\begin{table*}
    \caption{Light-curve fits with \program{MOSFiT}: models, parameters, priors and marginalised posteriors}
    \scriptsize
    \centering
        \begin{tabular}{lcrrrrrrrr|r}
             \toprule
             Parameter  & Prior  & Magnetar & Magnetar\,+   & Magnetar\,+        & $^{56}$Ni          & Fallback  & CSM   & CSM\,+             & CSM\,+             & $^{56}$Ni          \\
                        &        &          & $^{56}$Ni     & $^{56}$Ni          &                    &           &       & $^{56}$Ni          & $^{56}$Ni          & (red)              \\
                        &        &          &               & (fixed $\kappa$'s) & (fixed $\kappa$'s) &           &       &                    & (fixed $\kappa$'s) & (fixed $\kappa$'s) \\
             \midrule
             \multicolumn{11}{c}{\textbf{Fitted properties}}\\
             \midrule
             \multicolumn{11}{c}{{General}} \\
             \midrule
             ejecta mass $M_{\rm ej}$ $\left(M_\odot\right)$                         & $\log \mathcal{U}\left(1, 300\right)$              &$81^{+12}_{-9}       $&$54^{+29}_{-23}     $&$141^{+33}_{-29}      $&$ 83\pm4      $&$74\pm5             $&$46^{+11}_{-13}   $&$34^{+3}_{-2}         $&$34\pm6            $&$107^{+8}_{-9}$\\
             explosion date $t_{\rm exp}$ (day)                                      & $\mathcal{U}\left(-200, 0\right)$                  &$-19\pm3             $&$-22\pm2            $&$-22^{+2}_{-3}        $&$-70\pm5      $&$-11\pm2            $&$-27^{+4}_{-6}    $&$-17^{+2}_{-3}        $&$-14\pm2           $&$-97^{+7}_{-8}$\\
             `$\gamma$-ray' opacity $\kappa_\gamma$ $\left(\rm cm^2\,g^{-1}\right)$& $\log \mathcal{U}\left(10^{-2}, 10^{4}\right)$     &$0.013\pm0.002       $&$14^{+744}_{-14}    $&0.03                   &0.03           &$0.010\pm0.001       $&\dots             &$155^{+1751}_{-143}  $&0.03                &0.03\\
             optical opacity $\kappa$ $\left(\rm cm^2\,g^{-1}\right)$                & $\mathcal{U}\left(0.01, 0.2\right)$                &$0.18\pm0.02         $&$0.05\pm0.02        $&0.07                   &0.07           &$0.19\pm0.01        $&\dots              &$0.07\pm0.01          $&0.07                &0.07\\
             scaling velocity $v_{\rm scale}$ $\left({\rm km\,s}^{-1}\right)$        & $\mathcal{U}\left(1000, 10000\right)$              &$5130^{+210}_{-200}  $&$5660^{+170}_{-160} $&$ 5620^{+180}_{-170}  $&$4050\pm130   $&$5560^{+240}_{-200} $&$5510^{+200}_{-220}$&$5950^{+230}_{-200}   $&$5850^{+190}_{-160}$&$3600^{+260}_{-310}$\\
             white noise parameter $\sigma$                                          & $\log \mathcal{U}\left(10^{-3}, 100\right) $       & $0.25 \pm 0.01      $&$0.21\pm0.01        $&$0.25 \pm0.01         $&$0.25 \pm0.01 $&$0.26\pm0.01        $&$0.21 \pm0.01     $&$0.21\pm0.01          $&$0.21\pm0.01         $&$0.23\pm0.01$\\
             \midrule
             \multicolumn{11}{c}{{Magnetar model}} \\
             \midrule
             magnetic field $B_\perp$ $\left(10^{14}\,{\rm G}\right)$                & $\log \mathcal{U}\left(0.01, 20\right)$            &$0.72^{+0.05}_{-0.08}$&$1.32^{+1.25}_{-1.24}$&$1.05^{+0.47}_{-0.36}$& \dots         & \dots               & \dots          & \dots                 & \dots              &\dots\\
             neutron-star mass $M_{\rm NS}$ $\left(M_\odot\right)$                   & $\mathcal{U}\left(1, 2.2\right)$                   &$ 2.1\pm0.1          $&$ 1.4^{ +0.5}_{ -0.3}$&$1.6^{ +0.4}_{ -0.3} $& \dots         & \dots               & \dots          & \dots                 & \dots              &\dots\\
             initial spin period $P_0$ (ms)                                          & $\mathcal{U}\left(1, 20\right)$                    &$1.0^{+0.1}_{-0.0}   $&$15.5^{ +3.0}_{ -4.8}$&$14.6^{ +3.1}_{ -3.5}$& \dots         & \dots               & \dots          & \dots                 & \dots              &\dots\\
             \midrule               
             \multicolumn{11}{c}{{$^{56}$Ni model}} \\
             \midrule
             nickel fraction $f_{\rm Ni}$                                            & $\log \mathcal{U}\left(10^{-3}, 1\right)$          & \dots                &$0.6\pm0.2           $&$0.2\pm0.1           $&$0.50\pm0.02  $& \dots               &\dots           &$0.9\pm0.1$            &$0.6\pm0.1         $&$0.6^{+0.3}_{-0.2}$\\
             \midrule
             \multicolumn{11}{c}{{Fallback model}} \\
             \midrule
             luminosity $L_1$ $\left(10^{55}\rm erg\,s^{-1}\right)$                  & $\log \mathcal{U}\left(10^{-4}, 10^3\right)$       & \dots                & \dots                   & \dots                &\dots          &$4.5\pm0.1              $& \dots   & \dots & \dots& \dots\\
             transition time $t_{\rm tr}$ (day)                                      & $\log \mathcal{U}\left(10^{-4}, 10^4\right)$       & \dots                & \dots                   & \dots                &\dots          &$0.003^{+0.014}_{-0.003}$& \dots   & \dots & \dots& \dots\\
             \midrule
             \multicolumn{11}{c}{{CSM}} \\
             \midrule
             CSM mass $M_{\rm CSM}$ $\left(M_\odot\right)$                           & $\log \mathcal{U}\left(0.01, 300\right)$           & \dots                & \dots                   & \dots                & \dots         &\dots             &$58^{+13}_{-11}   $&$1.2^{+0.8}_{-0.5}$   &$13^{+3}_{-2}$ &\dots\\
             CSM density $\rho \left(10^{-14}\rm cm^{-3}\right)$                     & $\log \mathcal{U}\left(10^{-14}, 10^{2}\right)$    & \dots                & \dots                   & \dots                & \dots         &\dots             &$19^{+9}_{-6}     $&$14^{+57}_{-10}      $&$36300^{+18700}_{-15900}$&\dots\\
             power-law index of the CSM                                              & $\mathcal{U}\left(0, 2\right)$                     & \dots                & \dots                   & \dots                & \dots         &\dots             &$0.6\pm0.3        $&$1.8\pm0.1           $&$0.5\pm0.2    $&\dots\\
             density profile $s$\\   
             slope of the outer SN ejecta                                            & $\mathcal{U}\left(8, 12\right)$                    & \dots                & \dots                   & \dots                & \dots         &\dots             &$8.8^{+0.6}_{-0.4}$&$9.7^{+0.8}_{-0.9}$&$10.0^{+0.6}_{-0.7}$&\dots\\
             density profile $n$\\   
             slope of the inner SN ejecta                                            & fixed                                              & \dots                & \dots                   & \dots                & \dots         &\dots             &0                  &0               & 0              &\dots\\
             density profile $\delta$\\   
             progenitor radius $R_0$ (AU)                                            & $\log \mathcal{U}\left(0.1, 1000\right)$           & \dots                & \dots                   & \dots                & \dots         &\dots             &$617^{+160}_{-149}$&$59^{+67}_{-35}$& $10^{+6}_{-4}$ &\dots\\
             \midrule
             \multicolumn{11}{c}{\textbf{Fit quality}}\\
             \midrule
             log Bayesian evidence ($\log~Z$)                                        &                                                    & 516             & 640                   &  639                 & 541     & 497        & 644          & 642                     &638           & 287\\ 
             Number of free parameters                                               &                                                    & 11              & 12                    &  10                   & 7       & 10          & 11           & 14                      &13            & 7\\ 
             \midrule
             \multicolumn{11}{c}{\textbf{Derived properties}}\\
             \midrule
             $\gamma$-ray escape time $t_0$ (day)                                    &                                                    &$320^{+30}_{-20}$&$9200^{+76200}_{-8400}$&$630\pm80            $&$570\pm20$&$290\pm20$ &\dots         &$23700^{+64800}_{-17400}$&$300\pm30    $&$690^{+60}_{-50}$ \\
             nickel mass $M_{\rm Ni}$ $\left(M_\odot\right)$                         &                                                    &\dots            &$32^{+22}_{-18}$       &$32^{+11}_{-9}$       &$42\pm3  $&\dots      &\dots         &$31^{+4}_{-3}           $&$20^{+6}_{-5}$&$63^{+31}_{-19}$\\ 
             kinetic energy $E_{\rm kin}$ $\left(10^{51}\rm erg\right)$              &                                                    &$21\pm2         $&$17^{+9}_{-7}         $&$45^{+22}_{-19}$      &$14\pm1  $&$23\pm2   $&$14^{+3}_{-4}$&$12\pm1                 $&$12\pm2      $&$14^{+1}_{-2}$\\ 
             rotational energy $E_{\rm rot}$ $\left(10^{51}\rm erg\right)$           &                                                    &$33^{+5}_{-4}   $&$0.09^{+0.05}_{-0.04} $&$0.11^{+0.03}_{-0.02}$&\dots     &\dots      &\dots         &\dots                    & \dots        & \dots\\
             \bottomrule
        \end{tabular}
        \tablefoot{The model `$^{56}$Ni (red)' only fitted the data in the $r$ and redder bands. We used uniform ($\mathcal{U}$) and log uniform ($\log \mathcal{U}$) priors. The uncertainties of the marginalised posteriors are quoted at $1\sigma$ confidence. The explosion date is measured with respect to the date of the first detection. All marginalised posteriors are reported in linear units. The Bayesian evidence is reported in log units. The kinetic energy of the ejecta was computed via $E_{\rm kin}=1/2\, M_{\rm ej}\,v_{\rm scale}^{2}$ and the rotational energy of the magnetar via $E_{\rm rot}=2\times10^{52}\,\left(M_{\rm NS}/1.4\,M_\odot\right)^{3/2}\,\left(P_0/1\,\rm ms\right)^{-2}~\rm erg$.
        }
        \label{tab:mosfit}
\end{table*}

The CSM + $^{56}$Ni model might be viable. However, the inferred nickel fraction and CSM parameters are very sensitive to the \textit{a priori} unknown effective opacities $\kappa$ and $\kappa_\gamma$. The two parameters are different for each radiation component and come with their own uncertainties. Furthermore, the contribution of CSM interaction and $^{56}$Ni to the observed light curve changes with time resulting in a time-variable $\kappa$ and $\kappa_\gamma$. Capturing these complexities is not possible with that particular CSM + $^{56}$Ni model. Furthermore, the \citet{Chatzopoulos2012a} CSM model used in \program{MOSFiT} is debated \citep[see the discussion in][]{Sorokina2016a, Moriya2018a}, limiting the interpretability of the results. More sophisticated CSM + $^{56}$Ni models are needed to disentangle the two sources of energy \citep[e.g.][]{Chevalier2017a, Tolstov2017a, Suzuki2021a, Takei2022a}.

The pure magnetar model can be rejected on physical and statistical grounds. The aforementioned increased over-prediction of observed flux at late time makes this model unviable. Furthermore, the magnetar model requires extreme conditions ($M_{\rm NS}\sim2.1~M_\odot$, $P_0=1~\rm ms$ and $M_{\rm ej}\sim81~M_\odot$) to squeeze out as much energy as possible from the neutron star. The lower limit on the progenitor mass ($M_{\rm progenitor} > M_{\rm ej} + M_{\rm NS}$) exceeds $83~M_\odot$. Explosion models predict that such a massive star leaves behind a black hole but not a neutron star  \citep{Heger2002a}. From an empirical point of view, the H-poor SLSNe, which are thought to be powered by a magnetar, have ejecta masses of only $\sim5~M_\odot$. The most massive ejecta reach a few $10~M_\odot$ but never exceed $50~M_\odot$ \citep{Nicholl2017a, Blanchard2020a, Tinyanont2022a, Chen2023b, West2023a}, in stark contrast to the $\sim81~M_\odot$ required to power \sn. Previously, \citet{Eftekhari2021a} reported that \sn\ is powered by a magnetar (using the same magnetar model) but using a very sparse data set. We caution against this practice. Even with our comprehensive data set, only the data after \tmax+600~days broke the degeneracy between the central-engine models and the nickel-powered models. Furthermore, the best-fit magnetar properties presented here and in \citet{Eftekhari2021a} are significantly different, demonstrating that datasets with a large wavelength coverage and a wide time span are required to determine the powering mechanism of SLSNe. Our results echo the conclusions from \citet{Moriya2017a}, who performed a parameter study of magnetar and nickel models, that the two models could produce indistinguishable light curves if the data set covers only a limited time interval. Very late-time observations, such as those presented here, are critical to break the degeneracy in the light curve modelling \citep{Moriya2017a}.

The fallback model has implausible parameters. The total energy input from the fallback accretion required to power the light curve is given by $2.5\,L_1\times t^{-2/3}_{\rm tr}$ \citep{Moriya2018b}. For the best-fit values (Table \ref{tab:mosfit}), we infer a total energy input of 0.9--$4.4~M_\odot\,c^2$. Assuming a realistic conversion efficiency from the fallback accretion disk to the large-scale outflow of $10^{-3}$ \citep[i.e. the SN luminosity;][]{Dexter2013a}, the mass of the accretion disk would have to be 900--$4400~M_\odot$, which is unphysical. In the case of a jet forming during the accretion process, the conversion efficiency could be $\sim10\%$ \citep{McKinney2005a, Kumar2008a, Gilkis2016a} and the mass of the accretion disk would only have to be 9--$44~M_\odot$. Irrespective of the accretion efficiency, the fallback model always over-predicts the brightness by $\sim2$~mag at late times. To mitigate this issue, the accretion rate would also have to be fine-tuned to match observations. Owing to these issues, we do not deem the fallback model to be viable for \sn.

To strengthen our conclusions on the model selection, we also fit the observations using the software package \program{Redback} \citep{Sarin2023a}, which implements these different models. We fit the multi-band data in magnitude space with a Gaussian likelihood function and the exact same priors, and utilise the \program{nestle}\footnote{\href{ http://kylebarbary.com/nestle/}{http://kylebarbary.com/nestle/}} sampler implemented in \program{bilby} \citep{Ashton2019a, RomeroShaw2020a}. The fit parameters of the pure magnetar and nickel models as well as the magnetar + nickel model are consistent with the results from \program{MOSFiT} (Table \ref{tab:redback}). The pure nickel model is the only viable model. Furthermore, the necessity for a large ejecta is solely determined by the long rise. It does not depend on the availability of the data in the blue bands. We verified that by fitting the data in the $r$ band and in redder filters with \program{MOSFiT} and \program{Redback} (Tables \ref{tab:mosfit} and \ref{tab:redback}).

In summary, the multi-band light curve is consistent with being powered by radioactive $^{56}$Ni. The undulations, visible in several bands, are not captured by this nickel model. This could point to an additional source of energy, for example, interaction with circumstellar material. The required $^{56}$Ni mass of $\sim35~M_\odot$ can only be produced during a PISN explosion. Central engine models can be excluded with high confidence. The necessity for a large ejecta and nickel mass is solely determined by the long rise, the high peak luminosity and the slow decline. It does not depend on the availability of the data in particular filters.

Due to this extraordinary result, we want to briefly comment on whether the derived ejecta mass ($M_{\rm ej}\sim83~M_\odot$), nickel fraction ($\sim50\%$), velocity ($\sim4000\kms$) and kinetic energy ($14\times10^{51}~\rm erg$) are sensible for PISN models (values taken from Table \ref{tab:mosfit}). The PISN models by \citet{Kozyreva2017a}, \citet{Gilmer2017a} and \citet{Heger2002a} predict for a progenitor with a nickel yield of $\sim35~M_\odot$ an ejecta mass of $\sim130~M_\odot$, a nickel fraction of $\sim30\%$, an ejecta velocity of 9000--11000~km\,s$^{-1}$ and a kinetic energy of 80--$90\times10^{51}~\rm erg$. The model-predicted ejecta velocities, kinetic energy and nickel fractions appear to be in tension with our results, but that is not critical for the following reasons.
\textit{i}) The nickel model used in the light-curve fitting is an analytical one-dimensional model, but not a PISN model.
\textit{ii}) The velocity of the nickel model is a scaling velocity, which is not the ejecta velocity. In Section \ref{sec:velocities}, we showed that the ejecta velocity is $8500\kms$. The fastest portions of the ejecta move at $\sim12,500\kms$. These velocities yield a more realistic kinetic energy of 60--$130\times10^{51}~\rm erg$, consistent with PISN model predictions.
\textit{iii}) Different light-curve fitting codes give slightly different results.
\textit{iv}) PISN models have large uncertainties in predicting the energy release and the amount of $^{56}\mathrm{Ni}$. They depend on the nuclear network and assumptions of a given stellar evolution code, such as convection, details of the solution of the stellar structure equations and the dynamical phase of the pair-instability explosion, etc. For example, $^{56}\mathrm{Ni}$ masses calculated by \citet{Takahashi2018a} and \citet{Umeda2002a} differ significantly from \citet{Heger2002a}. In \citet{Umeda2002a}, a $270~M_\odot$ star would produce $10~M_\odot$ of $^{56}\mathrm{Ni}$ and a similar star in \citet{Takahashi2018a} would produce between a few $M_\odot$ masses and $>30~M_\odot$ of $^{56}\mathrm{Ni}$. In contrast to that, a $250~M_\odot$ star in \citet{Heger2002a} yields $\sim55~M_\odot$ of $^{56}\mathrm{Ni}$. As shown by \citet{Farmer2019a, Farmer2020a} and \citet{Kawashimo2023a}, the choice of the $^{12}\mathrm{C}(\alpha,\gamma)^{16}\mathrm{O}$ reaction rate leads to the scatter in the final $^{56}\mathrm{Ni}$ mass between $10 M_\odot$ and $70~M_\odot$ for the He-core stellar model of $130~M_\odot$, and a factor of 2 in the amount of the released explosion energy.

\subsubsection{Matching the light curve with PISN templates}\label{discussion:lightcurve:template_matching}

Motivated by the light-curve fits, we compare the bolometric light curve to the PISN templates from \citet{Kasen2011a}, \citet{Gilmer2017a} and \citet{Kozyreva2017a}. The grid of models from \citet{Kasen2011a} are the metal-free helium models from \citet{Heger2002a}. The \citet{Gilmer2017a} and \citet{Kozyreva2017a} models assume a metallicity of 7\% solar. Very massive stars in low-metallicity environments ($Z\sim 0.07~Z_\odot$) lose their hydrogen envelopes during the early evolution, assuming up-to-date wind mass-loss rates. The details about the used mass-loss rates are described in \citet{Ekstroem2012a} and \citet{Yusof2013a}. Therefore, these stars are hydrogen-free by the time of the pair-instability episode, and the helium-core models from \citet{Heger2002a} are a good representation of these explosions. This is in agreement with the models from \citet{Gilmer2017a}. Their suite of models, which were computed self-consistently, are initially hydrogen-rich models; however, owing to mass loss, the highest-mass models become hydrogen-free by the time of the explosion. 

\begin{figure*}[t]
    \centering
    \includegraphics[width=1\textwidth]{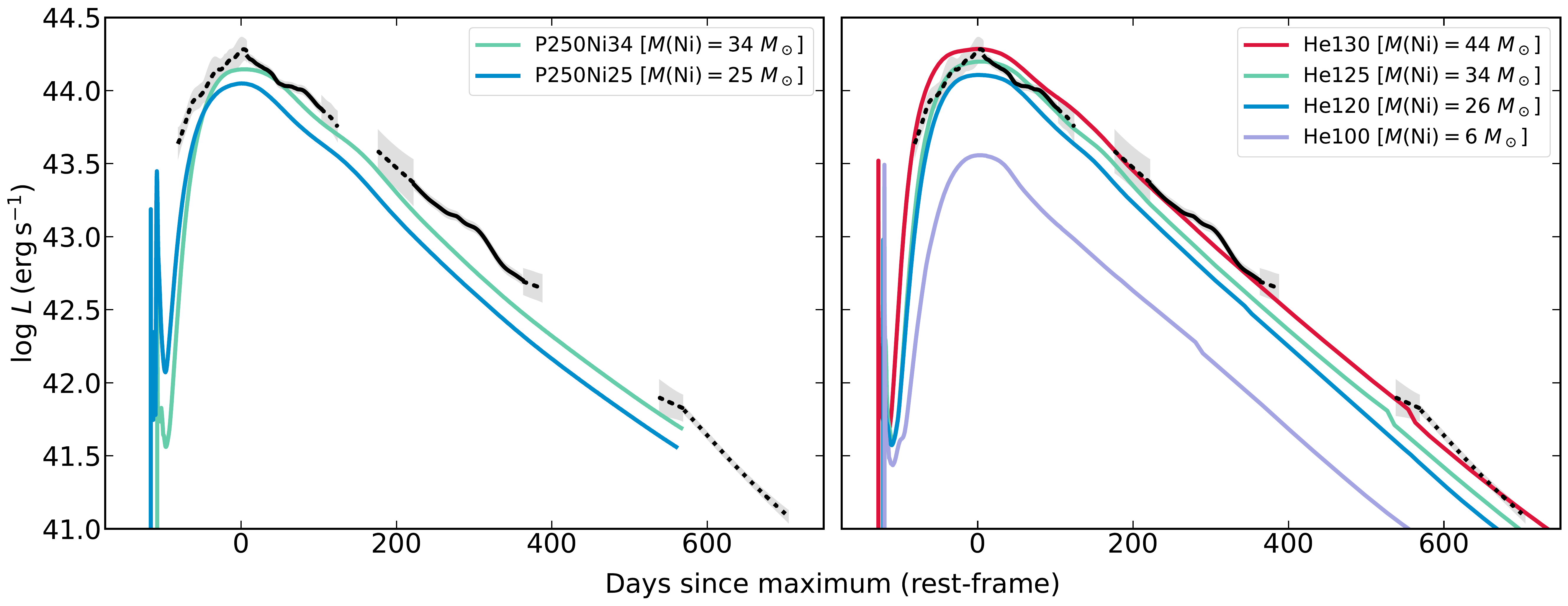}
    \caption{Comparison of \sn\ with the PISN models P250Ni25 and P250Ni34 from \citet{Kozyreva2017a} (left panel) and the 
    He100, He125, and He130 from \citet{Heger2002a} (right panel). Templates with nickel masses of 34--$44~M_\odot$ are required to describe the entire bolometric light curve from \tmax$-93$ to \tmax+706~days. Models with $M({\rm Ni})=25~M_\odot$ systematically underestimate the observed bolometric light curve, but they could still be viable if CSM interaction contributes significantly throughout the evolution.
    }
    \label{fig:lc:pisn_models}
\end{figure*}

Among the suitable models, we chose the P250Ni25 and P250Ni34 templates from \citet{Gilmer2017a}, and the He100, He120, He125 and He130 templates from \citet{Kasen2011a} (where the number stands for the helium core mass in $M_\odot$). The models have nickel yields between 5.8 and $40~M_\odot$. The vital properties of these models are presented in Table \ref{tab:pisn_properties}. The P250Ni25 model starts with an initial mass of $250~M_\odot$. At the time of the explosion, a helium core of $127~M_\odot$ has formed, which is similar to the helium models He125 and He130. We note that the P250Ni25 model not only loses its hydrogen envelope but also most of its helium layer (total mass of $2.6~M_\odot$ before the loss of the He layer) and ends up as a bare carbon-oxygen core with a tiny helium fraction by the time of the pair-instability explosion. In contrast to that, the He125 and He130 models evolve without mass losses and retain $2.4~M_\odot$ and $2.8~M_\odot$ of helium, respectively.

\begin{table}
    \caption{Summary of PISN model parameters}
    \label{tab:pisn_properties}
    \centering
    \scriptsize
    \begin{tabular}{ccccccc}
    \toprule
    Name    & $M(\rm ZAMS)$ & $M(\rm He)$   & $M(\rm Ni)$   & Metallicity   &  $v_{\rm ejecta}$  & $E_{\rm kin}$        \\
            & $\left(M_\odot\right)$   & $\left(M_\odot\right)$   & $\left(M_\odot\right)$   & $\left(Z/Z_\odot\right)$       & $\left(\rm km\,s^{-1}\right)$ & $\left(10^{51}\,\rm erg\right)$  \\
    \midrule
    He100     & 205           & 100           & 6             & 0.01          & 8400      & 42 \\
    He120     & 242           & 120           & 26            & 0.01          & 10000     & 71 \\
    He125     & 251           & 125           & 34            & 0.01          & 10300     & 79 \\
    He130     & 260           & 130           & 44            & 0.01          & 10600     & 87 \\
    P250Ni25  & 250           & 127           & 25            & 0.07          & 7500      & 82 \\
    P250Ni34  & 250           & 127           & 34            & 0.07          & 8850      & 82 \\
    \bottomrule
    \end{tabular}
    \tablefoot{The mass in column (3) lists the mass of the He core before the progenitor explodes. The values of the He100--He130 models were taken from \href{https://2sn.org/DATA/HW01/bulk_yields.txt}{https://2sn.org/DATA/HW01/bulk\_yields.txt} and are based on \citet{Heger2002a}. The values of the P250 models are from \citet{Kozyreva2017a} \missing{and \citet{Gilmer2017a}}. The ZAMS masses of the He100--He130 models are computed with $M({\rm He}) = 13/24 \times \left[M({\rm ZAMS}) - 20\,M_\odot\right]$ \citep{Heger2002a}.
    }
\end{table}

To build the bolometric light curves of the PISN models, we use the hydrodynamics radiative transfer code \program{STELLA} \citep{Blinnikov2006a}. The slight difference of the P250Ni34 light curve between our calculation and that in \citet{Gilmer2017a} and \citet{Kozyreva2017a} is caused by the different versions of \program{STELLA} used in the two studies. A relevant discussion can be found in \citet{Kozyreva2020a}. The re-calculated light curves of the helium models are consistent with those calculated with the spectral synthesis code \program{SEDONA} \citep{Kasen2011a}. In Figure \ref{fig:lc:pisn_models}, we compare the bolometric light curve of \sn\ to those computed for a series of PISN models. The PISN templates with nickel yields between 34 and $44~M_\odot$ (He125, He130, P250Ni34) provide excellent matches to the rise, the peak, and the fading parts of the bolometric light curve of \sn. While the He125 and P250Ni34 models describe the rise and peak well, they systematically underestimate the late-time flux. Such a deviation at the late epochs may not necessarily refute these two models since the observed bolometric flux of \sn\ may include a time-varying contribution from CSM interaction. As we show in Section \ref{sec:spectra_discussion}, this contribution is not negligible and could boost the luminosity by a few 0.1 dex.

\begin{figure*}
    \centering
    \includegraphics[width=1\textwidth]{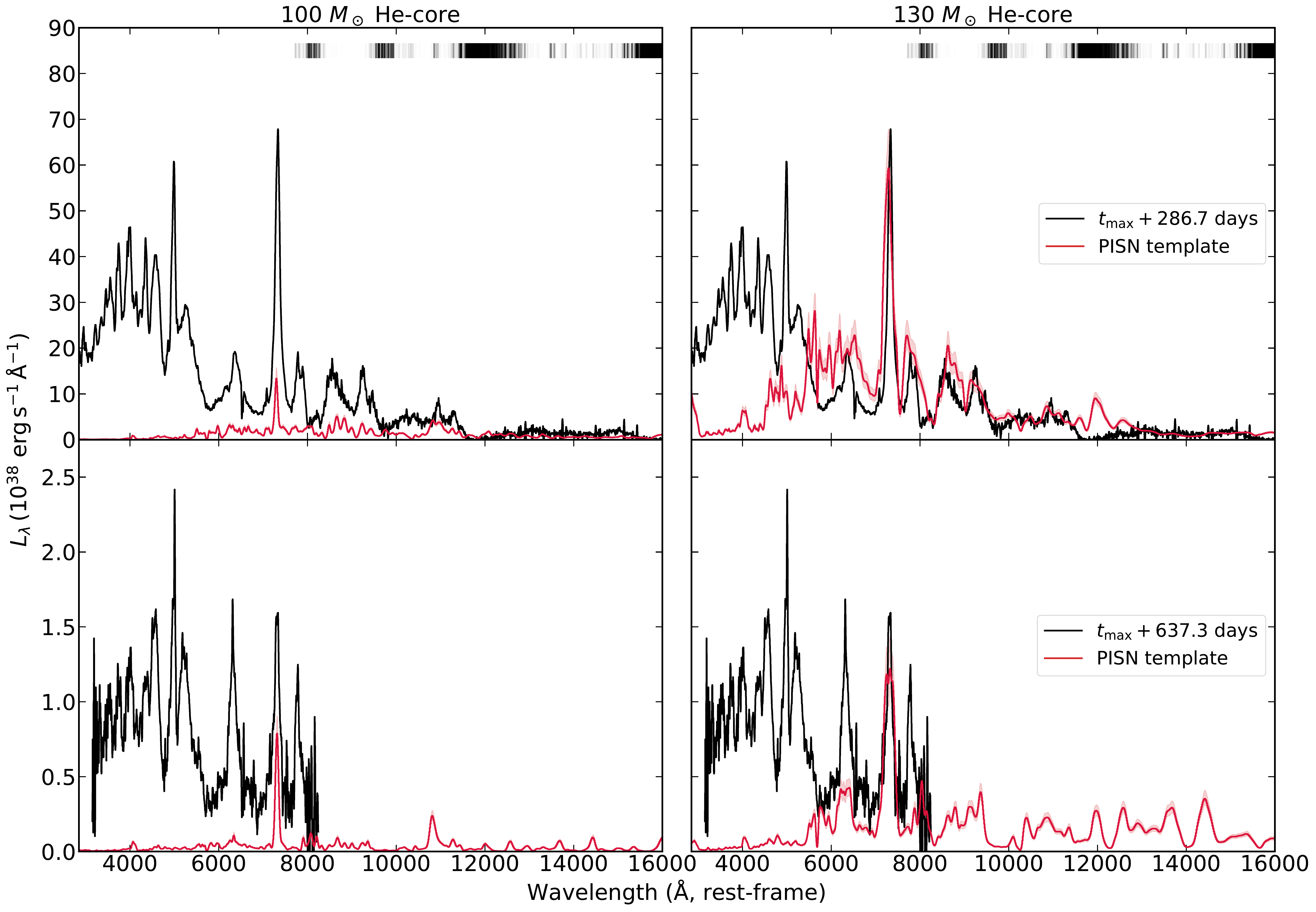}
    \caption{Late-time spectra of \sn\ at 287 and 637 days after its maximum. Overlaid are the computed PISN spectra from \citet{Jerkstrand2016a} scaled to these epochs. The shaded region indicates the uncertainty of the explosion time. The He130 model provides an adequate description of the emission redwards of 6000 \AA\ at \tmax+286.7~days, but a worse match for the second epoch. The observed spectra show a considerable excess at shorter wavelengths that is not expected from the model spectra. We argue that the blue excess is due to the interaction of the SN ejecta with circumstellar material, which is not included in existing PISN models. The He100 model matches the observation of neither epoch. The vertical bars at the top of each panel indicate the location of telluric features. }
    \label{fig:spec:pisn_comparison}
\end{figure*}

\begin{figure}
    \centering
    \includegraphics[width=1\columnwidth]{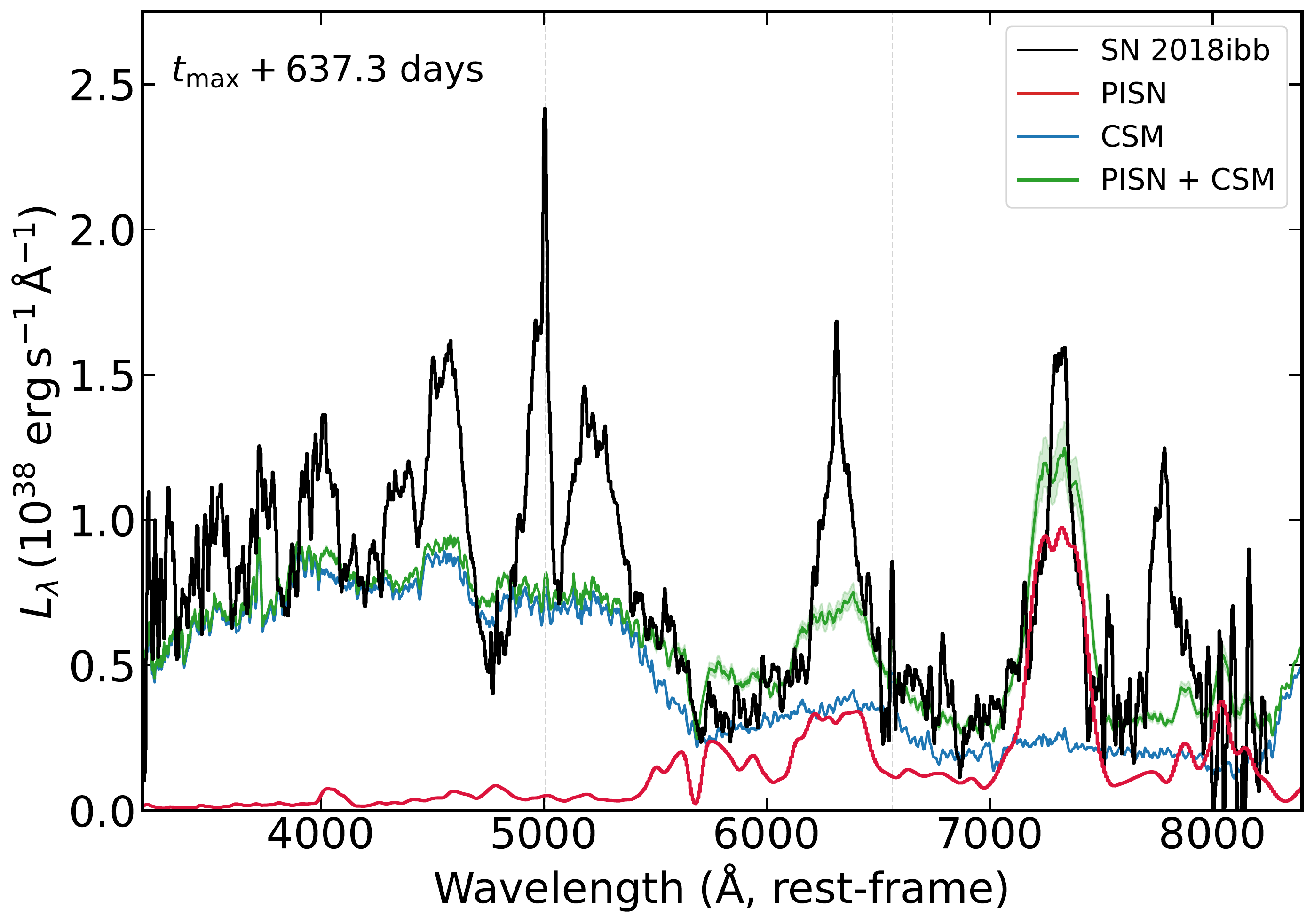}
    \caption{Nebular spectrum of \sn\ (black) at \tmax+637.3~days and its decomposition into a CSM interaction (blue) and PISN component (red). This decomposition reveals that the shape of the spectrum in the blue is similar to the pseudo-continuum seen in interaction-powered SNe (Type Ia-CSM, Ibn, Icn and IIn).    The emission lines in the blue arise either from CSM interaction or from material in the CSM shell excited by the SN light. The dotted vertical lines indicate the location of strong galaxy emission lines.
    }
    \label{fig:spec:decomposition}
\end{figure}

The templates with $M({\rm Ni}) \sim 25~M_\odot$ (He120 and P250Ni25) also provide reasonable matches to the data, even though they exhibit a faster rise and produce peak luminosities that are $\approx0.3$~dex fainter. In the case that CSM interaction contributes at all times, the apparent tension might be alleviated. The PISN model He100, which produces the smallest amount of Ni in our set, generates a light curve that is incompatible with observations. The peak bolometric luminosity of such a model is 0.8 dex lower compared to the observed value, implying that a different energy source must account for $>84\%$ of the observed peak luminosity.

\subsubsection{Late-time spectra of \sn\ compared to PISN models}\label{sec:spectra_discussion}

\citet{Jerkstrand2016a} computed spectra of the He100 [$M({\rm Ni})=5.8~M_\odot$] and He130 [$M({\rm Ni})=44~M_\odot$] PISN models at 400 and 700 days after the explosion.\footnote{The model spectra of the P250 templates will be presented in a forthcoming paper by Kozyreva et al. (in prep.).} To compare these spectra with the observations, we need to constrain the poorly measured explosion date of \sn. The bolometric light curve of the He100 and He130 models peaked at $\approx130$~rest-frame days. Assuming that \sn's bolometric light curve peaked up to 20 days before the peak in the $g$|$r$ band, we can scale the computed spectra to the epochs of the observed spectra via $\exp\left(\Delta t/t_{1/2,\rm Co}\right)$, where $\Delta t$ is the phase difference between the observed and computed spectra and $t_{1/2,\rm Co}$ is the mean lifetime of $^{56}$Co.

Figure \ref{fig:spec:pisn_comparison} shows the observed spectra of \sn\ at \tmax+286.7 (top row) and \tmax+637.3~days (bottom row) in black. The upper left and the bottom left panels compare the earlier and the later spectra with the phase-adjusted He100 model (red) at \tmax+400~days and \tmax+700~days, respectively. The right column presents the same comparison to the He130 model spectra. The phase-adjusted spectra have a shaded band to indicate the impact of the uncertain peak time of the bolometric light curve on the model flux. We selected the observed spectra at these specific epochs to minimise the phase correction and cover a wide wavelength range.

The He100 model fails to match the spectra of \sn\ at both \tmax+286.7 and \tmax+637.3~days. The predicted emission lines are significantly weaker compared to the data, and the relative strength of the features does not match the shape of the observed spectra. In addition, the model spectra also exhibit lines that are significantly narrower compared to the observed spectra since the He100 model yields a lower ejecta velocity (Table \ref{tab:pisn_properties}). The He130 model provides a better match. At \tmax+286.7~days, the model spectrum describes the observed spectrum redwards of 6000~\AA\ well, in terms of the absolute and relative strength of the features as well as the line widths. The computed spectrum also matches the observed NIR spectrum, albeit the strongest predicted feature at 1.2~$\mu$m (\ion{Fe}{i} and \ion{Si}{i}) is redshifted to a region that is strongly affected by atmospheric absorption (indicated by the black-shaded region in the upper half of the figure). The match at \tmax+637.3~days appears to be less plausible compared to that at \tmax+286.7~days. While the model reproduces the [\ion{O}{ii}]+[\ion{Ca}{ii}] at 7300~\AA, the observed spectrum shows an elevated continuum level and stronger [\ion{O}{i}]\,$\lambda\lambda$\,6300, 6364 in emission. The observed spectrum also shows prominent \ion{O}{i}\,$\lambda$\,7773 in emission that is not generated by the model. However, in Section \ref{sec:csm:oiii_oii} we showed that \ion{O}{i}\,$\lambda$\,7773 is produced by the CSM interaction.

Bluewards of 6000~\AA, the discrepancy between the observed and computed spectra is considerable in both epochs. A similar excess in the blue part of the spectrum was observed in other slow-evolving SLSNe \citep[e.g.][]{Jerkstrand2017a}, and it was used as a critical piece of evidence against the PISN interpretation \citep[e.g.][]{Dessart2013a, Nicholl2013a}. \citet{Mazzali2019a} performed detailed modelling of a nebular phase spectrum of the candidate PISN 2007bi and confirmed these conclusion from \citet{Dessart2013a}, \citet{Nicholl2013a} and \citet{Jerkstrand2017a}. In addition, these authors put forward the idea that if the core of the explosion was mixed, Ca, Mg and O would become efficient coolants and would produce conspicuous emission lines at 4570~\AA\ (Mg), 6300,6363~\AA\ (O), and 7291, 7324~\AA\ (Ca). However, PISN calculations do not support the notion of mixing in the inner ejecta \citep{Mazzali2019a, Chen2020a}.

This blue excess is not a critical piece of evidence against the PISN interpretation for \sn. In Sections \ref{sec:csm:shell}, \ref{sec:csm:spectrum} and \ref{sec:csm:oiii_oii}, we showed that \sn\ is not solely powered by $^{56}$Ni. \sn's progenitor had an eruptive mass-loss episode shortly before the explosion. The interaction of the SN ejecta with CSM contributes to the observed light curve via discrete emission lines, and it could even produce a blue pseudo-continuum similar to that seen in interaction-powered SNe \citep{Silverman2013a, Hosseinzadeh2017a, Perley2022a}\footnote{The pseudo-continuum in Type Ibn SNe is the product of the blending of thousands of iron emission lines \citep[e.g.][]{Dessart2022a}.}. This raises the questions of whether the blue excess in \sn\ is similar to that seen in interaction-powered SNe and how large the contribution of CSM interaction is to the bolometric light curve.

In Figure \ref{fig:spec:decomposition}, we further inspect the spectrum of \sn\ at \tmax+637~days against the phase-adjusted spectrum of the He130 model. We attempt to decompose the spectrum of \sn\ into two elements, namely a PISN and an ejecta-CSM interaction component. The CSM component is represented by a spectrum of the Type Icn SN\,2021csp obtained at $\sim52.7$~days after the explosion from \citet{Perley2022a}.\footnote{Using spectra of Type Ia-CSM or Ibn SNe would give similar results. We decided to use an Icn template because it shows no emission lines from H and He, which are present in Type Ia-CSM and Ibn SNe.} Its flux scale is scaled so that the sum of the PISN and CSM components (green) matches the shape of \sn's pseudo-continuum. This approach is similar to that in \citet{BenAmi2014a}, where these authors used a spectrum of a Type IIn SN to deduce that the ejecta of the Type Ic SN\,2010mb interacted with a large amount of H-free circumstellar material. Indeed, this toy model captures the general shape of \sn, suggesting that a considerable fraction of the flux bluewards of 6000~\AA\ is produced by the CSM interaction. Most of the emission lines in the blue and \ion{O}{i}\,$\lambda$\,7773 feature were not observed in the spectrum of SN\,2021csp. However, we have shown that some of the observed lines in \sn, for example, \ion{O}{i}\,$\lambda$\,7773,9262, [\ion{O}{ii}]\,$\lambda\lambda$\,7320,7330 and [\ion{O}{iii}]\,$\lambda\lambda$\,4959,5007, are generated by the CSM interaction. Others, such as [\ion{O}{i}]\,$\lambda\lambda$\,6300,6364 and \ion{Mg}{i}]\,$\lambda$\,4571, are likely formed in the unshocked SN ejecta or the contact discontinuity (cool-dense shell) between the SN ejecta and the CSM (Section \ref{sec:csm:oiii_oii}).

\begin{figure}
    \centering
    \includegraphics[width=1\columnwidth]{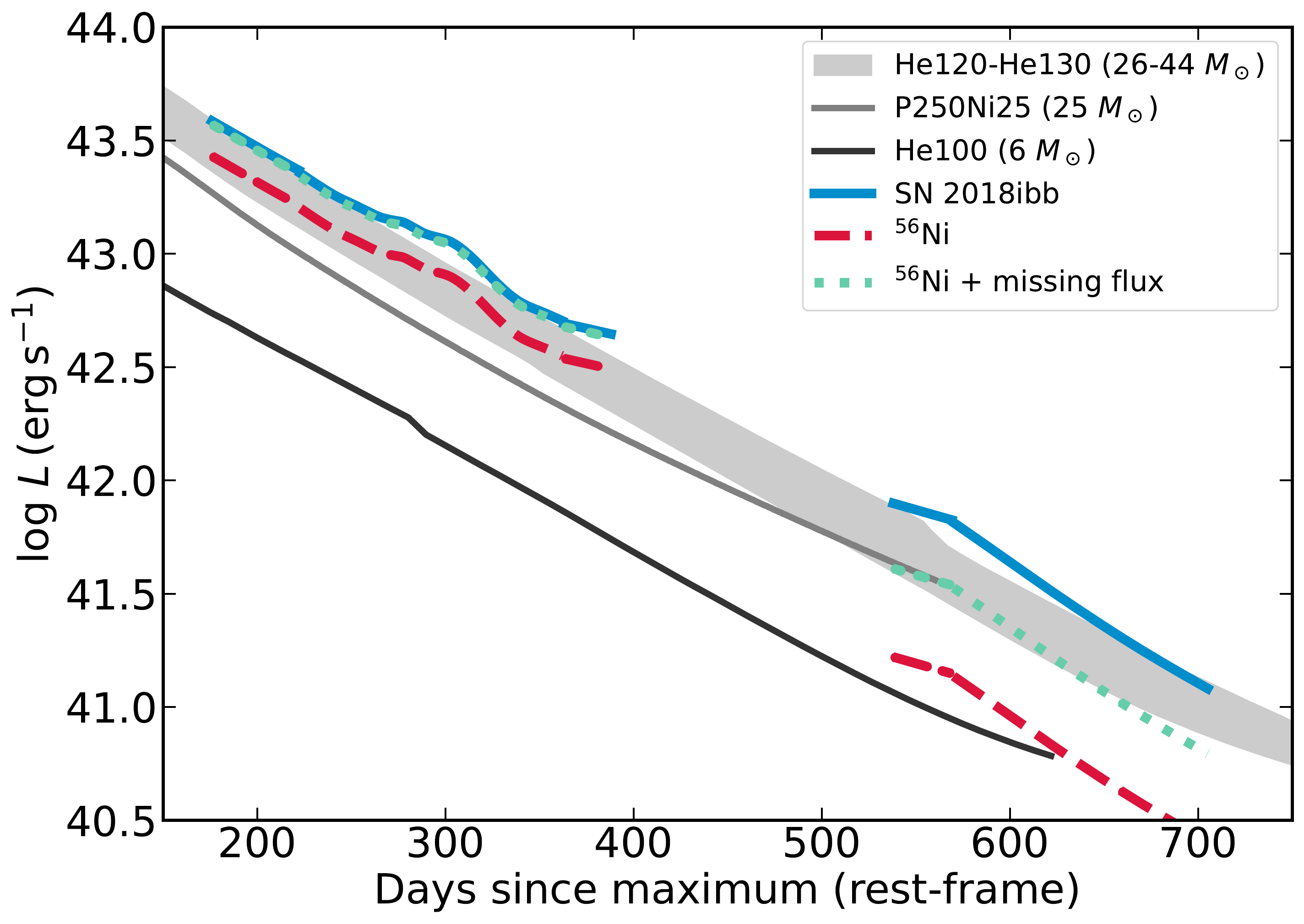}
    \caption{Observed late-time bolometric light curve (solid blue lines) and the fraction of light that could be attributed to $^{56}$Ni after accounting for CSM interaction (dashed, red). The dotted green curves show the $^{56}$Ni light curve after adding the missing IR flux (up to $5~\mu$m). The IR correction pushes the light curves back to the regime of PISN models that produce 25--$44~M_\odot$ of $^{56}$Ni. Even in the case of a substantial contribution from CSM interaction, a total amount of 25--$44~M_\odot$ of $^{56}$Ni appears to be essential to power \sn.
    }
    \label{fig:lc:pisn_csm_contribution}
\end{figure}

Assuming \sn's progenitor is similar to the He130 star model, we can roughly estimate the fractions of the observed bolometric flux that have been produced by the nickel decay and the CSM interaction. The bolometric luminosity calculated at \tmax+286.7~days covers the wavelength range from 3020 to 14,250~\AA\ (rest-frame). The phase-adjusted model spectrum from \citet{Jerkstrand2016a} accounts for 70\% of the bolometric flux. In other words, the nickel-powered light curve would be 0.1 dex fainter than the observed bolometric light curve. At \tmax+637.3~days, the observed bolometric luminosity covers the range from 3930~\AA\ to 8500~\AA\ (rest-frame). The phase-adjusted PISN spectrum accounts for only 21\% of the observed bolometric flux. This means that the Ni-powered light curve would be 0.7 dex fainter. To illustrate that, we show in Figure \ref{fig:lc:pisn_csm_contribution} the observed bolometric light curve (solid blue lines) and the fraction of the observed bolometric light curve that can be attributed to the He130 model (dashed red lines).

However, there is a critical detail that we need to take into account before drawing a conclusion. The bolometric light curves of the P250Ni25|34 and He100--He130 models extend to 50,000~\AA. Therefore, our observed bolometric light curve could miss a substantial fraction of the true bolometric flux. The \citet{Jerkstrand2016a} model spectra cover the wavelength range from the far UV to 25,000~\AA, and figure 13 in \citet{Jerkstrand2016a} shows the fraction of light emitted between 25,000~\AA\ and 50,000~\AA, allowing us to estimate the missing IR fractions. At \tmax+286.7~days, the missing IR fraction is 0.14~dex. This fraction increases to $\sim0.39$~dex at \tmax+637.3~days, due to the shorter wavelength coverage of the observed bolometric light curve and an increased mid-IR contribution from the PISN model. The dotted green line in Figure \ref{fig:lc:pisn_csm_contribution} shows the estimated Ni-powered light curve of \sn\ after correcting for the missing IR flux. The IR correction fortuitously compensates for most of the observed bolometric flux lost to CSM interaction, corroborating that even with significant CSM interaction a total mass of 25--44~M$_{\odot}$ of $^{56}$Ni is still needed to power the light curve and spectra.

A progressively increasing contribution from CSM interaction to the bolometric flux is not a contrived scenario. If the shock is radiative, as is expected for a high metallicity and dense CSM, then the luminosity from the shock is $\sim \dot M \Delta v_{\rm rel}^3/2$, where $\Delta v_{\rm rel}$ is the relative velocity of the ejecta and the CSM. If the density gradient, $n$, of the ejecta is steep, the shock velocity is only slowly decreasing, $v_{\rm s} \propto t^{-1/(n-s)}$ for a CSM with $r^{-s}$ density profile, and the shock luminosity will only be a slowly decreasing function of time \citep{Chevalier2017a}. Because the radioactive input decreases exponentially, it is expected that the shock contribution will increase relative to the radioactively powered input. 

\subsubsection{[\ion{Co}{ii}]\,$\lambda$\,1.025~$\mu$m}\label{disc:CoII}

\begin{figure}
    \centering
    \includegraphics[width=1\columnwidth]{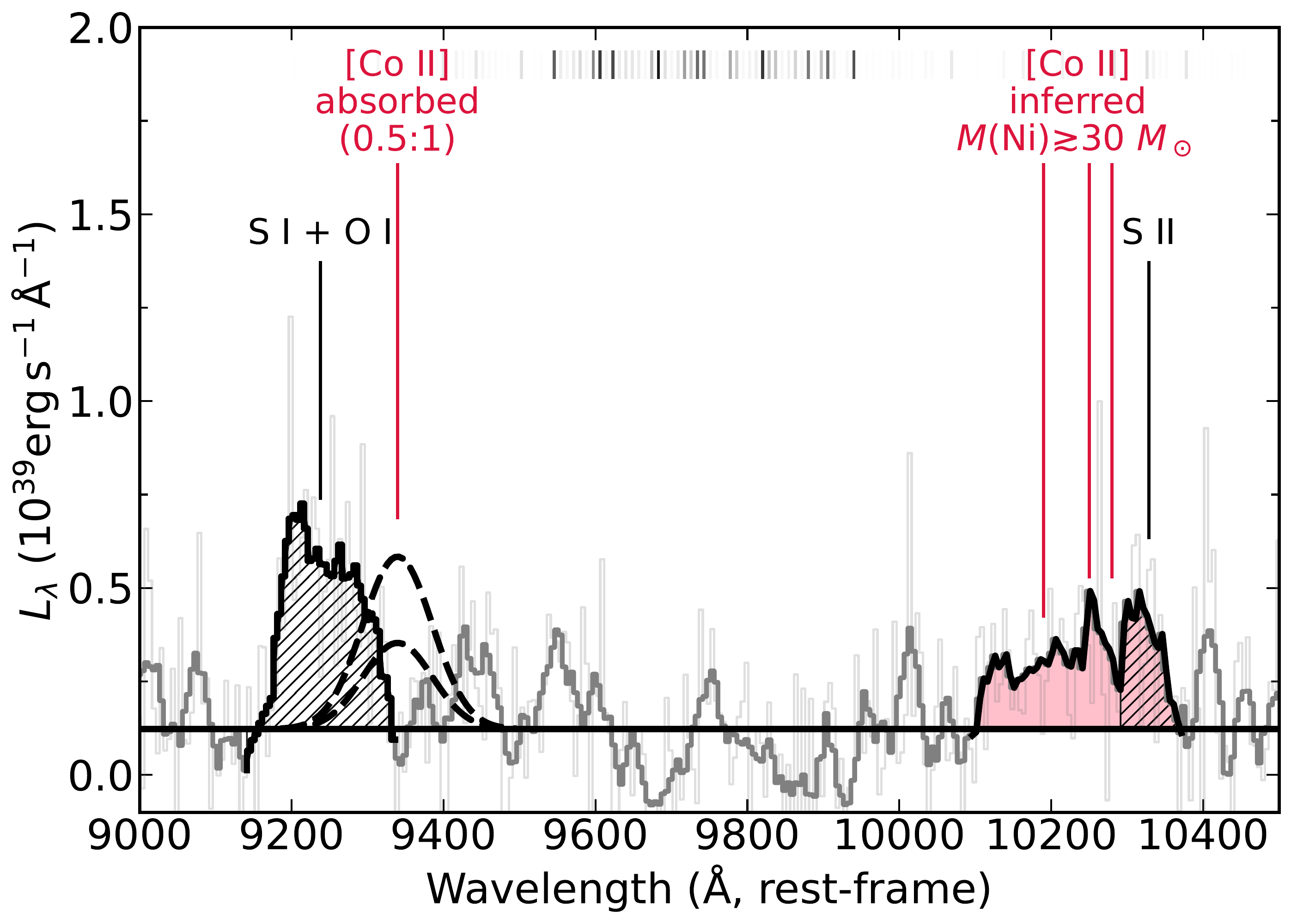}
    \caption{Zoom-in of the region from 9000~\AA\ to 10,500~\AA\ at \tmax+377.5~days. Cobalt has its strongest feature at 1.025~$\mu$m. Its tentative detection translates to $^{56}$Ni mass of $\gtrsim30~M_\odot$, consistent with the light curve modelling. The second strongest cobalt feature is at 9340~\AA. Its location is indicated by a fiducial Gaussian centred 9340~\AA. Its integrated luminosity is expected to be between 50 and 100\% of [\ion{Co}{ii}]\,$\lambda$\,1.025~$\mu$m. The absence of [\ion{Co}{ii}]$\lambda$\,9340 is not an argument against the discovery of [\ion{Co}{ii}]\,$\lambda$\,1.025~$\mu$m (see Section \ref{disc:CoII} for details).
    Lines from other elements that could blend with the [\ion{Co}{ii}] lines are marked. Regions of strong atmospheric absorption are indicated by vertical bars at the top of the figure.
    }
    \label{fig:spec:CoII}
\end{figure}

The NIR spectra of \sn\ after \tmax+300~days reveal an emission line at 1.025~$\mu$m that we interpret as [\ion{Co}{ii}] (a triplet of individual lines at 1.019, 1.025 and 1.028~$\mu$m, which result from the 9-1, 10-2, and 11-3 transitions as sorted from higher to lower energies, respectively; Figure \ref{fig:spec:CoII}). The line luminosities are $(2.9\pm0.8)\times10^{40}~\rm erg\,s^{-1}$ and $(5.4\pm1.4)\times10^{40}~\rm erg\,s^{-1}$ at \tmax+352.6 and \tmax+377.5~days, respectively. Assuming optically thin LTE, we can convert the line luminosity to a (temperature-dependent) \ion{Co}{ii} mass. The line luminosity of the [\ion{Co}{ii}]\,$\lambda$\,$1.025~\mu$m multiplet can be written as the sum of the individual transitions
\begin{equation}
    L ({\rm\ion{Co}{ii}}) = N_9\,A_{9-1}\,E_{9-1} + N_{10}\,A_{10-2}\,E_{10-2} + N_{11}\,A_{11-3}\,E_{11-3}\nonumber ,
\end{equation}
where $N_u$ is the total number of ions in the upper state $u$, $A_{u-l}$ the transition rate for spontaneous emission from the upper state $u$ to the lower state $l$, and $E_{u-l}$ the energy level of the transition. We use \citet{Quinet1998a} for the values of the Einstein coefficients and the energy levels. The partition function can be taken as 20, with an error less than a factor of 2 for reasonable temperatures. Then, using equation 42 in \citet{Jerkstrand2017b}, the cobalt mass is given by

\begin{equation}
    M\left({\rm \ion{Co}{ii}}\right) \gtrsim 0.5~M_\odot\,\times\,\left[\frac{L\left(\rm \ion{Co}{ii}\right)}{5\times 10^{40}~\rm erg/s}\right] \,\times\,\frac{\exp{\left(15410/T\right)}}{\exp{\left(15410/5000\right)}},\nonumber
\end{equation}
where $T$ is the temperature of the ejecta, in units of $K$. The temperature factor (the ratio of exponentials) varies from 0.2 at $T=10,000$ K to 100 at $T=2000$ K. 

To calculate the initial nickel mass, we need to account for the amount that has decayed over time. The initial nickel mass is a factor of $\exp\left(t/t_{1/2,\rm Co}\right)\gtrsim\exp\left(450/111\right)\simeq60$ larger, where $t$ is the time since explosion and $t_{1/2,\rm Co}$ is the mean lifetime of  $^{56}$Co. Averaging over the line luminosities of the two epochs, the inferred $^{56}$Ni mass is $\gtrsim30~M_\odot$ if $T \approx 5000$ K. This estimate is consistent with the inferred $^{56}$Ni mass from the light-curve modelling (Sections \ref{discussion:lightcurve:katz}, \ref{discussion:lightcurve:mosfit}, \ref{discussion:lightcurve:template_matching}). However, lower values of the nickel mass would be expected if the temperature is higher ($6~M_\odot$ at $T=10,000$ K). For temperatures below $\sim3500$ K, the nickel mass becomes unphysically large, $>100~M_\odot$.  We note that [\ion{Co}{ii}]\,$\lambda$\,1.025~$\mu$m can be blended with \ion{S}{ii}\,$\lambda$\,1.032~$\mu$m\footnote{This feature consists of six lines between 1.0287 and 1.0370~$\mu$m.}, indicated by the hatched region in Figure \ref{fig:spec:CoII}.

This Ni-mass estimate assumes that the transitions are optically thin. The Sobolev optical depth of the 9-1 transition line in LTE is \citep[][ignoring stimulated emission]{Jerkstrand2017a}:

\begin{align}
\tau_{9,1} &= A_{9,1}\,\lambda_{9,1}^3\,\frac{1}{8 \pi} \frac{g_9}{g_1}\,n_1 t\nonumber\\
 &\approx 0.08 \times \left(\frac{M\left({\rm \ion{Co}{ii}}\right)[450d]}{1~M_\odot}\right) \frac{x_1}{f} ,\nonumber
\end{align}
where $\lambda_{9,1}$ is the wavelength of the emitted photon, $g_n$ is the multiplicity of the $n$th state, $n_1$ is the number density of atoms in the ground state, $x_1$ is the fraction of Co II ions in the ground state, and $f$ is the filling factor for the $^{56}$Ni zone. In LTE at 5000 K $x_1$ is $\approx 0.5$, whereas at lower temperatures and/or in NLTE $x_1$ is typically higher (towards unity). A typical CCSN has a characteristic filling factor of $f \sim 0.1$ for any given zone, which means that 0.5 $M_\odot$ of $^{56}$Co are optically thin at $\sim$ 450 days.  For SLSNe, filling factors for the oxygen zones have been derived to be $f \approx 10^{-3}-10^{-2}$ \citep{Jerkstrand2017a}. If these filling factors also hold for the $^{56}$Ni zone of \sn, then the Co II lines would be optically thick at $\sim$ 450 days, and determining a mass from them would be impossible at that time \citep{Jerkstrand2017b}.  One may note that numerical simulations of PISNe show little clumping or mixing of the inner material \citep{Chen2020a}, and the low filling factors derived for other SLSNe may be due to mixing from the central engine \citep[e.g.][]{Suzuki2021a} or compression by circumstellar interaction \citep{vanMarle2010a}.

In the stripped-envelope-supernova models from \citet{Jerkstrand2015a}, [\ion{Co}{ii}]\,$\lambda$\,1.025~$\mu$m is the strongest predicted line from cobalt. The second strongest Co feature is a blend of two lines at 9338 and 9344~\AA. This [\ion{Co}{ii}] feature could be blended with the red wing of the \ion{O}{i}\,$\lambda$\,9263 recombination line. In optically thin LTE, the expected line ratio between [\ion{Co}{ii}]\,$\lambda$\,9340 and [\ion{Co}{ii}]\,$\lambda$\,1.025~$\mu$m is between $0.5-1$ for a wide range of plausible temperatures. To examine whether [\ion{Co}{ii}]\,$\lambda$\,9340 could be present in the spectrum at \tmax+387~days, we show in Figure \ref{fig:spec:CoII} a Gaussian centred at 9340~\AA\ that has either the same integrated luminosity as [\ion{Co}{ii}]\,$\lambda$\,1.025~$\mu$m or a luminosity that is 50\% smaller. Clearly, [\ion{Co}{ii}]\,$\lambda$\,9340 is not present in our data at those luminosities.

In the He130 model of \citet{Jerkstrand2016a} at 400~days after the explosion, neither of the [\ion{Co}{ii}] lines are present in any significant strength as they are absorbed by line blocking extending into the NIR. Under conditions with less line blocking (as in the \citealt{Jerkstrand2015a} CCSN models), the [\ion{Co}{ii}]\,$\lambda$\,1.025~$\mu$m line can still be visible, also as iron has few strong emission lines around this particular wavelength. The same cannot be said about the 9340~\AA\ region where iron is stronger. In a PISN ejecta, the densities are about 100-times higher for a given epoch, and the NIR region is still largely opaque at 400 days after the explosion. To explain the observed [\ion{Co}{ii}]\,$\lambda$\,1.025~$\mu$m line but the absence of the [\ion{Co}{ii}]\,$\lambda$\,9340 feature, we need to call upon absorption of the 9340~\AA\ line but not the 1.025-$\mu$m line. In other words, the He130 model reproduces the spectral shape near 9340~\AA~but not at 1.025 $\mu$m if \sn\ is a PISN. 

While the association of [\ion{Co}{ii}]\,$\lambda$\,1.025~$\mu$m could be the smoking gun signature that \sn\ is a PISN, we caution that the interpretation hinges on the detection of a single line. An IR spectrum with NIRSpec \citep{Jakobsen2022a} aboard the \textit{James Webb Space Telescope} could resolve such ambiguity. It is the only instrument that can provide an uncensored view from 1 to $5~\mu$m. Such a spectrum could reveal, for instance, Co, Fe and Ni lines at $>2.7~\mu$m as seen in the Type Ia SN 2021aefx \citep{Kwok2023a}.

\subsection{Velocity evolution}\label{discussion:velocity_evolution}

In Section \ref{sec:velocities}, we showed that the photospheric velocity is $\approx8500~\kms$ and remained constant between \tmax\ and \tmax+100~days. Here, we contrast our observations with predictions of PISN models.

From the observational point of view, the photospheric velocity is defined via the minimum of the \ion{Fe}{ii}\,$\lambda\lambda$\,4924,5018,5169 P-Cygni profiles, which serve as a good proxy of the photospheric velocity \citep{Dessart2005a}. In contrast to that, there is no specific location of the photosphere from the radiative transfer point of view, because its location is wavelength dependent. An average photosphere could be estimated, for instance, via the Rosseland mean optical depth \citep{Rybicki1986a} and electron scattering.

The photospheric velocity of the light curves simulated with \program{STELLA} is defined as the velocity of the mass shell where the integrated optical depth in the $B$ band is equal to 2/3. The integrated optical depth accounts for the total opacity, which includes contributions from continuum opacity (photoionisation, free-free absorption, and electron scattering processes assuming local thermodynamical equilibrium in the plasma) and line interactions. The location of the photosphere defined with this method coincides to a large extent with the location of the electron-scattering photosphere \citep{Kozyreva2022a} and permits a direct comparison with our observations.

\begin{figure}
    \centering
    \includegraphics[width=1\columnwidth]{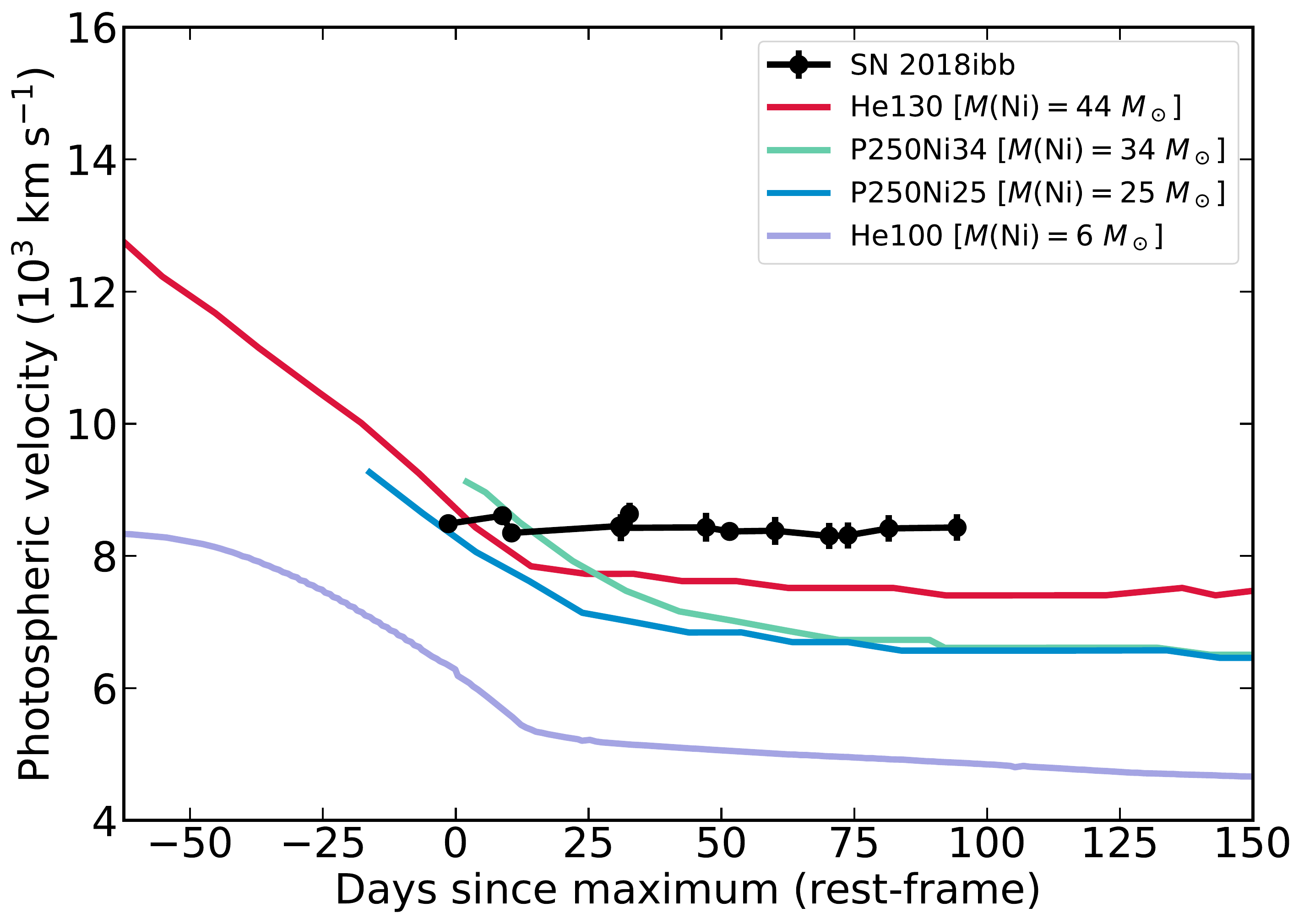}
    \caption{
    Photospheric velocity evolution of \sn\ (probed with \ion{Fe}{ii}) and the predicted evolution from different PISN models. The synthetic velocities of the P250Ni25|34 and He130 models are very similar to the observed velocities and velocity evolution. These are also the models that match the observed bolometric light curve well, in particular the He130 model.
    }
    \label{fig:spec:pisn_vel_comparison}
\end{figure}

The velocity evolution of the P250Ni25, P250Ni34, He100, and He130 PISN models together with the observed \ion{Fe}{ii} velocities are shown in Figure \ref{fig:spec:pisn_vel_comparison}. The match between the observed and predicted velocity tracks of the P250Ni25|34 and He130 models is very good in both the predicted velocities and the velocity evolution, in particular the He130 model. Small deviations exist in the absolute values of the velocities (He130: $|v_{\rm obs} - v_{\rm predicted}| \lesssim 1000~\kms$ and P250Ni25|34: $|v_{\rm obs} - v_{\rm predicted}| \sim +2000~\kms$) but not the velocity evolution. These discrepancies are in part due to differences in the definitions of the synthetic velocity and the physical model used in \program{STELLA}. Furthermore, none of the parameters of the PISN models were tuned to fit the observations. The velocity track of the He100 model is inconsistent with the observations ($|v_{\rm obs} - v_{\rm predicted}| \gtrsim +3000~\kms$). As shown in Section \ref{discussion:lightcurve:template_matching}, the predicted bolometric light curve of this model also does not match the observations.

\subsection{Comparison to other slow-evolving SLSNe}\label{discussion:comparison_w_slow_slsn}

Among the $\gtrsim200$ H-poor SLSNe known, only eight objects belong to such a phenomenological subclass of slow-evolving SLSNe: SN\,1999as \citep{Hatano2001a}, SN\,2007bi \citep{GalYam2009a}, PS1-11ap \citep{McCrum2014a}, PTF12dam \citep{Nicholl2013a}, LSQ14an \citep{Inserra2016a}, PS1-14bj \citep{Lunnan2016a}, SN\,2015bn \citep{Nicholl2016a} and \sn.\footnote{The ZTF-I SLSN sample contains a further possible slow-evolving SLSN. SN\,2018lzx has a rise ($t_{1/e,\rm rise}=60.5^{+8.2}_{-6.8}$~days) and a decline ($t_{1/e,\rm decline}=108.8^{+10.0}_{-13.2}$~days) time scale comparable to \sn\ (Table \ref{tab:lc_prop}). The peak absolute magnitude is 0.2 mag brighter than that of \sn\ \citep{Chen2023a}. Owing to its high redshift of $z=0.44$, the light curve spans a short time interval before SN\,2018lxz faded below the detection threshold, and the quality of the spectra is significantly lower compared to that of \sn. Therefore, we exclude this SLSN from the comparison.} In the following sections, we compare the photometric and spectroscopic properties of \sn\ to those of the historical slow-evolving SLSNe to comprehensively examine its exceptional properties. We omit SN\,1999as from this analysis because its light curve and spectra were never published.

We utilise the multi-band and bolometric light curves and host-subtracted spectra of LSQ14an presented in \citet{Inserra2017a} and \citet{Jerkstrand2017a}, PS1-11ap from \citet{McCrum2014a}, PS1-14bj from \citet{Lunnan2016a}, PTF12dam from \citet{Nicholl2013a}, \citet{Chen2015a} and \citet{Quimby2018a}, SN\,2007bi from  \citet{GalYam2009a}, \citet{Young2010a} and \citet{Jerkstrand2017a}, and SN\,2015bn from \citet{Nicholl2016a, Nicholl2016b, Nicholl2018a} and \citet{Jerkstrand2017a}. Furthermore, we use the \ion{Fe}{ii} velocity measurements from \citet{Liu2017a} and \citet{Lunnan2016a}.
All light curves and spectra were corrected for MW extinction. The spectrum of SN\,2015bn in \citet{Nicholl2018a} is not corrected for any host contribution. In Appendix \ref{app:SN2015bn}, we describe our approach to subtract the host contamination for SN\,2015bn.

\begin{table}
    \caption{Light-curve properties of slow-evolving SLSNe}
    \label{tab:slow_slsne}
    \footnotesize
    \centering
    \begin{tabular}{cccccc}
    \toprule
                & Redshift & $t_{1/e, \rm rise}$ & $t_{1/e, \rm decline}$ & $M_{g,\rm peak}$ & $(g-r)_{\rm peak}$ \\
                &          & (day)                 & (day)                 & (mag)            & (mag)              \\
    \midrule
    \sn           & 0.166  & 68                    & 102                   & $-21.8$          & $-0.12$            \\
    \midrule
    LSQ14an$^1$   & 0.163  & \nodata               & $\sim100$             & $<-20.8$         & $-0.21$            \\
    PS1-11ap$^2$  & 0.524  & $<25$                 & 38                    & $-21.8$          & \nodata            \\
    PS1-14bj$^2$  & 0.521  & $83$                  & 130                   & $-20.6$          & \nodata            \\
    PTF12dam      & 0.107  & $50$                  & 56                    & $<-21.7$         & $-0.20$            \\
    SN\,2007bi$^3$& 0.128  & $<23$                 & $<77$                 & $-21.3$          & 0                  \\
    SN\,2015bn    & 0.114  & $<31$                 & 56                    & $-22.0$          & $-0.17$            \\
    \midrule
    ZTF SLSNe   &          & $29$                  & 43                    & $-21.5$          & $-0.12$            \\
    \bottomrule
    \end{tabular}
    \tablefoot{The peak absolute magnitudes and the colours at $t_{\rm peak}$ are k-corrected and corrected for MW extinction. The uncertainty on the time scales is of the order of a few days. The peak magnitudes and colours have a statistical error of $\approx 0.1$~mag. The row `ZTF SLSNe' summarises the median values of the homogeneous ZTF SLSN sample \citep{Chen2023a}. \citet{Chen2023a} do not report the median rise and decline $1/e$ time-scale. We inferred that value from their median $1/10$ rise time-scale and the relationship between the rise and decline time scales reported in that paper. \\
    \tablefoottext{1}{The light curve of LSQ14an covers only the post-peak evolution.}\\
    \tablefoottext{2}{The redshifts of PS1-11ap and PS1-14bj are so high that the rest-frame $g-r$ colour cannot be inferred from spectra published in \citet{Lunnan2014a} and \citet{McCrum2014a}.}\\
    \tablefoottext{3}{The light curve of SN\,2007bi is only well-observed in $R$ band. We use the $R$-band data as an upper limit on the $g$-band time scale.}
    }
\end{table}

\subsubsection{Light curves}

Following the methodology of \citet{Chen2023a}, we measure for each slow-evolving SLSN the k-corrected peak absolute magnitude in the $g$ band, the k-corrected rest-frame $g-r$ colour, and the $1/e$ rise and decline time-scales of the $g$-band light curves.\footnote{Owing to the high redshift of PS1-11ap and PS1-14bj, we use their $i$-band light curves, which probe a rest-frame wavelength interval similar to that of the $g$ band of \sn.} All measurements are summarised in Table \ref{tab:slow_slsne}. We also report in that table the measurements of \sn\ and, for a broader comparison, the median values of the homogeneous ZTF SLSN sample \citep{Chen2023a}. 

Slow-evolving SLSNe, including \sn, have peak absolute magnitudes between $\sim-20.8$ and $-22$~mag in the $g$ band and k-corrected $g-r$ colours between $-0.2$ and 0 at peak (Table \ref{tab:slow_slsne}). Both their absolute peak magnitudes and the peak $g-r$-colours are comparable to the median values of the ZTF SLSN sample (median values being $M_{g,\rm peak}-21.5$~mag and $g-r=-0.12$~mag; Table \ref{tab:slow_slsne}). The rising parts of the light curves of the historical slow-evolving SLSNe are not well sampled, limiting the comparison with \sn\ and the ZTF SLSN-I sample. Only PS1-14bj has a rise time that is at least as long as that of \sn\ and even 30 days longer than that of \sn. The decline time scales of the historical slow-evolving SLSNe are well measured. They vary between 38 and 130 days, placing those events above the average of the ZTF-I sample (Table \ref{tab:slow_slsne}). Yet, only one historical slow-evolving SLSN had a decline time-scale as extreme as \sn. With a decline time scale of 130 days, PS1-14bj evolves even slower than \sn\, but its peak luminosity in the rest-frame $g$-band was 1.2~mag fainter than that of \sn. This makes \sn\ an unprecedented case even among the most extreme SLSNe known. LSQ14an also has a decline time scale of 100~days, but the observed light curve only covers the declining light curve, adding an unknown systematic error to its time scale measurement.

\begin{figure}[t]
    \centering
    \includegraphics[width=1\columnwidth]{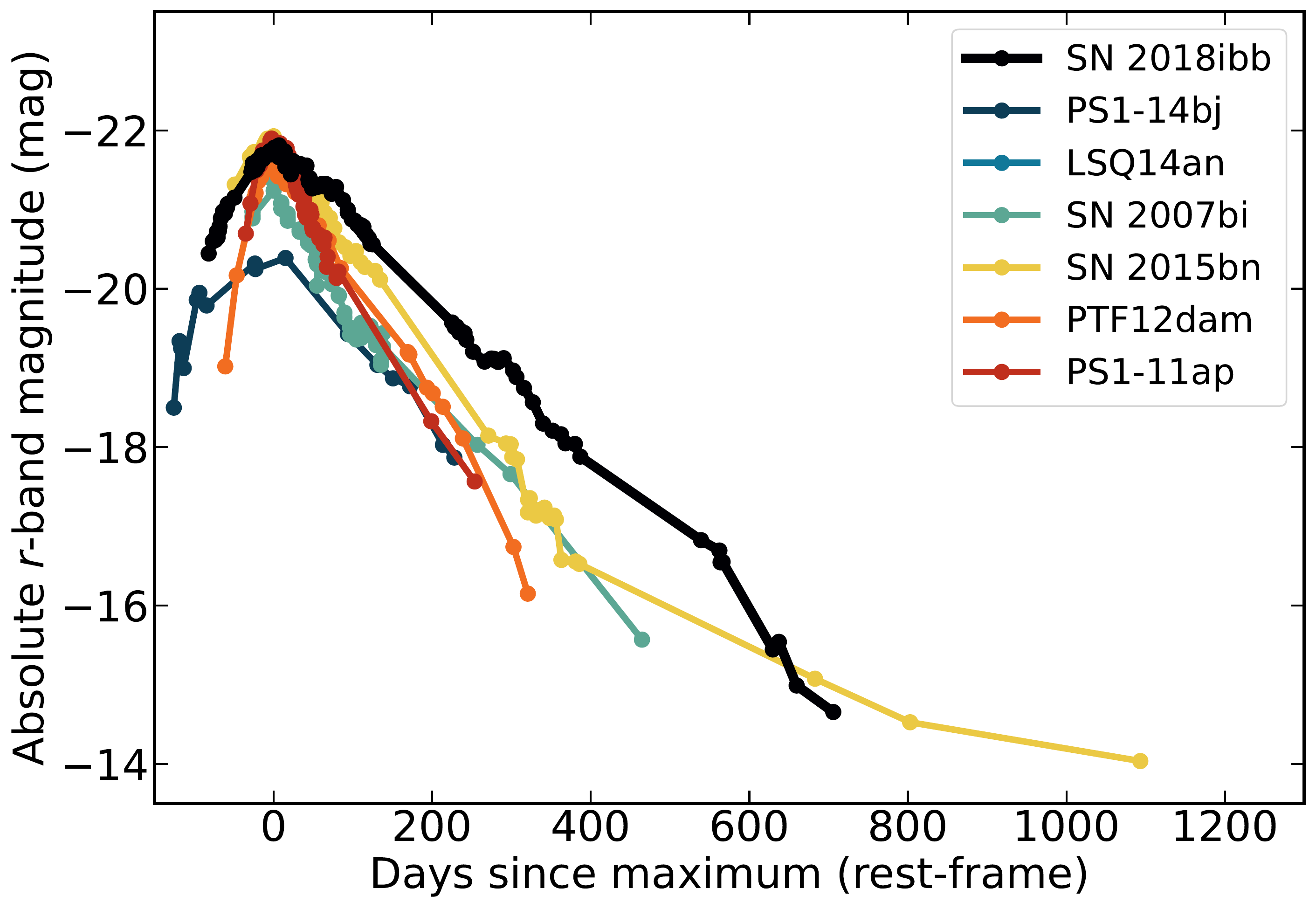}
    \caption{\sn\ in the context of the phenomenological sub-class of slow-evolving SLSNe. Even among this rare sub-class of SLSNe, \sn\, with its exceptionally broad light curve and high peak luminosity is an extreme object with unprecedented properties.
    }
    \label{fig:lc:comparison_slow_slsn}
\end{figure}

Figure \ref{fig:lc:comparison_slow_slsn} shows the $r$-band absolute magnitude light curves of all slow-evolving SLSNe. The supernovae 2015bn and 2018ibb are the only SLSNe with observations extending beyond 500 rest-frame days after maximum light. The light curve of SN\,2015bn faded much faster than that of \sn. At about 400~days after peak, the decline slowed down and became very gradual. In contrast, \sn's light curve faded linearly with a decline slope of $\sim1.1~\rm mag\,(100~days)^{-1}$ that steepened to $\sim1.5~\rm mag\,(100~days)^{-1}$ at 500 days after maximum. These differences translate into differences in the powering mechanisms. Magnetars lose their rotational energy efficiently through dipole radiation, which scales as $\dot{E}_{\rm rot}\propto t^{-2}$. The energy deposition (and henceforth the SN luminosity) evolves as a power law. Therefore, the light curve is expected to flatten at later times (in the time vs. magnitude space). Radioactive material has an exponentially declining energy deposition rate, which results in a linear decline in the time vs. magnitude space. The loss of $\gamma$-ray trapping accelerates the fading independently of the powering mechanism, but it only modifies the light curve without altering its general shape, i.e. the loss of gamma-ray trapping cannot convert a power-law decline into an exponential decline \citep[e.g.][]{Chen2015a, Wang2015a, Nicholl2018a}. Therefore, the power-law-shaped decline of SN\,2015bn could point to magnetar powering, as concluded in \citet{Nicholl2018a}. In return, \sn's continued linear decline excludes powering by a magnetar.

\begin{figure}[t]
    \centering
    \includegraphics[width=1\columnwidth]{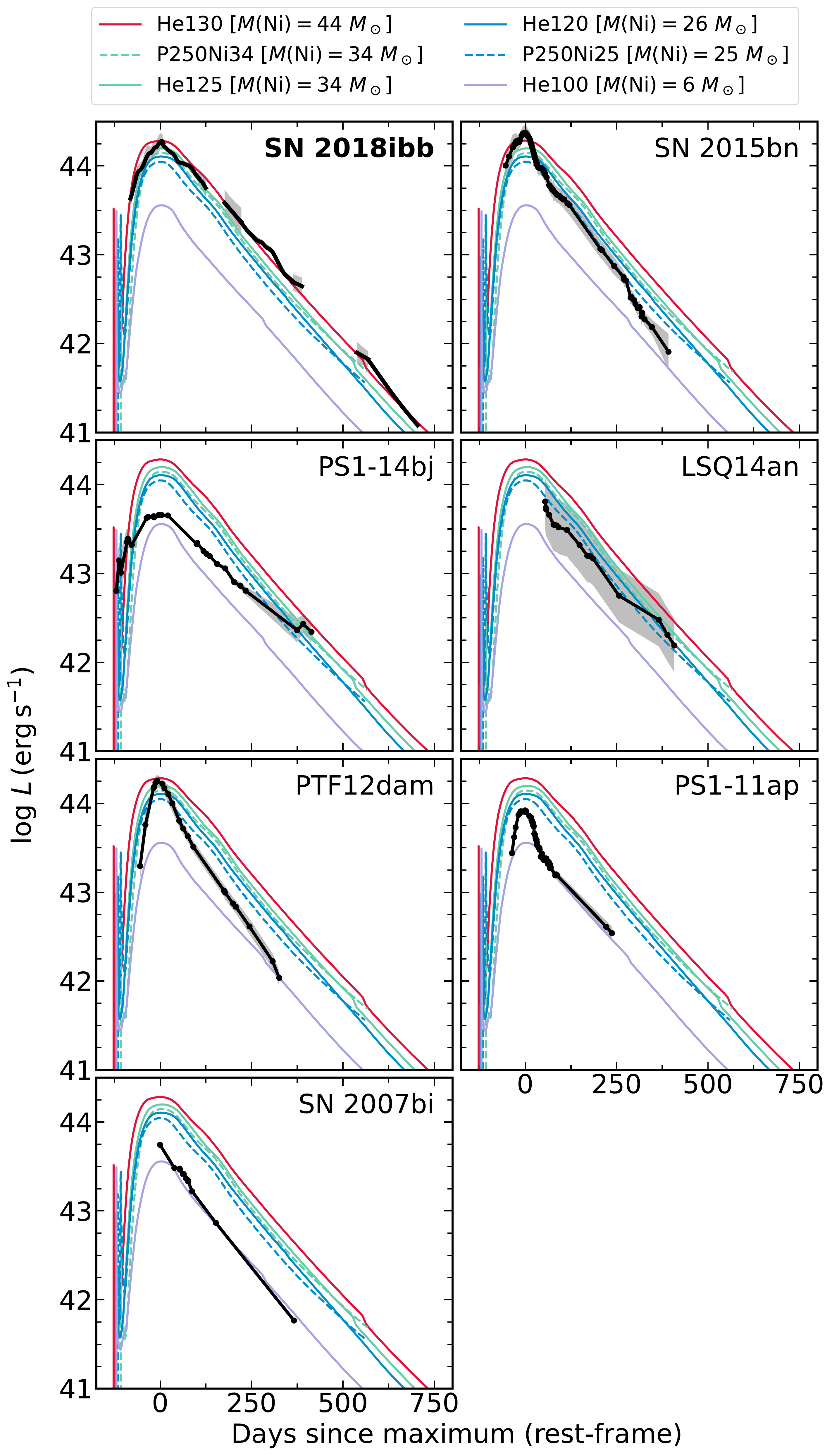}
    \caption{The bolometric light curves of \sn\ and historical slow-evolving SLSNe in the context of PISN models with nickel masses between 5 and $44~M_\odot$. \sn\ is the only SLSN whose entire light curve from \tmax$-93$ to \tmax+706~days is consistent with PISN templates. The other SLSNe have either too fast declining light curves, light curve shapes inconsistent with PISN models, or their light curves are poorly sampled, hindering a comparison with PISN templates. The grey-shaded region indicates the $1\sigma$ uncertainty.}
    \label{fig:lc:slow_slsn_pisn_comparison}
\end{figure}

\cite{Nicholl2017a} fitted the multi-band data of the slow-evolving SLSNe with the \program{slsn} magnetar model in \program{MOSFiT}. This model provides an adequate description of the observations, even to the data of SN\,2015bn at 1000~rest-frame days after maximum \citep{Nicholl2018a}. The best-fit parameters cover the range from 5.3 to $14~M_\odot$ for the ejecta mass $M_{\rm ej}$ (median being $6.3~M_\odot$), 0.1 to $0.8\times10^{14}$~G for the orthogonal component of the magnetic field strength $B$ (median being $0.3\times10^{14}$~G), and 2.3 to 3.9~ms for the initial spin period $P_0$ (median being 2.8~ms). These values are typical for SLSN light curves fitted with that particular magnetar model \citep{Nicholl2017a, Chen2023b}. \sn\ has starkly different values (Section \ref{discussion:lightcurve:mosfit}; Table \ref{tab:mosfit}). The best fit requires a magnetar with an initial spin period of 1~ms and an ejecta mass of $86~M_\odot$, to squeeze out as much energy as possible from the magnetar model. As alluded to in Section \ref{discussion:lightcurve:mosfit}, such massive stars do not have neutron star remnants. Furthermore, the magnetar model overpredicts the late-time flux significantly due to its power-law-shaped energy deposition.

In Figure \ref{fig:lc:slow_slsn_pisn_comparison}, we compare the bolometric light curves of the slow-evolving SLSNe to the suite of PISN models used for \sn\ in Section \ref{discussion:lightcurve:template_matching}. The bolometric light curves of all historical slow-evolving SLSNe are either inconsistent with the PISN models or the comparison is inconclusive: PTF12dam evolves too fast, the light curves of PS1-11ap and SN\,2015bn have a different shape to PISN templates, PS1-14bj shows a flattening at late times, and the bolometric light curves of SN\,2007bi and LSQ14an have no pre-max data. The lack of an estimate of the rising bolometric light curve for the latter two objects precludes concluding whether the two SLSNe could be PISNe or not. Dedicated studies on PS1-11ap, PS1-14bj, PTF12dam and SN\,2015bn revealed that the magnetar model provides an adequate description of the light curves \citep{Nicholl2013a, McCrum2014a, Lunnan2016a, Nicholl2018a, Vurm2021a}.

In conclusion, \sn\ is the \emph{only} SLSN among the hundreds of SLSNe known whose entire light curve is consistent with PISN models. This result is even more revelatory considering that the bolometric light curve covers an exceptionally wide time interval from \tmax$-93$ to \tmax+706~days.

\subsubsection{Spectra}

\begin{figure}[t]
    \centering
    \includegraphics[width=1\columnwidth]{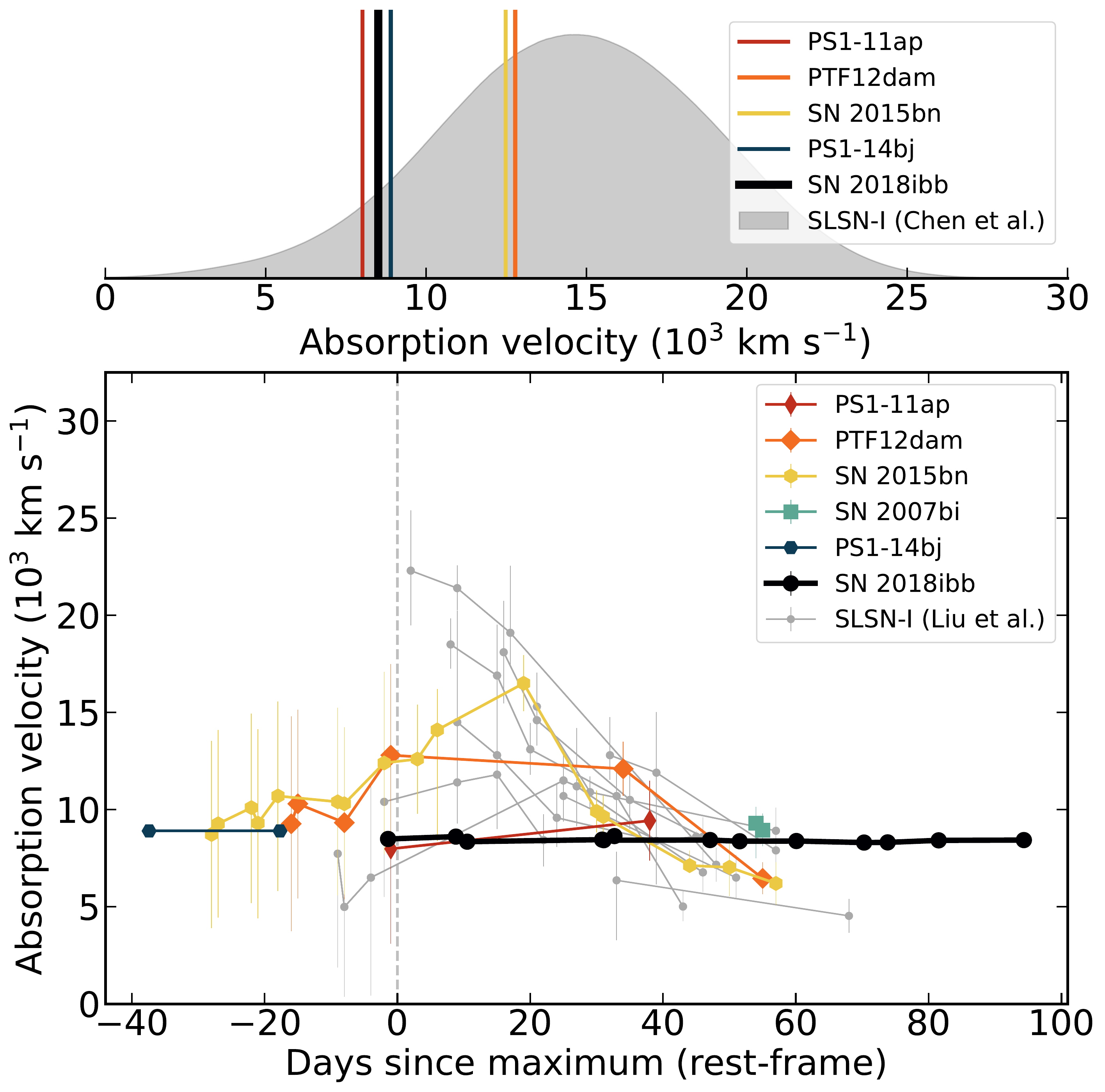}
    \caption{\ion{Fe}{ii} ejecta velocities of slow-evolving SLSNe (in colour) 
    and general SLSNe samples (grey) at the time of maximum (top panel) and as a function of time (bottom panel). \sn\ has a markedly low velocity at the time of maximum and a flat velocity evolution, which is in stark contrast to the bulk of the SLSN population. Its velocity at peak is similar to the slow-evolving SLSNe PS1-14bj and SN\,2007bi. However, both comparison objects lack spectra at earlier and later times.
    }
    \label{fig:spec:velocities_2}
\end{figure}

In this section, we compare the spectroscopic properties of \sn\ to those of other slow-evolving SLSNe. First, we compare the photospheric velocities measured with the \ion{Fe}{ii}\,$\lambda$\,5169 region. The top panel of Figure \ref{fig:spec:velocities_2} displays the photospheric velocities, measured from the \ion{Fe}{ii}\,$\lambda$5169 region, of \sn\ and other slow-evolving SLSNe\footnote{LSQ14an is omitted from this comparison. Conspicuous galaxy lines from [\ion{O}{iii}]\,$\lambda\lambda$\,4959,5007 contaminate the \ion{Fe}{ii}\,$\lambda$5169 region.} (in colour) and of the ZTF-I sample (kernel density estimate). The slow-evolving SLSNe have velocities between 8000 and $12,000~\rm km\,s^{-1}$ at peak, lower than the median value of the ZTF-I sample ($14,800~\rm km\,s^{-1}$). PTF12dam and SN\,2015bn have the fastest expanding ejecta ($\sim12,000~\rm km\,s^{-1}$ at peak), but their ejecta rapidly decelerate to $\sim6000~\rm km\,s^{-1}$ in $\sim60$~rest-frame days. Their velocities and velocity evolution is similar to those of other SLSNe \citep{Liu2017a}. In stark contrast to that, \sn\ has a velocity of merely $8500~\rm km\,s^{-1}$, comparable to those of PS1-14bj and SN\,2007bi. Furthermore, the velocity of \sn\ remains constant for 100 rest-frame days, which has not seen for any other SLSN before. Though the velocities of PS1-14bj and SN\,2007bi are very similar to SN\,2018ibb, the spectroscopic sequences of these two events are limited, precluding comparing their velocity evolution to that of \sn.

\begin{figure*}[t]
    \centering
    \includegraphics[width=1\textwidth, angle=0]{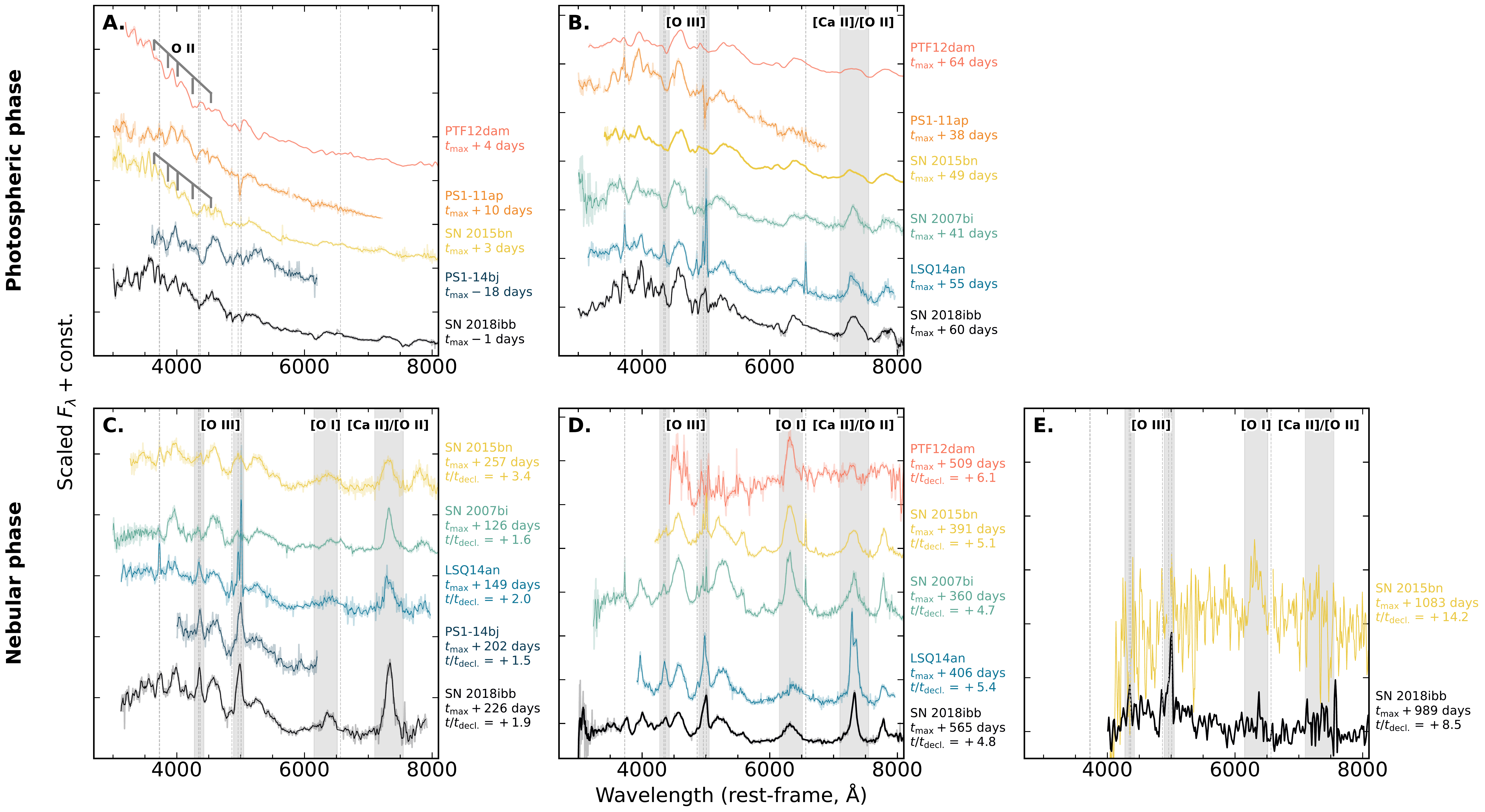}
    \caption{Comparison of the spectra of \sn\ to those of other slow-evolving SLSNe between \tmax+30 days and \tmax+1000~days (darker colour: 5~\AA\ binning; light shade: unbinned spectra).
    \textbf{Photospheric phase (Panels A--B)}: Around \tmax\ (Panel A), the spectra of PTF12dam and SN\,2015bn are characterised by a hot continuum with superimposed \ion{O}{ii} absorption lines as seen in many SLSNe at a similar epoch. SN\,2007bi, LSQ14an and \sn\ have cooler photospheres, and their spectra exhibit absorption lines from Ca, Fe, Mg, O and Si (see Figure \ref{fig:spec:line_id} for their locations) but not \ion{O}{ii}. At around \tmax+60 days (Panel B), all spectra appear similar, though differences exist. LSQ14an, PS1-14bj, and \sn\ are the only SLSNe showing [\ion{O}{iii}] in emission. Furthermore, SN\,2007bi, LSQ14an and \sn\ exhibit strong [\ion{Ca}{ii}] + [\ion{O}{ii}] in emission. This feature is also present in PTF12dam and SN\,2015bn but is less pronounced.
    \textbf{Nebular phase (Panels C--E)}: Differences start to emerge during the early nebular phase and become stronger with time. \sn, LSQ14an and PS1-14bj continue to show conspicuous [\ion{O}{iii}] in emission, in contrast to PTF12dam, SN\,2007bi and SN\,2015bn that have very strong [\ion{O}{i}] and \ion{O}{i} in emission. SNe\,2015bn and 2018ibb are the only SLSNe with spectra at $\sim$\tmax+1000~days (Panel E). \sn\ continues to show intermediate-width [\ion{O}{iii}], whereas the spectrum of SN\,2015bn exhibits [\ion{O}{i}].
    The elevated noise in the \sn\ spectrum at \tmax+989.2~days at $\lambda>6000~\rm\AA$ is due to residuals of the skyline subtraction. The dashed vertical lines indicate the expected locations of emission lines commonly seen from \ion{H}{ii} regions.
    }
    \label{fig:spec:comparison}
\end{figure*}

Next, we explore the spectroscopic properties of slow-evolving SLSNe during their photospheric and nebular phases. Panel A in Figure \ref{fig:spec:comparison} shows the photospheric spectra at the time of maximum light. PTF12dam and SN\,2015bn sustained a hot photosphere with a temperature of $\gtrsim12,000$~K \citep{Nicholl2016a, Vreeswijk2014a}. One of the strong features in their spectra is a comb of \ion{O}{ii} absorption lines, a characteristic feature of SLSNe, which are only seen in photospheres with $T>15,000$~K \citep{Quimby2018a} and probably also require non-thermal excitation \citep{Mazzali2019a}. The spectra of PS1-11ap, PS1-14bj and \sn\ are cooler (blackbody temperatures of 10,000 to 12,000 K). Their spectra do not show \ion{O}{ii} absorption lines but instead absorption lines from Ca, Fe, Mg, O and Si (see Figure \ref{fig:spec:line_id} for the locations). Common to PTF12dam, SN\,2015bn and \sn\ is the presence of [\ion{Ca}{ii}]\,$\lambda\lambda$\,7291,\,7323 in emission. It is one of the strongest features seen in nebular phase spectra of SNe \citep{Filippenko1997a, GalYam2017a} but is only seen during the photospheric phase in slow-evolving SLSNe \citep{GalYam2009a, Inserra2017a, Nicholl2019a}. It is also seen in SN\,2007bi and LSQ14an, but these SLSNe lack spectra at peak. 

Around the time of maximum light (Panel B), all objects have similar spectra. Due to the differences in the ejecta velocities, features appear sharper in LSQ14an, PS1-14bj and \sn\ that in PTF12dam and SN\,2015bn. Some clear differences are well visible though. LSQ14an, PS1-14bj and \sn\ reveal [\ion{O}{iii}] in emission. As we concluded in Section \ref{sec:csm:oiii_oii}, the 7300~\AA\ feature in \sn\ is not dominated by [\ion{Ca}{ii}], but [\ion{O}{ii}]\,$\lambda\lambda$\,7320,7330. In the other objects, the centre of the 7300~\AA\ feature is consistent with [\ion{Ca}{ii}]. Moreover, the 7300~\AA\ feature is well developed in SN\,2007bi, LSQ14an and \sn\ but still very weak in PTF12dam and SN\,2015bn. The line profiles also differ. In \sn\, the line profile is flat-topped but triangular and skewed to the blue for the other objects. 

During the early nebular phase ($t/t_{\rm decl.}\sim2$; Panel C), all objects show a blue pseudo-continuum with superimposed forbidden and allowed emission lines from calcium, magnesium and oxygen (for the line identifications see Figure \ref{fig:spec:line_id}). \sn\ and PS1-14bj are spectroscopically indistinguishable, though their overlap in wavelength is limited and PS1-14bj is significantly fainter than \sn\ (Figure \ref{fig:lc:comparison_slow_slsn}). The other objects reveal an increasing level of dissimilarities (LSQ14an $\rightarrow$ SN\,2015bn $\rightarrow$ SN\,2007bi). LSQ14an has a similar blue pseudo-continuum but its emission lines are not well developed. This is best seen in \ion{Ca}{ii}\,$\lambda\lambda$\,3933, 3968, [\ion{O}{iii}]\,$\lambda$\,4363 and [\ion{Ca}{ii}]\,$\lambda\lambda$\,7291,7324 + [\ion{O}{ii}]\,$\lambda\lambda$\,7320,7330. The SNe 2007bi and 2015bn have redder pseudo-continua and significantly weaker [\ion{Ca}{ii}]+[\ion{O}{ii}]. Moreover, SN\,2007bi has only a few features bluewards of 5000~\AA.

These differences develop further with time. During the late nebular phase ($t/t_{\rm decl.}\sim5$; Panel D), the pseudo-continuum of all objects fades. \sn\ and LSQ14an are characterised by a weaker [\ion{O}{i}]\,$\lambda\lambda$\,6300,6364 than SNe\,2007bi and 2015bn. The ratio between [\ion{Ca}{ii}]+[\ion{O}{ii}] and [\ion{O}{i}] is 2--3:1. Intriguingly, the emission lines of \sn\ evolved much slower than for LSQ14an. Now, LSQ14an exhibits more conspicuous emission lines than \sn, best seen in [\ion{O}{iii}] and [\ion{Ca}{ii}]+[\ion{O}{ii}]. The [\ion{O}{ii}] feature of \sn\ has a Lorentzian profile, whereas the profile of LSQ14an is double-peaked. In contrast to LSQ14an and \sn, PTF12dam, SN\,2007bi and SN\,2015bn have exceptionally strong [\ion{O}{i}]. It is, in fact, their strongest feature. Moreover, the [\ion{O}{i}] is markedly narrower than for \sn\ and LSQ14an: 6000--9000~km\,s$^{-1}$ vs. 16,000~km\,s$^{-1}$. The [\ion{Ca}{ii}]+[\ion{O}{ii}] to [\ion{O}{i}] ratio is 1:2--3 and inverted compared to LSQ14an and \sn.

Panel E shows spectra of \sn\ and SN\,2015bn at 1000--1100 rest-frame days after maximum ($t/t_{\rm decline}=9$--13). These are the only two SLSNe with such extensive spectroscopic observations. Despite the low signal-to-noise ratio, their spectra exhibit well-defined SN features. \sn\ continues to show intermediate-width [\ion{O}{iii}] with a similar width as in the spectral epochs before, whereas SN\,2015bn exhibits [\ion{O}{i}] like in the previous epochs.

\begin{figure}
    \centering
    \includegraphics[width=1\columnwidth]{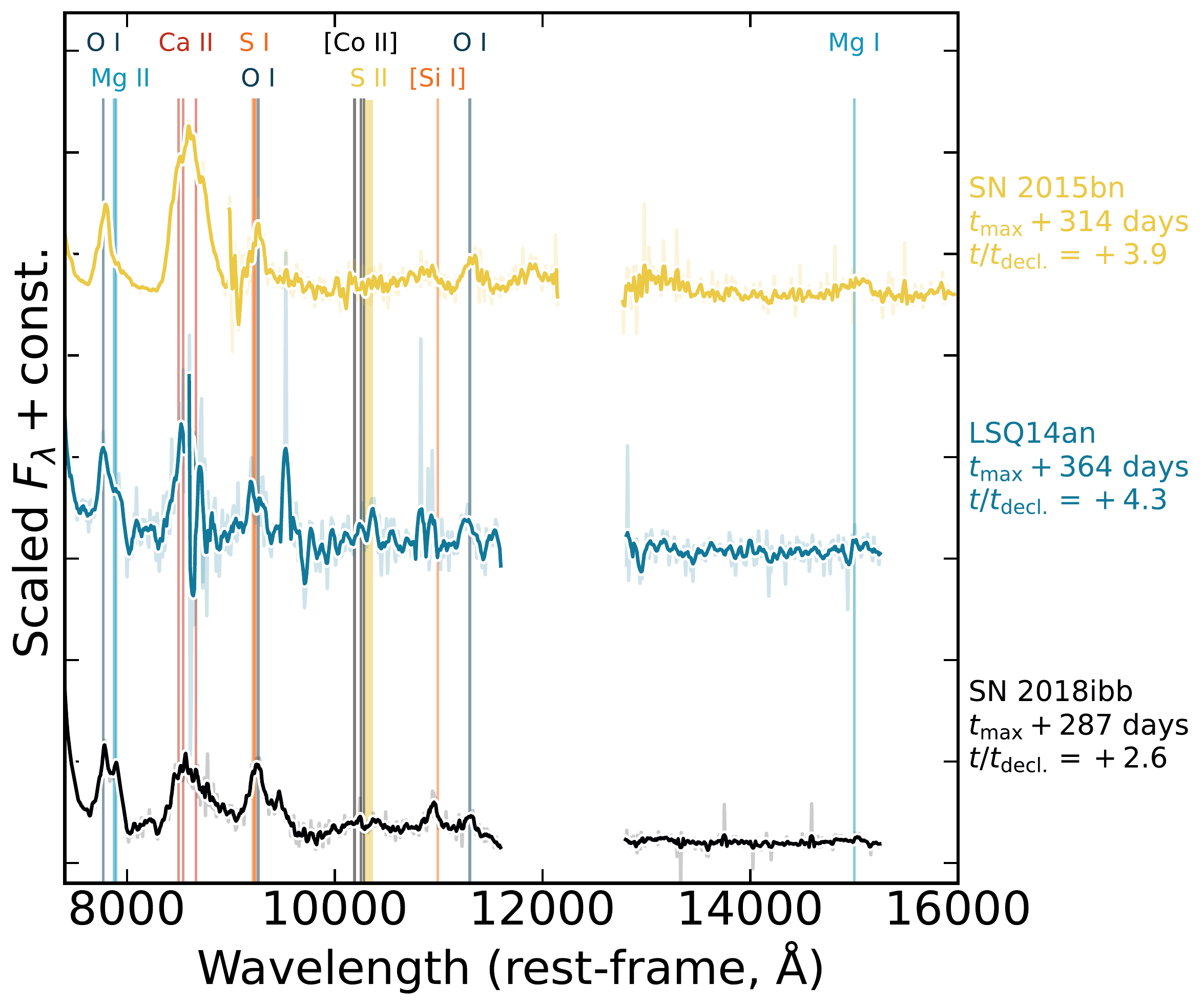}
    \caption{Late-time NIR spectra of \sn\ and the slow-evolving SLSNe LSQ14an and SN\,2015bn. The strongest features are labelled. \sn\ is the only SLSN that shows cobalt in emission, which has its strongest optical-NIR feature at 1.025~$\mu$m. Its luminosity translates to a nickel mass of $\gtrsim30~M_\odot$, consistent with the light curve modelling.  All spectra are scaled so that \ion{O}{i}\,$\lambda$\,7773 has the same amplitude in all objects. Regions of strong atmospheric absorption are cropped.
    }
    \label{fig:spec:comparison_nir}
\end{figure}

Figure \ref{fig:spec:comparison_nir} presents NIR spectra of LSQ14an, SN\,2015bn and \sn\ at 3--4-times their respective decline time scales. All spectra reveal only very few features beyond 1~$\mu$m, which is expected for models of PISNe \citep{Jerkstrand2016a}, SLSNe \citep{Jerkstrand2017a}, and regular stripped-envelope supernovae \citep{Jerkstrand2015a}. Some of the brightest expected features are redshifted to regions of strong atmospheric absorption at the average redshift of SLSNe. A feature that has been commonly seen among all known SLSNe is \ion{O}{i}\,$\lambda$\,1.13$\mu$m. \sn\ reveals an emission feature at 1.025~$\mu$m, which we identified as [\ion{Co}{ii}] (Section \ref{disc:CoII}). [\ion{Co}{ii}] is not present in any of the other spectra. The data quality of the spectra of LSQ14an and SN\,2015bn is higher compared to that of \sn, suggesting that if a substantial amount of $^{56}$Ni was also formed in these supernovae, the [\ion{Co}{ii}] line should have been visible. Instead, SN\,2015bn reveals \ion{Mg}{i}\,$\lambda$\,1.50~$\mu$m that is not visible in \sn\ but possibly in LSQ14an \citep{Jerkstrand2017a}.

In conclusion, \sn\ is spectroscopically similar to other SLSNe, including slow-evolving SLSNe. During the photospheric phase, \sn\ stands out by its low ejecta velocity and flat velocity evolution. The early nebular phase does not differ from other SLSNe. Very late-time observations ($t/t_{\rm decl.}>5$) show clear differences between \sn\ and other SLSNe, for example, the weak and broad [\ion{O}{i}] that stays optically thick throughout the entire evolution. Late-time NIR spectroscopy revealed the tentative detection of [\ion{Co}{ii}] in \sn. This feature is unprecedented for a SLSN and could be the smoking gun that \sn\ is powered by the decay of $^{56}$Ni. In Sections \ref{sec:csm:shell}, \ref{sec:csm:oiii_oii}, \ref{sec:csm:spectrum} and \ref{sec:spectra_discussion}, we argued that the blue pseudo-continuum in \sn\ is produced by the interaction of the SN ejecta with CSM. The prevalence of this feature in the other slow-evolving SLSNe raises the question of whether CSM interaction is also present in these objects. If this is the case, it is necessary to treat nebular spectra of SLSNe as the sum of at least two powering mechanisms, for example, magnetar + CSM or $^{56}$Ni + CSM, necessitating more complex SLSN models than the ones that currently exist. This also means that distinguishing between different powering mechanisms is more difficult and requires comprehensive data sets.

\subsection{Is \sn\ a pair-instability supernova?}

Models of H-poor PISNe make very clear predictions for PISNe in the regime of SLSNe ($M_{\rm peak}\lesssim-20$~mag) for their light curves \citep{Kasen2011a, Dessart2013a, Gilmer2017a, Kozyreva2017a}, ejecta velocities, spectra \citep{Dessart2013a, Jerkstrand2016a} and the environments in which their progenitors are formed \citep{Langer2007a}. In Sections \ref{res:host} and \ref{sec:discussion:lc_spec}, we tested the most critical predictions of the PISN models on the light curves, spectra and host galaxy. \sn\ passes most tests of PISN models with a nickel yield of 25--44~$M_\odot$. However, \sn\ did not comply with two predictions, although it could pass these tests with the interpretations that we propose. Table \ref{tab:pisn_test} summarises all tests.

The tentative detection of [\ion{Co}{ii}]\,$\lambda$\,1.025$\mu$m is unprecedented for a PISN candidate. However, existing PISN models do not predict significant emission from [\ion{Co}{ii}] because of line blocking. In Section \ref{disc:CoII}, we proposed that line blocking could be less severe than predicted by existing models.

The shape and the relative line intensities of the nebular spectra of \sn\ are compatible with those predicted by PISN models. Our observations reveal a significant excess at wavelengths shorter than 5000~\AA, which should not be present if 25--44~$M_\odot$ of iron group elements were formed. As we concluded in Section \ref{sec:spectra_discussion}, we propose that CSM interaction may account for some, if not all, of the excess. PISN models consider mass loss \citep{Kasen2011a, Gilmer2017a, Kozyreva2017a, Dessart2013a} in the evolution of the progenitor star. However, their light curves and spectra are computed assuming that any interaction between the SN ejecta and the circumstellar material is negligible. \citet{Kasen2011a} pointed out that the CSM interaction could actually have a non-negligible contribution. Furthermore, the CSM might not only be produced by stellar winds but also by eruptions similar to that seen in Eta Carinae in 1843. That this is indeed a non-negligible effect is corroborated by recent findings in \citet{Chen2023b}. These authors studied the light curves of 77 events from the homogenous ZTF SLSN-I sample, and concluded that CSM is common around H-poor SLSNe (in at least 25--44\% of the events) and that it contributes to the observed emission, albeit finding spectroscopic evidence in the spectra is difficult. Owing to the lack of predictions of PISN models on CSM interaction, we cannot firmly conclude that \sn\ is a PISN.

Our observations demonstrate that interactions between the SN ejecta and the ambient CSM play a non-negligible effect in the observed photometric and spectroscopic properties (Sections \ref{sec:csm:shell}, \ref{sec:csm:spectrum}, \ref{sec:csm:oiii_oii}, \ref{sec:spectra_discussion}). PISN models of H-poor progenitors with CSM are urgently needed. In the coming years, the Rubin Observatory, and the \textit{James Webb}, \textit{Euclid} and \textit{Roman Space Telescopes} will systematically explore the high-redshift Universe. Since PISNe require metal-poor stars and the early Universe was less chemically enriched than today, PISNe are thought to be more abundant at higher redshifts. Several teams have proposed search strategies to find PISNe with these new observing facilities \citep[e.g.][]{Wang2017a, Regoes2020a, Moriya2022a, Moriya2022b}. However, their search strategies are based on PISN models that, for instance, do not include CSM interaction. Considering that these new observing facilities either just started or will commence their science operations in the next years, it is critical to expand the suite of existing PISN models in order to find high-$z$ PISNe in real-time.

\subsection{Could \sn\ be a pulsational pair-instability supernova?}

\addtolength{\tabcolsep}{-2pt}   
\begin{table*}
\caption{Summary of the PISN tests applied on \sn}
\label{tab:pisn_test}
\small
\centering
\begin{tabular}{cccccc}
\toprule
Test            & Condition                 & Observation       & Section  & Pass & Reference\\
\midrule
\multicolumn{6}{c}{\textbf{Light curve}}\\
\midrule
Rise time       & 120\,--\,150~days             & $>93$~days                & \ref{sec:lc}  & \qmark & 1, 2\\
Decline rate    & 1~mag\,(100~day)$^{-1}$   & 1.1~mag\,(100~day)$^{-1}$ & \ref{sec:lc}  & \cmark & 1, 2\\
Peak absolute magnitude $M_{\rm bol}$ & $-20$\,--\,$-22.5$~mag        & $<-21.8$~mag                  & \ref{sec:lc}  & \cmark & 1, 2\\
Nickel mass     & 10\,--\,40~$M_\odot$ & 25\,--\,40~$M_\odot$ & \ref{discussion:lightcurve:katz}, \ref{discussion:lightcurve:mosfit}, \ref{discussion:lightcurve:template_matching} & \cmark & 1, 2\\
PISN template        & He100\,--\,He130,             & He120\,--\,He130,             & \ref{discussion:lightcurve:template_matching} & \cmark  & 1, 2, 3\\
                & P200\,--\,P250                & P250, P250Ni34            & \\
\midrule
\multicolumn{6}{c}{\textbf{Spectra}}\\
\midrule
Photospheric velocity        & 7000--11,000~km\,s$^{-1}$             & 8500~km\,s$^{-1}$                      & \ref{sec:velocities}         & \cmark & 3, 4\\
Photospheric velocity evolution        & flat              & flat                      & \ref{discussion:velocity_evolution}         & \cmark & here\\
Nebular spectra & He100, He130              & He130, but blue excess    & \ref{sec:spectra_discussion} & \xmark & 5\\
{[}\ion{Co}{ii}{]} \,$\lambda$\,1.025$\mu$m  & not predicted              & detected    & \ref{disc:CoII} & \xmark & 5\\
\midrule
\multicolumn{6}{c}{\textbf{Contribution from CSM interaction to the light curve and spectra}}\\
\midrule
CSM interaction & not explored & observed & \ref{sec:csm:oiii_oii}, \ref{sec:csm:spectrum}, \ref{sec:spectra_discussion} & \qmark & \\
\midrule
\multicolumn{6}{c}{\textbf{Host galaxy}}\\
\midrule
Metallicity     & $<Z_\odot/3$              & \missing{$0.25^{+0.07}_{-0.06}~Z_\odot$}                  & \ref{res:host}                & \cmark & 6\\
\bottomrule
\end{tabular}
\tablebib{
1) \citet{Kasen2011a};
2) \citet{Kozyreva2017a};
3) \citet{Gilmer2017a};
4) \href{https://2sn.org/DATA/HW01/bulk_yields.txt}{https://2sn.org/DATA/HW01/bulk\_yields.txt} based on \citet{Heger2002a}
5) \citet{Jerkstrand2016a};
6) \citet{Langer2007a}
}
\end{table*}

The massive eruptions in a pulsational pair-instability supernova (PPISN) with a large kinetic energy can, under the right conditions, be an ideal case for a luminous interacting SN, as demonstrated in several studies \citep[e.g.][]{Woosley2007a,Yoshida2016a,Woosley2017a,Leung2019,Marchant2019,Renzo2020a}. While the PPI mechanism is difficult to avoid for a He core in the mass range of 40--65~$M_\odot$ \citep{Woosley2007a}, the number of pulses and the interval between these, as well as the mass ejected and their kinetic energies, are more uncertain and differ between various studies. For a bright event to take place, the relative velocities between the shells of the different ejections, as well as their relative masses are important. The brightest event would result from the collision between a very fast, massive shell and a shell of low or zero velocity. The first shell must also be dense enough for the shocks to be radiative and massive enough for the second shell to be completely decelerated. Finally, the collision has to take place close enough to the star, on the order of $10^{15}-10^{16}$cm, so that it will radiate the energy on a timescale of approximately a year. This means that the interval between the pulses should not be more than approximately a year. However, a collision at a very small radius, and short time interval, will result in a very optically thick shell where most of the released energy will go into adiabatic expansion. In summary, there are a number of conditions which have to be fulfilled for a bright SN to result. This has been illustrated in detail by the different radiation-hydrodynamical models, for example, \citet{Woosley2017a}.

Below, we discuss the most extreme models in order to judge whether a pure PPISN could explain the large total radiated energy we find for SN 2018ibb. For a pure He core, \citet{Woosley2017a} finds an upper limit to the kinetic energy of $\sim 2 \times 10^{51}$ erg, with the highest energy from the highest He core mass, if no additional power source (e.g. magnetar or black hole accretion) is involved. The most extreme model with a $62~M_\odot$ He core resulted in a 36~$M_\odot$ ejecta with a total kinetic energy of $2.1 \times 10^{51}$ erg. This is distributed over several pulses, with most of the energy being dissipated in the first pulse. Without any previous strong mass loss this will, however, not be converted into radiation over a timescale of approximately a year. This is also confirmed by the light curve models in \citet{Woosley2017a}. 

Brighter light curves could be obtained for models with a remaining hydrogen envelope. The most extreme, T130D in \citet{Woosley2017a}, had three pulses, ejecting the 70~$M_\odot$ hydrogen-rich envelope with a kinetic energy of $1.5\times10^{51}$~erg. About 3300 years later a second pulse ejected a 7.7~$M_\odot$ shell with He, C and O and energy $1.1\times10^{51}$~erg, and after another 8 months a 13.5~$M_\odot$ shell and energy $1.5\times10^{51}$~erg. The last two shells were ejected close enough in time to collide and create a luminous SN with a total radiated energy of $4.5\times10^{50}$~erg.

Similar calculations have been done by \cite{Marchant2019} and \cite{Leung2019}, using the \program{MESA} code (while \citealt{Woosley2017a} used the \program{Kepler} code). Qualitatively, these models agree, especially in the higher energies and mass ejected, as well as the number of pulses with increasing He core mass. In particular, \citet{Leung2020} find a maximum kinetic energy of $2.8 \times 10^{51}$~erg for the highest PPI He core mass, similar to the corresponding model by \citet{Woosley2007a}. However, as discussed by \cite{Leung2019}, there are also substantial quantitative differences between the models, including ejected masses and time interval between the pulses. Some of the differences can be traced back to the treatment of shocks, and convection in both the hydrostatic and hydrodynamic phases. 

We note that a large kinetic energy in PPISN models has also been invoked to explain the light curves of other luminous SNe. For the FBOT AT2018cow, \citet{Leung2020} invoked a kinetic energy of $5 \times 10^{51}$~erg from a 42~$M_\odot$ He-core interacting with an ejected shell with mass of 0.5~$M_\odot$. An obvious solution to supply the extra energy is a hybrid model with a combination of a PPISN and the energy from a magnetar or accretion. This has been discussed by \citet{Woosley2017a} and for other energetic SLSNe including PTF12dam \citep{Tolstov2017a}, Gaia16apd \citep{Tolstov2017b} and iPTF16eh \citep{Lunnan2018b}. However, it remains unclear how a magnetar can be formed from the core collapse of the very massive He core in a PPISN.

In summary, a pure PPISN, close to the upper He core mass limit, may potentially explain the observed radiated energy of $>3\times10^{51}~\rm erg$ (Section \ref{sec:lc:bolometric}). The conversion of kinetic energy to radiative energy, however, requires rather special conditions in terms of pulse intervals, ejecta mass and velocities. The uncertainties in the models are, unfortunately, large, and it is difficult to draw any firm conclusions. Additional energy sources can not be excluded, such as a magnetar or a black hole. A contribution from a magnetar would result in the flattening of the late-time light curve which is in stark contrast to our observations. In the case a black hole was formed during the gravitational collapse of the progenitor star, the accretion rate would need to be well-tuned to be consistent with the exponentially declining light curve, making the PPISN scenario less likely.

\subsection{PISN rate constraint}\label{sec:rates}

Assuming that \sn\ is a genuine PISN and that the PISN and SLSN luminosity functions are roughly the same, we can measure the PISN rate as a function of SLSNe. As of 25 July 2023, the spectroscopically complete ZTF Bright Transient Survey \citep[BTS;][]{Fremling2020a, Perley2020a} includes 24 genuine SLSNe-I with an absolute peak magnitude of $<-21$~mag and an apparent peak magnitude of $<19$~mag.\footnote{The BTS has a spectroscopic completeness of 95\% down to $m=18.5$~mag (\href{https://sites.astro.caltech.edu/ztf/bts/explorer.php}{https://sites.astro.caltech.edu/ztf/bts/explorer.php}). Although, the completeness is lower at $m=19$, we can go down to that fainter magnitude limit to get a minimum number of spectroscopically confirmed SLSNe and place an upper limit on the PISN-to-SLSN rate.} None of these objects are photometrically similar to \sn. Using Binomial statistics, we place an upper limit of $14\%$ ($2\sigma$ confidence) on the PISN-to-SLSN rate. The latest version of the BTS catalogue reports a volumetric SLSN-I rate of $5\pm2~\rm Gpc^{-3}\,yr^{-1}$ between $z=0.1$ and $z=0.2$ (D. A. Perley; priv. comm.). This translates the PISN-to-SLSN rate to a volumetric rate of $<0.7~\rm yr^{-1}\,Gpc^{-3}$ in the same redshift interval ($2\sigma$ confidence).

With a peak magnitude of $r\sim17.7$~mag (Table \ref{tab:lc_prop}), \sn\ passes the brightness cut of the BTS survey. However, it failed the BTS quality cuts as it contaminated all ZTF reference images. Assuming that \sn\ would have passed the BTS quality cuts, we can place a lower limit on the volumetric rate. Following the methodology in \citet{{Perley2020a}}, the volumetric rate is
\begin{equation}
R =\frac{N}{T \times 4/3\,\pi\,D^3_L\times f_{\rm sky}\times f_{\rm retention}}~\nonumber\\
\end{equation}
where $N$ is the number of events, $T$ is the BTS survey duration of 4.9 years, $D_L$ is the maximum distance at which a transient with a given magnitude can be detected, $f_{\rm sky}$ is the average nightly BTS sky coverage of 35\%, and $f_{\rm retention}$ is the retention factor ($=1$, as \sn\ is assumed to be detected). For $N$ and $D_L$, we apply the following considerations. Firstly, a Poisson distribution with a mean of 1 ($=\rm one$ PISN detection in the entire survey) has a lower $2\sigma$ confidence interval of 0.025 objects  \citep{Gehrels1986a}. Secondly, at the BTS completeness limit of 18.5~mag, \sn\ would have been detected up to a distance of 0.8 Gpc which we assume for $D_L$. This yields a lower limit of $>9\times10^{-3}~\rm yr^{-1}\,Gpc^{-3}$ at $2\sigma$ confidence (i.e. $0.2\%$ of the SLSN rate).

Expected PISN rates vary between $10^{-5}$ and $10^{-2}$ of the core-collapse supernova rate \citep[e.g.][]{Langer2007a, Briel2022a}. They are very sensitive to the assumed shape of the IMF, the metallicity of the progenitor star \citep{Langer2007a}, stellar evolution models \citep[e.g. nuclear reaction rates, rotation, mixing, mass-loss history;][]{Takahashi2018a, duBuisson2020a, Farmer2020a}, the cosmic star-formation history \citep{Briel2022a}, redshift \citep{Langer2007a, Briel2022a}, and whether the progenitor system evolves as an isolated object or in a binary system \citep{Briel2022a, Tanikawa2023a}. This precludes a detailed comparison of the observed and expected rates. Our rate measurement of $0.009$--$0.7~\rm yr^{-1}\,Gpc^{-3}$ is significantly smaller than all but one prediction. \citet{Briel2022a} computed the PISN rate using various prescriptions for the cosmic star-formation history. For one of the prescriptions, these authors deduced a rate of $\sim0.8~\rm yr^{-1}\,Gpc^{-3}$, comparable with our upper limit of $<0.7~\rm yr^{-1}\,Gpc^{-3}$. A more in-depth comparison is beyond the scope of this paper.

\section{Conclusion}\label{sec:conclusion}

In this paper, we have presented observations of the slow-evolving H-poor SLSN 2018ibb covering an exceptionally long time interval from $-93$ to +989 rest-frame days after maximum. \sn\ shares many similarities with H-poor SLSNe, but its properties are extreme even for SLSNe. It is one of the slowest evolving SLSNe known. The slow evolution is apparent through the long rise of $>93$~rest-frame days from 10\% peak flux to peak, the slow decline of merely 1.1~mag (100~days)$^{-1}$, and the low photospheric velocity of 8500~km\,s$^{-1}$ that remains constant between the time of maximum and the following 100 rest-frame days. At peak, \sn\ reached an absolute magnitude of $M_r=-21.7$~mag, comparable to the bulk of the SLSN population. The bolometric light curve had a peak luminosity of $>2\times10^{44}~\rm erg\,s^{-1}$. During its lifetime, \sn\ radiated $>3\times10^{51}$~erg. The peak luminosity and total radiated energy are strict lower limits.

We compared \sn\ with PISN and SLSN models. \sn\ complies with most tests of PISN models with peak luminosity $<-20$~mag, and possibly all tests with the interpretations that we propose, making \sn\ the best PISN candidate, to date. Specifically, \sn\ passes the following tests:
\begin{enumerate}
\item[\textit{i})]   a rise time of $>93$~days (expected: 120--150 days)
\item[\textit{ii})]  a decline time scale of 1.1~mag\,(100~day)$^{-1}$ (expected: 1.1~mag\,(100~day)$^{-1}$)
\item[\textit{iii})] the modelling of the multi-band light curves with physical SLSN models and the \cite{Katz2013a} method point to the production of 25--$44~M_\odot$ $^{56}$Ni (expected: 10--$44~M_\odot$)
\item[\textit{iv})]  the bolometric light curve is consistent with PISN templates that produce 25 and $44~M_\odot$ of $^{56}$Ni
\item[\textit{v})]   a low ejecta velocity of $8500\kms$ (expected: 7000--$11,000\kms$)

\item[\textit{vi})]  a flat evolution of the velocity after maximum light during the photospheric phase (expected: flat evolution)
\item[\textit{vii})]  a low metallicity of $\sim0.25$ solar (expected: $<1/3$~solar)
\item[\textit{viii})] none of the $>200$ SLSNe has properties similar to \sn\ (expected: PISNe are rare).
\end{enumerate}
Such a huge amount of nickel of 25--$44~M_\odot$ can only be produced in a pair-instability-supernova explosion of a star with a He-core mass of 120--130~$M_\odot$ at the time of the explosion (ZAMS mass of approximately 240--$260~M_\odot$). However, \sn\ does not comply with the following tests:
\begin{enumerate}
\item[\textit{i})] the tentative detection of [\ion{Co}{ii}]\,$\lambda$\,1.025~$\mu$m in emission, implying $M(^{56}{\rm Ni})\gtrsim30~M_\odot$ (expected: no [\ion{Co}{ii}] in emission)
\item[\textit{ii})] the nebular spectra are similar to the He130 [$M(^{56}{\rm Ni})=44~M_\odot$] PISN model but show a substantial excess bluewards of 5000~\AA\ due to CSM interaction.
\end{enumerate}

The tentative detection of [\ion{Co}{ii}] is unprecedented for a PISN candidate and any SLSN. It could be the smoking-gun evidence of \sn\ being a PISN, though the line identification hinges on the detection of a single line. PISN models predict no significant [\ion{Co}{ii}]\,$\lambda$\,1.025~$\mu$m in emission because of line blocking extending by iron to the NIR. We propose that the line blocking might be over-estimated in existing models. 

While the late-time spectra are similar to PISN models, they also exhibit a blue excess that should not be present due to the massive line-blanketing of 25--44~$M_\odot$ iron-group elements. A similar blue excess was also observed in previous PISN candidates. Its presence was used as a critical piece of evidence against the PISN interpretation. We argue that this is not the case for \sn. Three lines of evidence reveal that \sn\ is not  solely powered by radioactivity and that CSM interaction is also at play:
\textit{i}) the detection of a slow-moving CSM shell around the progenitor star;
\textit{ii}) the presence of similarly slow \ion{O}{i}, [\ion{O}{ii}], [\ion{O}{iii}] emission lines; and
\textit{iii}) a blue pseudo-continuum similar to that of interaction-powered SNe. This suggests that some, if not all, of the blue excess is produced by CSM interaction. We stress that even after accounting for a substantial contribution of CSM interaction to the bolometric flux, 25--$44~M_\odot$ are still required to power the entire bolometric light curve.

PISN models consider mass-loss episodes (winds and to some level eruptions) to evolve their progenitors to the point of explosion. However, the SN light curves and spectra are computed in sterile environments, assuming that any interaction between the SN ejecta and the circumstellar material is negligible. Our observations demonstrate that CSM interaction is an important non-negligible effect that needs to be systematically explored in PISN models.
The lack of such PISN models is the reason why we cannot conclusively argue for \sn\ being a PISN.

Our data set disfavours central engine models (magnetar powering and fallback accretion onto a black hole), the magnetar+$^{56}$Ni model and pure CSM models. The continued linear decline out to \tmax+706~days and the absence of any light curve flattening, expected for magnetar models, are in conflict with existing analytical prescriptions of magnetar models. Furthermore, the inferred values of the physical parameters of the magnetar and magnetar+$^{56}$Ni models are in conflict with existing stellar evolution models. A model with a simple-power-law-shaped fallback accretion rate, the default assumption in fallback models, would also result in a flattening of the light curve in contradiction with our observations. Analytical CSM models did not provide an adequate description either.

The extensive, high-quality dataset of \sn\ is predestined to perform definitive tests with SLSN and PISN models, and to explore rare explosion mechanisms, such as axion-instability supernovae \citep[AISNe;][]{Sakstein2022a}. Simulations by \citet{Mori2023a} suggest that AISNe evolve faster and are bluer than PISNe for a given He-core mass. AISNe might also be more abundant than PISNe. Therefore, revealing the powering mechanism of \sn\ will have immediate consequences not only for SN science but also for stellar evolution theory. The final confirmation of a PISN would also have ramifications for the interpretation of the observed drop in the black hole mass function and, therefore, gravitational wave astronomy.

In the coming years, the Rubin Observatory, and the \textit{James Webb}, \textit{Euclid} and \textit{Roman Space Telescopes} will be used to search for SLSNe, PISNe, and the explosions of Population III stars in the high-redshift Universe. To make this leap forward, the community requires a significantly improved understanding of the powering mechanisms and the progenitors of SLSNe. This can be accomplished with 
\textit{i}) comprehensive data sets of low-$z$ SLSNe similar to the one presented here and
\textit{ii}) more complex theoretical models with clear predictions for light curves and spectra. 
The \textit{James Webb Space Telescope} could be transformative for studying low-redshift SLSNe. Its IR spectrograph NIRspec has the sensitivity to provide an uncensored view from 1 to 5~$\mu$m. Such an IR spectrum of a \sn-like event could reveal strong emission lines from cobalt, nickel and iron between 2 and 5~$\mu$m during the nebular phase, which would be the smoking-gun evidence for powering by $^{56}$Ni.

\begin{acknowledgements}

We thank the referee for a careful reading of the manuscript
and for helpful comments that improved this paper.

We thank
St\'ephane Blondin (Laboratoire d'Astrophysique de Marseille, France),  
Luc Dessart (Sorbonne Universit\'e, France),
Sebastian Gomez (Space Telescope Science Institute, USA),
Ryosuke Hirai (Monash University, Australia),
Boaz Katz (Weizmann Institute of Science, Israel),
Keiichi Maeda (Kyoto University, Japan),
Ilya Mandel (Monash University, Australia),
and
Kanji Mori (Fukuoka University, Japan) for fruitful discussions.
U.C. Berkeley undergraduate students Nachiket Girish, Andrew Hoffman, Evelyn Liu, Shaunak Modak, Jackson Sipple, Samantha Stegman, Kevin Tang, and Keto Zhang helped obtain data with the Lick/Nickel telescope.

\newline
Z. Chen acknowledges support from the China Scholarship Council.
A.~V. Filippenko's supernova group at U.C. Berkeley received financial support from the Christopher R. Redlich Fund, Gary \& Cynthia Bengier, Clark \& Sharon Winslow, Sanford Robertson, Frank and Kathleen Wood (T. G. Brink is a Wood Specialist in Astronomy), Alan Eustace (W. Zheng is a Eustace Specialist in Astronomy), and numerous other donors.
C. Fransson acknowledges support from the Swedish Research Council and the Swedish National Space Board.
J.~P.~U. Fynbo acknowledges support from the Carlsberg Foundation. The Cosmic Dawn Center (DAWN) is funded by the Danish National Research Foundation under grant No. 140.
M. Gromadzki is supported by the EU Horizon 2020 research and innovation programme under grant agreement No. 101004719.
A. Jerkstrand acknowledges support from the European Research Council (ERC) under the European Union's Horizon 2020 Research and Innovation Programme (ERC Starting Grant No. [803189]).
H. Kuncarayakti was funded by the Academy of Finland projects 324504 and 328898.
G. Leloudas and M. Pursiainen are supported by a research grant (19054) from VILLUM FONDEN.
R. Lunnan is supported by the European Research Council (ERC) under the European Union's Horizon Europe research and innovation programme (grant agreement No. 10104229 - TransPIre). 
T.~E. M\"uller-Bravo and L. Galbany acknowledge financial support from the Spanish Ministerio de Ciencia e Innovaci\'on (MCIN), the Agencia Estatal de Investigaci\'on (AEI) 10.13039/501100011033, the European Social Fund (ESF) `Investing in your future', and the European Union Next Generation EU/PRTR funds under the PID2020-115253GA-I00 HOSTFLOWS project, the 2019 Ram\'on y Cajal program RYC2019-027683-I, the 2021 Juan de la Cierva program FJC2021-047124-I, and from Centro Superior de Investigaciones Cient\'ificas (CSIC) under the PIE project 20215AT016, and the program Unidad de Excelencia Mar\'ia de Maeztu CEX2020-001058-M.
M. Nicholl is supported by the European Research Council (ERC) under the European Union's Horizon 2020 research and innovation programme (grant agreement No. 948381) and by a Fellowship from the Alan Turing Institute.
D. Polishook is grateful for the Wise Observatory staff.
A. Rossi acknowledges support from Premiale LBT 2013.
M. Rigault has received funding from the European Research Council (ERC) under the European Union's Horizon 2020 research and innovation programme (grant agreement No. 759194 - USNAC).
N. Sarin is supported by a Nordita Fellowship. Nordita is funded in part by NordForsk.
S. Schulze acknowledges support from the G.R.E.A.T. research environment, funded by {\em Vetenskapsr\aa det},  the Swedish Research Council, project number 2016-06012.
L.~J. Shingles acknowledges support by the European Research Council (ERC) under the European Union's Horizon 2020 research and innovation program (ERC Advanced Grant KILONOVA No. 885281).
L. Tartaglia acknowledges support from MIUR (PRIN 2017 grant 20179ZF5KS).
Y. Yang acknowledges support from a Benoziyo Prize Postdoctoral Fellowship and the Bengier-Winslow-Robertson Fellowship.

This work was funded by ANID, Millennium Science Initiative, ICN12\_009.

\newline

Based in part on observations at the European Southern Observatory, Program IDs 199.D-0143, 0105.D-0380, 0106.D-0524, 1103.D-0328, 2102.D-5026, and 2104.D-5006 (PIs C. Inserra, S. Schulze, and S.~J. Smartt);
Gemini-South, Program ID 2021B-Q-901 (PI A. Gal-Yam);
Hubble Space Telescope, Program ID GO-16657 (PI C. Fremling);
Keck, Program IDs C323, U023, U025 (PIs S. R. Kulkarni, A. V. Filippenko); 
Large Binocular Telescope, Program ID DDT\_2019B\_13 (PI E. Palazzi);
Las Cumbres Observatory, Program IDs FTPEPO2017AB-001, KEY2017AB-001, SUPA2019A-001, SUPA2019A-002, SUPA2019B-007, and NOAO2020B-012 (PIs P.~J. Brown, K. De);
Liverpool Telescope, Program IDs JL18B06, JL18B07, JL19A24, JL19B11, and JL20B15  (PI D.~A. Perley);
Nordic Optical Telescope, Program IDs 57-502, 58-802, and 61-606, (PIs G. Leloudas, J. Sollerman);
P200 (PI L. Yan); and
\xmm, Program ID 08221501 (PI R. Margutti).

\newline

We thank the staffs of the many observatories at which we conducted observations.
\newline
This work has made use of data from the European Space Agency (ESA) mission {\it Gaia}\footnote{\href{https://www.cosmos.esa.int/gaia}{https://www.cosmos.esa.int/gaia}}, processed by the \gaia\ Data Processing and Analysis Consortium\footnote{\href{https://www.cosmos.esa.int/web/gaia/dpac/consortium}{https://www.cosmos.esa.int/web/gaia/dpac/consortium}} (DPAC). Funding for the DPAC has been provided by national institutions, in particular the institutions participating in the \gaia\ Multilateral Agreement.
\newline
Part of the funding for GROND (both hardware as well as personnel) was generously granted from the Leibniz-Prize to Prof. G. Hasinger (DFG grant HA 1850/28-1). 
\newline
This work is based in part on observations made with the Large Binocular Telescope (LBT). The LBT is an international collaboration among institutions in Italy, the United States, and Germany. LBT Corporation partners are Istituto Nazionale di Astrofisica, Italy; The University of Arizona on behalf of the Arizona university system; LBT Beteiligungsgesellschaft, Germany, representing the Max Planck Society, the Astrophysical Institute Potsdam, and Heidelberg University; The Ohio State University; and The Research Corporation on behalf of The University of Notre Dame, University of Minnesota, and University of Virginia. 
\newline
Some of the observations with the Las Cumbres Observatory data have been obtained via OPTICON proposals and as part of the Global Supernova Project. The OPTICON project has received funding from the European Union's Horizon 2020 research and innovation programme under grant agreement No 730890.
\newline
This work made use of data supplied by the UK Swift Science Data Centre at the University of Leicester.
\newline
CRTS is supported by the U.S. National Science Foundation (NSF) under grants AST-0909182, AST-1313422, and AST-1413600. The Catalina Sky Survey (CSS) is a NASA-funded project supported by the Near Earth Object Observation Program (NEOO) under the Planetary Defense Coordination Office (PDCO).
\newline
This publication makes use of data products from the Two Micron All Sky Survey, which is a joint project of the University of Massachusetts and the Infrared Processing and Analysis Center/California Institute of Technology, funded by NASA and the U.S. NSF.
\newline
Based in part on observations obtained with the Samuel Oschin Telescope 48-inch and the 60-inch Telescope at the Palomar Observatory as part of the Zwicky Transient Facility project. ZTF is supported by the U.S. NSF under grant AST-1440341 and a collaboration including Caltech, IPAC, the Weizmann Institute of Science, the Oskar Klein Center at Stockholm University, the University of Maryland, the University of Washington, Deutsches Elektronen-Synchrotron and Humboldt University, Los Alamos National Laboratories, the TANGO Consortium of Taiwan, the University of Wisconsin at Milwaukee, and Lawrence Berkeley National Laboratories. Operations are conducted by COO, IPAC, and UW.
The SED Machine is based upon work supported by the U.S. NSF under grant 1106171.
\newline
Partially based on observations made with the Nordic Optical Telescope, owned in collaboration by the University of Turku and Aarhus University, and operated jointly by Aarhus University, the University of Turku and the University of Oslo, representing Denmark, Finland and Norway, the University of Iceland and Stockholm University at the Observatorio del Roque de los Muchachos, La Palma, Spain, of the Instituto de Astrofisica de Canarias.
This work makes use of observations from the Las Cumbres Observatory network. The Las Cumbres Observatory team is supported by NSF grants AST-1911225 and AST-1911151.
\newline
Some of the data presented herein were obtained at the W. M. Keck
Observatory, which is operated as a scientific partnership among the
California Institute of Technology, the University of California, and
NASA; the observatory was made possible by the generous financial
support of the W. M. Keck Foundation.
KAIT, and its ongoing operation were made possible by donations from Sun Microsystems, Inc., the Hewlett-Packard Company, AutoScope Corporation, the Lick Observatory, the U.S. NSF, the University of California, the Sylvia \& Jim Katzman Foundation, and the TABASGO Foundation.    
A major upgrade of the        
Kast spectrograph on the Shane 3\,m telescope at Lick Observatory was        
made possible through generous gifts from William and Marina Kast as          
well as the Heising-Simons Foundation. Research at Lick Observatory is        
partially supported by a generous gift from Google.

\end{acknowledgements}

\bibliographystyle{aa}

\begin{appendix}

\section{Photometric observations}\label{appendix:photometry}

\subsection{All-sky surveys}\label{sec:obs:surveys}

\paragraph{ATLAS}---~
Between 2018 and 2019, the ATLAS survey utilised two 0.5\,m telescope systems on Haleakala and Mauna Loa, Hawaii (USA). Each of the telescopes has a field of view of 29 square degrees. The two telescopes work in tandem to survey the entire visible sky from $-45^\circ < \delta < 90^\circ$ with a 2-day cadence. ATLAS observes in two wide filters, called `cyan' or `c', which roughly covers the SDSS/Pan-STARRS $g$ and $r$ filters (4200--6500~\AA), and `orange' or `o', which covers the SDSS/PanSTARRS $r$ and $i$ (5600--8200~\AA) to a depth of $\sim19$~mag \citep[$5\sigma$, averaged over both telescope sites and weather conditions;][]{Tonry2011a}. All data immediately go through an automatic data-processing pipeline, described in \citet{Stalder2017a}.

After the field of \sn\ appeared from behind the sun, the ATLAS survey obtained the first image on 7 August 2018 (\tmax$-$107~days). On 10 September, \sn\ got brighter than the detection threshold of the ATLAS pipeline. To recover emission from \sn\ between 7 August and 10 September and boost the quality of all data, we ran the ATLAS Forced Photometry service\footnote{\href{https://fallingstar-data.com/forcedphot}{https://fallingstar-data.com/forcedphot}} \citep{Shingles2021a} on all images starting from summer 2018. We clipped and binned the data using the python script \program{plot\_atlas\_fp.py}\footnote{\href{https://gist.github.com/thespacedoctor/86777fa5a9567b7939e8d84fd8cf6a76}{https://gist.github.com/thespacedoctor/\\gi86777fa5a9567b7939e8d84fd8cf6a76}}. We used a bin size of 2 days for the first observing season and 5 days for the second observing season. Measurements with a significance of $<3\sigma$ were removed from the final light curve. To identify the epoch of first light, we built custom-made nightly stacks and visually inspected each coadded image. The earliest $3\sigma$ detection was recovered from data obtained on 31 August 2018 (\tmax$-$80.1~days). The brightness was $o=19.13\pm0.21$~mag. The last non-detection before the first detection was on 17 August 2018 (\tmax$-$92.1~days) and had a  depth of $o\approx19.9$~mag at $3\sigma$ confidence. The full light curve covers the period up until \tmax+386~days.

\paragraph{ZTF}---~
The Zwicky Transient Facility uses the Samuel Oschin 48-inch (1.22\,m) Schmidt telescope at Mount Palomar (USA) equipped with a 47-square-degree camera  \citep{Dekany2020a}. Since 17 March 2018, the public ZTF Northern Sky Survey monitors the northern hemisphere every 3 days in $g$ and $r$ band to a depth of $\sim20.7$~mag \citep[$5\sigma$;][]{Bellm2019b, Bellm2019a}. The field of \sn\ was observed for the first time on 29 August 2018. 

ZTF started its science operations only in March 2018. This resulted in all reference images between September and November 2018 being contaminated by \sn. Hence, the difference photometry in the ZTF alert packages only provided a partial event history, and the measurements were unusable. Owing to that, we obtained the science-ready but not-template-subtracted images from the NASA/IPAC Infrared Science Archive\footnote{\href{https://irsa.ipac.caltech.edu/}{https://irsa.ipac.caltech.edu/}}. (Their data reduction is described in \citealt{Masci2019a}.) We measured the brightness of \sn\ using the aperture photometry tool\footnote{\href{https://github.com/steveschulze/Photometry}{https://github.com/steveschulze/Photometry}} presented in \citet{Schulze2018a}. Once an instrumental magnitude was established, it was calibrated against the brightness of several stars from a cross-matched PanSTARRS-\gaia\ catalogue. The full light curve covers the interval from \tmax$-$81.9 to \tmax+306~days.

\paragraph{Gaia} ---~
The \gaia\ satellite has two $\rm 1.45\,m\,\times\,0.5\,m$ telescopes pointing in two directions separated by an angle of 106\fdg5 and merged into a single focal plane. It monitors the sky with a $\gtrsim30$~day cadence in the unfiltered, white-light \gaia\ `G' band \citep[3320--10515~\AA;][]{Gaia2016a}. We retrieved the light curve from the Gaia Photometric Science Alerts database\footnote{\href{http://gsaweb.ast.cam.ac.uk/alerts/home}{http://gsaweb.ast.cam.ac.uk/alerts/home}}. The light curve covers the period from \tmax$-$93 to \tmax+259~days. 

\begin{figure}
    \centering
    \includegraphics[width=1\columnwidth]{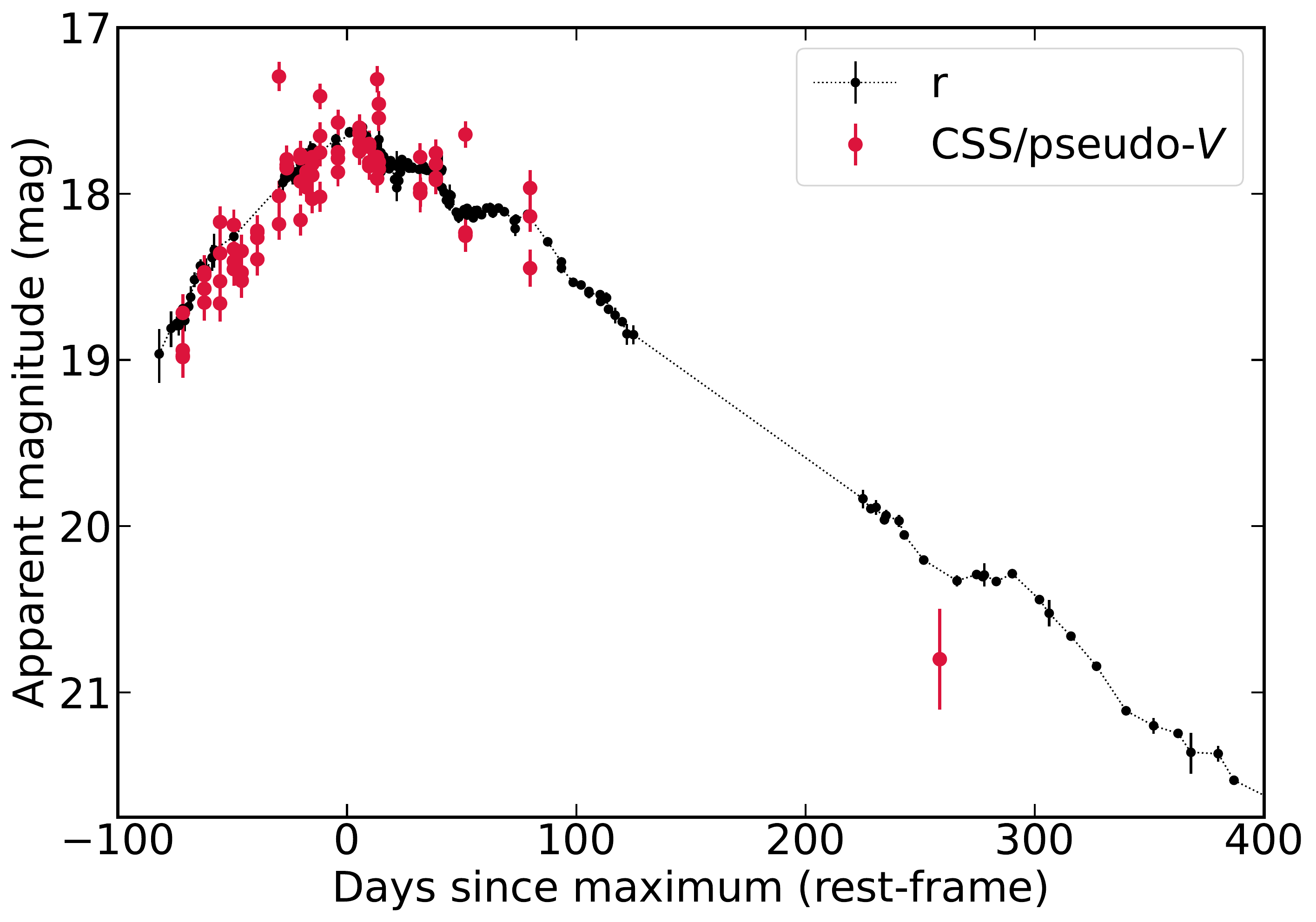}
    \caption{
    The light curve from the Catalina Sky Survey in pseudo-$V$ band together with the $r$-band light curve of \sn.
    }
    \label{fig:lc:css}
\end{figure}

\paragraph{Catalina Sky Survey} ---~ The data were taken by the Catalina Sky Survey \citep{Larson2003a} using the 0.7\,m Schmidt telescope\footnote{\href{https://catalina.lpl.arizona.edu/telescopes}{https://catalina.lpl.arizona.edu/telescopes}}. All observations were taken unfiltered using 30\,s exposures and typically reach $V\sim19.5$~mag. The photometry was performed using Source Extractor \citep{Bertin1996a} and is calibrated to a pseudo-$V$ based on a preselected set of calibrator stars as described in \citet{Drake2011a}. The light curve covers the period from \tmax$-$72 to \tmax+259~days. Owing to the scatter in the data, we show the data in Figure \ref{fig:lc:css}.

\begin{table}
    \caption{Log of photometric observations\label{tab:phot}}
    \centering
    \begin{tabular}{cccc}
    \toprule
    Telescope/ & Filter & MJD & Brightness \\ 
    Instrument &  & (day) & (mag) \\
    \midrule
    Swift/UVOT & $uvw2$ & 58464.766 & $21.07\pm0.16$ \\
    Swift/UVOT & $uvw2$ & 58472.770 & $21.36\pm0.19$ \\
    Swift/UVOT & $uvw2$ & 58476.387 & $21.26\pm0.19$ \\
    Swift/UVOT & $uvw2$ & 58481.176 & $21.40\pm0.24$ \\
    Swift/UVOT & $uvw2$ & 58484.828 & $21.76\pm0.23$ \\
    Swift/UVOT & $uvw2$ & 58493.922 & $21.69\pm0.25$ \\
    Swift/UVOT & $uvw2$ & 58497.102 & $21.55\pm0.22$ \\
    Swift/UVOT & $uvw2$ & 58500.664 & $21.84\pm0.36$ \\
    Swift/UVOT & $uvw2$ & 58501.531 & $21.92\pm0.34$ \\
    Swift/UVOT & $uvw2$ & 58505.371 & $21.75\pm0.21$ \\
    Swift/UVOT & $uvw2$ & 58507.598 & $21.80\pm0.21$ \\
    Swift/UVOT & $uvw2$ & 58516.656 & $22.02\pm0.25$ \\
    Swift/UVOT & $uvw2$ & 58521.641 & $21.80\pm0.22$ \\
    Swift/UVOT & $uvw2$ & 58528.613 & $22.22\pm0.34$ \\
    \bottomrule
\end{tabular}
\tablefoot{All measurements are reported in the AB system. An s-correction was applied to all measurements but no correction for reddening. A machine-readable table is available online on WISeREP. 
}
\end{table}

\subsection{Large observing campaigns}\label{sec:obs:large_campaigns}

\paragraph{2.2\,m MPG} ---~Between 16 December 2018 and 29 October 2019 (\tmax+11 -- \tmax+283 days), we monitored the light-curve evolution with the seven-channel imager GROND \citep[Gamma-Ray Burst Optical/Near-Infrared Detector;][]{Greiner2008a} from 4500 to 22,000~\AA\ at the 2.2\,m Max Planck Gesellschaft telescope at La Silla Observatory (Chile) as a part of the GREAT survey \citep{Chen2018a}.

The images were reduced with the GROND pipeline \citep{Kruehler2008a}, which applies bias and flat-field corrections, stacks images and provides astrometric calibration. The photometry was extracted similarly to the ZTF data. To establish the absolute flux scale, we used a local sequence of stars from a cross-matched PanSTARRS-\gaia\ catalogue ($griz$) and 2MASS \citep[$JHK$;][]{Skrutskie2006a}. \sn\ evaded detection in $K_s$ band. These images are very shallow, and we omit them in this paper.

\paragraph{3.58\,m ESO/NTT} ---~
At the end of October 2019, we switched from the 2.2\,m MPG/ESO telescope to the 3.58\,m ESO New Technology Telescope at La Silla Observatory (Chile). We obtained photometry in $BVgriz$ between 22 October 2019 and 11 November 2020 (\tmax+277 -- \tmax+619 days) with the EFOSC2 instrument \citep[ESO Faint Object Spectrograph and Camera version 2;][]{Buzzoni1984a}. Furthermore, we obtained a few epochs of $JHK$ photometry with the SOFI \citep{Moorwood1998a}. All observations were carried out as a part of the ePESSTO survey \citep{Smartt2015a}. On 1 March 2022 (\tmax+1023 days), we obtained a final image in $B$ band. At this phase, the brightness of \sn\ was well below the host level in the $B$ band. We use these data to expand the host galaxy observations to shorter wavelengths.

The data were reduced with the PESSTO pipeline\footnote{\href{https://github.com/svalenti/pessto}{https://github.com/svalenti/pessto}} \citep{Smartt2015a}, which applies bias and flat-field corrections, and provides astrometry calibration.
We utilised the tool from \citet{Schulze2018a} to extract the photometry. For the SN photometry, we used circular apertures and an elliptical aperture for the host galaxy whose size was matched to that of the VLT/FORS2 data. The $BVgriz$ photometry was calibrated with a local sequence of stars from a cross-matched PanSTARRS-\gaia\ catalogue and the $JHK$ photometry with the stars from the 2MASS point source catalogue. 

\paragraph{8.2\,m ESO/VLT} ---~
Between September 2020 and March 2021, we obtained photometry in $gRIz$ with the FOcal Reducer/low dispersion Spectrograph 2 \citep[FORS2,][]{Appenzeller1998a} at the 8.2\,m Very Large Telescope at Paranal Observatory (Chile), covering the time interval from \tmax+564~days to \tmax+706~days. These data suffered from a progressively increasing contamination by the host galaxy. To remove the host contribution, we obtained a final set of $gRIz$ photometry in January/February 2022 (\tmax+991~days). These reference images were obtained under similar observing conditions, but their integration time was a factor of 2 larger to ensure a reliable host subtraction.

We reduced the data with \program{IRAF} in the same way as the GROND and NTT data. The world coordinate system was calibrated with the software package \program{astrometry.net} \citep{Lang2010a}. To remove the host contribution, we aligned all images in a given filter to the corresponding template image and subtracted the images with the \program{High Order Transform of Psf ANd Template Subtraction} code version 5.11 \citep[\program{HOTPANTS};][]{Becker2015a}. We measured the brightness in the difference images using aperture photometry. The photometry was calibrated against a set of stars identified in archival $griz$ CTIO/DECam images from the Dark Energy Survey \citep{DES2016a} and DESI Legacy Imaging Surveys \citet{Dey2018a} 

\paragraph{Neil Gehrels Swift Observatory} ---~ We observed the field with the 30\,cm Ultraviolet/Optical Telescope \citep[UVOT;][]{Roming2005a} aboard the \swift\ satellite \citep{Gehrels2004a} between \tmax+8.4 and \tmax+224~days in $w2$, $m2$, $w1$, $u$, $b$, $v$. While \sn\ was not observable from the ground between April and August 2019, we made strategic use of \swift's orbit to reduce the seasonal gap of our ground-based campaigns from 100 to 54 rest-frame days.

We retrieved the science-ready data from the \swift\ archive\footnote{\href{https://www.swift.ac.uk/swift_portal}{https://www.swift.ac.uk/swift\_portal}}. We co-added all sky exposures for a given epoch and filter to boost the S/N using \program{uvotimsum} in \program{HEAsoft}\footnote{\href{https://heasarc.gsfc.nasa.gov/docs/software/heasoft/}{https://heasarc.gsfc.nasa.gov/docs/software/heasoft/}} version 6.26.1. Afterwards, we measured the brightness of \sn\ with the \swift\ tool \program{uvotsource}. The source aperture had a radius of $5''$ while the background region had a significantly larger radius. The photometry was calibrated with the latest calibration files from September 2020.

\subsection{Supplementary observations}\label{sec:obs:small_campaigns}

We augmented the campaigns mentioned above with targeted observations using
\begin{itemize}
    \item the Alhambra Faint Object Spectrograph and Camera (ALFOSC)\footnote{\href{http://www.not.iac.es/instruments/alfosc}{{http://www.not.iac.es/instruments/alfosc}}} on the 2.56\,m Nordic Optical Telescope (NOT) at the Observatorio del Roque de los Muchachos on La Palma (Spain) 
    in $griz$
    \item the 0.76\,m Katzman Automatic Imaging Telescope (KAIT) in $BVRI$ and `clear' as part of the Lick Observatory Supernova Search \citep[LOSS;][]{Filippenko2001a}
    \item the Las Cumbres Observatory (LCO) in $ugriz$ and $BVRI$
    \item the optical imager (IO:O) on the robotic 2\,m Liverpool Telescope \citep[LT;][]{Steele2004a} located at the Observatorio del Roque de los Muchachos in $griz$
    \item the Low Resolution Imaging Spectrometer (LRIS, \citealt{Oke1995a}) on the 10\,m Keck I telescope at Maunakea (USA) in $gVRi$
    \item the Multi-Object Double Spectrographs MODS-1 and 2 cameras in $ugriz$ and the near-IR LUCI camera in $JHK$ at the 8.4\,m Large Binocular telescope at Mt. Graham (USA)
    \item 1\,m Nickel telescope at Lick Observatory in $BVRI$ \citep{Li2003a}
    \item the SED Machine \citep{Blagorodnova2018a} at the Palomar 60-inch telescope at Mount Palomar (USA) in $gri$
    \item the 28-inch telescope at the Wise Observatory (Israel) in $gri$
    \item the Wide Field Camera 3 in $F336W$ aboard the \textit{Hubble Space Telescope} to obtain a late-time image in August 2022
\end{itemize}

\noindent All data were reduced in a similar fashion with instrument-specific pipelines, for example, \program{FPipe} for the SEDm data \citep{Fremling2016a}, the \program{LOSSPhotPypeline}\footnote{\href{https://github.com/benstahl92/LOSSPhotPypeline}{https://github.com/benstahl92/LOSSPhotPypeline}} for the KAIT and Lick data \citep{Stahl2019a} and \program{IRAF}. We applied aperture photometry to extract the photometry similar to the method described in Section \ref{sec:obs:large_campaigns}. In the case of KAIT and Lick data Point-spread function photometry \citep{Stetson_1987} using DAOPHOT was applied.

The Keck observations were performed when the host contribution was non-negligible but smaller than 10\% in all filters. Owing to the size of the host contribution, we omit any host correction but added an error of 0.2 mag in quadrature to those measurements. The NOT data were obtained in adverse observing conditions. We omit to report their photometry owing to issues in obtaining a reliable photometric calibration. The LCOGT $z$-band images suffer from fringing, resulting in photometry with large systematic errors. We omit to report these measurements in this paper.

\section{Spectroscopic observations}\label{appendix:spectroscopy}

\subsection{10\,m Keck telescope}

We obtained 3 epochs with LRIS between \tmax$-$1.4~days and \tmax+562.3~days. The first two observations, acquired on 1 December 2018 (\tmax$-$1.42 days) and 29 August 2019 (\tmax+231.2~days), used the B400/3400 blue-side grism and the R400/8500 red-side grating, the dichroic 560 and a 1\farcs0 wide slit. The integration times were 300 s for each epoch. The third epoch was obtained with the B600/4000 blue-side grism and R400/8500 red-side grating and a 1\farcs0 wide slit on 18 September 2020 (\tmax+562.3~days). The integration time was 4935 s. All spectra were reduced in a standard fashion with the data reduction pipeline \program{LPipe} \citep{Perley2019a}.

\subsection{Palomar 60-inch telescope}

We acquired 3 epochs of spectroscopy with the SED Machine between \tmax+5.4 and \tmax+21.6~days. The SED Machine is a very low resolution ($R\sim100$) integral field unit covering the wavelength range from 3650 to 10,000~\AA. The first two epochs (9 and 12 December 2018) are of sufficient quality and are reported to WISeREP but are not presented in the paper owing to the availability of higher resolution and higher S/N data. The final epoch obtained on 28 December 2018 is of very poor quality and is reported here only for completeness. All observations were reduced using the pipeline described by \citet{Rigault2019a}.

\subsection{Palomar 200-inch telescope}

We obtained one epoch of spectroscopy with the DouBle-SPectrograph \citep[DBSP;][]{Oke1982a} on 13 December 2018 (\tmax+8.8~days). The observations were taken using the D-55 dichroic beam splitter, a blue grating with 600 lines per mm blazed at 4000~\AA, a red grating with 316 lines per mm blazed at 7500~\AA, and a 1\farcs5 wide slit. The data are reduced using the python package \program{DBSP\_DRP}\footnote{\href{https://github.com/finagle29/dbsp_drp}{https://github.com/finagle29/dbsp\_drp}} that is primarily based on \program{PypeIt} \citep{Prochaska2020b, Prochaska2020a}.

\subsection{Extended Public ESO Spectroscopic Survey of Transient Objects}

Between 15 December 2018 and 20 October 2019 (\tmax+10.5 -- \tmax+275.6~days), we obtained 6 epochs of spectroscopy using EFOSC2. The observations were performed with grisms \#11, \#13 and \#16. Depending on weather conditions, we used either a 1\farcs0 or 1\farcs5 wide slit. The integration times were between 1800 and 5400 s. We reduced the data in a standard fashion using the PESSTO pipeline \citep{Smartt2015a}.

\subsection{2.56\,m Nordic Optical Telescope}

We acquired 3 epochs of low-resolution spectroscopy with ALFOSC between 7 January and 22 February 2019 (\tmax+30.8 and \tmax+70.3~days). The spectra were obtained with a 1\farcs3 wide slit and grism \#4. For the second epoch, we used the second-order blocking filter WG345.

The data were reduced using \program{IRAF} (epoch 2), the pipeline \program{foscgui}\footnote{\href{http://sngroup.oapd.inaf.it/foscgui.html}{http://sngroup.oapd.inaf.it/foscgui.html}} (epochs 1 and 3). The reduction includes cosmic-ray rejection, bias corrections, flat fielding, and wavelength calibration using HeNe arc lamps imaged immediately after the target. The relative flux calibration was done with spectrophotometric standard stars observed during the same night.

\subsection{8.2\,m ESO Very Large Telescope}

\paragraph{X-shooter} ---~
We obtained 7 medium-resolution spectra with the X-shooter instrument \citep{Vernet2011a} between 10 January 2019 and 16 February 2020
(\tmax+32.7 -- \tmax+377.5~days). All observations were performed in nodding mode and with 1\farcs0/0\farcs9/0\farcs9 wide slits (UVB/VIS/NIR). The first four epochs covered the full spectral range from 3000 to 24,800~\AA. For the other epochs, we used the $K$-band blocking filter \citep[cutting the wavelength coverage at 20,700~\AA;][]{Vernet2011a} to increase the signal-to-noise ratio (S/N) in the $H$ band. The integration times were varied between 1800 and 3600 s.

The data were reduced following \citet{Selsing2019a}. In brief, we first removed cosmic-rays with the tool  \program{astroscrappy}\footnote{\href{https://github.com/astropy/astroscrappy}{https://github.com/astropy/astroscrappy}}, which is based on the cosmic-ray removal algorithm by \citet{vanDokkum2001a}. Afterwards, the data were processed with the X-shooter pipeline v3.3.5 and the ESO workflow engine ESOReflex \citep{Goldoni2006a, Modigliani2010a}. The UVB and VIS-arm data were reduced in stare mode to boost the S/N. In the background limited case, this can increase the S/N by a factor of $\sqrt{2}$ compared to the standard nodding mode reduction. The individual rectified, wavelength- and flux-calibrated two-dimensional spectra files were co-added using tools developed by J. Selsing\footnote{\href{https://github.com/jselsing/XSGRB_reduction_scripts}{https://github.com/jselsing/XSGRB\_reduction\_scripts}}. The NIR data were reduced in nodding mode to ensure a good sky-line subtraction. In the third step, we extracted the one-dimensional spectra of each arm in a statistically optimal way using tools by J. Selsing. Finally, the wavelength calibration of all spectra were corrected for barycentric motion. The spectra of the individual arms were stitched by averaging the overlap regions.

\paragraph{FORS2} ---~
We obtained 3 low-resolution spectra with the FORS2 spectrograph between 20 September and 1 February 2022 (\tmax+565.3 -- \tmax+989.2~days). Each observation was performed with the 300V grism and a 1\farcs0 wide slit. The first epoch comprised of $6\times1100$~s (executed on 20 and 23 September), and the latter two epochs consisted of $12\times1200$~s each. The data were reduced with \program{IRAF} similar to the datasets mentioned above.

\subsection{3\,m Shane telescope}

We obtained a series of four spectra with the Kast double spectrograph\footnote{\href{https://mthamilton.ucolick.org/techdocs/instruments/kast/Tech\%20Report\%2066\%20KAST\%20Miller\%20Stone.pdf}{https://mthamilton.ucolick.org/techdocs/instruments/kast/\\Tech\%20Report\%2066\%20KAST\%20Miller\%20Stone.pdf}} mounted on the Shane 3\,m telescope at Lick Observatory.  These spectra span the time period tmax+24.3 through tmax+89.3 days. The Kast observations utilised the $2"$ slit, 600/4310 grism, and 300/7500 grating. This instrument configuration has a combined wavelength range of $\sim3500$-10,500~\AA. To minimise slit losses caused by atmospheric dispersion \citep{Filippenko1982a}, the Kast spectra were acquired with the slit oriented at or near the parallactic angle.

The Kast data were reduced following standard techniques for CCD processing and spectrum extraction \citep{Silverman2012a} utilising \program{IRAF} routines and custom \program{Python} and \program{IDL} codes\footnote{\href{https://github.com/ishivvers/TheKastShiv}{https://github.com/ishivvers/TheKastShiv}}. Low-order polynomial fits to comparison-lamp spectra were used to calibrate the wavelength scale, and small adjustments derived from night-sky lines in the target frames were applied. The spectra were flux calibrated using observations of appropriate spectrophotometric standard stars observed on the same night, at similar airmasses, and with an identical instrument configuration.

The spectrum from \tmax+47.3~days was obtained in very poor conditions. The quality of the spectrum was insufficient to extract a useful spectrum.

\subsection{8.4\,m Large Binocular Telescope}

We observed \sn\ in the optical with the Multi-Object Double Spectrographs MODS-1 and MODS-2 \citep{Pogge2010a} and in the near-IR with the two LUCI \citep[LBT Utility Camera in the Infrared;][]{Ageorges2010a, Seifert2010a} cameras. In the optical, we used MODS-1 and -2 in dual-grating mode (grisms G400L and G670L) providing a wavelength coverage from 3200--9500~\AA\ and a slit mask with a width of $1\farcs2$ for each camera. The integration time was 0.8~hours for the observation on 22 October 2019 (\tmax+276.1~days) and on 29/30 January 2020 (\tmax+361.6~days). In the NIR, we used the G200 grating and a 1\farcs0 wide slit. The first epoch was obtained with the $zJspec$ bandpass filter (0.90--1.25 $\mu$m) on 19 October 2019 (integration time 0.8 hours). A second epoch was obtained on 28 January 2020 in $zJspec$ and $HKspec$ (1.47--2.35 $\mu m$). The integration of each setup was 1 hour. 

The MODS data were reduced first with \program{modsCCDRed} \citep{Pogge2019a} version 2.04 to remove the bias and flat-field the data using a slit-less pixel flat. After that, we used custom \program{IRAF} scripts to extract the science spectrum using a nearby star for tracing. The observations of the spectrophotometric standard were combined in order to measure the trace of the dispersion along the entire slit. This trace was used along with the wavelength calibration from arc-lamp lines to rectify the tilt in the direction of the dispersion and the cross-dispersion axes for the full-frame ($8192 \times 3072$ pixel). The final wavelength calibration was cross-checked with known strong auroral skylines in the blue ([\ion{O}{i}]\,$\lambda$\,5577.338) and red ([\ion{O}{i}]\,$\lambda$\,6300.3) channels. The one-dimensional spectrum was extracted from each channel using a $1\farcs2$-wide aperture. The spectra were flux-calibrated using the spectrophotometric standard stars. Telluric features were removed from the red channels using the normalised spectrophotometric standard spectrum. The data from the four channels were combined into a single spectrum and re-binned to a common scale of $\Delta\lambda=0.85$~\AA\ pixel$^{-1}$. All data were corrected for heliocentric motion.

The LUCI spectroscopic data were reduced at the Italian LBT Spectroscopic Reduction Center\footnote{\href{http://www.iasf-milano.inaf.it/software}{http://www.iasf-milano.inaf.it/software}.} by means of scripts optimised for LBT data adopting the standard procedure for long-slit spectroscopy with bias subtraction, flat-fielding,  bad-pixel correction, sky subtraction, and cosmic-rays rejection. The wavelength calibration was obtained from sky-lines achieving an r.m.s. of $<0.3$~\AA\ in zJ and 0.9~\AA\ in HK. We flux-calibrated the spectra with telluric standard stars.

\subsection{8.1\,m Gemini-South telescope}

We obtained one epoch of spectroscopy with the Gemini Multi-Object Spectrograph \citep[GMOS;][]{Hook2004a,Gimeno2016a} at the Gemini-South telescope at Cerro Pachon, Chile, starting on 28 January 2022 (\tmax+988.1~days). We used the R150 grating with the GG455 blocking filter and a $1''$-wide slit. We divided the observation into three sets of $4\times1200$~s each. Owing to technical problems, the campaign had to be aborted after the first set was obtained.

We reduced the data using the \program{Gemini IRAF} package\footnote{\href{https://www.gemini.edu/observing/phase-iii/understanding-and-processing-data/data-processing-software}{https://www.gemini.edu/observing/phase-iii/understanding-and-processing-data/data-processing-software}} version 1.14. We detect the very faint trace of the \sn, but due to the significantly shorter total integration time, the quality of the spectrum was insufficient to detect features of \sn\ or its host galaxy.

\section{Bolometric light curve}

The tabulated version of the bolometric light curve is shown in Table \ref{tab:bolometric}.

\begin{table}[h!]
    \caption{Bolometric luminosity and blackbody properties\label{tab:bolometric}
    }
    \centering
    \begin{tabular}{cccc}
    \toprule
    Phase & $\log\,L_{\rm bol}$ & $\log\,R$ & $T$ \\ 
    (day) & $\left(\rm erg\,s^{-1}\right)$ & (cm) & (K)  \\ 
    \midrule
    -81.4	&$ 43.64 \pm 0.11 $& \nodata & \nodata\\
    -71.4	&$ 43.79 \pm 0.08 $& \nodata & \nodata\\
    -61.4	&$ 43.93 \pm 0.08 $& \nodata & \nodata\\
    -51.4	&$ 43.97 \pm 0.08 $& \nodata & \nodata\\
    -41.4	&$ 44.06 \pm 0.12 $& \nodata & \nodata\\
    -31.4	&$ 44.14 \pm 0.09 $& \nodata & \nodata\\
    -21.4	&$ 44.17 \pm 0.07 $& \nodata & \nodata\\
    -11.4	&$ 44.21 \pm 0.08 $& \nodata & \nodata\\
    -1.4	&$ 44.27 \pm 0.10 $& \nodata & \nodata\\
    -0.4	&$ 44.27 \pm 0.09 $& \nodata & \nodata\\
    8.4		&$ 44.23 \pm 0.04 $&$ 15.64 \pm 0.02 $&$ 11869 \pm 421 $\\
    18.4	&$ 44.19 \pm 0.03 $&$ 15.68 \pm 0.02 $&$ 10513 \pm 306 $\\
    28.4	&$ 44.16 \pm 0.03 $&$ 15.70 \pm 0.01 $&$ 10092 \pm 245 $\\
    38.4	&$ 44.12 \pm 0.03 $&$ 15.70 \pm 0.02 $&$ 9892 \pm 314 $\\
    48.4	&$ 44.05 \pm 0.03 $&$ 15.68 \pm 0.02 $&$ 9567 \pm 346 $\\
    58.4	&$ 44.03 \pm 0.03 $&$ 15.70 \pm 0.02 $&$ 9281 \pm 357 $\\
    68.4	&$ 44.02 \pm 0.03 $&$ 15.70 \pm 0.02 $&$ 9256 \pm 359 $\\
    78.4	&$ 44.00 \pm 0.03 $&$ 15.68 \pm 0.04 $&$ 9428 \pm 599 $\\
    88.4	&$ 43.96 \pm 0.03 $&$ 15.67 \pm 0.03 $&$ 9089 \pm 512 $\\
    98.4	&$ 43.90 \pm 0.04 $&$ 15.64 \pm 0.03 $&$ 8980 \pm 475 $\\
    103.4	&$ 43.87 \pm 0.10 $&$ 15.63 \pm 0.04 $&$ 9006 \pm 518 $\\
    \bottomrule
\end{tabular}
\tablefoot{The bolometric luminosity and the properties of the blackbody spectrum are reported in 10-day bins. The blackbody radius and temperature are reported between \tmax+8.4 and \tmax+100.3~days when multi-band photometry from the $u$-band to the NIR are available and \sn\ is during its photospheric phase. Errors quote the statistical uncertainties. A machine-readable table is available online. 
}
\end{table}

\section{VLT/FORS2 spectrum from January 2022}

\begin{figure}
    \centering
    \includegraphics[width=1\columnwidth]{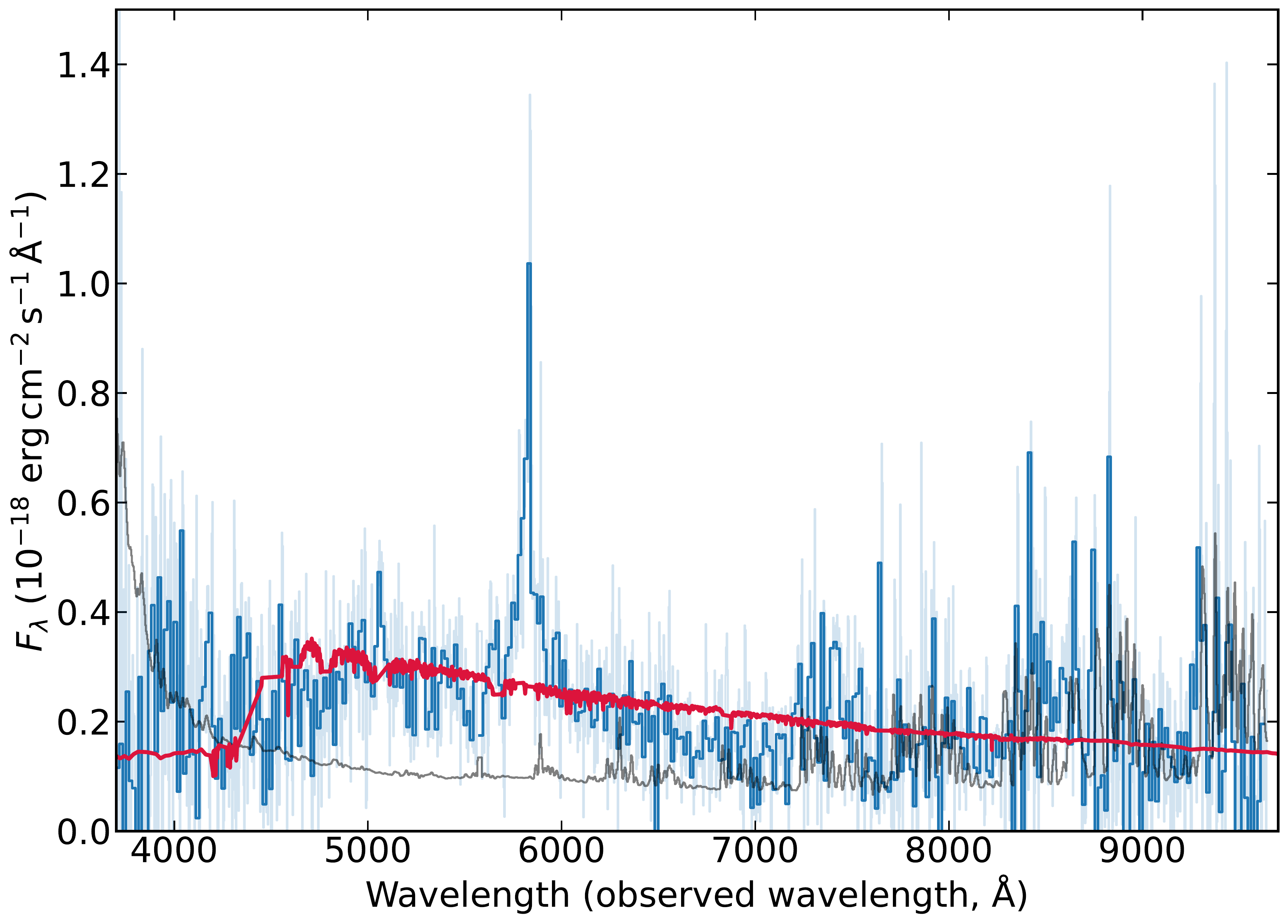}
    \caption{
    Observed spectrum of \sn\ at \tmax+989.2 days (unbinned: light blue, 30~\AA\ binning: blue; error spectrum: grey) and the scaled galaxy SED (red). The only remaining SN feature is the broad [\ion{O}{iii}]\,$\lambda\lambda$\,4949,5007 emission line.
    }
    \label{fig:host_contamination}
\end{figure}

Figure \ref{fig:host_contamination} shows the observed VLT/FORS2 spectrum from January 2022, the unbinned spectrum light-blue and a binned version in a darker shade (bin size 30~\AA). We scaled the best fit of the host galaxy SED to the brightness at the SN explosion site (shown in red, galaxy emission lines are clipped). The shape of the continuum of the VLT spectrum is consistent with the host galaxy SED. The only remaining SN feature is the broad [\ion{O}{iii}] at 5000~\AA.

\section{Modelling the SN light curve with \program{Redback}}

Table \ref{tab:redback} summarises the results of fitting the multi-band light curve with the software package \program{Redback}.

\begin{table*}[h!]
    \caption{Light-curve fits with Redback: models, parameters, priors and marginalised posteriors}
    \scriptsize
    \centering
    \begin{tabular}{lcrrrr|r}
    \toprule
    Parameter  & Prior  & Magnetar & Magnetar\,+    &Magnetar\,+        & $^{56}$Ni          & $^{56}$Ni \\
               &        &          & $^{56}$Ni      &$^{56}$Ni          &(fixed $\kappa$'s)  & (red)     \\
               &        &          &                &(fixed $\kappa$'s) &                    & (fixed $\kappa$'s)\\
    \midrule
    \multicolumn{7}{c}{\textbf{Fitted properties}}\\
    \midrule
    \multicolumn{7}{c}{{General}} \\
    \midrule
    ejecta mass $M_{\rm ej}$ $\left(M_\odot\right)$                         & $\log \mathcal{U}\left(1, 300\right)$          &$104^{+21}_{-15}$        &$140^{+37}_{-34}      $&$139^{+39}_{-40}       $&$100\pm5      $&$119\pm7$\\
    explosion date $t_{\rm exp}$ (day)                                      & $\mathcal{U}\left(-200, 0\right)$              &$-16^{+2}_{-3}$          &$-5\pm1               $&$-5\pm1                $&$-68\pm5      $&$-79\pm7$\\
    `$\gamma$-ray' opacity $\kappa_\gamma$ $\left(\rm cm^2\,g^{-1}\right)$& $\log \mathcal{U}\left(10^{-2}, 10^{4}\right)$ &$0.013\pm0.0002         $&$9.0^{+769.7}_{-8.9}  $&$                      $&$    $&\\
    optical opacity $\kappa$ $\left(\rm cm^2\,g^{-1}\right)$                & $\mathcal{U}\left(0.01, 0.2\right)$            &$0.16^{+0.02}_{-0.03}   $&$0.11^{+0.06}_{-0.07} $&$                      $&$    $&\\
    scaling velocity $v_{\rm scale}$ $\left({\rm km\,s}^{-1}\right)$        & $\mathcal{U}\left(1000, 10000\right)$          &$5810^{+250}_{-240}     $&$6390\pm140           $&$6390\pm140            $&$4250^{+140}_{-150}  $&$3640^{+210}_{-200}$  \\
    white noise parameter $\sigma$                                          & $\log \mathcal{U}\left(10^{-3}, 100\right) $   &$0.25\pm0.01$            &$0.22\pm0.01          $&$0.21\pm0.01           $&$0.25\pm0.01         $&$0.23^{+0.01}_{-0.01}$\\
    $V$-band total extinction $A_V$ (mag)                                   & $\mathcal{U}\left(0, 1\right)$                 &$0.01\pm0.01$            &$0.22^{+0.06}_{-0.05} $&$0.22^{+0.06}_{-0.05}  $&$0.01^{+0.02}_{-0.01}$&$0.52^{+0.07}_{-0.08}$\\
    \midrule
    \multicolumn{7}{c}{{Magnetar model}} \\
    \midrule
    magnetic field $B_\perp$ $\left(10^{14}\,{\rm G}\right)$                & $\log \mathcal{U}\left(0.01, 20\right)$        &$0.98^{+0.05}_{-0.08}$   &$0.09^{+10.65}_{-0.07}$&$0.06^{+11.50}_{-0.04}$&\dots & \dots \\
    neutron-star mass $M_{\rm NS}$ $\left(M_\odot\right)$                   & $\mathcal{U}\left(1, 2.2\right)$               &$2.1^{0.0}_{-0.1}    $   &$1.6\pm0.4            $&$1.6\pm0.4            $&\dots & \dots \\
    initial spin period $P_0$ (ms)                                            & $\mathcal{U}\left(1, 20\right)$              &$1.02^{+0.03}_{-0.01}$   &$13^{+5}_{-6}         $&$14^{+4}_{-5}         $&\dots & \dots \\
    \midrule 
    \multicolumn{7}{c}{{$^{56}$Ni model}} \\
    \midrule
    nickel fraction $f_{\rm Ni}$                                            & $\log \mathcal{U}\left(10^{-3}, 1\right)$      &\dots                     &$0.3\pm0.1           $&$0.3\pm0.1             $&$0.52\pm0.02   $&$0.93^{+0.05}_{-0.09}$\\
    \midrule
    \multicolumn{7}{c}{\textbf{Fit quality}}\\
    \midrule
    log Bayesian evidence ($\ln~Z$)                                         &                                                &$-105.4$                  &$-4.3                $&$-4.8                      $&$-99.0            $&$-21.5$\\
    Number of free parameters                                               &                                                &11                         & 12                  & 10                          & 7                & 7\\
    \midrule
    \multicolumn{7}{c}{\textbf{Derived properties}}\\
    \midrule
    $\gamma$-ray escape time $t_0$ (day)                                    &                                                &$730\pm60                $&$800\pm100           $&$760^{+100}_{-120}  $&$980\pm40        $&$1240^{+80}_{-70}$\\
    nickel mass $M_{\rm Ni}$ $\left(M_\odot\right)$                         &                                                &\dots                     &$48\pm3              $&$48\pm3             $&$52\pm2          $&$109^{+9}_{-10}$\\
    kinetic energy $E_{\rm kin}$ $\left(10^{51}\rm erg\right)$              &                                                &$59^{+14}_{-11}          $&$57^{+12}_{-14}      $&$57^{+16}_{-17}     $&$30^{+3}_{-2}    $&$16\pm2$\\
    rotational energy $E_{\rm rot}$ $\left(10^{51}\rm erg\right)$           &                                                &$35.7^{+2.0}_{-3.3}      $&$ 0.1^{+0.3}_{-0.1}  $&$0.1^{+0.3}_{-0.1}  $&\dots             &\dots\\
    \bottomrule
    \end{tabular}
    \tablefoot{The model `$^{56}$Ni (red)' only fitted the data in the $r$ and redder bands. We used uniform ($\mathcal{U}$) and log uniform ($\log \mathcal{U}$) priors. The uncertainties of the marginalised posteriors are quoted at $1\sigma$ confidence. The explosion date is measured with respect to the date of the first detection. All marginalised posteriors are reported in linear units. The kinetic energy of the ejecta was computed via $E_{\rm kin}=1/2\, M_{\rm ej}\,v_{\rm scale}^{2}$ and the rotational energy of the magnetar via $E_{\rm rot}=2\times10^{52}\,\left(M_{\rm NS}/1.4\,M_\odot\right)^{3/2}\,\left(P_0/1\,\rm ms\right)^{-2}~\rm erg$. 
    }
    \label{tab:redback}
    \end{table*}

\section{Galaxy photometry of SN\,2015bn}\label{app:SN2015bn}

We use science-ready coadded images from PanSTARRS DR1 and archival science-ready images obtained with MegaCAM at the 3.58\,m Canada-France-Hawaii Telescope (CFHT). We augmente this data set with archival data from \swift/UVOT in $w2$, $m2$ and $w1$ obtained between January 2016 and February 2017 after the SN faded, and Subaru/Suprime-Cam \citep{Miyazaki2002a} in $B$ and $V$ bands.

The \swift/UVOT data is processed and analysed as described in Appendix \ref{appendix:photometry}. The Subaru data is reduced with the software package \program{SDFRED2} \citep{Ouchi2004a}. 

The photometry is extracted using sufficiently large apertures to encompass the entire host galaxy. The instrumental magnitudes of the ground-based images is calibrated against stars from PanSTARRS. We apply known colour equations between PS1, SDSS and Bessell-like filters to account for differences in the filter response function as described in Section \ref{sec:obs:imaging}. A summary of the host photometry is reported in Table \ref{tab:host_phot_SN2015bn}.

\begin{figure}
    \centering
    \includegraphics[width=1\columnwidth]{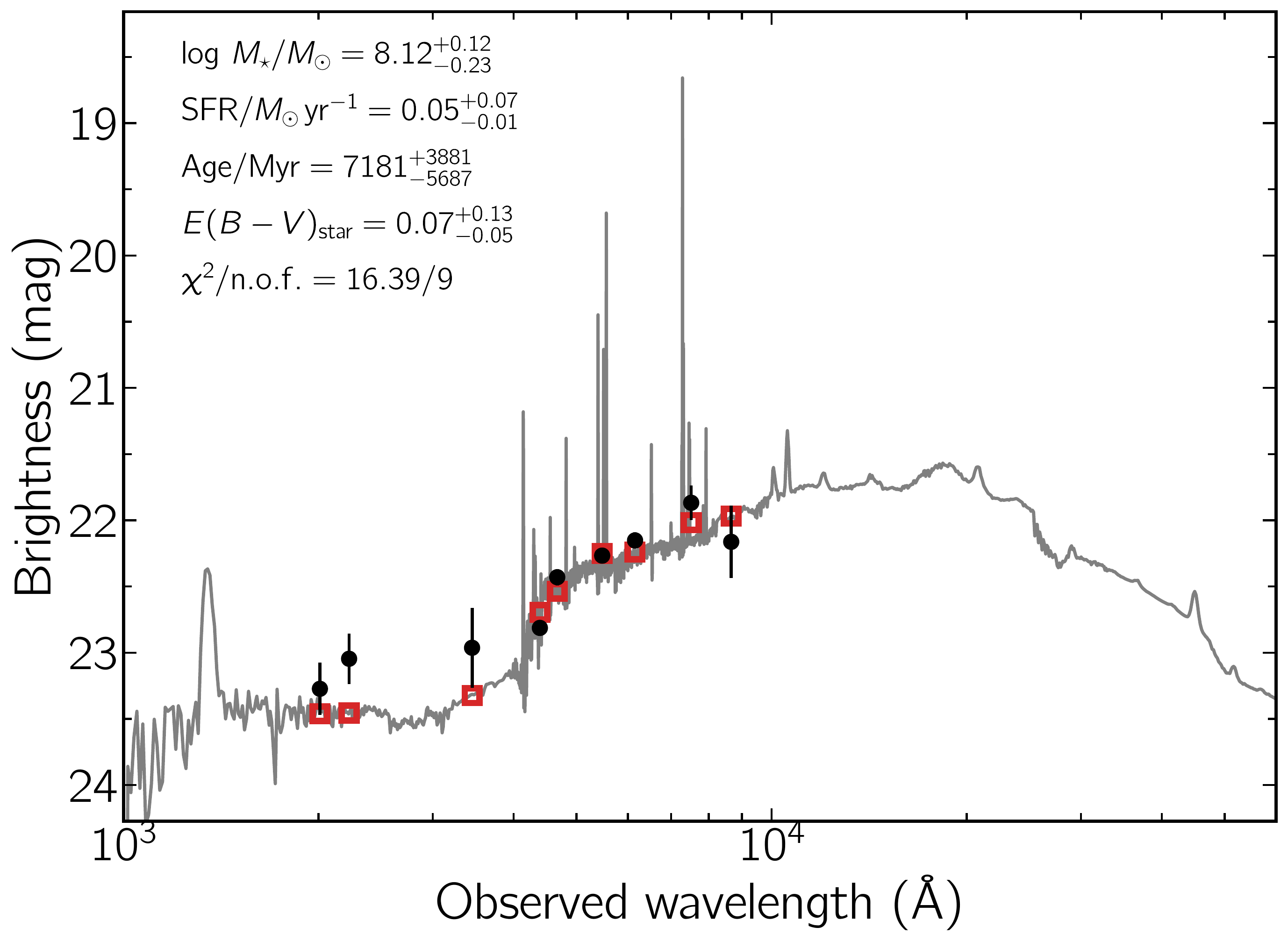}
    \caption{
    Spectral energy distribution (SED) of the host galaxy of SN\,2015bn (black data points). The solid line displays the best-fitting model of the SED. The red squares represent the model-predicted magnitudes. The fitting parameters are shown in the upper-left corner. The abbreviation `n.o.f.' stands for the number of filters.
    }
    \label{fit:host:SN2015bn}
\end{figure}

We model the host SED with \program{Prospector} as described in Section \ref{res:host}. The observed SED is adequately described by a galaxy model with the stellar mass of $\log\,~M_\star/M_\odot=8.12^{+0.12}_{-0.23}$ and a star-formation rate of $0.05^{+0.07}_{-0.01}~M_\odot\,\rm yr^{-1}$ (Figure \ref{fit:host:SN2015bn}). We use the best-fit SED to remove the host contribution of the SN spectrum from \citet{Nicholl2018a} in Section \ref{discussion:comparison_w_slow_slsn}.

\newpage

\begin{table}[h!]
    \caption{Photometry of SN2015bn's host galaxy}\label{tab:host_phot_SN2015bn}
    \centering
    \begin{tabular}{lllc}
    \toprule
    Telescope & Instrument    & Filter    & Brightness\\
    \midrule
    \swift    & UVOT          &$w2  $& $23.47 \pm 0.17$\\
    \swift    & UVOT          &$m2  $& $23.26 \pm 0.16$\\
    \swift    & UVOT          &$u   $& $23.07 \pm 0.28$\\
    Subaru    & Suprime-Cam   &$B   $& $22.99 \pm 0.03$\\
    Subaru    & Suprime-Cam   &$V   $& $22.32 \pm 0.02$\\
    CFHT      & MegaCam       &$ g  $& $22.51 \pm 0.04$\\
    CFHT      & MegaCam       &$ r  $& $22.21 \pm 0.04$\\
    PanSTARRS &               &$ i  $& $21.91 \pm 0.12 $\\
    PanSTARRS &               &$ z  $& $22.20 \pm 0.28 $\\
    \bottomrule
    \end{tabular}
    \tablefoot{All measurements are reported in the AB system and not corrected for reddening. Non-detections are reported at $3\sigma$ confidence.
    }
    \end{table}

\end{appendix}

\end{document}